\documentclass{report}[amsfonts,amsmath,floatfix]

\usepackage{physics}
\usepackage{fullpage}
\usepackage[applemac]{inputenc}  
\usepackage[T1]{fontenc}       
\usepackage{amsmath}
\usepackage{amssymb}
\usepackage{dsfont}
\usepackage{graphicx}
\usepackage[dvipsnames]{xcolor} 
\usepackage{braket}
\usepackage{pict2e}
\usepackage{xspace}
\usepackage{subfig}
\usepackage{titlesec}
\usepackage{pdfpages}
\usepackage{tikz}
\usepackage{marvosym}

\newcommand{\lgr}{\left\{}
\newcommand{\rgr}{\right\}}

\newcommand{\sse}{\subsection}
\newcommand{\ssse}{\subsubsection}

\newcommand{\lt}{\left(}
\newcommand{\rt}{\right)}
\newcommand{\ba}{\begin{eqnarray}}
\newcommand{\ea}{\end{eqnarray}}
\newcommand{\fr}{\frac}
\newcommand{\nn}{\nonumber}
\newcommand{\be}{\begin{equation}}
\newcommand{\ee}{\end{equation}}
\newcommand{\bc}{\begin{center}}
\newcommand{\ec}{\end{center}}
\newcommand{\beq}{\begin{equation}}
\newcommand{\eeq}{\end{equation}}
\newcommand{\beqq}{\begin{equation*}}
\newcommand{\eeqq}{\end{equation*}}
\newcommand{\beqa}{\begin{align}}
\newcommand{\eeqa}{\end{align}}
\newcommand{\barr}{\begin{array}}
\newcommand{\earr}{\end{array}}
\newcommand{\bi}{\begin{itemize}}
\newcommand{\ei}{\end{itemize}}
\newcommand{\module}[1]{\ensuremath{\vert#1\vert}}

\newcommand{\R}{\mathbb{R}}

\newcommand{\Z}{\ensuremath{\mathds{Z}}}
\newcommand{\Id}{\ensuremath{\mathds{1}}}

\newcommand{\sinc}{\ensuremath{\mathrm{sinc}}}

\usepackage[authoryear]{natbib}
\usepackage[colorlinks]{hyperref}
\hypersetup{%
        plainpages=true,
        breaklinks=true,
        hypertexnames=false,
        pageanchor=true,
        colorlinks=true,
        linkcolor={blue},
        citecolor={blue},
        urlcolor={blue},
        anchorcolor={black}
      }

\usepackage{enumerate}
\usepackage{units}
\usepackage{url}

\usepackage{mleftright} 

\newcommand{\figref}[1]{\mbox{Fig.~\ref{#1}}}

\newcommand{\secref}[1]{\mbox{Sec.~\ref{#1}}}
\newcommand{\chpref}[1]{\mbox{Chapter~\ref{#1}}}

\renewcommand{\eqref}[1]{\mbox{Eq.~(\ref{#1})}}

\renewcommand{\braket}[2]{\langle #1|#2\rangle}
\newcommand{\brakket}[3]{\langle #1 | #2 | #3 \rangle}

\newcommand{\abssq}[1]{\mleft| #1 \mright|^2}

\newcommand{\bea}{\begin{eqnarray}}
\newcommand{\eea}{\end{eqnarray}}

\usepackage{amsthm}
\usepackage{circuitikz}
\usetikzlibrary{quantikz2}
\usepackage{float}

\usepackage{listings} 

\definecolor{codegreen}{rgb}{0,0.6,0}
\definecolor{codegray}{rgb}{0.5,0.5,0.5}
\definecolor{codepurple}{rgb}{0.58,0,0.82}
\definecolor{backcolour}{rgb}{0.95,0.95,0.92}

\lstdefinestyle{mystyle}{
    backgroundcolor=\color{backcolour},  
    commentstyle=\color{NavyBlue},
    keywordstyle=\color{BrickRed},
    numberstyle=\tiny\color{codegray},
    stringstyle=\color{codepurple},
    basicstyle=\footnotesize,
    breakatwhitespace=false,         
    breaklines=true,                 
    captionpos=b,                    
    keepspaces=true,                 
    numbers=left,                    
    numbersep=5pt,                  
    showspaces=false,                
    showstringspaces=false,
    showtabs=false,                  
    tabsize=2
}
\DeclareMathOperator{\CNOT}{CNOT}
\DeclareMathOperator{\CZ}{CZ}
\newtheorem*{trick}{Trick}
\newcounter{tutorial}
\newtheorem{ex}{Exercise}[tutorial]
\newcommand{\Mod}{\mkern-10mu\mod}

\usepackage{authblk}
\begin{document}

\title{Lecture Notes on Quantum Computing}

\renewcommand{\thefootnote}{\fnsymbol{footnote}}

\author[1]{Anton Frisk Kockum\thanks{anton.frisk.kockum@chalmers.se}}
\affil[1]{Department of Microtechnology and Nanoscience, Chalmers University of Technology, 412 96 Gothenburg, Sweden}

\author[1]{Ariadna Soro}

\author[1]{Laura Garc\'{i}a-\'{A}lvarez}

\author[1]{Pontus Vikst{\aa}l}

\author[2]{Tom Douce}
\affil[2]{Laboratoire Mat\'{e}riaux et Ph\'{e}nom\'{e}nes Quantiques, 
Univ.~Paris Diderot, CNRS UMR 7162, 75013, Paris, France}

\author[1]{G{\"o}ran Johansson}

\author[1]{Giulia Ferrini\thanks{ferrini@chalmers.se}}

\maketitle

\begin{abstract}

These are the lecture notes of the master's course ``\href{https://www.chalmers.se/en/education/your-studies/find-course-and-programme-syllabi/course-syllabus/MCC155/?acYear=2023/2024}{Quantum Computing}'', taught at Chalmers University of Technology every fall since 2020, with participation of students from RWTH Aachen and Delft University of Technology. The aim of this course is to provide a theoretical overview of quantum computing, excluding specific hardware implementations. Topics covered in these notes include quantum algorithms (such as Grover's algorithm, the quantum Fourier transform, phase estimation, and Shor's algorithm), variational quantum algorithms that utilise an interplay between classical and quantum computers [such as the variational quantum eigensolver (VQE) and the quantum approximate optimisation algorithm (QAOA), among others], quantum error correction, various versions of quantum computing (such as measurement-based quantum computation, adiabatic quantum computation, and the continuous-variable approach to quantum information), the intersection of quantum computing and machine learning, and quantum complexity theory. Lectures on these topics are compiled into 12 chapters, most of which contain a few suggested exercises at the end, and interspersed with four tutorials, which provide practical exercises as well as further details. At Chalmers, the course is taught in seven weeks, with three two-hour lectures or tutorials per week. It is recommended that the students taking the course have some previous experience with quantum physics, but not strictly necessary.

\end{abstract}

\begin{figure}[ht!]
\centering
\includegraphics[width=\linewidth]{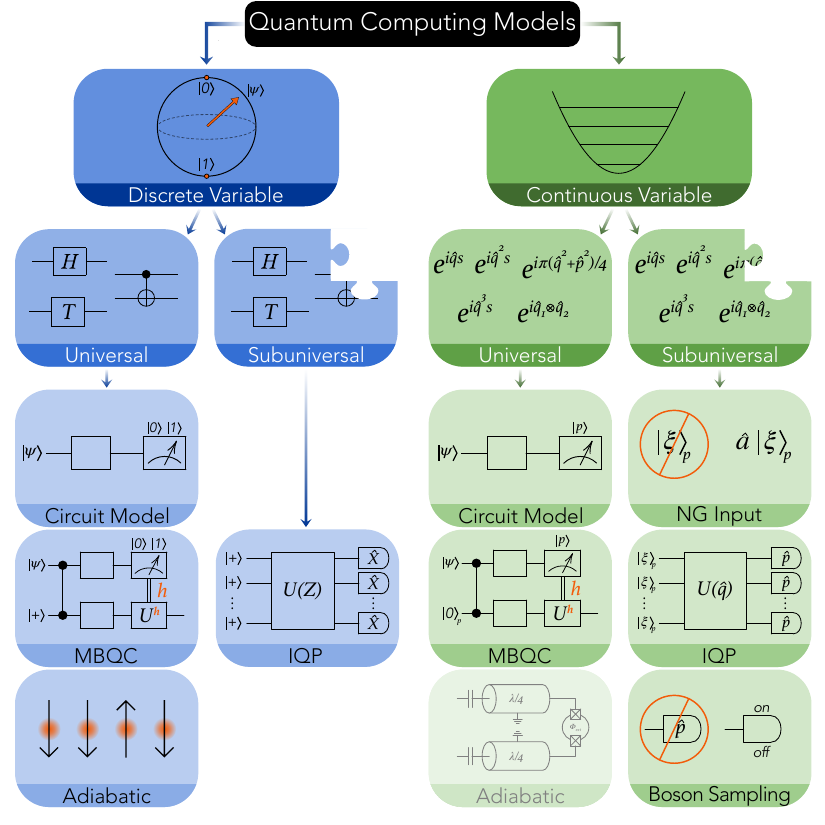}
\caption{Course roadmap -- by A.S.
\label{fig:course-roadmap}}
\end{figure}

\newpage

\section*{Acknowledgements}
\addcontentsline{toc}{chapter}{Acknowledgements}

We thank Oliver Hahn, Marika Svensson, and Alexandru Gheorghiu for proofreading some of the sections in these notes, as well as Timo Hillmann for giving ideas for Chapter 12 and Tutorial 4. We also thank all the students of previous editions of this course who have pointed out typos or suggested other improvements for the notes.

Quantum circuits were drawn using the Quantikz package~\citep{quantikz}. Numerical simulations were performed using QuTiP~\citep{Johansson2012, Johansson2013}.

A.F.K, A.S., L.G.-\'{A}., P.V., G.J., and G.F.~acknowledge support from the Knut and Alice Wallenberg Foundation through the Wallenberg Centre for Quantum Technology (WACQT). A.F.K.~and A.S.~acknowledge support from the Swedish Research Council (grant number 2019-03696). A.F.K.~also acknowledges support from the Swedish Foundation for Strategic Research (grant numbers FFL21-0279 and FUS21-0063) and the Horizon Europe programme HORIZON-CL4-2022-QUANTUM-01-SGA via the project 101113946 OpenSuperQPlus100. G.F.~also acknowledges support from the Swedish Research Council (Grant QuACVA) and the HORIZON-EIC-2022-PATHFINDERCHALLENGES-01 programme under Grant Agreement Number 101114899 (VeriQuB).

Views and opinions expressed are however those of the authors only and do not necessarily reflect those of the European Union. Neither the European Union nor the granting authority can be held responsible for them.

\section*{Author contributions}
\addcontentsline{toc}{chapter}{Author contributions}

A.F.K.~wrote Chapters 1, 3, 5, and 10. G.F.~wrote Chapters 9, 11, and 12; the latter with contributions from T.D.~and A.S. G.J.~wrote the first version of Chapter 2 and A.F.K.~and G.F.~edited it. G.J.~wrote the first version of Chapter 4 and G.F.~edited it. G.F.~and A.F.K.~wrote Chapter 6. A.F.K., G.F., and P.V.~wrote Chapter 7. G.F.~and A.F.K.~wrote Chapter 8 with contributions from T.D.~and Alexandru Gheorghiu. L.G.-\'{A}.~wrote Tutorials 1 and 3, and A.S.~edited them. A.S.~wrote Tutorials 2 and 4. All the quantum circuits in the notes were drawn by A.S.

\tableofcontents


\chapter{The circuit model for quantum computation}
\label{chp:CircuitModel}

In this course, we will give an overview of various approaches to quantum computation, reflecting many of the latest developments in the field. We will cover several different models of quantum computation, from the foundational circuit model through measurement-based and adiabatic quantum computation to boson sampling. We will discuss quantum computation with both discrete and continuous variables. When it comes to the algorithms that we study, they include both classics like Shor's algorithm and newer, heuristic approaches like the quantum approximate optimization algorithm (QAOA). We will also see how quantum computing can be combined with machine learning.

We assume that the students taking this course already have some familiarity with quantum physics (superposition, entanglement, etc.) and some basic concepts in quantum computation. We will repeat some of these basic concepts at the beginning of the course, but perhaps give a more thorough justification for why they can be used in quantum computation.

In this first chapter, we will study the circuit model of quantum computation. This introduces quantum bits, quantum gates, and other components in close similarity with concepts in classical computing and gives us the tools to begin investigating whether quantum computers can ever outperform classical computers. For this chapter, we have borrowed parts from Refs.~\citep{Nielsen2000, Aaronson2018, Kockum2019}.

\section{Components of the circuit model}

Loosely speaking, a computation requires a system that can represent data, a way to perform manipulation of that data, and a method for reading out the result of the computation. In the circuit model of quantum computation, we use:
\begin{itemize}
\item \textbf{Quantum bits (qubits)} to represent the data.
\item \textbf{State preparation} to initialize the qubits in the input state we need to begin the computation.
\item \textbf{Quantum gates} on the qubits to manipulate the data.
\item \textbf{Measurements} on the qubits to read out the final result.
\end{itemize}

Below, we first say a few words about what qubits are. We then discuss various quantum gates, and what is required of such gates to allow us to perform any quantum computation we would like. We assume for now that it is possible to initialize our quantum computer in some simple state, and that we can read out the state of the qubits at the end of a computation.

\section{Quantum bits}

In a classical computer, the most basic unit of information is a \textit{bit}, which can take two values: 0 and 1. In a quantum computer, the laws of quantum physics allow phenomena like superposition and entanglement. When discussing information processing in a quantum world, the most basic unit is therefore a \textit{quantum bit}, usually called \textit{qubit}, a two-level quantum system with a ground state $\ket{0}$ and an excited state $\ket{1}$. Unlike a classical bit, which only has two possible states, a quantum bit has infinitely many states: all superpositions of $\ket{0}$ and $\ket{1}$,
\be
\ket{\psi} = \alpha \ket{0} + \beta \ket{1} = 
\alpha 
\begin{pmatrix} 1 \\ 0 \end{pmatrix}
+
\beta
\begin{pmatrix} 0 \\ 1 \end{pmatrix}
=
\begin{pmatrix} \alpha \\ \beta \end{pmatrix},
\ee
where $\alpha$ and $\beta$ are complex numbers satisfying $\abssq{\alpha} + \abssq{\beta} = 1$. A measurement on this qubit state (in the basis of $\ket{0}$ and $\ket{1}$) gives the result 0 with probability $\abssq{\alpha}$ and the result 1 with probability $\abssq{\beta}$.

\begin{figure}
\centering
\includegraphics[width=0.7\linewidth]{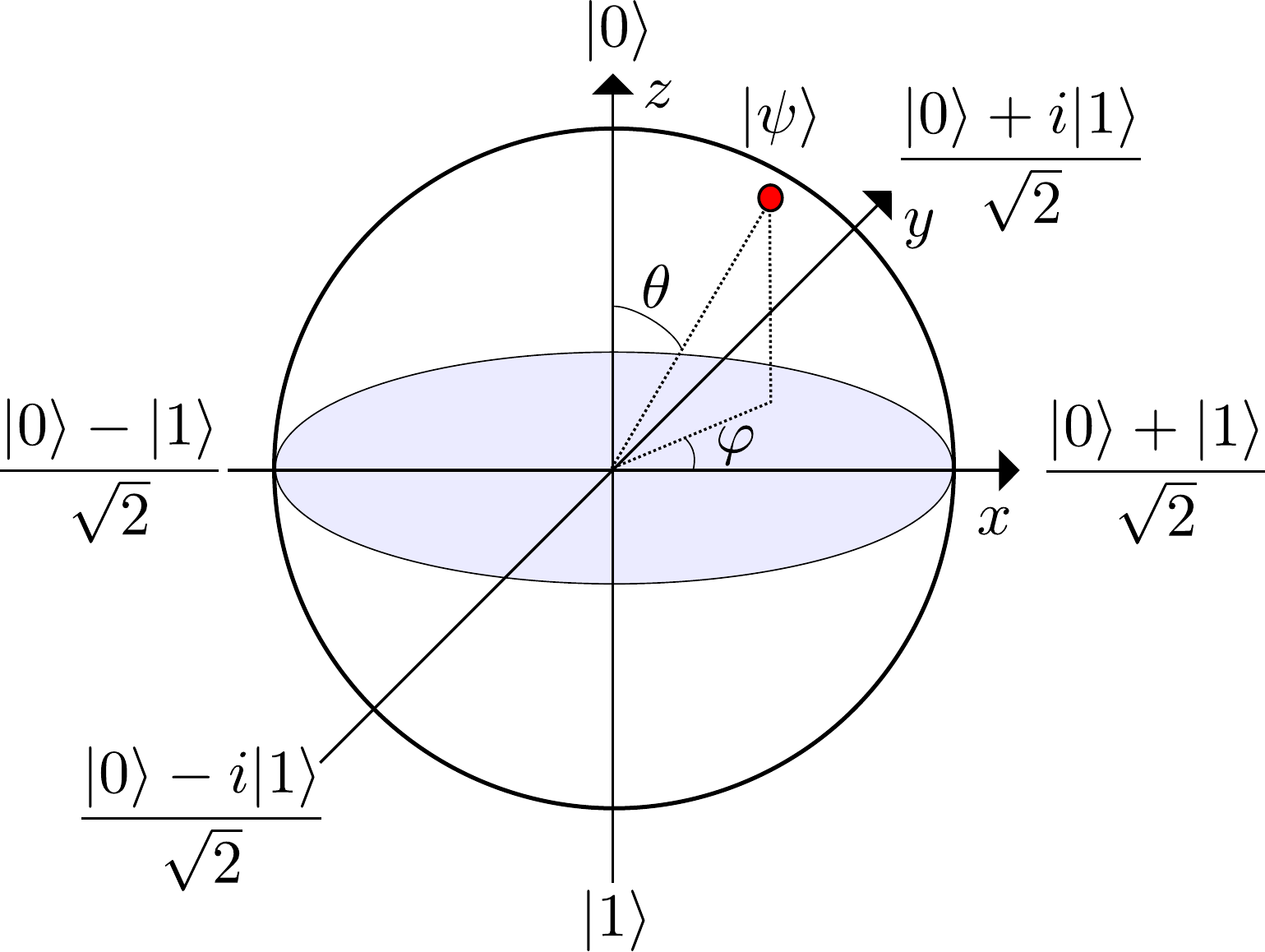}
\caption{The Bloch-sphere representation of a qubit state. The north pole is the ground state $\ket{0}$ and the south pole is the excited state $\ket{1}$. To convert an arbitrary superposition of $\ket{0}$ and $\ket{1}$ to a point on the sphere, the parametrization $\ket{\psi} = \cos \frac{\theta}{2} \ket{0} + e^{i \varphi} \sin \frac{\theta}{2} \ket{1}$ is used.
\label{fig:BlochSphere}}
\end{figure}

A useful tool for visualizing a qubit state is the \textit{Bloch sphere} shown in \figref{fig:BlochSphere}. A state of the qubit is represented as a point on the surface of the sphere, which has radius 1. The two states of a classical bit correspond to the north and south poles on the sphere. The two states on opposite ends of the $x$ axis are often denoted
\be
\ket{+} = \frac{\ket{0} + \ket{1}}{\sqrt{2}}, \qquad \ket{-} = \frac{\ket{0} - \ket{1}}{\sqrt{2}}.
\ee
%

If there are $N$ qubits in a system, the total state of that system can be a superposition of $2^N$ different states: $\ket{000 \ldots 00}$, $\ket{100 \ldots 00}$, $\ket{010 \ldots 00}$, $\ldots$, $\ket{111 \ldots 10}$, $\ket{111 \ldots 11}$. Note that $N$ classical bits can be in \textit{one} of these $2^N$ states, but not in a superposition of several of them. To store all the information about a general $N$-qubit state, one needs to keep track of the $2^N$ amplitudes in the superposition. This means that at least $2^N$ classical bits may be required to represent the quantum system. This explains why it is hard for classical computers to simulate some quantum systems, and gives a first hint that quantum computers can be more powerful than classical ones (at least when it comes to simulating quantum systems).

There are many physical implementations of qubits, e.g., superconducting qubits, trapped ions, natural atoms, etc. These implementations are a topic for another course. In the following, we assume that we have access to qubits, but do not care much about how they are made.

\section{Single-qubit gates}

An operation on a single qubit changes its state from $\ket{\psi} = \alpha \ket{0} + \beta \ket{1}$ to $\ket{\psi'} = \alpha' \ket{0} + \beta' \ket{1}$ while preserving the norm $\abssq{\alpha} + \abssq{\beta} = 1 = \abssq{\alpha'} + \abssq{\beta'}$. This preservation of the norm is mandated by the fact that the probabilities of different measurement results must sum up to 1. Such single-qubit operations (gates) can be described by $2 \times 2$ unitary matrices. A general such matrix is commonly denoted $U$. 

Here we list some of the most common single-qubit gates. First are the Pauli matrices:
\bea
X &=& \begin{pmatrix} 0 & 1 \\ 1 & 0 \end{pmatrix} =
\begin{quantikz}
    & \gate{X} & 
\end{quantikz}, \label{eq:PauliX}\\
Y &=& \begin{pmatrix} 0 & -i \\ i & 0 \end{pmatrix} =
\begin{quantikz}
& \gate{Y} & 
\end{quantikz}, \\
Z &=& \begin{pmatrix} 1 & 0 \\ 0 & -1 \end{pmatrix} = 
\begin{quantikz}
    & \gate{Z} & 
\end{quantikz}
.
\eea
In these equations, we have shown on the right how these gates are represented when drawing a quantum circuit (the instructions for a quantum algorithm). One line represents one qubit. The circuit progresses from left to right, so the qubit arrives on the left in some quantum state $\ket{\psi}$, the gate $U$ acts on it, and the qubit leaves on the right in a new quantum state $U \ket{\psi}$. 

The Pauli matrices generate rotations around the corresponding axes on the Bloch sphere when exponentiated, e.g.,
\be
R_x (\theta) = \exp (- i \theta X / 2) = \cos (\theta / 2) I - i \sin (\theta / 2) X = \begin{pmatrix} \cos (\theta / 2) & - i \sin (\theta / 2) \\ - i \sin (\theta / 2) & \cos (\theta / 2) \end{pmatrix} =
\begin{quantikz}
    & \gate{R_x (\theta)} & 
\end{quantikz}
,
\label{eq:RotX}
\ee
where $I$ is the identity matrix.

The $X$ gate is the quantum equivalent of the classical NOT gate:
\be
X \mleft( \alpha \ket{0} + \beta \ket{1} \mright) = \begin{pmatrix} 0 & 1 \\ 1 & 0 \end{pmatrix} \begin{pmatrix} \alpha \\ \beta \end{pmatrix} = \begin{pmatrix} \beta \\ \alpha \end{pmatrix} = \beta \ket{0} + \alpha \ket{1}.
\ee

By adding the $X$ and $Z$ gates, one obtains the Hadamard gate
\be
H = \frac{1}{\sqrt{2}} \begin{pmatrix} 1 & 1 \\ 1 & -1 \end{pmatrix} = \frac{X + Z}{\sqrt{2}} = \begin{quantikz}
    & \gate{H} & 
\end{quantikz}.
\ee
This gate transforms the qubit state from the $\ket{0}, \ket{1}$ basis to the $\ket{+}, \ket{-}$ basis, and vice versa. The Hadamard gate is often used to create superposition states at the beginning of a quantum algorithm.

The $Z$ gate applies a phase factor $-1$ to the $\ket{1}$ part of the qubit state. Two gates that apply other phase factors are often given their own names. One is the $T$, or $\pi / 8$, gate:
\be
T = 
\begin{pmatrix} 
1 & 0 \\
0 & \exp (i \pi / 4) 
\end{pmatrix} 
= 
\exp (i \pi / 8) 
\begin{pmatrix} 
\exp (- i \pi / 8) & 0 \\
0 & \exp (i \pi / 8)
\end{pmatrix}
= 
\begin{quantikz}
 & \gate{T} & 
\end{quantikz}.
\ee
The other is the phase, or $S$, or $P$, gate
\be
S = \begin{pmatrix} 1 & 0 \\ 0 & i \end{pmatrix} = T^2 = \begin{quantikz}
    & \gate{S} & 
\end{quantikz}.
\ee

\section{Multi-qubit gates}

To realize useful quantum algorithms, we also need to be able to make two or more qubits interact through multi-qubit gates. To discuss this topic, we first need to establish some notation for states of multiple qubits and operators acting on them. If you are not familiar with the tensor product ($\otimes$), we recommend you to follow the dedicated Qiskit tutorial~\citep{TensorProduct} or chapter 2 in Ref.~\citep{Nielsen2000}. Briefly, if we have two qubits $a$ and $b$ with states $\ket{\psi_a} = a_0 \ket{0}_a + a_1 \ket{1}_a$ and $\ket{\psi_b} = b_0 \ket{0}_b + b_1 \ket{1}_b$, their total state is
\be
\ket{\psi_a} \otimes \ket{\psi_b} =
\begin{pmatrix}
a_0 \\
a_1
\end{pmatrix}
\otimes
\begin{pmatrix}
b_0 \\
b_1
\end{pmatrix}
=
\begin{pmatrix}
a_0 
    \begin{pmatrix}
    b_0 \\
    b_1
    \end{pmatrix}
\\
a_1
    \begin{pmatrix}
    b_0 \\
    b_1
    \end{pmatrix}
\end{pmatrix}
=
\begin{pmatrix}
a_0 b_0 \\
a_0 b_1 \\
a_1 b_0 \\
a_1 b_1 \\
\end{pmatrix}
=
a_0 b_0 \ket{00} + a_0 b_1 \ket{01} + a_1 b_0 \ket{10} + a_1 b_1 \ket{11},
\ee
where $\ket{01}$ denotes $\ket{0}_a\otimes\ket{1}_b$, etc.
If an X gate acts on qubit $a$ (and does nothing, i.e., identity, to qubit $b$) and another X gate acts on qubit $b$ (and does nothing, i.e., identity, to qubit $a$) in this state, we can write it as
\be
(X \otimes I_2) (I_2 \otimes X) \mleft(\ket{\psi_a} \otimes \ket{\psi_b} \mright) = (X \otimes X) \mleft(\ket{\psi_a} \otimes \ket{\psi_b} \mright) = \mleft( X \ket{\psi_a} \mright) \otimes \mleft( X \ket{\psi_b} \mright) = X_a X_b \ket{\psi_a \psi_b} ,
\label{eq:TensorNotation}
\ee
where $I_2$ is the $2 \times 2$ identity matrix. Note that we will mostly omit tensor product signs and use abbreviated notation like in the rightmost expression in \eqref{eq:TensorNotation}. Also note that entangled states of qubits, like the ones produced by many two-qubit gates, are not separable, i.e., cannot be written as the tensor product of individual single-qubit states (like $\ket{\psi_a} \otimes \ket{\psi_b}$ above).

Now, one way to achieve an interaction between two qubits is to let the state of one qubit control whether a certain single-qubit gate is applied to another qubit. One such gate is the controlled-NOT (CNOT) gate:
\be
\text{CNOT} 
= 
\begin{pmatrix}
1 & 0 & 0 & 0  \\
0 & 1 & 0 & 0 \\
0 & 0 & 0 & 1 \\
0 & 0 & 1 & 0
\end{pmatrix} 
= 
\begin{quantikz}
& \ctrl{1} &\\
& \targ{} & 
\end{quantikz}
= 
\begin{quantikz}
& \ctrl{1} &\\
& \gate{X} & 
\end{quantikz}
.
\label{eq:CNOTdef}
\ee
Note that the two-qubit Hilbert space is spanned by the basis vectors $\ket{00}$, $\ket{01}$, $\ket{10}$, and $\ket{11}$, in that order. This is the tensor product of the two single-qubit Hilbert spaces. These four states are the computational basis, and this is the basis we will use when writing all the following multi-qubit gates. However, one could also choose another basis for the two-qubit Hilbert space, e.g., $\ket{++}$, $\ket{+-}$, $\ket{-+}$, and $\ket{--}$.

The action of the CNOT gate in \eqref{eq:CNOTdef} is thus to do nothing if the first qubit is in state $\ket{0}$ ($\ket{00}$, $\ket{01}$ changes to $\ket{00}$, $\ket{01}$), and to apply NOT (X) to the second qubit if the first qubit is in state $\ket{1}$ ($\ket{10}$, $\ket{11}$ changes to $\ket{11}$, $\ket{10}$). The dot in the circuit diagram denotes that the first qubit acts as the control.

The controlled-Z (CZ) gate can be defined in the same manner:
\be
\text{CZ} 
= 
\begin{pmatrix}
1 & 0 & 0 & 0  \\
0 & 1 & 0 & 0 \\
0 & 0 & 1 & 0 \\
0 & 0 & 0 & -1
\end{pmatrix} 
= 
\begin{quantikz}
 & \ctrl{1} & \\
 & \control{} & 
\end{quantikz}
= 
\begin{quantikz}
& \ctrl{1} &\\
& \gate{Z} & 
\end{quantikz}
.
\ee
Here, both qubits can be seen as controls, since something only happens in the $\ket{11}$ state (it gets a factor $-1$).
More generally, the controlled application of a single-qubit unitary $U$ to the second qubit takes the form
\be
\begin{pmatrix}
I_2 & 0_2 \\
0_2 & U
\end{pmatrix}
=
\begin{quantikz}
& \ctrl{1} &\\
& \gate{U} & 
\end{quantikz}.
\ee

There are also two-qubit gates that are not controlled operations. For example, the SWAP gate
\be
\text{SWAP} = \begin{pmatrix}
1 & 0 & 0 & 0  \\
0 & 0 & 1 & 0 \\
0 & 1 & 0 & 0 \\
0 & 0 & 0 & 1
\end{pmatrix} = 
\begin{quantikz}
 & \swap{1} & \\
 & \targX{} & 
\end{quantikz}
\ee
swaps the states $\ket{01}$, $\ket{10}$ to $\ket{10}$, $\ket{01}$.

It is also possible to define gates involving more than two qubits. For three-qubit gates, the most well-known ones are the Toffoli and Fredkin gates. The Toffoli gate is a controlled-controlled-NOT (CCNOT), i.e., the state of the third qubit is flipped if and only if both the first two qubits are in state $\ket{1}$:
\be
\text{Toffoli} = \begin{pmatrix}
1 & 0 & 0 & 0 & 0 & 0 & 0 & 0  \\
0 & 1 & 0 & 0 & 0 & 0 & 0 & 0  \\
0 & 0 & 1 & 0 & 0 & 0 & 0 & 0  \\
0 & 0 & 0 & 1 & 0 & 0 & 0 & 0  \\
0 & 0 & 0 & 0 & 1 & 0 & 0 & 0  \\
0 & 0 & 0 & 0 & 0 & 1 & 0 & 0  \\
0 & 0 & 0 & 0 & 0 & 0 & 0 & 1  \\
0 & 0 & 0 & 0 & 0 & 0 & 1 & 0  \\
\end{pmatrix} = 
\begin{quantikz}
 & \ctrl{1} & \\
 & \ctrl{1} & \\
 & \targ{} & 
\end{quantikz}.
\ee
The Fredkin gate is a controlled-SWAP (CSWAP) gate, swapping the states of the second and third qubits if and only if the state of the first qubit is $\ket{1}$:
\be
\text{Fredkin} = \begin{pmatrix}
1 & 0 & 0 & 0 & 0 & 0 & 0 & 0  \\
0 & 1 & 0 & 0 & 0 & 0 & 0 & 0  \\
0 & 0 & 1 & 0 & 0 & 0 & 0 & 0  \\
0 & 0 & 0 & 1 & 0 & 0 & 0 & 0  \\
0 & 0 & 0 & 0 & 1 & 0 & 0 & 0  \\
0 & 0 & 0 & 0 & 0 & 0 & 1 & 0  \\
0 & 0 & 0 & 0 & 0 & 1 & 0 & 0  \\
0 & 0 & 0 & 0 & 0 & 0 & 0 & 1  \\
\end{pmatrix} = 
\begin{quantikz}
 & \ctrl{1} & \\
 & \swap{1} & \\
 & \targX{} & 
\end{quantikz}.
\ee

Most practical implementations of quantum computing have limited connectivity between qubits, only allowing for pairwise interactions between nearest-neighbour qubits. This can prohibit direct implementations of multi-qubit gates with three or more qubits, but it turns out that any multi-qubit gate can be decomposed into a number of single- and two-qubit gates.

\section{Universal quantum computation}
\label{sec:UniversalQComp}

Are all the gates we specified above enough to carry out any quantum computation that we would like? Could we do any quantum computation using just a small subset of the gates above? These are questions about \textit{universality}. 

For classical computers, a set of gates is called universal if, by applying enough gates from this set in a sequence, it is possible to express any Boolean function on any number of bits. The classical NAND gate turns out to be universal all on its own, but there are other sets of gates that are not universal. For example, the set $\{ \text{AND}, \text{OR} \}$ is not universal, because these gates cannot express non-monotone Boolean functions; changing an input bit from 0 to 1 in a circuit with these gates will never result in an output bit changing from 1 to 0.

For quantum computers, a gate set is called \textit{universal} if the gates therein can be used to approximate any unitary transformation on any number of qubits to any desired precision. To understand what makes a gate set universal for quantum computing, let us first see how a gate set can fail to be universal:

\begin{itemize}
\item \textbf{Inability to create superposition states} \\
Some gates, e.g., X and CNOT, only change states in the computational basis ($\ket{0}$ and $\ket{1}$) into other states in the computational basis (e.g, $\text{CNOT} \ket{11} = \ket{10}$). These gates can maintain superpositions, but they cannot create new ones.
\item \textbf{Inability to create entanglement} \\
The Hadamard gate can create a superposition (e.g., $H \ket{0} = \ket{+}$), but it only acts on a single qubit, so it cannot create entanglement between two or more qubits. Clearly, no single-qubit gate can. Since an unentangled state of $N$ qubits can be written $(\alpha_1 \ket{0} + \beta_1 \ket{1}) \otimes \ldots \otimes (\alpha_N \ket{0} + \beta_N \ket{1})$, it can be specified using only $2N$ amplitudes, which a classical computer would be able to simulate efficiently.
\item \textbf{Inability to create non-real amplitudes} \\
A gate set like $\{ \text{CNOT}, H \}$ would be able to create both entanglement and superposition states. However, the matrices specifying these gates only have real entries. Thus, they would never be able to create states with complex amplitudes.
\item \textbf{The Gottesman--Knill theorem} \\
If we take our gate set to be $\{ \text{CNOT}, H, S \}$, we circumvent all the objections above. However, it turns out that this still is not enough to achieve universal quantum computation. This is the \\

\textbf{Gottesman--Knill theorem~\citep{Gottesman1999}:}
A quantum circuit using only the following elements can be simulated efficiently on a classical computer:
\begin{enumerate}
\item Preparation of qubits in computational basis states,
\item Quantum gates from the Clifford group $\{ \text{CNOT}, H, S \}$, and
\item Measurements in the computational basis.
\end{enumerate}
%
The Clifford group is the group of operations that map Pauli matrices onto (possibly different) Pauli matrices, and is generated by the gates $\{ \text{CNOT}, H, S \}$.
Quantum circuits with only these gates are also called stabilizer circuits (we will return to this concept in \chpref{chp:QEC}, where we will give a proof of the Gottesman--Knill theorem and discuss quantum error correction). If one starts in the computational basis, the single-qubit states that can be reached using these circuits are the $\pm 1$ points on the $x$, $y$, and $z$ axes of the Bloch sphere. This restriction of the state space turns out to make the circuit classically simulatable even though the state space includes states with both superposition, entanglement, and complex amplitudes.
\end{itemize}

What is then a universal gate set? Actually, replacing the Hadamard gate in the last gate set considered above, $\{ \text{CNOT}, H, S \}$, with almost any other possible single-qubit gate makes the set universal. We can also replace S with some other gate. One universal gate set is $\{ \text{CNOT}, H, T \}$. Almost any two-qubit gate on its own is also universal.

We thus now know that we can implement any unitary operation on $N$ qubits in a realistic experimental architecture that allows for a few single- and two-qubit gates. Does that mean that we can run quantum algorithms that are faster than classical algorithms for some problems? This will be discussed below and in the next chapters.

\section{The Solovay--Kitaev theorem}
\label{sec:SolovayKitaev}

Above, we saw that there are many gate sets that are universal for quantum computing, i.e., they can approximate any unitary transformation on any number of qubits to any desired precision $\epsilon$. However, it is a different question whether that approximation is fast and efficient. For example, if the number of gates needed would scale exponentially in $1 / \epsilon$, there would probably not be any practical use for such a universal gate set. Luckily, there is a theorem telling us that the scaling for any universal gate set is much more benign: \\

{\bf Solovay--Kitaev theorem:} Let $G$ a finite subset of $SU(2)$ and $U \in SU(2)$. If the group generated by $G$ is dense in $SU(2)$, then for any $\varepsilon>0$ it is possible to approximate $U$ to precision $\varepsilon$ using $O \mleft(\log^4(1/\varepsilon) \mright)$ gates from $G$.\\ 

The proof can be found in Appendix 3 of Ref.~\citep{Nielsen2000}. What the theorem essentially says is that if we have a gate set that can approximate any single-qubit rotation, then we only need relatively few gates from that set to do the approximation, so we can do the approximation fast. By adding some two-qubit gate to make the gate set universal, it can be shown that the total number of gates needed to approximate $U$ on $N$ qubits is at most $O \mleft( 4^N \log^4 \mleft[ 1/\varepsilon \mright] \mright)$. More recent results have shown that for certain gate sets, the number of gates can be brought down from $O \mleft( \log^4 \mleft[ 1/\varepsilon \mright] \mright)$ to $O \mleft( \log \mleft[ 1/\varepsilon \mright] \mright)$, and that the exponent can be reduced below 4 for general gate sets; see, e.g., Ref.~\citep{Kuperberg2023} and references therein. There are good algorithms for finding the gate sequences needed to achieve the Solovay-Kitaev bound.

This theorem, together with the observations above, is what gives us confidence that quantum computing has potential. We now know how to find a universal gate set to approximate any unitary transformation, and we know that we can implement that approximation efficiently.

\newpage

\section*{Exercises}

\begin{enumerate}

\item (Ref.~\citep{Nielsen2000}, hereafter abbreviated NC, exercise 4.7) Show that $XYX = - Y$ and use this to prove that $X R_y (\theta) X = R_y (- \theta)$.

\item (NC 4.13) It is useful to be able to simplify circuits by inspection, using well-known identities. Prove the following three identities:
\be
HXH = Z; \quad HYH = -Y; \quad HZH = X. \nn
\ee
%

\item (NC 4.14) Use the previous exercise to show that $HTH = R_x (\pi / 4)$, up to a global phase.

\item (NC 4.17) Construct a CNOT gate from one CZ gate and two Hadamard gates, specifying the control and target qubits.

\item Construct a Fredkin gate using three Toffoli gates.

\item Starting from the two-qubit state $\ket{00}$, can you create the following state using the gate set $\{ \text{CNOT}, H, S \}$? If yes, show how. If no, explain why.
\be
\frac{1}{\sqrt{2}} \mleft( \ket{00} + e^{i \pi /4} \ket{11} \mright) \nn
\ee

\end{enumerate}


\chapter{Grover's algorithm}
\label{chp:Grover}

Could a quantum computer find the answer to a problem that \textit{cannot} be solved on a classical computer, even if there are no restrictions on the resources available to the classical computer? The answer is no. The classical computer could always simulate the quantum computation by storing, to enough precision, the $2^N$ amplitudes for the whole state of the quantum computer's $N$ qubits, and changing these amplitudes according to the gates applied in the quantum algorithm. However, there is clearly a possibility that this classical simulation requires a lot more resources than the quantum computation itself. So a better question to ask is: are there problems that a quantum computer could solve \textit{faster} than a classical computer? In this chapter, we begin to answer that question by introducing one of the most famous quantum algorithms: Grover's algorithm. For a more formal treatment of how to compare the ability of classical and quantum computers when it comes to solving different types of problems, see the discussion of complexity classes in \chpref{chp:complexity-classes}.

\section{Quantum parallelism}
\label{quantum-parallelism}

As a prelude to Grover's algorithm, we will first look at one of the simplest examples showing how a quantum algorithm can work faster than any classical algorithm. The problem may seem somewhat contrived, but the purpose is to give a clear demonstration of the special type of parallelism you can achieve with {\em superposition} states.

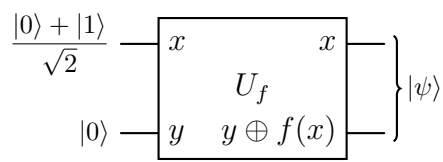
\begin{figure}[H]
\centering
\begin{quantikz}
    \lstick{$\dfrac{\ket{0}+\ket{1}}{\sqrt{2}}$} & \gate[2][2.5cm]{\text{\large $U_f$}}\gateinput{\large $x$}\gateoutput{\large $x$} & \rstick[2]{$\ket{\psi}$}\\
    \lstick{$\ket{0}$} & \gateinput{\large $y$}\gateoutput{\large $y\oplus f(x)$} & 
\end{quantikz}
\caption{A circuit demonstrating quantum parallelism. From Ref.~\citep{Nielsen2000}, Fig.~1.17.}
\label{QuantumParallellismFig}
\end{figure}

Consider the circuit in \figref{QuantumParallellismFig}, where the box named $U_f$ performs the two-qubit unitary transformation $\ket{x}\ket{y} \rightarrow \ket{x}\ket{y \oplus f(x)}$. Here, $f(x)$ is a function taking a single bit as input and giving a single bit as output. This box could be regarded as a subroutine of some sort, performing time-consuming calculations; we therefore want to run it as few times as possible.

When we send in the input states shown in \figref{QuantumParallellismFig}, the output state becomes 
\be
\ket{\psi} = \frac{\ket{0}\ket{f(0)} + \ket{1}\ket{f(1)}}{\sqrt{2}},
\ee
which contains information about both the function values $f(0)$ and $f(1)$ after only a {\em single} run of the function $U_f$. However, measuring this output state only gives a single outcome, {\em either} $(0,f(0))$ {\em or} $(1,f(1))$, each with probability $1/2$.

\section{The Deutsch-Jozsa algorithm}
\label{sec:DeutschAlgorithm}

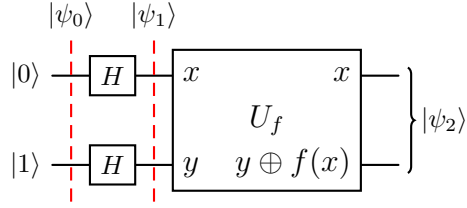
\begin{figure}[t]
\centering
\begin{quantikz}
    \lstick{$\ket{0}$}\slice{$\ket{\psi_0}$} & \gate{H}\slice{$\ket{\psi_1}$} & \gate[2][2.5cm]{\text{\large $U_f$}}\gateinput{\large $x$}\gateoutput{\large $x$} & \rstick[2]{$\ket{\psi_2}$}\\
    \lstick{$\ket{1}$} & \gate{H} & \gateinput{\large $y$}\gateoutput{\large $y\oplus f(x)$} & 
\end{quantikz}
\caption{The beginning of Deutsch's algorithm. A final Hadamard gate on the first qubit completes the algorithm. From Ref.~\citep{Nielsen2000}, Fig.~1.19.}
\label{DeutschsAlgorithmCircuitFig}
\end{figure}

Deutsch began showing~\citep{Deutsch1985} in 1985 that with a slightly more elaborate circuit, a {\em global} property of $f(x)$ may be deduced in a single run of $U_f$, which is classically impossible. Later refinements by Jozsa and others led to what is now known as the Deutsch-Josza algorithm, but the simple version we present here is often just called Deutsch's algorithm. In that case, we want to check whether the function $f(x)$ is constant or not, i.e., whether or not $f(0) = f(1)$. To do so, we use the circuit shown in \figref{DeutschsAlgorithmCircuitFig}.

When the input state
\be
\ket{\psi_0} = \ket{0} \ket{1}
\ee
in \figref{DeutschsAlgorithmCircuitFig} is affected by a Hadamard gate on each qubit, the resulting state is
\be
\ket{\psi_1} = \mleft( \frac{\ket{0} + \ket{1}}{\sqrt{2}} \mright) \mleft( \frac{\ket{0} - \ket{1}}{\sqrt{2}} \mright).
\ee
Applying $U_f$ to the state $\ket{x} (\ket{0} - \ket{1}) / \sqrt{2}$ gives
\be
\ket{x} \frac{\ket{f(x)} - \ket{1 \oplus f(x)}}{\sqrt{2}} = \mleft\{
\begin{array}{cc}
\frac{\ket{x} (\ket{0} - \ket{1})}{\sqrt{2}}, & f(x) = 0 \\
\frac{\ket{x} (\ket{1} - \ket{0})}{\sqrt{2}}, & f(x) = 1
\end{array} \mright\} 
= (-1)^{f(x)} \frac{\ket{x} (\ket{0} - \ket{1})}{\sqrt{2}}.
\label{eq:DeutschOracle}
\ee
This expression can be generalised to the case where $\ket{x}$ denotes a set of multiple qubits. For the circuit here in \figref{DeutschsAlgorithmCircuitFig}, the resulting output state is
\bea
\ket{\psi_2} &=& \frac{1}{2} \mleft[ (-1)^{f(0)} \ket{0} (\ket{0} - \ket{1}) + (-1)^{f(1)} \ket{1} (\ket{0} - \ket{1}) \mright] \nn\\
&=& (-1)^{f(0)} \frac{1}{2} \mleft( \ket{0} + (-1)^{f(0) \oplus f(1)} \ket{1} \mright) \mleft( \ket{0} - \ket{1} \mright)
\eea
Ignoring the global phase and only focussing on the state of the first qubit, we see that it is
\be
\ket{\psi_2^{(1)}} = \frac{1}{\sqrt{2}} \mleft( \ket{0} + (-1)^{f(0) \oplus f(1)} \ket{1} \mright).
\ee
Applying a Hadamard gate to this qubit then yields
\bea
\ket{\psi_3^{(1)}} &=& \frac{1}{2} \mleft( \ket{0} + \ket{1} + (-1)^{f(0) \oplus f(1)} \ket{0} - (-1)^{f(0) \oplus f(1)} \ket{1} \mright) \nn\\
&=& \frac{1}{2} \mleft[ \mleft( 1 + (-1)^{f(0) \oplus f(1)} \mright) \ket{0} + \mleft( 1 - (-1)^{f(0) \oplus f(1)} \mright) \ket{1} \mright]. 
\eea
So if $f(0) = f(1)$, i.e., if $f(0) \oplus f(1) = 0$, we will measure the first qubit to be in $\ket{0}$, and if $f(0) \neq f(1)$, i.e., if $f(0) \oplus f(1) = 1$, we will measure the first qubit to be in $\ket{1}$. So from just running this circuit once and measuring the first qubit, we will know for sure whether or not $f(0) = f(1)$.

The generalization to the case when the function $f$ takes multiple bits as input is the Deutsch-Jozsa algorithm. Even in that complicated case, it turns out that a single run of $U_f$ is sufficient to determine whether $f(x)$ is constant (gives the same result for all possible inputs) or balanced (gives the result 0 for half of the inputs and 1 for the other half), if we know that $f(x)$ is restricted to being only one of these two options. 

\section{Quantum search algorithms}

Is it possible to use the quantum parallelism to faster find what you are looking for, e.g., in a database? The answer is yes, but the speedup is ``only'' from a classical $O(N)$ runtime to a quantum $O(\sqrt{N})$ runtime, where $N$ is the number of items you have to look through. The performance is measured in the number of times you have to consult the ``oracle'', i.e., the subroutine which recognizes the item you are looking for. The quantum search algorithm is optimal in the sense that you may prove [see Sec.~6.6 in Ref.~\citep{Nielsen2000}] that {\em no quantum algorithm} can find the item using fewer than $O(\sqrt{N})$ consultations of the oracle. This implies that you generally {\em cannot} get an exponential speedup using a quantum algorithm for solving a classical problem, unless the problem has some specific structure that can be used.

\section{Grover's algorithm}

We now present Grover's algorithm~\citep{Grover98}, which achieves the $O(\sqrt{N})$ runtime bound for quantum search algorithms.

\subsection{Action of the oracle}

Grover's algorithm generalizes and puts to use the concept of quantum parallelism seen in \secref{quantum-parallelism}. In general, the black-box function that we introduced therein is called an oracle. Formally, the oracle $O$ is a subroutine (unitary operator) taking an $n$-bit input $x$ and flipping an output bit if and only if $x$ is the item you are looking for. More explicitly,
\be
\ket{x} \ket{q} \rightarrow \ket{x} \ket{q \oplus f(x)},
\ee
where $f(x)=1$ if $x$ is an item you are looking for and zero otherwise.

\subsection{The algorithm}

We assume that we have $N=2^n$ items to search. As we have seen in Deutsch's algorithm in \secref{sec:DeutschAlgorithm}, it is useful to place the oracle qubit $\ket{q}$ in a superposition such that
\be
\ket{x} \mleft( \frac{\ket{0} - \ket{1}}{\sqrt{2}} \mright) \rightarrow (-1)^{f(x)} \ket{x} \mleft( \frac{\ket{0} - \ket{1}}{\sqrt{2}} \mright),
\ee
as shown in \eqref{eq:DeutschOracle}. In this way, the action of the oracle can be written
\be
\ket{x} \rightarrow (-1)^{f(x)} \ket{x},
\ee
suppressing the oracle qubit which does not change state. The oracle thus marks the solution by shifting the phase of the solution state.

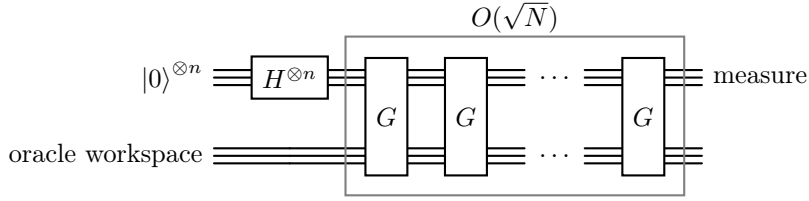
\begin{figure}
\centering
\begin{quantikz}[wire types={b,b}, classical gap=0.1cm]
\lstick{$\ket{0}^{\otimes n}$} & \gate{H^{\otimes n}}& \gate[2]{G}\gategroup[2,steps=4,style={gray, inner
    sep=4pt}]{$O(\sqrt{N})$} & \gate[2]{G} & \push{\;\;\dots\;\;} & \gate[2]{G} & \rstick{measure}\\
\lstick{oracle workspace} & & & & \push{\;\;\dots\;\;} & &
\end{quantikz}
\caption{An overview circuit figure of Grover's search algorithm. From Ref.~\citep{Nielsen2000}, Fig.~6.1.}
\label{GroverOverviewFig}
\end{figure}

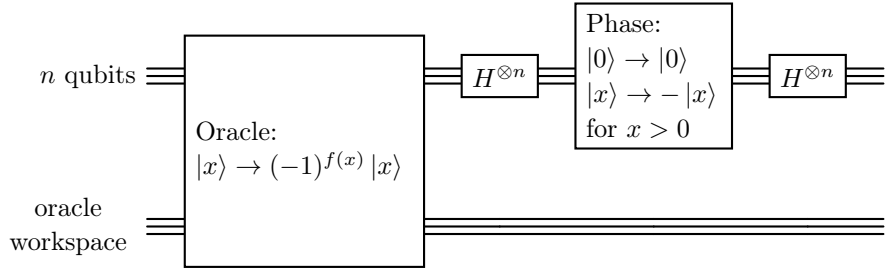
\begin{figure}
\centering
\begin{quantikz}[wire types={b,b}, classical gap=0.1cm]
\lstick{$n$ qubits} & \gate[2]{\parbox{2.9cm}{Oracle:\\ $\ket{x} \to (-1)^{f(x)}\ket{x}$}} & \gate{H^{\otimes n}} & \gate{\parbox{1.8cm}{Phase:\\$\ket{0}\to\ket{0}$\\$\ket{x}\to-\ket{x}$\\for $x>0$}} & \gate{H^{\otimes n}} & \\
\lstick{\parbox{1.8cm}{\centering oracle\\workspace}} & & & & &
\end{quantikz}
\caption{Detailed circuit figure showing the Grover operator $G$. From Ref.~\citep{Nielsen2000}, Fig.~6.2.}
\label{GroverDetailFig}
\end{figure}

The search algorithm shown in \figref{GroverOverviewFig} starts by putting the input register in an equal superposition $\ket{\psi}$ of all
input states using the Hadamard $H^{\otimes n}$ operation:
\be
\ket{\psi} = \frac{1}{\sqrt{N}} \sum_{x=0}^{N-1} \ket{x} .
\ee
Then the oracle is called inside the Grover operator $G$, shown in \figref{GroverDetailFig}. This operator is called an $O(\sqrt{N})$ number of times, after which the input register is read out, giving the sought element $x_0$.

\subsection{Grover's operator}

The action of the Grover operator $G$ in \figref{GroverDetailFig} can be expressed in four steps:
\begin{enumerate}
\item Apply the oracle $O$.
\item Apply the Hadamard transform $H^{\otimes n}$.
\item Give all states a phase shift of $\pi$, except the $\ket{0}$ state, which is left untouched:
\[ \ket{x} \rightarrow -(-1)^{\delta_{x,0}} \ket{x} .\]
\item Apply the Hadamard transform $H^{\otimes n}$.
\end{enumerate}

More compactly, one may write the Grover operator as
\be
G = H^{\otimes n} \mleft( 2 \ketbra{0}{0} - \hat{1} \mright) H^{\otimes n} O = \mleft( 2 \ketbra{\psi}{\psi} - \hat{1} \mright) O ,
\ee
where $\ket{\psi}$ is the equal superposition of all input states.

\subsection{Geometric visualization of the state evolution}

\begin{figure}
\centering
\includegraphics[width=0.45\textwidth]{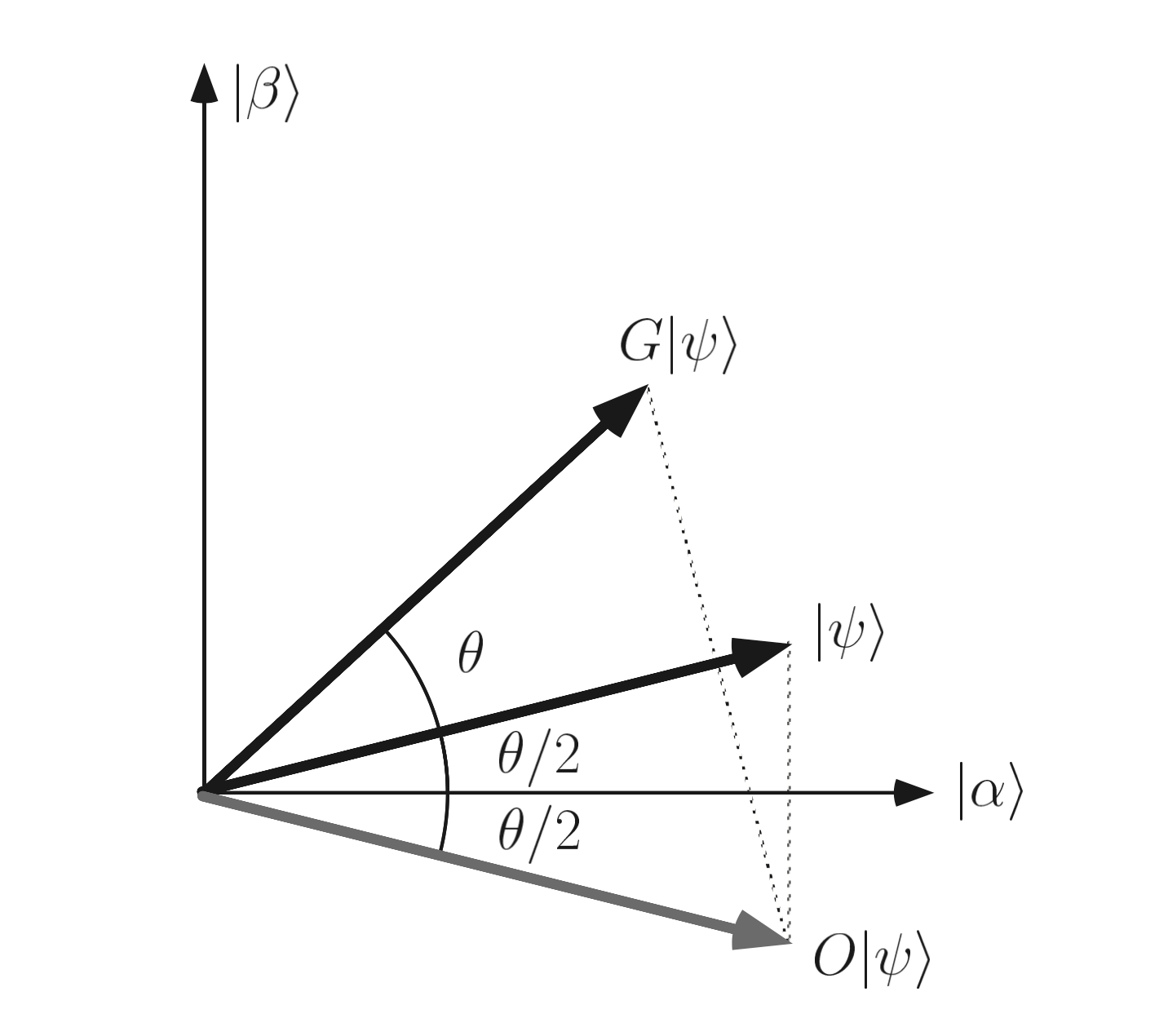}
\caption{One iteration of the Grover operator $G$ visualized as two reflections in a two-dimensional vector space. From Ref.~\citep{Nielsen2000}, Fig.~6.3.} \label{GroverVectorsFig}
\end{figure}

Limiting ourselves to the case when we are looking for only one state $\ket{x_0}$, we see that the algorithm evolves in a two-dimensional vector space spanned by the solution vector
\be
\ket{\beta} = \ket{x_0}
\ee
and the state consisting of an equal superposition of all the other states,
\be
\ket{\alpha} = \frac{1}{\sqrt{N-1}} \sum_{x \neq x_0} \ket{x} .
\ee
The initial state can then be written
\be
\label{geometric-grover}
\ket{\psi} = \frac{1}{\sqrt{N}} \sum_{x \neq x_0}^{N-1} \ket{x} + \frac{1}{\sqrt{N}} \ket{x_0} 
= \sqrt{\frac{N-1}{N}} \ket{\alpha} + \sqrt{\frac{1}{N}} \ket{\beta}.
\ee
The oracle operation $O$ changes the sign of the state $\ket{\beta}$ such that the new state is
\be
\ket{\psi'} = \sqrt{\frac{N-1}{N}} \ket{\alpha} - \sqrt{\frac{1}{N}} \ket{\beta} .
\ee
The operator $\ketbra{\psi}{\psi}$ is a projection operator on the vector $\ket{\psi}$, and thus the operator $\mleft( 2 \ketbra{\psi}{\psi} - \hat{1} \mright)$ changes the sign of the component orthogonal to $\ket{\psi}$, i.e., it performs a reflection with respect to the vector $\ket{\psi}$. Indeed, it acts as:
\bea
\ket{\psi} &\rightarrow& 2 \ket{\psi} - \ket{\psi} =  \ket{\psi}, \\
\ket{\psi}_\perp &\rightarrow& - \ket{\psi}_\perp.
\eea
Two consecutive reflections is equal to a rotation. From the geometric construction in \figref{GroverVectorsFig}, the angle $\theta$ of the rotation can be extracted:
\be
\sin{\frac{\theta}{2}} = \frac{1}{\sqrt{N}} ,
\ee
where we have used that the projection of $\ket{\psi}$ on $\ket{\beta}$ is $\frac{1}{\sqrt{N}}$ according to \eqref{geometric-grover}. Therefore the state becomes
\be
G \ket{\psi} = \cos \frac{3 \theta}{2} \ket{\alpha} + \sin \frac{3 \theta}{2} \ket{\beta}.
\ee
%

After $k$ applications of $G$, the state makes an angle
$\theta_k = k \theta + \frac{\theta}{2} = \mleft( k + \frac{1}{2} \mright) \theta$
with the $\ket{\alpha}$ state. For $\theta_k\approx \frac{\pi}{2}$, this is close to the desired state $\ket{\beta}$, namely
\be
G \ket{\psi}_k = \cos \mleft[ \mleft( k + \frac{1}{2} \mright) \theta \mright] \ket{\alpha} + \sin \mleft[ \mleft( k + \frac{1}{2} \mright) \theta \mright] \ket{\beta} .
\ee
Therefore a measurement in the computational basis identifies the solution $\ket{\beta} = \ket{x_0}$ with a high probability. In the interesting
case $N \gg 1$, we can estimate the optimal number $k_{\rm opt}$ of iterations:
\be
\frac{\pi}{2} \approx \mleft( k_{\rm opt} + \frac{1}{2} \mright) \theta \approx \mleft( k_{\rm opt} + \frac{1}{2} \mright) \frac{2}{\sqrt{N}}
\rightarrow k_{\rm opt} \approx \frac{\pi \sqrt{N}}{4} - \frac{1}{2} .
\ee
%
The maximum error in angle is $\theta/2 \simeq 1/{\sqrt{N}}$ (otherwise we could perform a further iteration of the algorithm), which translates into a probability of error
\be
P_{\rm err} \leq \cos^2{\mleft( \frac{\pi}{2} - \frac{\theta}{2} \mright) } = \sin^2{\frac{\theta}{2}} \simeq \mleft( \frac{\theta}{2} \mright)^2 = \frac{1}{N} .
\ee
Therefore, the correct state is read out with probability
\be
P_{\rm succ} = 1 - P_{\rm err} \geq 1 - \frac{1}{N} .
\ee
%
It is straightforward to analyze the situation when more than one item fulfil the search criteria.



\chapter{Quantum error correction}
\label{chp:QEC}

We have now seen that universal quantum computation can be performed with a rather small and simple set of quantum gates, and that there is good reason to believe that a universal quantum computer will be able to solve some problems faster than classical computers can. However, a crucial question remains to answer before we go ahead and invest large resources into actually building a quantum computer: can a quantum computer deal with errors? 

If even a small error can wreck a quantum computation beyond repair, a practical quantum computer can never be realized. In today's classical computers, the probability $p$ of an error occurring during a logical operation is amazingly low. Numbers on the order of $p \approx 10^{-18}$ are often quoted. For state-of-the-art quantum computers, however, $p$ is rather on the order of $10^{-2}$ or, in the best case, $10^{-3}$. In this chapter, we show that quantum error correction is indeed possible and feasible, even with error rates close to what we have today. For a more detailed account, se Chapter 10 in Ref.~\citep{Nielsen2000}.

\section{Challenges for quantum error correction}
\label{sec:ChallengesForQEC}

Correcting for errors on classical bits is quite straightforward. The basic idea is to encode the state of one bit redundantly using several bits such that an error on one of the latter does not change the encoded information. The simplest example is a three-bit code with majority voting. The state of one bit, 0 or 1, is encoded into three bits as 000 or 111. The encoded information is read out by measuring all three bits and going with the majority vote. For example, if the third bit is flipped, 000 changes into 001 (and 111 into 110), but the majority vote among 001 (110) still tells us that the encoded bit was 0 (1). For the error correction to fail, two bits need to be flipped. This means that while the error probability for the single unencoded bit is $p$, the error probability for the encoded bit is $\sim p^2$.

Correcting for errors on quantum bits is more complicated. Indeed, this was a major headache for researchers in the early days of quantum computing. Before Peter Shor showed in 1995~\citep{Shor1995} how to achieve quantum error correction, it was hard to be optimistic about the prospects for quantum computing. The major obstacles thought to prevent quantum error correction were three:
\begin{itemize}
\item The no-cloning theorem~\citep{Wootters1982, Dieks1982}. Quantum mechanics prohibits the existence of a unitary operation $U$ that can change a known state $\ket{\phi}$ into a copy of an unknown arbitrary quantum state $\ket{\psi}$ without perturbing the latter, i.e., $U \ket{\phi} \ket{\psi} = \ket{\psi} \ket{\psi}$. For the classical error correction described above, such cloning was essential for encoding.
\item Measuring a quantum state causes it to collapse into an eigenstate of the measured observable. How to correct errors on an arbitrary superposition state $\alpha \ket{0} + \beta \ket{1}$ without disturbing the state when performing a measurement on it?
\item In a classical bit, there is only one possible error: a bit flip, taking the bit from 0 to 1 or vice versa. For a quantum bit, there are infinitely many possible errors: any single-qubit operation, i.e., any rotation around any axis of the Bloch sphere by any angle, could possibly be induced by some external error source. How to make an error-correction procedure general enough to be able to deal with all these possible errors?
\end{itemize}

\section{The three-qubit bit-flip code}
\label{sec:QEC-3q-bit-flip-code}

\begin{figure}
\centering
\begin{quantikz}[row sep={0.8cm,between origins}]
    \lstick{$\ket{\psi}$} & \ctrl{1}\gategroup[3,steps=2,style={gray, dashed, inner sep=1pt}]{Encoding} & \ctrl{2} & \gate[3, disable auto
    height]{\parbox{1cm}{\centering bit\\flip\\error}\vphantom{0}} & \gate[2]{M_{12}}\gategroup[3,steps=2,style={gray, dashed, inner sep=1pt}]{Measurements} & & \gate{\text{Flip iff } M_{12} = -1, M_{23} = 1}\gategroup[3,steps=1,style={gray, dashed, inner sep=1pt}]{Correction} \\
    \lstick{$\ket{0}$} & \targ{} & & & & \gate[2]{M_{23}} & \gate{\text{Flip iff } M_{12} = -1, M_{23} = -1}\\
    \lstick{$\ket{0}$} & & \targ{} & & & & \gate{\text{Flip iff } M_{12} = 1, M_{23} = -1}
\end{quantikz}
\caption{The three-qubit bit-flip error-correction code. In the first step, CNOT gates are applied to produce the state $\ket{\psi_3}$ from $\ket{\psi}$ [see \eqref{EqPsi3}]. After a bit-flip error occurs, parity measurements are done on pairs of qubits and correcting bit flips are applied to the qubits depending on the measurement results (-1 means the qubits are in opposite states, +1 that they are in the same state). Figure from Ref.~\citep{Kockum2014}.
\label{Fig3QbCode}}
\end{figure}
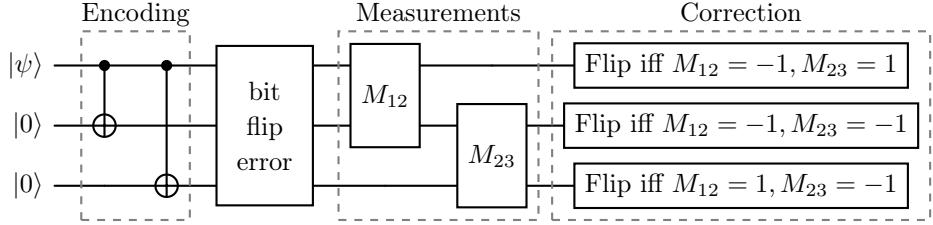

In this section, we will look at the simplest quantum error-correction code, which demonstrates that all three obstacles above can be dealt with. The code is the three-qubit bit-flip code. The encoding and error-correction process is shown schematically in \figref{Fig3QbCode}.

We begin with an arbitrary one-qubit state 
\be
\ket{\psi} = \alpha \ket{0} + \beta \ket{1}.
\ee
To protect it from bit-flip errors, i.e., from an unwanted application of the $X$ gate, we encode it using three qubits as
\be
\ket{\psi_3} = \alpha\ket{000} + \beta\ket{111}.
\label{EqPsi3}
\ee
Note that this state can be created by application of CNOT gates, which is different from cloning the state; that would have resulted in $\ket{\psi_3} = \mleft( \alpha \ket{0} + \beta \ket{1} \mright)^{\otimes 3}$ (if cloning had been possible).

Now, let us assume that the third qubit is flipped. This gives
\be
\ket{\psi_{3,\text{err}}} = X_3 \ket{\psi_3} = \alpha \ket{001} + \beta \ket{110}.
\label{EqPsi3err}
\ee
If we first perform a parity measurement on qubits 1 and 2, i.e., measure the product $Z_1 Z_2$, and then measure the parity of qubits 2 and 3, i.e., $Z_2 Z_3$, we would not affect the state $\ket{\psi_3}$, since $Z_1 Z_2 \ket{\psi_3} = \ket{\psi_3} = Z_2 Z_3 \ket{\psi_3}$. Similarly, performing these measurements on $\ket{\psi_{3,\text{err}}}$ does not change the coefficients $\alpha$ and $\beta$ determining the superposition; it just adds a global phase factor. However, the result of the measurements lets us draw the conclusion that qubit 3 has been flipped (assuming that the probability of more than one qubit flipping is negligible). We can then apply an $X$ gate to this qubit, flipping it back to its original state and recovering $\ket{\psi_3}$. As shown in \figref{Fig3QbCode}, the same set of measurements also lets us correct the state if a bit-flip error instead occurs on qubit 1 or 2. We have thus managed to circumvent the problem of quantum measurements being projective.

What happens if the error is an arbitrary $X$ rotation $R_x(\theta)$ on the third qubit instead of just $X$? As we can see from \eqref{eq:RotX}, the resulting state would then be
\be
R_x(\theta) \ket{\psi_3} = \cos(\theta/2) \ket{\psi_3} - i \sin (\theta/2) \ket{\psi_{3,\text{err}}}.
\ee
Measuring the observables $Z_1 Z_2$ and $Z_2 Z_3$ will either project this state into $\ket{\psi_3}$ or $\ket{\psi_{3,\text{err}}}$ (modulo a global phase). In both cases, we will know what operation to apply to recover $\ket{\psi_3}$. This demonstrates how quantum error-correcting codes deal with the continuum of possible errors: measurements are used to project the perturbed state into a finite set of states from which we know how to recover the original state.

\section{The three-qubit phase-flip code}

The three-qubit bit-flip code can only correct for $X$ errors on a single qubit. However, the idea of the code can be extended to instead deal with $Z$ errors, i.e., phase flips. The crucial observation is that just as $X$ flips $\ket{0}$ to $\ket{1}$ and vice versa, $Z$ flips $\ket{+}$ to $\ket{-}$ and vice versa. So by encoding the one-qubit state we want to protect in the $\{ \ket{+}, \ket{-} \}$ basis instead of the $\{ \ket{0}, \ket{1} \}$ basis,
\be
\ket{\psi_3} = \alpha\ket{+++} + \beta\ket{---}.
\label{EqPsi3Z}
\ee
we achieve an encoded three-qubit state where a $Z$ error flips one of the qubits from one of its basis states to the other. This encoding is implemented by adding a Hadamard gate on each qubit at the end of the encoding step in \figref{Fig3QbCode}.

To detect a phase-flip error in one of the three qubits, we measure the products $H^{\otimes 3} Z_1 Z_2 H^{\otimes 3} = X_1 X_2$ and $H^{\otimes 3} Z_2 Z_3 H^{\otimes 3} = X_2 X_3$. Just as above, these measurements do not change the coefficients $\alpha$ and $\beta$, but let us conclude whether a qubit has suffered a phase flip (and which qubit it was), allowing us to apply a $Z$ gate to correct. And just as before, this procedure also works for arbitrary $Z$ rotations by projecting the state into either $\ket{\psi_3}$ or a state that can be corrected by applying a $Z$ gate.

\section{The nine-qubit Shor code}
\label{sec:9qbShor}

The breakthrough of Shor in 1995, which showed that quantum error correction could correct arbitrary single-qubit errors, was a nine-qubit code~\citep{Shor1995} that combines the three-qubit bit-flip and phase-flip codes into one. The two codes are concatenated such that the single-qubit state $\ket{\psi}$ first is encoded with the three-qubit phase-flip code and then each of these three qubits are encoded with the three-qubit bit-flip code into three more qubits each. The resulting state is
\be
\ket{\psi_9} = \alpha \frac{\mleft( \ket{000} + \ket{111} \mright) \mleft( \ket{000} + \ket{111} \mright) \mleft( \ket{000} + \ket{111} \mright)}{2\sqrt{2}} + \beta \frac{\mleft( \ket{000} - \ket{111} \mright) \mleft( \ket{000} - \ket{111} \mright) \mleft( \ket{000} - \ket{111} \mright)}{2\sqrt{2}}.
\ee
To find out whether a bit flip has occurred, we can treat each of the three groups of three qubits separately, measuring the product of $Z$s on the first two and the last two qubits in each group. For phase-flip errors, the procedure is slightly more complicated. We first determine in which of the three groups of three qubits that a phase flip has occurred. This is done by measuring the product of $X$s on all qubits in two groups of qubits at a time, i.e., measuring $X_1 X_2 X_3 X_4 X_5 X_6$ and $X_4 X_5 X_6 X_7 X_8 X_9$. The results of these measurements will tell us if a phase flip has occurred in one of the three groups, but it will not tell us which of these three qubits has been flipped. However, we do not need to know which qubit it was, because we can correct the error to the encoded state by applying $Z$s to all three qubits in the group.

Studying the procedure for error-correction in the nine-qubit code, it becomes clear that the code also is able to correct a combined $Z$ and $X$ error on one qubit. The check for bit-flip errors will locate the bit flip without being affected by the presence of the phase-flip error. Once the bit flip has been corrected, only the phase-flip error remains and will thus be detected and corrected as before.

Since the nine-qubit code thus can correct for both $X$, $Z$, and $XZ$ errors, it is able to correct for any single-qubit error. This is seen by noting that any rotation of a qubit on the Bloch sphere can be decomposed into $R_z (\gamma) R_x (\beta)  R_z (\alpha)$. Applying this operation to a qubit gives
\be
R_z (\gamma) R_x (\beta)  R_z (\alpha) \ket{\psi} = \mleft[ \cos(\gamma/2) - i \sin (\gamma/2) Z \mright) \mleft( \cos(\beta/2) - i \sin (\beta/2) X \mright)  \mleft( \cos(\alpha/2) - i \sin (\alpha/2) Z \mright] \ket{\psi},
\ee
which results in terms where either the identity operator, $X$, $Z$, or $XZ$ has been applied to $\ket{\psi}$ (note that $ZXZ = -X$). Measuring the observables for error correction will thus project any single-qubit error to identity, a bit-flip error, a phase-flip error, or a combined bit- and phase-flip error on that qubit. Since each of these errors can be corrected, any single-qubit error can be corrected.

\section{Stabilizers}
\label{sec:Stabilizers}

The error-correction codes above, and many others, can be understood in terms of \textit{stabilizers}. If a state $\ket{\psi}$ is unchanged under the action of a unitary operator $U$, i.e., $U \ket{\psi} = \ket{\psi}$, we say that the state $\ket{\psi}$ is stabilized by $U$. We can extend this to a collection of operators and states. Let $S$ be a group of $N$-qubit operators and $V_S$ the set of $N$-qubit states that are stabilized by every element in $S$. Then we say that $S$ is the stabilizer of the vector space $V_S$. For $V_S$ to be non-trivial (i.e., not just zero), it can be seen quite easily that the elements of $S$ need to commute and that $-I$ cannot be in $S$.

Recall that for the three-qubit bit-flip code in \secref{sec:QEC-3q-bit-flip-code}, we measured the operators $Z_1 Z_2$ and $Z_2 Z_3$, noting that they left the encoded state, a superposition of $\ket{000}$ and $\ket{111}$, unchanged. Here, the stabilized vector space $V_S$ is spanned by $\ket{000}$ and $\ket{111}$ and the stabilizer $S$ is the group generated by $Z_1 Z_2$ and $Z_2 Z_3$, i.e., the group $\{ I, Z_1 Z_2, Z_2 Z_3, Z_1 Z_3\}$ (products of the elements in the group are elements in the group; the elements $I$ and $Z_1 Z_3$ can be constructed by multiplying together the generators $Z_1 Z_2$ and $Z_2 Z_3$ in various ways).

So by measuring the generators of $S$, we were able to correct certain errors on the stabilized states. We can imagine constructing other error-correction codes by finding a stabilizer, its generators, and the corresponding stabilized states. But how do we know which errors the code protects from? If we work with a stabilizer $S$ and errors $\{E_j\}$ that are subgroups of the $N$-qubit Pauli group $G_N$ (products of single-qubit Pauli matrices and factors $\pm 1, \pm i$), there is a theorem [Theorem 10.8 in Ref.~\citep{Nielsen2000}] that tells us that these errors can be corrected if $E_j^\dag E_k \notin N(S) - S \ \forall j, k$.
Here, the normalizer of $S$, $N(S)$, leaves the set $S$ fixed under conjugation, i.e., it consists of all elements $g\in G_N$ such that $gsg^\dagger\in S$ for all $s\in S$.
This means that elements of $N(S)$ commute with $S$ as a set.
For our example with the three-qubit bit-flip code, it is quite straightforward to see that any product of two elements in the error set $\{ I, X_1, X_2, X_3\}$ anti-commutes with at least one of the elements in the stabilizer group generated by $Z_1 Z_2$ and $Z_2 Z_3$ (except $I$, but $I \in S$, so $I \notin N(S) - S$), so all errors in the set can be corrected.

\section{Proving the Gottesman--Knill theorem}

Here we outline how to prove the Gottesman--Knill theorem (which we discussed in \secref{sec:UniversalQComp}) using stabilizers. For a more detailed description of the proof, see Sections 10.5.2--10.5.4 in Ref.~\citep{Nielsen2000}.

If we have a vector space $V_S$ stabilized by $S$, the action of any unitary $U$ on a state $\ket{\psi} \in V_S$ can, for any element $g \in S$, be written
\be
U \ket{\psi} = U g \ket{\psi} = U g U^\dag U \ket{\psi}.
\ee
This means that $U \ket{\psi}$ is stabilized by $U g U^\dag$, and thus $U V_S$ is stabilized by $U S U^\dag$, which is generated by the operators $U g_1 U^\dag, ..., U g_n$ if $g_1, ..., g_n$ are the generators of S.

The point of this is that it sometimes allows a compact representation of qubit states under transformations. Consider, for example, the state $\ket{0}^{\otimes N}$. Applying a Hadamard gate to each qubit in this state transforms it to $\ket{+}^{\otimes N}$, which requires $2^N$ amplitudes to represent in the computational basis. However, the state $\ket{0}^{\otimes N}$ is the only state (up to a global phase) that is stabilized by the stabilizer generated by $\{ Z_1, Z_2, ..., Z_N \}$. After the transformation, $\ket{+}^{\otimes N}$ is uniquely determined by $ H \{ Z_1, Z_2, ..., Z_N \} H^\dag = \{ X_1, X_2, ..., X_N \} $, which only is $N$ generators.

It turns out that the gate set in the Gottesman--Knill theorem, $\{ \text{CNOT}, H, S \}$, is such that any unitary that transforms elements of the $N$-qubit Pauli group to other elements of the $N$-qubit Pauli group can be composed from $\mathcal{O} (N^2)$ gates in that set. Thus, starting in a computational basis state, specified by $N$ generators in the $N$-qubit Pauli group, we can keep track of changes to the state by keeping track of how these generators change under the action of $\{ \text{CNOT}, H, S \}$. This only requires $\mathcal{O} (N^2)$ steps on a classical computer. The same goes for measurements of these states, so the quantum circuit is classically simulatable.

\section{Fault-tolerant quantum computing}

The error-correction codes we have looked at so far have helped us reduce the probability that a single-qubit error will result in an actual error in an encoded single-logical-qubit state formed by several physical qubits. However, this is just the first step towards quantum computing that works in the presence of errors. Such \textit{fault-tolerant quantum computing} requires schemes to perform logical operations and measurements on logical qubits in a way that itself is robust against errors. Explaining how this is done is beyond the scope of this course, but it can be done.

The most important concept in fault-tolerant quantum computing is the \textit{error threshold}. This is the single-qubit error probability which can be tolerated in practice and still allow fault-tolerant quantum computing. The concept can be intuitively understood already from the three-qubit bit-flip code. For an unencoded qubit, the error probability is $p$. For the encoded three-qubit state, we can correct for one single-qubit error, but two single-qubit errors will result in an error for the logical qubit. This occurs with a probability $c p^2$ for some $c$ that is determined by how the code works (in this case, $c = 3$). For it to make sense to invest qubit resources into the three-qubit encoding, we must have $ c p^2 < p$, i.e., $ p < 1/c$. The error threshold is thus $1/c$. To be precise, this is actually called a pseudo-threshold, since it arose from considering a particular size (three qubits) of the error-correction code. The threshold is the value you find for a calculation like this in the limit of a large code (for some codes, it will equal the pseudo-thresholds you find for smaller instances of the code).

If we are below the error threshold, we can reduce the logical error rate further by concatenating the code at more levels. For the bit-flip case, each of the three qubits making up the logical qubit could in turn be encoded into three qubits each to protect against bit-flip errors. And each of those qubits could be encoded into three more, and so on. But will this not lead to having to invest too many resources to lower the logical error rate enough? Fortunately, the answer is no. This is the \textit{threshold theorem for quantum computation}: A quantum circuit containing $s$ gates may be simulated with error probability of at most $\epsilon$ using $\mathcal{O} \mleft( \mathrm{poly} \mleft( \log s / \epsilon \mright) s \mright)$ gates on hardware whose components fail with probability at most $p$, provided $p$ is below some constant threshold, $p < p_{\rm th}$, and given reasonable assumptions about the noise in the underlying hardware. Here, ``poly'' is a polynomial of fixed degree.

\section{The surface code}
\label{sec:SurfaceCode}

In practice, only small examples of error-correcting codes have been demonstrated in experiments so far. There are several practical challenges for experimental implementations of error correction at larger scale. One is that some concatenated codes may require high qubit connectivity. Recall, as a simple example, that the nine-qubit Shor code required measuring two six-qubit stabilizers to identify the phase-flip errors.

The error-correction code attracting most attention for large-scale implementation today is the surface code. For a detailed explanation of how the surface code works, we refer to Ref.~\citep{Fowler2012}. Here, we will only explain some basic points of the code.

\begin{figure}
\centering
\includegraphics[width=0.7\linewidth]{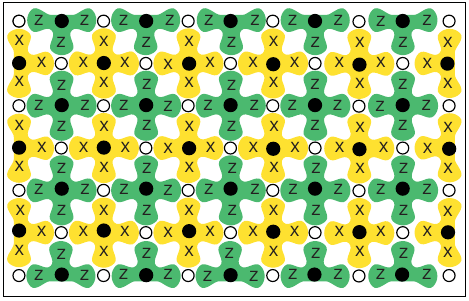}
\caption{The layout for the surface code. The open circles represent qubits that are part of the encoded logical state. The solid circles are qubits that are not part of the logical state, but are used for measurements of the four-qubit $XXXX$ and $ZZZZ$ stabilizers of the code. From Ref.~\citep{Fowler2012}.
\label{fig:SurfaceCode}}
\end{figure}

The surface code is adapted to a square grid of qubits, where each qubit can interact (perform two-qubit gates) with its four nearest neighbours. The layout is depicted in \figref{fig:SurfaceCode}. The surface code only uses four-qubit stabilizers on the form $XXXX$ and $ZZZZ$, which can be measured using the nearest-neighbour interactions on the grid to perform sequential CNOT or CZ gates between the four ``data qubits'' to be measured and a ``measurement qubit'' connected to them, as shown in the figure.

Once the system state has been projected into a state that is stabilized by the four-qubit measurements, a $Z$ error on a data qubit anti-commutes with the $XXXX$ stabilizer measurements that involve one of the two $X$-measurement qubits connected to this data qubit. Similarly, an $X$ error on a data qubit anti-commutes with the $ZZZZ$ stabilizer measurements that involve one of the two $Z$-measurement qubits connected to this data qubit. In this way, it is possible to identify when single-qubit errors (also $XZ$ errors) occur on data qubits. However, if the error probability increases such that several data qubits close to each other experience errors, it can be hard to deduce which error configuration actually caused the measurement results, even if a logical error has not occurred.

\begin{figure}
\centering
\includegraphics[width=0.7\linewidth]{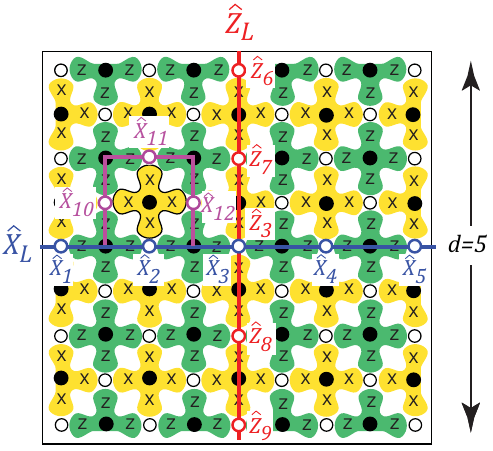}
\caption{Logical operations on the surface code. The depicted setup has distance 5, i.e., $d = 5$ data qubits along each side of the square. There are 41 data qubits in the setup and together with 40 measurement qubits for $X$ and $Z$ stabilizers. The difference leaves room for encoding one logical qubit. To perform logical operations on this encoded qubit, a chain of operators must be applied across the square grid, as shown. A chain of $X$ operators connecting the two opposite sides with $X$ measurements at the boundaries implement a logical $X$ operation. Likewise, a chain of $Z$ operators connecting the two opposite sides with $X$ measurements at the boundaries implement a logical $Z$ operation. From Ref.~\citep{Fowler2012}.
\label{fig:SurfaceCodeLogic}}
\end{figure}

To flip the state of the encoded qubit, a chain of $X$ or $Z$ operations, stretching from one side to the other of the square grid, is required, as shown in \figref{fig:SurfaceCodeLogic}. The larger the distance $d$ across the grid of data qubits, the better protected the encoded qubit is, provided the error probability is below the threshold. For the surface code, the threshold is on the order of \unit[1]{\%}, which is better than many other error-correcting codes. This contributes to the great interest in implementing the surface code.

It is possible to perform two-qubit operations on encoded qubits in the surface code. However, explaining how this works is beyond the scope of these notes. See Ref.~\citep{Fowler2012} for details.

A comprehensive list of possible quantum error correcting codes can be found at \url{https://errorcorrectionzoo.org}.

Experimental progress on implementing the surface code and other quantum error-correcting codes has been impressive in the past few years. Two prominent recent examples are Google demonstrating improved error correction with increasing distance (up to $d=7$, using 101 superconducting qubits) for a surface-code single-qubit memory, which only is possible when all components have error rates below the threshold~\citep{Acharya2025}, and Quera running algorithms with up to 48 encoded logical qubits, using up to 280 physical qubits on a neutral-atom quantum processor~\citep{Bluvstein2024}. 

\newpage

\section*{Exercises}

\begin{enumerate}

\item (NC 10.3) For the three-qubit bit-flip code, show by explicit calculation that measuring $Z_1 Z_2$ followed by $Z_2 Z_3$ is equivalent, up to labeling of the measurement outcomes, to measuring the four projectors defined by 
\bea
&&P_0 \equiv \ketbra{000}{000} + \ketbra{111}{111} \quad \text{no error} \nn \\
&&P_1 \equiv \ketbra{100}{100} + \ketbra{011}{011} \quad \text{bit flip on qubit one} \nn \\
&&P_2 \equiv \ketbra{010}{010} + \ketbra{101}{101} \quad \text{bit flip on qubit two} \nn \\
&&P_3 \equiv \ketbra{001}{001} + \ketbra{110}{110} \quad \text{bit flip on qubit three} , \nn
\eea
in the sense that both procedures result in the same measurement statistics and post-measurement states.

\item (NC 10.6) For the nine-qubit Shor code, show that recovery from a phase flip on any of the first three qubits may be accomplished by applying the operator $Z_1 Z_2 Z_3$.

\item (NC 10.8) Verify that the three-qubit phase-flip code $\ket{0_L} = \ket{+++}$, $\ket{1_L} = \ket{- - -}$ satisfies the quantum error-correction conditions for the set of operators $\{I, Z_1, Z_2, Z_3\}$.

\end{enumerate}


\chapter*{Tutorial 1: Pauli and Clifford groups, stabilizers, and QEC}
\addcontentsline{toc}{chapter}{Tutorial 1} 
\renewcommand{\thesection}{\arabic{section}} 
\setcounter{section}{0} 
\setcounter{tutorial}{1}
\setcounter{ex}{0}
\setlength{\parindent}{0pt} 

\setcounter{equation}{0} 
\renewcommand{\theequation}{T1.\arabic{equation}} 

Some of the most common single-qubit gates are  Pauli matrices (X, Y, Z), so we need to be agile when manipulating them. That is why we start with a very basic exercise, but that will come in handy later on.\\

Let us first get everyone on the same page about tensor-product notation:
\be
\underbrace{P}_{\text{qb}1}\otimes \underbrace{Q}_{\text{qb}2} \longrightarrow (P\otimes Q) (R\otimes S) = PR\otimes QS .
\ee

\section{Pauli matrices}

We define Pauli matrices by $\sigma_a,\; a\in\{x, y, x\}$, where
\begin{equation}
X=\sigma_x=\mqty(0 & 1\\1 & 0), \quad Y=\sigma_y=\mqty(0 & -i\\i & 0),\quad Z=\sigma_z=\mqty(1 & 0\\0 & -1).
\end{equation}

\subsection{Properties}

From the definition, three important properties follow:
\begin{align}
&\bullet \sigma_a^2=\Id,\;\forall a\\
&\bullet \comm{\sigma_a}{\sigma_b} =  \sigma_a\sigma_b - \sigma_b\sigma_a = 2 i \varepsilon_{ab} \sigma_c,\quad
\varepsilon_{ab}=\begin{cases} +1 & (a,b) = \{(x,y),\; (y,z),\; (z,x)\}\\
-1 & (a,b) = \{(y,x),\; (z,y),\; (x,z)\}\\
0 & a=b
\end{cases}\\
&\bullet \acomm{\sigma_a}{\sigma_b} = \sigma_a\sigma_b + \sigma_b\sigma_a = 2 \delta_{ab} \Id,\qquad
\delta_{ab}=\begin{cases}1 & a=b\\ 0 & a\neq b \end{cases}
\end{align}
Remember that, from the definition of commutation ($\comm{}{}$) and anticommutation ($\acomm{}{}$), it follows that
\be
\comm{\sigma_a}{\sigma_b} = 0 \Leftrightarrow \sigma_a\sigma_b = \sigma_b\sigma_a
\qquad\qquad\qquad
\acomm{\sigma_a}{\sigma_b} = 0 \Leftrightarrow \sigma_a\sigma_b = -\sigma_b\sigma_a.
\ee

\begin{ex}
    Show that $[XXX, XYZ] = 0$ (here $XXX=X\otimes X\otimes X$).
\end{ex}
\begin{proof}[Solution]
    \begin{align}
        [XXX, XYZ] &= XX\otimes XY\otimes XZ - XX\otimes YX \otimes ZX \stackrel{\acomm{}{}}{=} \nn\\
        &= XX \otimes XY \otimes XZ - XX\otimes (-XY)\otimes(-XZ) = 0
    \end{align}
\end{proof}

Note that there is a trick to skip the proof that will come in very handy in the future:
\begin{trick}
    If the number of Pauli matrices that anticommute is even, the commutator is zero.
\end{trick}

\subsection{Pauli group}
The Pauli group for one qubit is generated by the Pauli matrices:
%
\be
\mathcal{P}_1 = \expval{X,Y,Z} = \{\pm\Id, \pm i\Id, \pm X, \pm i X, \pm Y, \pm iY, \pm Z, \pm i Z\} .
\ee
Similarly, the Pauli group for $n$ qubits is generated by the same operators above applied to each of the $n$ qubits:
\be
\mathcal{P}_n = \expval{X_1, \dots, X_n, Y_1, \dots, Y_n, Z_1, \dots, Z_n},
\ee
where $X_1 = X\otimes \overbrace{\Id \otimes \dots \otimes \Id}^{(n-1)}$, etc.

\section{Clifford group}

The Clifford group is the normalizer of the Pauli group, i.e., it transforms elements of the Pauli group into elements of the Pauli group via conjugation:
\be
C = \{U\;|\;UPU^\dagger \in \mathcal{P}_n \quad\forall P\in\mathcal{P}_n\} .
\ee
This means that if we know how the transformation acts on the generators of the Pauli group, we know how it acts on all the elements in the Pauli group. Take, for example, an element $W\in \mathcal{P}_n$ that is the multiplication of two generators of the Pauli group, $P,Q\in \mathcal{P}_n$, i.e., $W=PQ$. Then, since $U\in C$ is unitary, 
\be
U W U^\dagger = U(PQ)U^\dagger = (UPU^\dagger)(UQU^\dagger).
\ee
The Clifford group is generated by $C=\expval{\CNOT, H, S}$. So it is enough to see how these three gates transform the Pauli group. This is very useful, for instance, when calculating what a certain quantum circuit does.

\begin{ex}
How does CNOT transform $X_1$?
\end{ex}
Note that if you do this also with $S$ and $H$ and see the transformations for all the generators of the Pauli group, you obtain a characterization on how the Clifford group acts on the Pauli group.

\begin{proof}[Solution]
We could either solve it in matrix notation or Dirac notation, but let us go with the latter.
We begin by refreshing some definitions:
\begin{align}
&\CNOT = \ketbra{0} \otimes \Id + \ketbra{1}\otimes X\\
&\mqty{\Id = \ketbra{0} + \ketbra{1}\\
Z = \ketbra{0} - \ketbra{1}} \implies
\mqty{\ketbra{0} = (\Id + Z)/2\\ \ketbra{1} = (\Id-Z)/2}
\end{align}
Then,
\begin{align}
    \CNOT (X\otimes\Id) \CNOT^\dagger =&
    \underbrace{\left[\frac{(\Id + Z)}{2}\otimes \Id + \frac{(\Id - Z)}{2}\otimes X\right]
    (X\otimes \Id)} \CNOT = \nn\\[8pt]
    =& \left[\frac{(X + ZX)}{2}\otimes \Id + \frac{(X - ZX)}{2}\otimes X\right]\left[\frac{(\Id + Z)}{2}\otimes \Id + \frac{(\Id - Z)}{2}\otimes X\right] =\nn\\[8pt]
    =&\frac{1}{4}\underbrace{(X + ZX)(\Id + Z)}_{X+XZ+ZX+ZXZ}\otimes \Id \;+\; \frac{1}{4}\underbrace{(X + ZX)(\Id - Z)}_{X -XZ +ZX - ZXZ}\otimes X + \nn\\
    &+ \frac{1}{4}\underbrace{(X - ZX)(\Id + Z)}_{X+XZ-ZX-ZXZ}\otimes X \;+\; \frac{1}{4}\underbrace{(X - ZX)(\Id - Z)}_{X-XZ-ZX+ZXZ}\otimes \underbrace{X^2}_\Id \stackrel{\text{regroup}}{=}\nn\\[8pt]
    =& \frac{2X+2ZXZ}{4} \otimes \Id + \frac{2X-2ZXZ}{4} \otimes X \stackrel{ZXZ=-X}{=}\nn\\[8pt]
    =& X\otimes X.
\end{align}
Thus, CNOT: $X\otimes \Id \longrightarrow X\otimes X$.
\end{proof}
\begin{trick}
    Check out Table 1 in Ref.~\citep{Gottesman1999} to see how all the Clifford-group generators transform all the Pauli-group generators.
\end{trick}

\section{Stabilizers}

A stabilizer $U$ of a state $\ket{\psi}$ is a unitary operator that, when applied on the state $\ket{\psi}$, leaves the state unchanged, i.e.,
\be
U\ket{\psi} = \ket{\psi}.
\ee

Let us illustrate why they are useful through a couple of examples.

\begin{ex}
In the circuit below, find the state $\ket{\psi}$ using stabilizers.
\end{ex}

\begin{figure}[H]
\centering
\begin{quantikz}
\lstick{$\ket{0}$} & \gate{H}\slice{A} & \ctrl{1}\slice{B} & \rstick[wires=2]{$\ket{\psi}$}\\
\lstick{$\ket{0}$} & & \targ{} & 
\end{quantikz}
\end{figure}

Note that when we have a circuit like this, we always have two options when it comes to finding the final state:
\begin{enumerate}
    \item Tracking states:\\
    We calculate the state after each gate, i.e., $\ket{\psi} = \CNOT\left(H\ket{00}\right)$. This is practical for small circuits like this one, but when you have multiple ($n$) qubits and many gates, it becomes very tedious to calculate, as we have to keep track of $2^n$ amplitudes.
    \item Tracking stabilizers:\\
    We find the generators of the stabilizer group of the initial state. Then, we calculate how they are transformed after each gate. The final stabilizers determine the final state. This is what we do in this exercise, and it is very practical for larger circuits, as the number of generators scales linearly with the number of qubits.
\end{enumerate}

\begin{proof}[Solution]
We denote the initial state of the circuit by $\ket{00}$. The stabilizers of this state are
\be
\langle Z\otimes\Id, \;\Id\otimes Z \rangle = \langle Z_1, \,Z_2 \rangle = \{\Id,\, Z_1,\, Z_2,\, Z_1Z_2\},
\ee
because $Z$ leaves $\ket{0}$ unchanged and adds a $-$ to $\ket{1}$. Note that these stabilizers uniquely determine the state $\ket{00}$.\\

The Hadamard gate transforms our stabilizers, such that the state in step A is stabilized by $\langle X_1, \,Z_2 \rangle$. Similarly, the CNOT gate transforms them again. Using the result in Exercise 2, we obtain that the stabilizers in step B are $\langle X_1X_2, \, Z_1Z_2 \rangle$. The state stabilized by these is a Bell state:
\be
\ket{\psi} = \frac{1}{\sqrt{2}} (\ket{00} +\ket{11}),
\ee
which you can check by calculating $X_1X_2\ket{\psi}$ and $Z_1Z_2\ket{\psi}$.\\

To sum up:
\be
\ket{00}:\; \langle Z_1, \,Z_2 \rangle \;\stackrel{H_1}{\longrightarrow}\; \langle X_1, \,Z_2 \rangle \;\stackrel{\CNOT_{1\to2}}{\longrightarrow}\; \langle X_1X_2, \,Z_1Z_2 \rangle:\; \ket{\psi}
\ee
\end{proof}

What we saw here is a \textit{stabilizer circuit}, i.e., a circuit that only contains Clifford gates. Our circuit also had the initial state in the computational basis, so if we were to measure the final state also in the computational basis, this circuit would be efficiently simulatable on a classical computer. This is what the Gottesman--Knill theorem dictates.\\

But maybe you are wondering: \emph{Why? I do not see why it is efficiently simulatable}. Well, it has to do with what we have seen until now. With Pauli matrices and Clifford gates, we only need to track how the generators of the stabilizer groups are transformed, not the full group. The advantage lies in the fact that the number of generators scales linearly with the number of qubits, not exponentially.\\

While the Gottesman--Knill theorem implies that quantum computation is only more powerful than classical computation when it uses gates outside the Clifford group, Clifford operations are still important for applications in quantum-error correction (QEC) and quantum communication.
As we will see in the next Section, stabilizer codes use only Clifford operations for encoding and decoding, yet are extremely useful for overcoming the effects of errors and decoherence. Other important communication problems, such as quantum teleportation, also use only Clifford group gates and measurements.\\

Note: There are many states in the Hilbert space which are not \textit{stabilizer states}, i.e., they cannot be completely described by specifying a stabilizer from the Pauli group. In fact, there are only a finite number of stabilizer states of a given size. But as long as the input of a stabilizer circuit is a stabilizer state, the output of the circuit will also be a stabilizer state.

\newpage
\section{Quantum error correction (QEC)}

\subsection{Formalism}

We use $[[n, k, d]]$ to describe a QEC code (QECC), where $n$ is the number of physical qubits, $k$ is the number of encoded qubits, and $d$ is the distance of the code. We typically use double brackets for quantum codes and single brackets for classical codes.\\

The distance of a code, $d$, is the minimum weight of the logical operators. The weight is the number of qubits on which it acts nontrivially. For instance, let us take the 3-qubit code, [[3,1,1]], which you have seen in class (\secref{sec:QEC-3q-bit-flip-code}).
In this case, the logical 0 and 1 are
\be
\ket{\overline{0}} = \ket{000}, \quad \ket{\overline{1}} = \ket{111}.
\ee
The logical $X$ and $Z$ are defined such that
\be
\overline{X}\ket{\overline{0}} = \ket{\overline{1}}, \quad \overline{X}\ket{\overline{1}} = \ket{\overline{0}}, \qquad \overline{Z}\ket{\overline{0}} = \ket{\overline{0}},  \quad \overline{Z}\ket{\overline{1}} = -\ket{\overline{1}}.
\ee
Therefore, we can define $\overline{X}=X_1X_2X_3$. The weight of this operator is 3, as it acts nontrivially on three different qubits. However, we said that the distance of this code is 1, so there must be another logical representation that has weight 1. Indeed, $\overline{Z}=Z_1$.\\
The problem with the logical operators not being uniquely defined ($\overline{Z}=Z_1$, but also $\overline{Z}=Z_2$, or $\overline{Z}=Z_3$), is that the circuit confuses errors with logical operations. That is why the 3-qubit code cannot detect nor correct all types of errors, only bit flips, which are $X$-type errors.\\

So when can a code detect or correct random errors?\\
A stabilizer code with distance $d$ can detect $d-1$ errors.
Any QECC can correct $t$ unknown errors, or up to $2t$ erasure errors if the location of the erased qubit is known. To be able to correct, the distance has to have distance $d\ge 2t+1$. Therefore, to correct a random error, we need a minimum distance of $d=3$. Codes that attain the Hamming bound ($d=2t+1$) are called perfect codes. \\

\subsection{The 9-qubit Shor code}

The code seen in class to correct a random error (the 9-qubit Shor code, \secref{sec:9qbShor}) is [[9,1,3]].
This code is a concatenation of the bit-flip code (left) and the phase-flip code (right):
\begin{figure}[H]
\centering
\begin{quantikz}
\lstick{$\ket{\psi}$} & \ctrl{1} & \ctrl{2} & \\
\lstick{$\ket{0}$} & \targ{} & &\\
\lstick{$\ket{0}$} &  & \targ{} & 
\end{quantikz}
\qquad
\begin{quantikz}
\lstick{$\ket{\psi}$} & \ctrl{1} & \ctrl{2} & \gate{H} &\\
\lstick{$\ket{0}$} & \targ{} &  & \gate{H} & \\
\lstick{$\ket{0}$} & & \targ{} & \gate{H} &
\end{quantikz}
\end{figure}

So the encoding in Shor's code looks like this:
\begin{figure}[H]
\centering
\begin{quantikz}
\lstick{$\ket{\psi}$} & \ctrl{3} & \ctrl{6} & \gate{H} & & \ctrl{1} & \ctrl{2} & \\
\wireoverride{n} & \wireoverride{n} & \wireoverride{n} &\wireoverride{n} & \lstick{$\ket{0}$}\wireoverride{n} & \targ{} & & \\
\wireoverride{n} & \wireoverride{n} & \wireoverride{n} & \wireoverride{n} &\lstick{$\ket{0}$}\wireoverride{n} & & \targ{} &\\
\lstick{$\ket{0}$} & \targ{} & & \gate{H} & &  \ctrl{1} & \ctrl{2} & \\
\wireoverride{n} & \wireoverride{n} & \wireoverride{n} & \wireoverride{n} & \lstick{$\ket{0}$}\wireoverride{n} & \targ{} & & \\
\wireoverride{n} & \wireoverride{n} & \wireoverride{n} & \wireoverride{n} &\lstick{$\ket{0}$}\wireoverride{n} & & \targ{} & \\
\lstick{$\ket{0}$} & & \targ{} & \gate{H} & &  \ctrl{1} & \ctrl{2} & \\
\wireoverride{n} & \wireoverride{n} & \wireoverride{n} & \wireoverride{n} & \lstick{$\ket{0}$}\wireoverride{n} & \targ{} & & \\
\wireoverride{n} & \wireoverride{n} & \wireoverride{n} & \wireoverride{n} & \lstick{$\ket{0}$}\wireoverride{n} & & \targ{} &
\end{quantikz}
\end{figure}
where the phase-flip encoding maps
\be
\ket{0}\to\ket{+++}, \qquad \ket{1}\to\ket{---}
\ee
and the bit-flip encoding maps each of those $\pm$ to
\be
\ket{+} \to \frac{\ket{000}+\ket{111}}{\sqrt{2}},\qquad 
\ket{-} \to \frac{\ket{000}-\ket{111}}{\sqrt{2}}.
\ee
Therefore, the logical $\ket{0}$ and $\ket{1}$ read
\be
\ket{\overline{0}} =  \frac{\left(\ket{000}+\ket{111}\right)^{\otimes 3}}{2\sqrt{2}}, \qquad 
\ket{\overline{1}} = \frac{\left(\ket{000}-\ket{111}\right)^{\otimes 3}}{2\sqrt{2}}.
\ee
By definition, the logical $Z$ can be written as
\be
\overline{Z} = X_1X_2X_3X_4X_5X_6X_7X_8X_9,
\ee
since it satisfies $\overline{Z}\ket{\overline{0}}=\ket{\overline{0}}$ and $\overline{Z}\ket{\overline{1}}=-\ket{\overline{1}}$.\\

Now, in class you learned stabilizers after these error-correction codes, so maybe you have not made the connection yet. In the lectures, it was shown that in Shor's code, you do parity checks with these 8 operators:
$$
Z_1Z_2, \quad Z_2Z_3, \quad Z_4Z_5, \quad Z_5Z_6, \quad Z_7Z_8, \quad Z_8Z_9, \quad X_1X_2X_3X_4X_5X_6, \quad X_4X_5X_6X_7X_8X_9.
$$
These are, in fact, the generators of the stabilizer group for this code. Note that it is not random that there are 8 generators: in a stabilizer code $[[n,k,d]]$, the stabilizer group is generated by $n-k$ operators.\\

Let us circle back to $\overline{Z}$. We know that Shor's code has distance $d=3$, so there must be a definition of $\overline{Z}$ that is different than the one provided above that has weight 3 instead of 9. Indeed, logical gates that differ by a stabilizer are equivalent:
\be
\overline{Z}\cdot X_4X_5X_6X_7X_8X_9 \stackrel{X^2=\Id}{=} X_1X_2X_3 \equiv \overline{Z}.
\ee

\renewcommand{\thesection}{\thechapter.\arabic{section}} 
\setlength{\parindent}{15pt} 
\renewcommand{\theequation}{\thechapter.\arabic{equation}} 

\chapter{Fast quantum algorithms}

Having reviewed the basics of quantum computing, seeing that it should be possible to live with some errors and limited universal gates sets, and having gotten a first taste of quantum speed-ups in Grover's algorithm, we are now ready to meet some really fast quantum algorithms. In this chapter, we first learn about a quantum version of the Fourier transform, then study an algorithm called phase estimation, and, finally, Shor's algorithm. For this chapter, we follow Ref.~\citep{NielsenChuang}.

\section{The quantum Fourier transform}
\label{sec:QFT}

You should all have seen the discrete Fourier transform (DFT), where a set of $N$ complex numbers $\{x_0, \dots ,x_{N-1}\}$ are transformed into $N$ new complex numbers $\{y_0, \dots ,y_{N-1}\}$ according to
\be
y_k=\frac{1}{\sqrt{N}}\sum_{m=0}^{N-1} e^{i \frac{2\pi m k}{N}} x_m .
\ee
The Fourier transformation is extremely useful, e.g., to detect periods in a signal where $\{x_0, \dots ,x_{N-1}\}$ could be the amplitude of some signal as a function of discretized time. The Fourier-transformed signal $\{y_0, \dots ,y_{N-1}\}$ then describes the frequency content. Solving the Schr{\"o}dinger equation on a lattice, DFT is what takes you between real space and momentum ($k$) space. 

The {\em quantum} Fourier transform (QFT) is a unitary $n$-qubit operation, transforming the initial $N=2^n$ basis states $\{|0\rangle, \dots ,|N-1\rangle\}$ into a new basis in a way which looks mathematically identical to the DFT,
\begin{equation}
\label{QFTdefEq}
|j\rangle \rightarrow \frac{1}{\sqrt{N}}\sum_{k=0}^{N-1} e^{i
\frac{2\pi j k}{N}} |k\rangle.
\end{equation}
The action on an arbitrary state is
\be
\sum_{j=0}^{N-1} x_j |j\rangle \rightarrow \sum_{k=0}^{N-1} y_k
|k\rangle,
\ee
where the amplitudes $y_k$ are the DFT transforms of the amplitudes $x_m$. One may easily verify that the new states are normalized and form an orthogonal set, and thus that the QFT is a unitary transform.

The QFT can be used to find periods and also to extract eigenvalues of unitary operators to a high precision. But before discussing these issues in more detail, let us see if we can find an effective implementation of the QFT. Remember that there are operators that need exponentially many single- and two-qubit gates for implementation, so what about the QFT?

\subsection{Another definition}

We will now rewrite the definition of the QFT in a way that is more transparent for constructing a circuit. First we need a simple way to number the basis states. We number the $n$-qubit state $|j\rangle$ using the binary $n$-bit representation of
$j=j_1 2^{n-1}+j_2 2^{n-2} + \dots + j_{n-1} 2^1 + j_n 2^0$. 
For example, in the 4-qubit case, the state
$|5\rangle=|0101\rangle=|0_1\rangle|1_2\rangle|0_3\rangle|1_4\rangle$.

\ssse{Alternative notation: the binary fraction}

Before starting, we introduce a notation that is used in Ref.~\citep{NielsenChuang} to perform calculations on the QFT and phase-estimation algorithms. However, in these notes, we will make use of both notations --- the standard one and the one based on the binary fraction --- leaving to the reader the choice of which one is preferable.

The definition of binary fractions is
\be
0.j_1j_2j_3 \dots j_n=j_1/2+j_2/2^2+j_3/2^3\dots+j_n/2^n,
\ee%
e.g., $0.101=0.5+0.125=0.625$, and more generally
\be
0.j_l j_{l+1}  \dots j_m = j_l /2 +j_{l+1}/2^{2}+ \dots+j_m/2^{m-l+1}.
\ee
Using this notation we can write the QFT in \eqref{QFTdefEq} as
\begin{equation}
|j\rangle = |j_1,j_2,\dots,j_n\rangle \rightarrow
\frac{\mleft( |0\rangle+e^{i 2\pi 0.j_n}|1\rangle \mright) \mleft(|0\rangle + e^{i 2\pi 0.j_{n-1}j_n}|1\rangle \mright) \dots \mleft( |0\rangle + e^{i 2\pi 0.j_1j_2 \dots j_n}|1\rangle \mright)}{2^{n/2}} .
\label{QFTproductFormEq}
\end{equation}
%


\ssse{Rewriting the output state of the quantum Fourier transform}

The algebraic manipulations connecting the two expressions are straightforward, but need some afterthought. Observing that
\be
\frac{k}{2^n} = \frac{k_1 2^{n-1}}{2^n} + \dots +  \frac{k_n 2^{0}}{2^n} = k_1 2^{-1} + \dots +  k_n 2^{-n} = \sum_{l=0}^{n} k_l 2^{-l},
\ee
from Eq.~(\ref{QFTdefEq}) we have
\begin{eqnarray}
\label{eq:steps-TF}
|j\rangle & \rightarrow & \frac{1}{\sqrt{N}}\sum_{k=0}^{N-1} e^{i \frac{2\pi j k}{N}} |k\rangle \nn \\
& = & \frac{1}{\sqrt{N}} \sum_{k_1=0}^{1} \dots \sum_{k_n=0}^{1} e^{i 2\pi j \sum_{l=1}^n k_l 2^{-l}} |k_1 \dots k_n \rangle \nn \\
& = & \frac{1}{\sqrt{N}} \sum_{k_1=0}^{1} \dots \sum_{k_n=0}^{1} \bigotimes_{l=1}^{n} e^{i 2\pi j  k_l 2^{-l}} |k_l  \rangle \nn \\ 
& = &  \frac{1}{\sqrt{N}} \bigotimes_{l=1}^{n}  \sum_{k_l=0}^{1} e^{i 2\pi j  k_l 2^{-l}} |k_l  \rangle \nn \\
& = &  \frac{1}{\sqrt{N}} \bigotimes_{l=1}^{n} ( |0 \rangle + e^{i 2 \pi j   2^{-l }} |1 \rangle ) \nn \\
& = &  \frac{1}{\sqrt{N}} (|0 \rangle + e^{i 2 \pi j   2^{-1 }} |1 \rangle)  (|0 \rangle + e^{i 2 \pi j 2^{-2 }} |1 \rangle) \dots (|0 \rangle + e^{i 2 \pi j   2^{-n }} |1 \rangle) \nn \\ 
& = &  \frac{1}{\sqrt{N}} (|0 \rangle + e^{i 2 \pi 0.j_n} |1 \rangle)  (|0 \rangle + e^{i 2 \pi 0.j_{n-1} j_{n}} |1 \rangle)\dots (|0 \rangle + e^{i 2 \pi   0.j_1...j_n } |1 \rangle) ,
\end{eqnarray}
where in the last step we have used that
\begin{eqnarray}
j 2^{-n} & =&   j_1 / 2 +  j_2 / 4 + \dots j_n / 2^{n} = 0.j_1 \dots j_n \nn \\
& & ...  \\
 j 2^{-1} & = &  j_1 2^{n-2}  + j_2 2^{n-3}+ \dots j_n / 2  =  j_1 2^{n-2}    + j_2 2^{n-3} +  \dots + 0.j_n \nn
\end{eqnarray}
and that the integer part of $j \cdot 2^l$ disappears in the exponent since it is multiplied by $2 \pi$.

\subsection{An efficient implementation}

Using the form of the QFT in Eq.~(\ref{QFTproductFormEq}), it is straightforward to implement the desired transformation with a quantum circuit. We realize that we have to implement conditional phase shifts on each
qubit; therefore we define the single-qubit operator
\be
R_k = 
\begin{bmatrix}
1 & 0 \\
0 & e^{i2\pi/2^k}
\end{bmatrix}.
\ee
The controlled-$R_k$ gate can be decomposed in terms of single qubit gates and CNOT gates (Exercise \ref{ex:TF}).

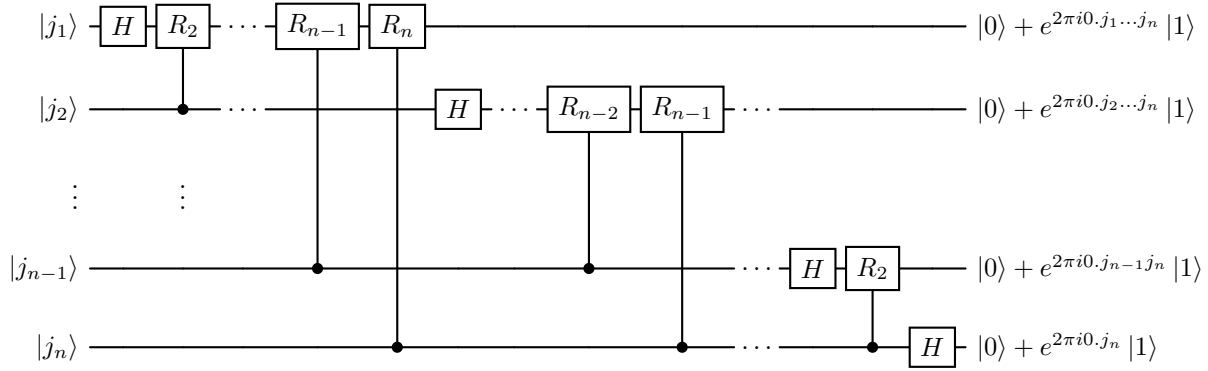
\begin{figure}
\begin{quantikz}[column sep = 4pt]
    \lstick{$\ket{j_1}$} & \gate{H} & \gate{R_2} & \push{\;\dots\;} & \gate{R_{n-1}} & \gate{R_n} & & & & & & & & & \rstick{$\ket{0}+e^{2\pi i 0.j_{1}\dots j_{n}}\ket{1}$}\\
    \lstick{$\ket{j_2}$} & & \ctrl{-1} & \push{\;\dots\;} & & & \gate{H} & \push{\;\dots\;} & \gate{R_{n-2}} & \gate{R_{n-1}} & \push{\;\dots\;} & & & & \rstick{$\ket{0}+e^{2\pi i 0.j_{2}\dots j_{n}}\ket{1}$} \\
    \lstick{\vdots}\setwiretype{n} & & \push{\vdots} & & & & & & &\\
    \lstick{$\ket{j_{n-1}}$} & & & & \ctrl{-3} & & & & \ctrl{-2} & & \push{\;\dots\;} & \gate{H} & \gate{R_2} & &\rstick{$\ket{0}+e^{2\pi i 0.j_{n-1}j_{n}}\ket{1}$}\\
    \lstick{$\ket{j_n}$} & & & & & \ctrl{-4} & & & & \ctrl{-3} & \push{\;\dots\;} & & \ctrl{-1} & \gate{H} & \rstick{$\ket{0}+e^{2\pi i 0.j_{n}}\ket{1}$}
\end{quantikz}
\caption{An efficient circuit to perform the quantum Fourier transform, from Ref.~\citep{NielsenChuang} (Fig. 5.1 therein). Here $0.j_1...j_n = j 2^{-n}, 0.j_2...j_n = j 2^{-(n-1)}, ..., 0.j_{n-1}j_n = j/2^{-2}$  and $0.j_n = j/2^{-1}$.}
\label{QFTcircuitFig}
\end{figure}

Now let us see what happens to an input state $|j_1, j_2, \dots, j_n\rangle$ when it passes through the circuit in \figref{QFTcircuitFig}. The first Hadamard gate produces the state $\mleft( |0\rangle + |1\rangle \mright) / \sqrt{2}$ if $j_1 = 0$ and $\mleft( |0\rangle - |1\rangle \mright) / \sqrt{2}$ if $j_1 = 1$, i.e.,
\be
\frac{1}{\sqrt{2}} \mleft( |0\rangle + e^{i 2\pi j_1/2} |1\rangle \mright) |j_2, \dots, j_n\rangle = \frac{1}{\sqrt{2}} \mleft( |0\rangle + e^{i 2\pi 0.j_1}|1\rangle \mright) |j_2, \dots, j_n\rangle,
\ee
since $e^{i 2\pi 0.j_1} = e^{i 2\pi j_1/2} = -1$ for $j_1=1$ and $+1$ otherwise. The controlled-$R_2$ gate rotates the component $|1\rangle$ of the first qubit by $e^{i 2 \pi /2^2}$ if $j_2 = 1$, i.e., it applies the phase $e^{i2\pi j_2/2^2} $. Therefore, it produces the state
\be
\frac{1}{\sqrt{2}} \mleft( |0\rangle + e^{i 2\pi j_2/ 2^2} e^{i 2\pi j_1/ 2}|1\rangle \mright) |j_2, \dots, j_n\rangle =
\frac{1}{\sqrt{2}} \mleft( |0\rangle + e^{i 2\pi 0.j_1j_2}|1\rangle \mright) |j_2, \dots, j_n\rangle.
\ee
After all the controlled-$R_k$ operations on the first qubit, the
state is
\be
\frac{1}{\sqrt{2}} \mleft( |0\rangle + e^{i 2 \pi \lt \fr{j_1}{2} + ... + \fr{j_n}{2^n} \rt}|1\rangle \mright) |j_2, \dots, j_n\rangle = \frac{1}{\sqrt{2}} \mleft(|0\rangle + e^{i 2\pi 0.j_1j_2 \dots j_n}|1\rangle \mright) |j_2, \dots, j_n\rangle.
\ee
The Hadamard on the second qubit produces
\ba
& \frac{1}{\sqrt{2^2}} \mleft( |0\rangle + e^{i 2 \pi \lt \fr{j_1}{2} + ... + \fr{j_n}{2^n} \rt}|1\rangle \mright) \mleft(|0\rangle + e^{i 2\pi \fr{j_2}{2}}|1\rangle \mright) |j_3, \dots, j_n\rangle \nn \\
& = \frac{1}{\sqrt{2^2}} \mleft( |0\rangle + e^{i 2\pi 0.j_1j_2 \dots j_n}|1\rangle \mright) \mleft( |0\rangle + e^{i 2\pi 0.j_2}|1\rangle \mright) |j_3, \dots, j_n\rangle,
\ea
and the controlled-$R_2$ to -$R_{n-1}$ gates yield the state
\ba
& \frac{1}{\sqrt{2^2}} \mleft( |0\rangle + e^{i 2 \pi \lt \fr{j_1}{2} + ... + \fr{j_n}{2^n} \rt}|1\rangle \mright) \mleft( |0\rangle + e^{i 2\pi \lt \fr{j_2}{2} + ... + \fr{j_n}{2^{n-1}} \rt}|1\rangle \mright)|j_3, \dots, j_n\rangle \nn \\
& = \frac{1}{\sqrt{2^2}} \mleft( |0\rangle + e^{i 2\pi 0.j_1j_2 \dots j_n}|1\rangle \mright) \mleft( |0\rangle + e^{i 2\pi 0.j_2...j_n}|1\rangle \mright) |j_3, \dots, j_n\rangle.
\ea
We continue in this fashion for all qubits, obtaining the final state
\ba
& \frac{1}{\sqrt{2^n}} \mleft( |0\rangle + e^{i 2 \pi j 2^{-n} }|1\rangle \mright) \mleft( |0\rangle + e^{i 2\pi j 2^{-(n-1)} }|1\rangle \mright) ... \mleft( |0\rangle + e^{i 2\pi j 2^{-1} }|1\rangle \mright) \nn \\
& = \frac{1}{\sqrt{2^n}} \mleft( |0\rangle + e^{i 2\pi 0.j_1j_2 \dots j_n}|1\rangle \mright) \mleft( |0\rangle + e^{i 2\pi 0.j_2...j_n}|1\rangle \mright) ... \mleft( |0\rangle + e^{i 2\pi 0.j_n}|1\rangle \mright).
\ea

We now need to reverse the order of the qubits, which can be achieved using a series of SWAP gates. The number of gates needed are $n$ on the first qubit, $n-1$ on the second qubit, and so on, adding up to $n(n+1)/2=O(n^2)$ gates. Then we need on the order of $n$ SWAP gates, not changing the scaling. Thus we can implement the QFT for $n$ qubits using on the order of $O(n^2)$ gates. The best classical algorithm (FFT) needs $O(n2^n)$ gates, indicating why the QFT could be used for speedup. This does not translate in an immediate speed-up for computing classical FFT, because we cannot access the amplitudes when measuring the Fourier-transformed quantum state, and we do not even know how to efficiently prepare the input state to be transformed. However, in the next
section, we will see one problem where the quantum Fourier transform is useful.

\section{Phase estimation}
\label{sec:PhaseEstimation}

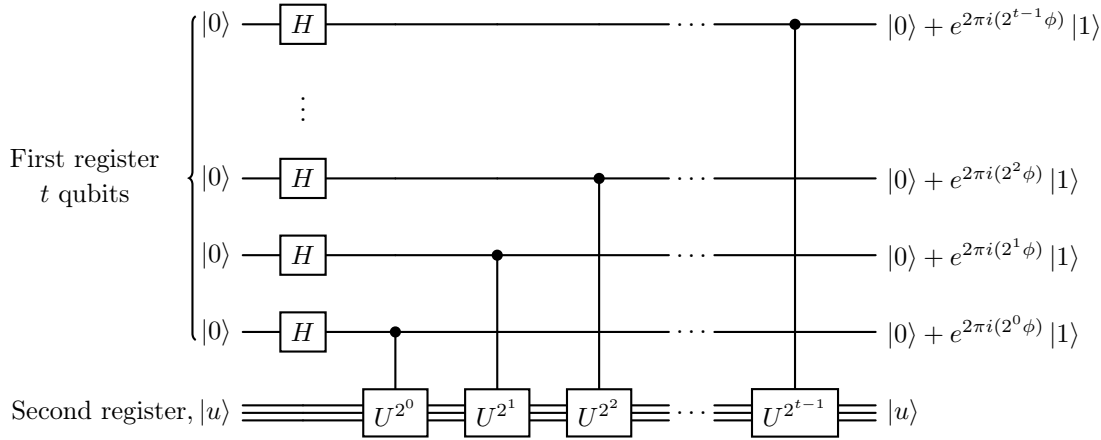
\begin{figure}[!ht]
\begin{quantikz}[wire types = {q, n, q, q, q, b}, classical gap = 0.1cm]
    \lstick[5]{\parbox{2.5cm}{\centering First register\\$t$ qubits}} & \lstick{$\ket{0}$}\wireoverride{n} & \gate{H} & & & & \push{\;\dots\;} & \ctrl{5} & \rstick{$\ket{0}+e^{2\pi i (2^{t-1}\phi)}\ket{1}$}\\
    & & \push{\vdots} & & & & \\
    & \lstick{$\ket{0}$}\wireoverride{n} & \gate{H} & & & \ctrl{3} & \push{\;\dots\;} & & \rstick{$\ket{0}+e^{2\pi i (2^{2}\phi)}\ket{1}$}\\
    & \lstick{$\ket{0}$}\wireoverride{n} & \gate{H} & & \ctrl{2}& & \push{\;\dots\;} & & \rstick{$\ket{0}+e^{2\pi i (2^{1}\phi)}\ket{1}$}\\
    & \lstick{$\ket{0}$}\wireoverride{n} & \gate{H} & \ctrl{1} & & & \push{\;\dots\;} & & \rstick{$\ket{0}+e^{2\pi i (2^{0}\phi)}\ket{1}$}\\
    \lstick[1]{Second register,} & \lstick{$\ket{u}$}\wireoverride{n} & & \gate{U^{2^0}} & \gate{U^{2^1}} & \gate{U^{2^2}} & \push{\;\dots\;} & \gate{U^{2^{t-1}}} & \rstick{$\ket{u}$}
\end{quantikz}
\caption{A circuit performing the first step of the phase estimation algorithm, from Ref.~\citep{NielsenChuang} (Fig. 5.2 therein).}
\label{PhaseEstCircuitDetailedFig}
\end{figure}

\begin{figure}[!ht]
\centering
\begin{quantikz}
    \lstick{$\ket{0}$} & \qwbundle{t} & \gate{H}\wire[r][1]["\ket{j}"{above,pos=0.3}]{a} & \ctrl{1} & \gate{FT^\dagger} & \meter{} \\
    \lstick{$\ket{u}$} & \qwbundle{} & & \gate{U^j} & & \rstick{$\ket{u}$}
\end{quantikz}
\caption{An overview circuit figure of the phase estimation circuit, from Ref.~\citep{NielsenChuang} (Fig. 5.3 therein).}
\label{PhaseEstCircuitOverviewFig}
\end{figure}
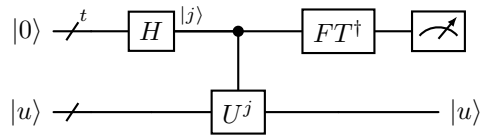

The aim of this algorithm is to estimate the eigenvalue $\lambda$ corresponding to an eigenvector $|u\rangle$ of a unitary operator $U$. Since the matrix $U$ is unitary, the eigenvalue can be expressed as $e^{i 2 \pi \phi}$ (Exercise \ref{ex:eigenvalue-unitary}). The vector $|u\rangle$ is given, as well as a circuit (black box, oracle) effectively implementing controlled-$U^n$ operations. A circuit solving a first stage of this problem is shown in \figref{PhaseEstCircuitDetailedFig} and an overview circuit showing the whole algorithm is given in \figref{PhaseEstCircuitOverviewFig}. Two qubit registers are needed; the first has $t$ qubits, which are initialized to zero. The number of qubits $t$ is determined by the required accuracy in the estimate of $\phi$. The second register is large enough to represent the eigenvector $|u\rangle$, and it is also initialized to $|u\rangle$ and remains in this state throughout the computation.

The initial set of Hadamard gates puts all qubits of register 1 in an equal superposition of $|0\rangle$ and $|1\rangle$. If the $k$th control qubit is in the state $|1\rangle$, a unitary operation $U^{2^k}$ will be performed on the second register, picking up a phase
$\mleft(e^{i 2 \pi \phi}\mright)^{2^k} = e^{i 2 \pi \phi {2^k}}$. For example, the first step ($k = 0$) gives
\begin{equation}
C_U \frac{1}{\sqrt{2}} \mleft( |0\rangle + |1\rangle \mright) |u\rangle = 
\frac{1}{\sqrt{2}} \mleft( |0\rangle + e^{i 2\pi \phi} |1\rangle \mright) |u\rangle.
\end{equation}
The final state of the first register in this first step is
\begin{equation}
\frac{1}{\sqrt{2^t}}
\mleft(|0\rangle + e^{i 2\pi 2^{t-1} \phi} |1\rangle \mright)
\mleft(|0\rangle + e^{i 2\pi 2^{t-2} \phi} |1\rangle \mright) \dots
\mleft(|0\rangle + e^{i 2\pi 2^{1} \phi} |1\rangle \mright)
\mleft(|0\rangle + e^{i 2\pi 2^{0} \phi} |1\rangle \mright) =
\frac{1}{\sqrt{2^t}} \sum_{k=0}^{2^t-1} e^{i 2\pi k \phi} |k\rangle.
\label{PhaseEstMidStateEq}
\end{equation}
By comparison with \eqref{eq:steps-TF}, we see that this state is nothing else than the Fourier transform of the state $|2^t \phi \rangle = | \phi_1 \phi_2\dots \phi_t \rangle$, where in the last step we have assumed that the phase $\phi$ has an exact representation in $t$ bits as $\phi = 0.\phi_1\phi_2\dots\phi_t$ (with a slight abuse of notation). The final step is hence to make an inverse quantum Fourier transform of the first register (Exercise \ref{ex:FT-inverse}). This allows recovering the latter state. Register 1 is then read out and $\phi$ is recovered.

If the phase is not an exact binary fraction in $t$ qubits, there will be some finite probability of reading out some other state ``close'' to the best approximation. A careful analysis gives that if we want $n$ bits of precision, with a success probability of at least $1 - \epsilon$, we need a register of size
\begin{equation}
\label{bits:phase-estimation}
t = n + \mleft\lceil \log{\mleft( 2 + \frac{1}{2\epsilon} \mright)} \mright\rceil.
\end{equation}

Note that if we are not able to prepare an eigenstate $\ket{u}$ of $U$, it is still possible to do phase estimation. Suppose that the input state is not an eigenstate $\ket{u}$, but a state $\ket{\psi} = \sum_u c_u \ket{u}$, where each component $\ket{u}$ has eigenvalue $2 \pi i \phi_u$. The result of running the phase-estimation algorithm will be an output state close to $\sum_u c_u \ket{\tilde{\phi}_u} \ket{u}$, where $\tilde{\phi}_u$ is a good approximation of the phase $\phi_u$. Therefore, reading out the first register will give a good approximation of $\tilde{\phi}_u$, where $u$ is chosen at random with probability $|c_u|^2$. This procedure allows one to avoid preparing a (possibly unknown) eigenstate, at the cost of introducing some additional randomness into the algorithm. 

Phase estimation is interesting in its own right. However, it is also a primitive to other algorithms, such as those for solving linear systems of equations in linear algebra (the HHL algorithm, which will be discussed in \chpref{chp:QML}), and those minimizing the number of features required in machine-learning applications (principal component analysis; again, see \chpref{chp:QML}). We will now see how it enters Shor's algorithm for factoring.

\section{Factoring -- Shor's algorithm}

We now know how to efficiently determine the phase of an eigenvalue to a unitary operator. In this section we will see how this enables us to efficiently solve a number-theoretical problem which is considered hard on classical computers: order finding. Finally, we show how factoring can be reduced to order finding.

\subsection{Modular arithmetics}

Order finding is defined in {\em modular arithmetics}. Modular arithmetics is based on the fact that, given any two positive integers $x$ and $N$, $x$ can be uniquely written as
\be
x = k \cdot N + q,
\ee
where $k$ is a non-negative integer and $0 \leq q < N$ is the remainder
\be
x = q \mod{N}.
\ee
As an example,
\be
2 = 5 = 8 = 11 \mod{3}.
\ee

The {\em greatest common divisor} $gcd(a,b)$ of two integers $a$ and $b$ is the largest integer dividing both $a$ and $b$. If $gcd(a,b) = 1$, then $a$ and $b$ are called {\em co-prime}.

\subsubsection{Multiplicative inverse}

Now let us look at modular multiplication by considering the series
\be
m_k = k \cdot x \mod{N}, \ 0 < k < N.
\ee
As an example, take $x=6$ and $N=15$ with $gcd(x,N)=3$, giving
\be
m_k = \{6, 12,  3,  9,  0,  6, 12,  3,  9,  0, 6, 12, 3, 9 \},
\ee
showing that the equation $k \cdot 6 = y \mod{15}$ has no solution for $y \in \{ 1, 2, 4, 5, 7, 8, 10, 11, 13, 14 \}$. Note that, in particular, it has no solution for $y = 1$. Then take $x = 7$ and $N = 15$, which are co-prime, giving
\be
m_k = \{ 7, 14, 6, 13, 5, 12, 4, 11, 3, 10, 2, 9, 1, 8 \},
\ee
showing that the equation $k \cdot 7 = y \mod{15}$ may be solved for all $y$.

The {\em multiplicative inverse} $x^{-1}$ of an integer $x$ modulus $N$ is another integer, which fulfils
\be
x^{-1} \cdot x = 1 \mod{N},
\ee
and it exists if and only if $x$ and $N$ are co-prime. If the inverse exists, we can solve the equation
\be
k \cdot x = y \mod{N}
\ee
for all integers $y$ through
\be
k = y \cdot x^{-1} \mod{N}.
\ee

Another way of formulating this is the following: all the integers between 1 and $(N-1)$ appear once and only once in $\{m_k\}$ if and only if $x$ and $N$ are co-prime. If not, we can write $x = k \cdot gcd(x,N)$ and $N = y \cdot gcd(x,N)$, where $0 < y < N$. So we obtain
\be
m_y = y \cdot k \cdot gcd(x,N) \mod{(y\cdot gcd(x,N))} = 0,
\ee
and then the series $\{m_k\}$ will just repeat from the start. In Exercises \ref{ex:mod1} and \ref{ex:mod2}, you get to show that the modular inverse of $7 \Mod{15}$ is 13, and that the solution of the equation $k \cdot 7 = 4 \mod{15}$ is $k = 7$.

\subsection{Order finding}

Consider the equation
\be
x^r = 1 \mod{N},
\ee
which has solutions for the integers $x$ and $N$ being co-prime, and $x < N$. The lowest positive integer $r$ solving the equation is called the {\em order} of $x$ modulo $N$. One straightforward method to calculate $r$ is to evaluate the series $m_k = x^k \mod{N}$ for $0 < k < N$. Then it is clear that the series $\{m_k\}$ is periodic with period $r$ since $x^{r+a} = x^r \cdot x^a = x^a \mod N$. In other words, the order is the period of the modular exponentiation function $m_k = x^k \mod{N}$. As an example let us take $x=5$ and $N=21$, giving
\be
m_k = \{ 5, 4, 20, 16, 17, 1, 5, 4, 20, 16, 17, 1, 5, 4, 20, 16, 17, 1, 5, 4\},
\ee
and we have the order $r=6$. There is no classical algorithm for finding $r$ which scales polynomially in the number of bits $L$ needed to represent the input, i.e., the integers $x$ and $N$.

\subsection{Factoring as order finding}

Factoring can be reduced to order finding as follows.
Suppose we want to factor $N = p q$. Consider the period $r$ of the function of $k$ defined as $x^k \mod N$ for some $x$. $x$ is chosen such that $r$ is even and $x^{r/2} \neq N - 1 \mod N$. Then $r$ allows us to find $p$ and $q$ as follows. Define $y = x^{r/2}$. Then $y^2 = x^r = 1 \mod N$ by the definition of the period $r$. Therefore we have
\be
y^2 - 1 = (y - 1 )  (y + 1 ) = 0 \mod N.
\ee
Therefore $(y - 1) (y + 1)$ is a multiple of $N$, i.e., it contains the two factors $p q$ in its decomposition. However, neither $(y - 1)$ nor $(y + 1)$ are multiples of $N$:
\begin{align}  
(y - 1 ) & \neq 0 \mod N, \mbox{ because otherwise the period would be smaller than r. }\\
(y + 1 ) & \neq 0 \mod N, \mbox{ by construction}.
\end{align}
Therefore $(y - 1)$ and $(y + 1)$ must split the two factors $q$ and $p$ appearing in the decomposition of $N$. Say, for instance,
\be
(y-1) = l p; \hspace{0.5cm} (y+1) = l' q.
\ee
We therefore finally obtain
\be
p = gcd (N, x^{r/2} - 1); \hspace{0.5cm} q = gcd (N, x^{r/2} + 1).
\ee
In other words, determining the order $r$ of the modular exponentiation function $x^k \mod N$ yields the determination of the two factors $p$ and $q$ such that $N = pq$. For $N$ having $L$ bits, this common factor can be found using Euclid's algorithm in $O(L^3)$ steps.
For uniformly chosen $x$, such that $0 < x < N$ and $x$ and $N$ are co-prime, one may calculate a lower bound for the probability of $r$ being even and that $y=x^{r/2}$ is non-trivial,
\be
p(r\ {\rm is\ even\ and }\ x^{r/2}\neq-1 \Mod{N} ) \geq 1-\frac{1}{2^m},
\ee
where $m$ is the number of different prime factors in $N$, i.e., $m \geq 1$. 

In the following, we are going to derive an efficient quantum algorithm for order finding. 

\subsection{A quantum algorithm for order finding}

Suppose that the integer $N$ that we want to factor consists of $L$ digits.
Given an integer $x$, for which we want to find the order mod $N$, consider the $L$-qubit unitary operation
\be
U|y\rangle \equiv 
\mleft\{ 
\begin{array}{ll}
|x\cdot y \Mod{N}\rangle, & 0 \leq y \leq N-1 \\
|y\rangle, & N \leq y \leq 2^L-1 
\end{array} 
\mright. 
.
\ee
The unitarity follows since $U$ basically permutes the basis states and $x$ has a multiplicative inverse modulus $N$ since $x$ and $N$ are co-prime. The states
\be
\ket{u_s} = \frac{1}{\sqrt{r}}\sum_{k=0}^{r-1} \exp{\mleft[ \frac{-i 2\pi s k}{r} \mright]} \ket{x^k \Mod{N}},
\ee
defined for integers $0 \leq s \leq r-1$, are eigenstates of $U$, since
\begin{align}
U \ket{u_s} 
&=\frac{1}{\sqrt{r}}\sum_{k=0}^{r-1} \exp{\mleft[\frac{-i 2\pi s k}{r} \mright]} \ket{x^{k+1} \mod{N}} = \nn \\
&= \frac{1}{\sqrt{r}}\sum_{k=1}^{r} \exp{\mleft[ \frac{-i 2\pi s
(k-1)}{r} \mright]} \ket{x^k \mod{N}}
= \exp{\mleft[ \frac{i 2\pi s}{r} \mright]} \ket{u_s},
\end{align}
since $x^r = x^0 \mod{N}$ and 
$\exp{\mleft[ \frac{-i 2\pi s (r-1)}{r} \mright]} = \exp{\mleft[ \frac{i 2\pi s}{r} \mright]}$. 

Using the phase-estimation algorithm, we can now efficiently determine $s/r$ with high accuracy. One requirement is that we can implement the operators $U^{2^k}$ efficiently, which can be done using a procedure known as modular exponentiation [see Box 5.2.~on page 228 in Ref.~\citep{NielsenChuang}], needing $O(L^3)$ gates. Furthermore, we need to produce one or more of the eigenstates $|u_s\rangle$, which is done by noting that
\be
\frac{1}{\sqrt{r}}\sum_{s=0}^{r-1}|u_s\rangle =
\frac{1}{r}\sum_{s=0}^{r-1} \sum_{k=0}^{r-1} \exp{\mleft[ \frac{-i 2\pi s k}{r} \mright]} |x^k \mod{N}\rangle
= |1\rangle .
\ee
To have enough accuracy in the phase estimation, we should use $t = 2L + 1 + \mleft\lceil
\log{\mleft( 2 + \frac{1}{2\epsilon} \mright)} \mright\rceil$ qubits in the first register and prepare the second register in the $|1\rangle$ state\footnote{In the expression of the number of digits, $2L + 1$ appears instead of $L$ [as one would expect from \eqref{bits:phase-estimation}] due to convergence requirements of the continued-fraction expansion algorithm.}. We then obtain the phase $\phi=s/r$, for a random $0 \leq s < r$, with $2L + 1$ bits precision, with a probability of at least $(1-\epsilon)$. Knowing that the phase $\phi = s/r$ is a rational number, where $s$ and $r$ are integers not larger than $L$ bits, we can classically determine $s$ and $r$. The appropriate algorithm is called the continued-fraction expansion and needs $O(L^3)$ gates.

\subsection{Performance}

The algorithm fails if $s = 0$, and also if $s$ and $r$ have common factors so that they cannot be extracted from $s/r$. Note that if instead $x$ and $N$ happen to not be co-prime, which can be checked efficiently by computing the gcd, the algorithm can just return the common factor, yielding one factor of the decomposition of $N$. Also note that the case of odd $r$ can be addressed, with no need to re-run the algorithm~\citep{ekeraa2021completely}. The success probability can be shown to be constant, 
and one needs only to repeat the procedure a constant number of times to obtain $r$ in order to obtain as high success probability as one wishes. In fact, the algorithm can be improved to succeed with a probability of $1 - 10^{-4}$ in a single run for moderate to large $r$~\citep{ekeraa2022success}. 

The number of gates needed are $O(L)$ for the initial Hadamards, inverse Fourier transform needs $O(L^2)$ gates, implementing $U^{2^k}$ through modular exponentiation requires $O(L^3)$ gates, and the classical continued-fraction algorithm needs $O(L^3)$ (classical) gates. The total scaling is therefore $O(L^3)$. Note that a more recent algorithm for modular exponentiation was also developed, which requires $\tilde{O}(L^2)$, where the tilde indicates that we are neglecting logarithmic contributions to the order~\citep{gidney1905factor}. This brings down the count of the gates to be run on the quantum processors to $\tilde{O}(L^2)$. A different algorithm for factoring has been introduced in Ref.~\citep{regev2023efficient}, where they bring down the quantum runtime to $\tilde{O}(L^{1.5})$, and conjecture that it could be further brought down to $\tilde{O}(L)$.

\subsubsection{The algorithm for factoring}

Find a factor of the composite $L$-bit integer $N$.

\begin{enumerate}

\item If $N$ is even, return the factor 2. 

\item Determine whether $N = a^b$, i.e., if there is only one prime factor. This can be done with $O(L^3)$ operations. If so, return the factor $a$.

\item Randomly choose $1 < x < (N-1)$, and check whether $x$ and $N$ are co-prime ($O(L^3)$ operations). If not co-prime, return the factor $gcd(x,N)$.

\item Find the order $r$ of $x$ modulo $N$, which can be done using $O(L^3)$ quantum gates (quantum subroutine!). 

\item If $r$ is even and $x^{r/2}\neq -1 \mod{N}$, then compute $gcd(x^{r/2}+1,N)$ and $gcd(x^{r/2}-1,N)$, and check if one is a non-trivial factor of $N$. Return this factor. If $r$ is odd, or $x^{r/2}= -1 \mod{N}$, the algorithm fails.

\end{enumerate}

\newpage

\section*{Exercises}

\begin{enumerate}

\item \label{ex:TF} (NC 5.5) Give a decomposition of the controlled-$R_k$ gate in terms of single-qubit gates and CNOT gates.

\item \label{ex:FT-inverse} (NC 5.5) Give a quantum circuit to perform the inverse quantum Fourier transform. 

\item Show that the quantum Fourier transform for a single qubit is the Hadamard gate. 

\item \label{ex:eigenvalue-unitary} (From Quic seminar 4) Let $\lambda$ be an eigenvalue of a unitary matrix $U$. Prove that $\abs{\lambda} = 1$.

\item  (From Quic seminar 4) \label{ex:phase-estimation $X$} We know that the Pauli matrix $X$ has eigenvalues $\pm 1$ and eigenvectors $\ket{\pm}$.
Show that the eigenvalue of $X$ corresponding to the state $\ket{+}$ is 1, by using phase estimation. Hint: start by writing the circuit and then compute its output.

\item (From Quic seminar 4) The Hadamard test is an algorithm for computing the expectation value of an operator that is similar to the phase-estimation algorithm. Let $\ket{\psi}$ be a state on $m$ qubits and $Q$ a unitary operator on $m$ qubits, and consider the algorithm described by \figref{fig:Hadamard-test}.
Compute the probabilities $p(0)$ and $p(1)$ of measuring outcomes $0$ and $1$, respectively, at the output of the circuit. Show that $p(0) - p(1) = \text{Re}[\brakket{\psi}{Q}{\psi}]$. 
Prove that implementing an $S$ gate before the final Hadamard gate on the first qubit allows one to compute the imaginary part of the expectation value, $\text{Im}[\brakket{\psi}{Q}{\psi}]$.

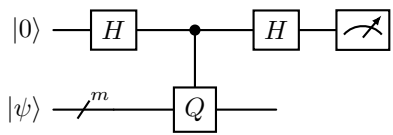
\begin{figure}[h!]
\centering
\begin{quantikz}
    \lstick{$\ket{0}$} & \gate{H} & \ctrl{1} & \gate{H} & \meter{}\\
    \lstick{$\ket{\psi}$} &  \qwbundle{m} & \gate{Q} & 
\end{quantikz}
\caption{Circuit implementing the Hadamard test.
\label{fig:Hadamard-test}}
\end{figure}

\item \label{ex:mod1} Compute the modular inverse of $7 \mod{15}$.

\item \label{ex:mod2} Solve the equation $k \cdot 7 = 4 \mod{15}$.

\end{enumerate}


\chapter*{Tutorial 2: Shor's algorithm}
\addcontentsline{toc}{chapter}{Tutorial 2} 
\renewcommand{\thesection}{\arabic{section}} 
\setcounter{section}{0} 
\setcounter{tutorial}{2}
\setcounter{ex}{0}
\setlength{\parindent}{0pt} 

\setcounter{equation}{0} 
\renewcommand{\theequation}{T2.\arabic{equation}} 

This tutorial tackles the circuit implementation of the quantum subroutine of Shor's algorithm. How do the gates actually look like? How do we know which gates we need to factorize a certain number? These questions are answered here.

\section{Shor's algorithm: toolkit}
    
\begin{ex}\label{Shor:ex1}
What controlled function $f_x(y)$ does the below gate array correspond to? Note that $x$ is a control bit, $y_i$ are bits of $y$ in binary ($y=y_3y_2y_1y_0$) and $f_i$ are bits of $f_x(y)$ in binary ($f_x(y) = f_3f_2f_1f_0$).
\end{ex}

\begin{figure}[H]
\centering
\begin{quantikz}
\lstick{$\ket{x}$} & \ctrl{2} & \ctrl{3} & \ctrl{4} & \\
\lstick{$\ket{y_3}$} & \swap{1} & & & \rstick{$\ket{f_3}$}\\
\lstick{$\ket{y_2}$} & \targX{}& \swap{1} & & \rstick{$\ket{f_2}$}\\
\lstick{$\ket{y_1}$} & & \targX{} & \swap{1} & \rstick{$\ket{f_1}$}\\
\lstick{$\ket{y_0}$} & & & \targX{} &  \rstick{$\ket{f_0}$}
\end{quantikz}
\end{figure}

\begin{proof}[Solution]
First of all, we see that $x$ is a control qubit, so when $x=0$, $f_0(y)=y$. However, when $x=1$, all the SWAP gates will be activated. The transformation undergone by each SWAP gate is the following:
\be
\ket{y_3y_2y_1y_0} \stackrel{\text{SWAP}}{\longrightarrow} \ket{y_2y_3y_1y_0} \stackrel{\text{SWAP}}{\longrightarrow} \ket{y_2y_1y_3y_0} \stackrel{\text{SWAP}}{\longrightarrow} \ket{y_2y_1y_0y_3} \equiv \ket{f_3f_2f_1f_0}.
\ee

Now that we know the effect of the circuit, we can write down the input-output map to calculate the function $f_1(y)$. But first, for those who are not familiar with binary code, we convert binary to decimal as follows:
\be
y =\sum_{i=0}^{i=3} 2^iy_i = y_0 + 2\cdot y_1 + 4\cdot y_2 + 8\cdot y_3.
\ee
So, for instance, $y=7$ in decimal is $y_3y_2y_1y_0=0111$ in binary.\\

After computing Table~\ref{tab:4mod15} (left), we deduce that $f_1(y)=2y\,\Mod 15$. Hence,
\be
f_x(y) = 2^xy\Mod 15.
\ee
\end{proof}


\begin{table}[H]
    \centering
    \begin{tabular}{c|c|c|c}
        $y$ & $y_3y_2y_1y_0$ & $f_3f_2f_1f_0$ & $f_1(y)$  \\\hline
        0 & 0000 & 0000 & 0\\ 
        1 & 0001 & 0010 & 2\\ 
        2 & 0010 & 0100 & 4\\ 
        3 & 0011 & 0110 & 6\\ 
        4 & 0100 & 1000 & 8\\ 
        5 & 0101 & 1010 & 10\\ 
        6 & 0110 & 1100 & 12\\ 
        7 & 0111 & 1110 & 14\\ 
        8 & 1000 & 0001 & 1\\ 
        9 & 1001 & 0011 & 3\\ 
        10 & 1010 & 0101 & 5\\ 
        11 & 1011 & 0111 & 7\\ 
        12 & 1100 & 1001 & 9\\ 
        13 & 1101 & 1011 & 11\\ 
        14 & 1110 & 1101 & 13\\ 
        15 & 1111 & 1111 & 15\\  
    \end{tabular}
    \qquad\qquad
     \begin{tabular}{c|c|c|c}
        $y$ & $y_3y_2y_1y_0$ & $f_3f_2f_1f_0$ & $f_1(y)$  \\\hline
        0 & 0000 & 0000 & 0\\ 
        1 & 0001 & 0100 & 4\\ 
        2 & 0010 & 1000 & 8\\ 
        3 & 0011 & 1100 & 12\\ 
        4 & 0100 & 0001 & 1\\ 
        5 & 0101 & 0101 & 5\\ 
        6 & 0110 & 1001 & 9\\ 
        7 & 0111 & 1101 & 13\\ 
        8 & 1000 & 0010 & 2\\ 
        9 & 1001 & 0110 & 6\\ 
        10 & 1010 & 1010 & 10\\ 
        11 & 1011 & 1110 & 14\\ 
        12 & 1100 & 0011 & 3\\ 
        13 & 1101 & 0111 & 7\\ 
        14 & 1110 & 1011 & 11\\ 
        15 & 1111 & 1111 & 15\\ 
    \end{tabular}
    \caption{Input-output maps for the circuits in Exercise \ref{Shor:ex1} (left) and Exercise \ref{Shor:ex2} (right).}
    \label{tab:4mod15}
\end{table}

\begin{ex}\label{Shor:ex2}
Construct a gate array for the function $f_x(y) = 4^x y \mod{15}$, for a single control bit $x$.
\end{ex}

\begin{proof}[Solution]
Since $x$ is again the control bit, we need a function that satisfies $f_0(y)=y$ and $f_1(y)=4y$, i.e., that does nothing when $x=0$ and multiplies by 4 when $x=1$.
Note that multiplying by 4 is simply multiplying twice by 2, an operation that we already computed in the previous exercise. Therefore, we want a circuit that implements the following map:
\be
\ket{y_3y_2y_1y_0} \stackrel{\times 2}{\longrightarrow} \ket{y_2y_1y_0y_3} \stackrel{\times 2}{\longrightarrow} \ket{y_1y_0y_3y_2} \equiv \ket{f_3f_2f_1f_0}.
\ee
If we look at the ordering of the bits, this circuit swaps $3 \leftrightarrow 1$ and $2 \leftrightarrow 0$. Hence, it is straightforward to build the circuit using two Fredkin (controlled-SWAP) gates:

\begin{figure}[H]
\centering
\begin{quantikz}
\lstick{$\ket{x}$} & \ctrl{3} & \ctrl{4} &  \\
\lstick{$\ket{y_3}$} & \swap{2} &   &  \rstick{$\ket{f_3}$}\\
\lstick{$\ket{y_2}$} &   & \swap{2} &  \rstick{$\ket{f_2}$}\\
\lstick{$\ket{y_1}$} & \targX{} &   &  \rstick{$\ket{f_1}$}\\
\lstick{$\ket{y_0}$} &   &  \targX{} &  \rstick{$\ket{f_0}$}
\end{quantikz}
\end{figure}

We can verify that our circuit performs as we want to by computing the input-output map [see Table~\ref{tab:4mod15} (right)].\\

\end{proof}

\begin{ex}
Construct a gate array for the function $f_x(y) = 2^x y \mod{15}$, for a two-bit control register $x$.
\end{ex}
\begin{proof}[Solution]
Since the control $x$ now has two bits $x_0$ and $x_1$, we can write it as $x = x_0 + 2 x_1$. Therefore, 
\be
2^x = 2^{x_0+2x_1} = 2^{x_0}4^{x_1}.
\ee
Hence, the function $f_x(y) = 2^{x_0} 4^{x_1} y\,\Mod{15}$ can be built as a concatenation of the functions in the two previous exercises. That is,

\begin{figure}[H]
\centering
\begin{quantikz}
\lstick{$\ket{x_1}$} & \ctrl{3} & \ctrl{4} & & & & \\
\lstick{$\ket{x_0}$} & & & \ctrl{2} & \ctrl{3} & \ctrl{4} & \\
\lstick{$\ket{y_3}$} & \swap{2} & & \swap{1} & & & \rstick{$\ket{f_3}$}\\
\lstick{$\ket{y_2}$} & & \swap{2} & \targX{} & \swap{1} & & \rstick{$\ket{f_2}$}\\
\lstick{$\ket{y_1}$} & \targX{} & & & \targX{} & \swap{1} & \rstick{$\ket{f_1}$}\\
\lstick{$\ket{y_0}$} & & \targX{} & & & \targX{} & \rstick{$\ket{f_0}$}
\end{quantikz}
\end{figure}
\end{proof}

\section{Shor's algorithm: factorizing 15}

Assuming a number $N$ can be factorized into two coprimes $p,q$ (i.e., $N=pq$), Shor's algorithm is designed to find those factors. Let us see how it works by factorizing $N=15$.

\subsection{Classical part}

\begin{enumerate}
    \item \textbf{Pick a number $x<N$.}\\
    We pick $x=2$.
    \item \textbf{If $\gcd(x, N)\neq 1$, $\gcd(x, N)$ is a factor of $N$ and we are done. Else, we continue.}\\
    We note that $\gcd(2,15)=1$, so we continue.
    \item \textbf{Calculate $x^j \Mod N$ to find the smallest integer $r>0$ such that $x^r = 1\Mod N$.}\\
    This is done by the quantum subroutine, but we do it here by hand to prepare for a later exercise. As shown in the table below, the order (a.k.a.~period) is $r=4$.
        \begin{table}[H]
        \centering
        \begin{tabular}{c|c|c}
            $j$ & $2^j$ & $2^j\Mod 15$ \\\hline
            0 & 1 & 1\\
            1 & 2 & 2\\
            2 & 4 & 4\\
            3 & 8 & 8\\
            4 & 16 & {\color{red}1}\\
            5 & 32 & 2\\
            6 & 64 & 4\\
            7 & 128 & 8\\
            8 & 256 & 1
        \end{tabular}
    \end{table}
    \item \textbf{If $r$ is odd, go back to step 1. Else, continue.}\\
    We note that $r=4$ is even, so we continue.
    \item \textbf{If $x^{r/2}-1 = 0\Mod N$, go back to step 1. Else, continue.}\\
    We note that $2^2-1 = 3 \neq 0 \Mod 15$, so we continue.
    \item \textbf{Both $\gcd(x^{r/2}\pm 1, N)$ are nontrivial factors of $N$.}\\
    We note that $\gcd(2^2 + 1, 15)=5$ and $\gcd(2^2 - 1, 15)=3$, and that $5\cdot 3 = 15$, just like we wanted.
\end{enumerate}

\subsection{Quantum part}

\begin{figure}[H]
\centering
\begin{quantikz}[column sep=0.4cm]
\lstick{$\ket{0}$} & \qwbundle{t}\slice{A} & \gate{H}\slice{B}  & \ctrl{1}\slice{C} & & \gate{QFT^\dagger}\slice{D} & \meter{}\\
\lstick{$\ket{1}$} & \qwbundle{L} &  & \gate{x^j \mod N} & & &   
\end{quantikz}
\end{figure}

\subsubsection*{Step A}
The quantum circuit to run the order-finding subroutine is composed of two registers. The first one, initialized to $\ket{0}$, has $t$ bits, whereas the second one, initialized to $\ket{1}$, has $L$ bits ($2^L>N$). In our case, we take $t=2, L=4$, so that in step $A$ (see circuits above and below), we have
%
\be
\ket{\psi_A} \stackrel{\text{decimal}}{=} \ket{0}\ket{1} \stackrel{\text{binary}}{=} \ket{00}\ket{0001}.
\ee

\subsubsection*{Step B}
Then, we apply Hadamard gates to the bits in the first register, to create a uniform superposition:
\be
\ket{\psi_B}
\stackrel{\text{decimal}}{=}
    \frac{1}{2^{t/2}}\sum_{j=0}^{2^t-1}\ket{j}\ket{1} 
\stackrel{t=2}{=}
    \frac{1}{2}\sum_{j=0}^3 \ket{j}\ket{1}
\stackrel{\text{binary}}{=} 
    \frac{1}{2} \mleft( \ket{00}+\ket{01}+\ket{10}+\ket{11} \mright) \ket{0001}.
\ee

\subsubsection*{Step C}
On the second register, we compute $x^j\Mod N$. In our case, we compute $2^j\Mod 15$, which is achieved by the circuit we built in Exercise 3. The state at step C thus becomes
\be
\ket{\psi_C} 
\stackrel{\text{decimal}}{=}
\frac{1}{2^{t/2}}\sum_{j=0}^{2^t-1}\ket{j}\ket{x^j \Mod N}
\stackrel{t,x=2, N=15}{=}
\frac{1}{2} \mleft( \ket{0}\ket{1} + \ket{1}\ket{2} + \ket{2}\ket{4} + \ket{3}\ket{8} \mright).
\ee

\subsubsection*{Step D}
Now, we apply the inverse quantum Fourier transform (QFT$^\dag$) on the first register. From the lectures, we know that it transforms the states as follows:
\be
QFT^\dag \ket{j} = \frac{1}{2^{t/2}} \sum_{k=0}^{2^t-1} e^{-i2\pi jk/(2^t)} \ket{k}.
\ee

Therefore, the state in step D becomes
\begin{align}
\ket{\psi_D} \stackrel{\text{decimal}}{=}&
\frac{1}{2^{t/2}}\sum_{j=0}^{2^t-1} \mleft( QFT^\dag \ket{j} \mright) \ket{x^j \Mod N} 
=
\frac{1}{2^{t}}\sum_{j,k=0}^{2^t-1} e^{-i 2\pi jk/(2^t)} \ket{k} \ket{x^j \Mod N}
\stackrel{t,x=2, N=15}{=} \nn \\
= \;\;\;& \frac{1}{2} \mleft[ \mleft( QFT^\dag \ket{0} \mright) \ket{1} + \mleft( QFT^\dag \ket{1} \mright) \ket{2} + \mleft( QFT^\dag \ket{2} \mright) \ket{4} + \mleft( QFT^\dag \ket{3} \mright) \ket{8} \mright] = \dots = \nn \\
= \;\;\;& \frac{1}{4} \mleft( \ket{0}\ket{1} + \ket{1}\ket{1} + \ket{2}\ket{1} + \ket{3}\ket{1} + \ket{0}\ket{2} - i\ket{1}\ket{2} - \ket{2}\ket{2} + i\ket{3}\ket{2} \mright. \nn \\
&+ \mleft. \ket{0}\ket{4} - \ket{1}\ket{4} + \ket{2}\ket{4} - \ket{3}\ket{4} + \ket{0}\ket{8} + i\ket{1}\ket{8} - \ket{2}\ket{8} - i\ket{3}\ket{8} \mright).
\end{align}
From the expression above, we see that we can measure four possible outcomes in the first register, $k={0, 1, 2, 3}$, all with the same probability:
\begin{align}
    P_0 = \abs{\frac{1}{4}}^2 + \abs{\frac{1}{4}}^2 + \abs{\frac{1}{4}}^2 + \abs{\frac{1}{4}}^2 = \frac{1}{4} \qquad
    P_1 = \abs{\frac{1}{4}}^2 + \abs{\frac{-i}{4}}^2 + \abs{\frac{-1}{4}}^2 + \abs{\frac{+i}{4}}^2 = \frac{1}{4} \nn \\
    P_2 = \abs{\frac{1}{4}}^2 + \abs{\frac{-1}{4}}^2 + \abs{\frac{1}{4}}^2 + \abs{\frac{-1}{4}}^2 = \frac{1}{4} \qquad
    P_3 = \abs{\frac{1}{4}}^2 + \abs{\frac{i}{4}}^2 + \abs{\frac{-1}{4}}^2 + \abs{\frac{-i}{4}}^2 = \frac{1}{4}.
\end{align}
The measurement outcome $k$ relates to the order $r$ as
\be
\frac{k}{2^t} = \frac{s}{r},
\ee
where $s$ is an integer between 0 and $r-1$.\footnote{See the end of the tutorial for where this expression comes from.} In our case,
\be
\frac{k}{4}  = \begin{cases}
\dfrac{0}{4} = \dfrac{s}{r} \implies \text{rerun algorithm}\\[8pt]
\dfrac{1}{4} = \dfrac{s}{r} \implies r=4 \implies \text{success}\\[8pt]
\dfrac{2}{4} = \dfrac{s}{r} \implies r=2 \implies \text{wrong order*}\\[8pt]
\dfrac{3}{4} = \dfrac{s}{r} \implies r=4 \implies \text{success}.
\end{cases}
\ee
In reality, the way to obtain the order $r$ from the measurement outcome $k$ is through the \textit{continued-fractions algorithm}, which you can find in Ref.~\citep{NielsenChuang} (Box 5.3), but, for simplicity, here we do it by hand.\\

Finally, the probability of success is technically $0.5$, although in reality, obtaining the wrong order, in this case, still allows us to find the factors. Let us say we obtain $r=2$ from the quantum subroutine. Then, when we compute the last step of the classical part of Shor's algorithm, we obtain
\be
\gcd(2^1+1, 15) = 3, \quad \gcd(2^1-1, 15) = 1.
\ee
Note that we have successfully obtained one nontrivial factor. Then, we can divide 15 by the factor and obtain the other nontrivial factor: $15/3=5$. So, in this case, the probability of success is actually $0.75$.\\

It is important to remark that there is a lower bound on the success probability, which means that by running the algorithm enough times, it will always be successful. This is key in demonstrating the polynomial runtime of the algorithm. \\

\begin{figure}[H]
\centering
\begin{quantikz}
\lstick{$\ket{x_1} = \ket{0}$} \slice{A} & \gate{H}\slice{B} & \ctrl{4} & \ctrl{5} &  &  & \slice{C} &[5mm] \gate{H}\gategroup[wires=2,steps
    =3,style={rounded corners, inner sep=1pt}]{semiclassical FT$^\dagger$} & \meter{}\vcw{1} &  & \slice{D} & \rstick{$\ket{k_0}$} \\
\lstick{$\ket{x_0}=\ket{0}$}& \gate{H} &  &  &\ctrl{2} & \ctrl{3} & \ctrl{4} & & \gate{R_2} & \gate{H} &  & \meter{}\rstick{$\ket{k_1}$}\\
\lstick{$\ket{y_3}=\ket{0}$}& & \swap{2} &  & \swap{1} &  &  & & &  &  &  \\
\lstick{$\ket{y_2}=\ket{0}$}& &  & \swap{2} & \targX{} & \swap{1} &  & & &  & &   \\
\lstick{$\ket{y_1}=\ket{0}$}& & \targX{} &  &  & \targX{} & \swap{1} & & &  & &    \\
\lstick{$\ket{y_0}=\ket{1}$}& &  & \targX{} &  &  & \targX{} & & &  &  &  
\end{quantikz}
\caption{Circuit implementation of the quantum subroutine of Shor's algorithm for factorizing 15 with base 2. Note that the FT changes the positions of the bits, but that $\ket{k}=\ket{k_1k_0}$, with $k_0$ being the least significant bit.}
\end{figure}
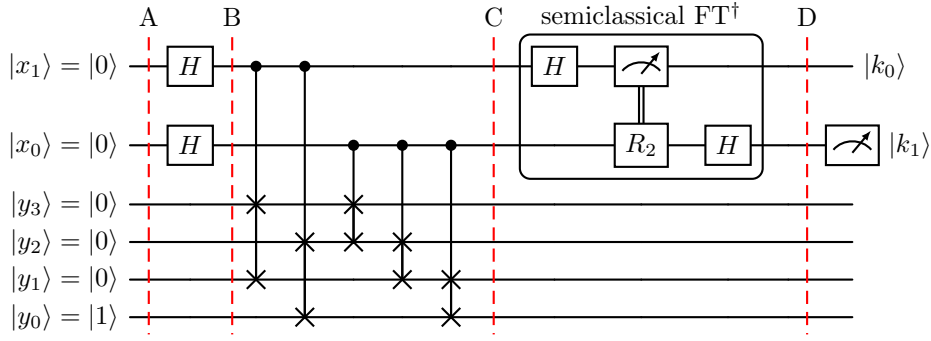

\section{Additional material}
$\bullet$ For more on circuit implementation, check Ref.~\citep{Vedral96}.\\

$\bullet$ In class, we use \href{https://phasespacecomputing.com/pdf/Product_sheet_Shors.pdf}{Shor's algorithm toolkit} from Phase Space Computing to simulate all the circuits shown throughout the tutorial. This toolkit is a modular set of electronic circuit boards that approximate the behavior of quantum gates via quantum simulation logic. To learn more about how these gates actually work, check Chapter 3 in Ref.~\citep{Johansson2019}.\\

$\bullet$ To see other circuit implementations of Shor's quantum subroutine (e.g., factorizing 15 with base 7 instead of 2), check Chapter 9 in Ref.~\citep{Johansson2019}.\\

$\bullet$ \textbf{Why do we measure the phase $s/r$ in the order-finding subroutine?}

The state in step C can also be written as
\be
\ket{\psi_C} = 
\frac{1}{\sqrt{2^{t}}}\sum_{j=0}^{2^t-1}\ket{j}\ket{x^j \Mod N}
\approx 
\frac{1}{\sqrt{2^{t}r}}\sum_{s=0}^{r-1}\sum_{j=0}^{2^t-1}e^{2\pi i sj/r}\ket{j}\ket{u_s},
\ee
where 
\be
\ket{u_s} = \frac{1}{\sqrt{r}}\sum_{k=0}^{r-1} e^{2\pi i sk/r}\ket{x^k\Mod N}.
\ee
Then, when applying the QFT$^\dag$ to the first register,
\be
\ket{\psi_D} = QFT^\dagger\ket{\psi_C} = \frac{1}{\sqrt{r}} \sum_{s=0}^{r-1}\ket{s/r}\ket{u_s}. 
\ee
Thus, measuring the first register, we read out $s/r$.
For a longer explanation, check Section 5.3.1 of Ref.~\citep{NielsenChuang}.

\renewcommand{\thesection}{\thechapter.\arabic{section}} 
\setlength{\parindent}{15pt} 
\renewcommand{\theequation}{\thechapter.\arabic{equation}} 


\chapter{Quantum machine learning}
\label{chp:QML}

Quantum computing and machine learning are arguably two of the ``hottest'' topics in science at the moment. Here in Sweden, this is, for example, reflected in the fact that the two largest programs supported by the Knut and Alice Wallenberg foundation are centered around quantum technologies and artificial intelligence. In this chapter, we will discuss efforts to combine the two fields into quantum machine learning. Since this is a course about quantum algorithms, we do not cover applications of classical machine learning to simulating and understanding quantum systems [even though that is a fascinating topic in itself; see, e.g., Ref.~\citep{Krenn2023}], but focus instead on how machine learning can be enhanced by quantum computation. We begin with a brief overview of classical machine learning and then study some examples of quantum machine learning algorithms. 

In the limited time we have available in this course, it is hard to do more than scratch the surface of quantum machine learning and give some basic ideas that are important in this field. For the reader who wants to go deeper into the topic than this chapter does, a good starting point is the reviews in Refs.~\citep{Biamonte2017, Cerezo2022}.

\section{A brief overview of classical machine learning}

What is machine learning? With the success of, and hype around, machine learning in recent years, there are examples of companies and researchers calling many things ``machine learning'' that would have been called something else a few years ago. According to Wikipedia, ``Machine learning is the scientific study of algorithms and statistical models that computer systems use to perform a specific task without using explicit instructions, relying on patterns and inference instead''. We like the following definition (source unknown): ``Machine learning studies algorithms whose performance improves with data (`learning from experience')''.

\subsection{Types of machine learning}

Broadly speaking, there are three paradigms in machine learning for extracting meaning from some data:
\begin{itemize}
\item \textbf{Unsupervised learning}: Learning structure in $P(\text{data})$ given samples from $P(\text{data})$. Here, the machine learning is used to \textit{generate} knowledge by analyzing unlabelled data. Examples of unsupervised learning are clustering (grouping data points), density estimation (estimating a probability density function giving rise to the data), and much of what is called data mining.
\item \textbf{Supervised learning}: Learning structure in $P(\text{labels} | \text{data})$ given samples from $P (\text{data}, \text{labels})$. Here, the machine learning is used to \textit{generalize} knowledge gained from studying labelled data to predict correct labels for unlabelled data. Examples of supervised learning are foremost various classification tasks, e.g., image recognition. 
\item \textbf{Reinforcement learning}: Learning from (possibly rare) rewards. Here, an agent learns by acting on its environment, observing the results of its actions (the results can include rewards), and updating its policy for actions based on the outcomes. Examples of reinforcement learning include the superhuman-level game-playing programs for go, chess, StarCraft, etc.~by Google's DeepMind.
\end{itemize}

\subsection{Neural networks}

\begin{figure}
\centering
\includegraphics[width=0.8\linewidth]{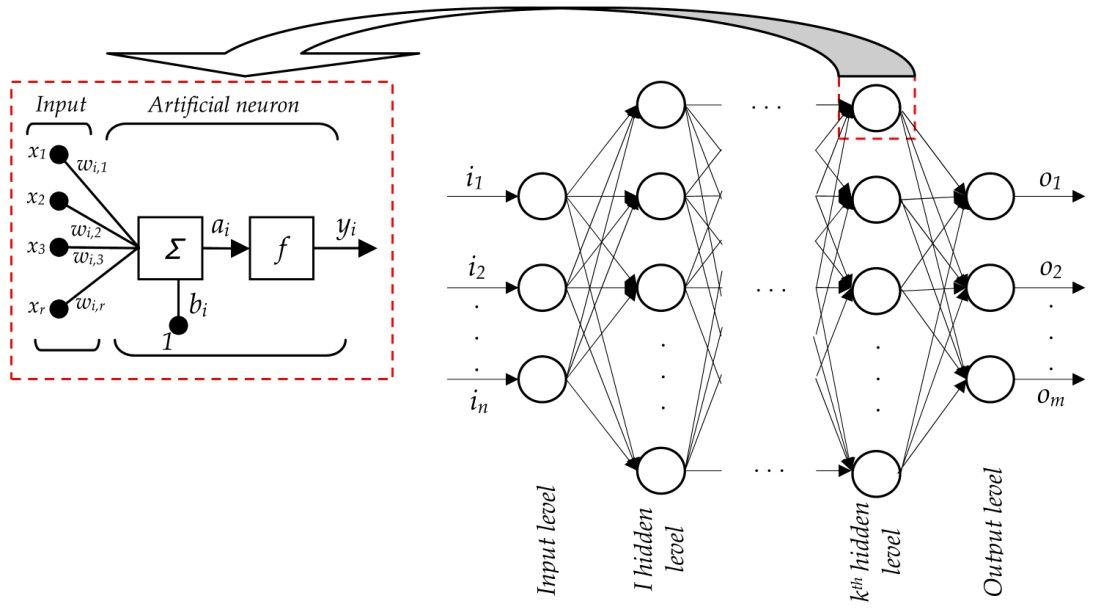}
\caption{Structure of a feed-forward neural network and a single neuron in that network. From Ref.~\citep{Tanikic2012}.
\label{fig:NeuralNetwork}}
\end{figure}

One common way to implement machine learning is using neural networks. The neurons in such a network can be connected in different layouts. Figure~\ref{fig:NeuralNetwork} shows a feed-forward neural network, where some input (data) is fed into the neurons in the input layer on the left. The output from these neurons becomes the input for neurons in the next layer, and so forth. In the end, some result is defined by the output from the last layer on the right. The layers between the input and output layers are called hidden layers.

The structure of a single neuron is shown on the left in \figref{fig:NeuralNetwork}. The inputs $x_j$ to neuron $i$ are weighted by weights $w_{i,j}$. To the weighted sum of the inputs, a ``bias'' $b_i$ can be added. The result $a_i$ is fed into a nonlinear activation function $f$, which usually is some smoothed version of a step function. The output is then
\be
y_i = f \mleft( \sum_j w_{i,j} x_j + b_i \mright).
\ee

A neural network can thus be said to be a complicated function, parameterized by all the weights and biases in the network, which transforms an input into an output. What functions can a neural network represent? The answer, provided by Cybenko in 1989~\citep{Cybenko1989}, gives a hint of why neural networks are powerful. It turns out that any arbitrary smooth function with vector input and vector output can be approximated to any desired precision by a feedforward neural network with only a single hidden layer. In practice, deep neural networks, i.e., networks with many hidden layers, have turned out to be more efficient at representing various functions. See, e.g., Ref.~\citep{Lin2017}.

\subsection{Training neural networks}

Training a neural network to perform a specific task boils down to adjusting the weights $w$ and biases $b$ such that the network approximates a function that solves the task. To do this, we first need to be able to say how close the output of the current network is to the desired output. The difference between the actual and the desired output is measured by some cost function. One example is the mean square error
\be
C (w, b) = \frac{1}{2n} \sum_x \abssq{y(x) - a},
\ee
where $n$ is the number of examples $x$ (data points), $y$ is the desired output, and $a$ is the actual output. To find good weights and biases for the task is to find weights and biases that minimize $C (w,b)$.

How does one minimize $C$ in a smart way? Clearly, there are too many unknowns to simply find the extremum by setting the gradient of $C$ to zero and solving the resulting equation. Therefore, gradient descent is used, with the update of the parameters being proportional to minus the gradient (the proportionality constant is called the \textit{learning rate}). However, the na\"ive approach to calculating the gradient of $C$ is time-consuming: to obtain each partial derivative of $C$, the network would need to be run once for each data point $x$ and for each variable in $w$ or $b$, to see how a small change in the input changes the output. One important reason for the prevalence of neural networks today is that two tricks have been found that can reduce the necessary calculations considerably.

The first trick is to use \textit{stochastic gradient descent}, which means that the partial derivatives are not calculated for all data points $x$ in each step of the gradient descent, but only for a subset, a \textit{mini-batch}, of $x$. The next step uses another subset, and so on until all subsets have been used (marking the end of an \textit{epoch} of training), whereupon the selection of mini-batches starts over. The use of stochastic gradient descent will only give an approximation to the actual gradient, but if the mini-batches are large enough and the learning rate is low enough, this approximation is usually sufficient.

The second trick is to calculate the partial derivatives not one by one by running the network many times, but by using \textit{back-propagation}. Essentially, back-propagation allows one to calculate all partial derivatives by running the network once, noting the result for each neuron, finding the derivative of the cost function with respect to the output from the last layer, and then applying the chain rule repeatedly, going backwards through the network to extract each partial derivative. For more details on how this works, see, e.g., Ref.~\citep{Nielsen2015}.

With the network having so many parameters, often orders of magnitude more than the number of training examples, a valid concern is whether the training will result in overfitting. Over the years, various strategies have been developed to counter this, but that is beyond the scope of this short introduction to classical machine learning.

\subsection{Reasons for the success of classical machine learning}

In explanations of the current success of classical machine learning, three factors are usually brought forward:
\begin{itemize}
\item \textbf{Data}: There are now a large number of huge data sets that can be used for training of neural networks.
\item \textbf{Computational power}: We now have much more powerful, and increasingly custom-built, hardware to run machine-learning algorithms on.
\item \textbf{Algorithms}: Clever algorithms like back-propagation, which enable efficient training, together with a number of other clever tricks discovered in the past decade or so, have led to large improvements.
\end{itemize}

\section{Quantum machine learning using qBLAS}
\label{sec:QML-using-qBLAS}

To see how quantum computers can aid or enhance machine learning, a first entry point is to note that many machine learning algorithms rely heavily on linear algebra. For classical computation of linear-algebra operations, there are optimized low-level routines called basic linear algebra subprograms (BLAS). For quantum computers, there are several algorithms that deal with linear-algebra problems. Together, these algorithms are sometimes referred to as quantum BLAS (qBLAS).

Some examples of qBLAS are the HHL algorithm for solving systems of linear equations~\citep{Harrow2009}, the quantum Fourier transform (see \secref{sec:QFT}), and quantum phase estimation for finding eigenvalues and eigenvectors (see \secref{sec:PhaseEstimation}). All of these examples have exponential speed-ups compared to their classical counterparts. However, it is important to ``read the fine print''~\citep{Aaronson2015} for these algorithms. They all rely on the problem being encoded in a quantum random access memory (QRAM). In a QRAM, data is encoded in the probability amplitudes of a large superposition state. For example, a vector $\mathbf{b}$ with $n$ entries can be stored in $\log_2 n$ qubits as $\sum_j b_j \ket{j}$, where $b_j$ are the entries in the vector and $\ket{j}$ are the computational basis states of the qubits.

The problem with the QRAM is that no efficient way is known to encode the data in the QRAM in the first place. The time it takes to encode the problem can therefore negate the exponential speed-up from the qBLAS algorithms. This is sometimes called the \textit{input problem}. There is also an \textit{output problem}: the output of the qBLAS algorithms is not necessarily the direct answer sought, but a state which lets you sample properties of the answer. For example, solving the system of linear equations $A \mathbf{x} = \mathbf {b}$ does not give the solution vector $\mathbf{x}$ as an easily measurable output, but just enables sampling properties of $\mathbf{x}$.

\section{Quantum support vector machines}

We will now look at an example of a classic machine-learning problem that can be tackled with quantum algorithms: support vector machines (SVMs). Before going to the quantum algorithm, we first define SVMs and see how they are implemented classically.

\subsection{Support vector machines}

\begin{figure}
\centering
\includegraphics[width=0.6\linewidth]{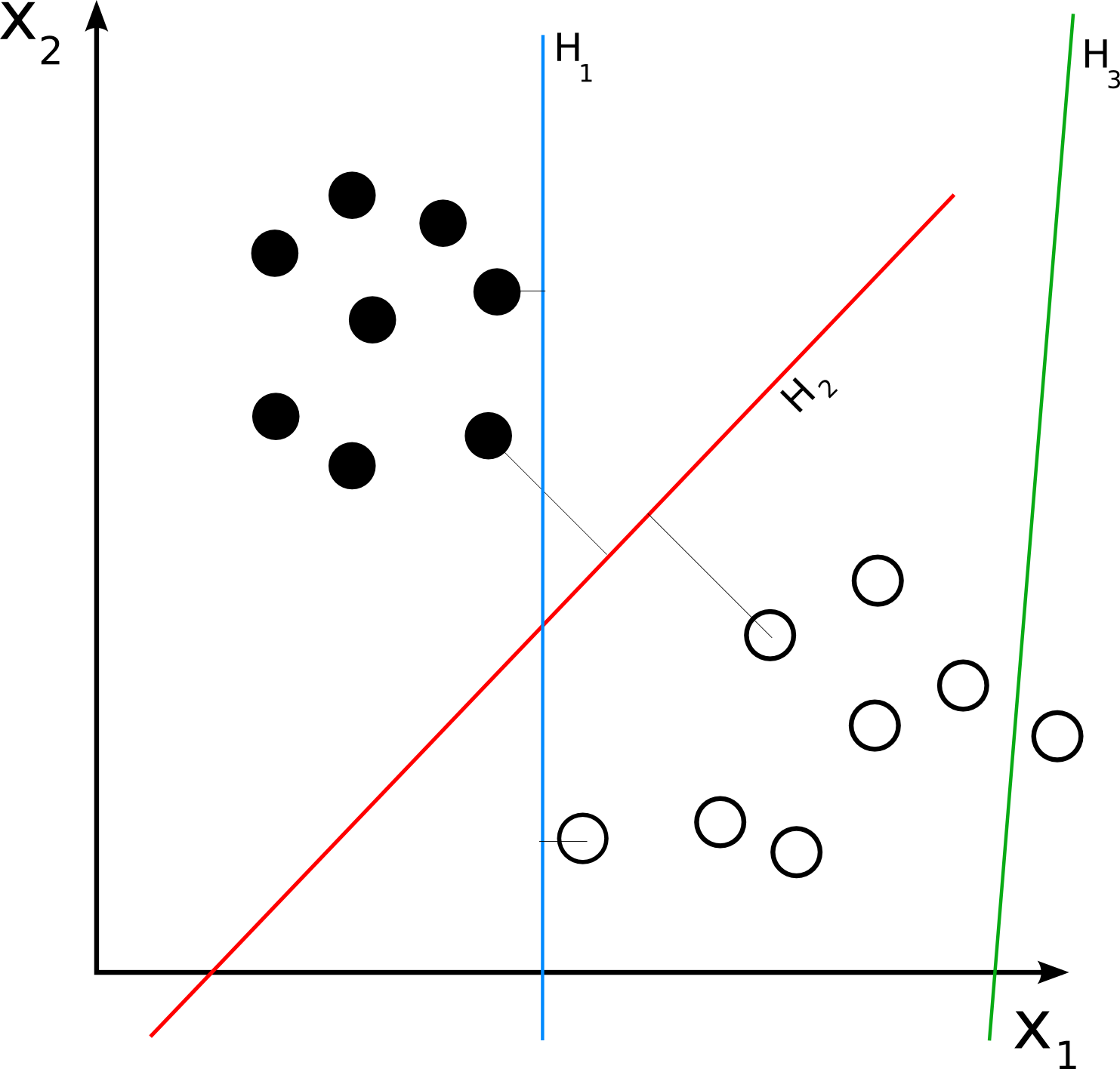}
\caption{Illustration of support vectors. From Wikipedia.
\label{fig:SVM}}
\end{figure}

A support vector machine is a classifier that divides data points into categories based on some boundary (a hyperplane) in space. To train an SVM is to show it labelled data points such that it can identify the optimal boundary. New unlabelled examples can then be classified by checking on which side of the boundary they fall. This is illustrated in \figref{fig:SVM}. The green line ($H_3$) does not separate the two categories of points (black and white). The blue and red lines ($H_1$ and $H_2$) both separate the points correctly, but the red line ($H_2$) is optimal, since the distance between $H_2$ and the nearest points in the training data is maximized. The points nearest to the optimal hyperplane thus define this hyperplane. These points are called support vectors.

\subsection{Classical computation}

Mathematically, the problem can be formulated as follows. We are given a set of data points $\mathbf{x}_j$ with labels $y_j \in \{-1, 1\}$. We want to find the equation $\mathbf{w} \cdot \mathbf{x} - b = 0 $ defining the optimal separating hyperplane. Here, $\mathbf{w}$ is the normal vector to the hyperplane, normalized such that the closest points in the two classes lie on the hyperplanes $\mathbf{w} \cdot \mathbf{x} - b = \pm 1 $. The distance from the optimal separating hyperplane to the closest point is $1/ \abs{\mathbf{w}}$. Since we wish to maximize this distance, we should minimize $\abs{\mathbf{w}}$, or, equivalently, $\frac{1}{2} \abssq{\mathbf{w}}$, under the constraints $y_j (\mathbf{w} \cdot \mathbf{x}_j - b) \geq 1 \: \forall j$.

To perform this minimization under constraints, we introduce Lagrange multipliers, forming the Lagrangian
\be
L (\mathbf{w}, b, \mathbf{\lambda}) = \frac{1}{2} \abssq{\mathbf{w}} - \sum_j \lambda_j \mleft[ y_j (\mathbf{w} \cdot \mathbf{x}_j - b) - 1 \mright].
\ee
Setting the partial derivatives of $L$ with respect to $\lambda_j$ equal to zero gives the constraints (the $\lambda_j$ not corresponding to support vectors will become zero). Setting the partial derivative of $L$ with respect to $\mathbf{w}$ to zero leads to
\be
0 = \mathbf{w} - \sum_j \lambda_j y_j \mathbf{x}_j \quad \Rightarrow \quad \mathbf{w} = \sum_j \lambda_j y_j \mathbf{x}_j,
\ee
so we see that $\mathbf{w}$ will be determined by the support vectors. Finally, we also use
\be
0 = \frac{\partial L}{\partial b} = \sum_j \lambda_j y_j,
\label{eq:ConstraintSVM}
\ee
and substitute these results back into the Lagrangian to obtain
\be
\sum_j \lambda_j - \frac{1}{2} \sum \lambda_i \lambda_j y_i y_j \mathbf{x}_i \cdot \mathbf{x}_j.
\label{eq:SVMLagrangian2}
\ee
The objective now is to find $\lambda_j$ that maximize this expression under the constraint given by \eqref{eq:ConstraintSVM}.

\begin{figure}
\centering
\includegraphics[width=0.8\linewidth]{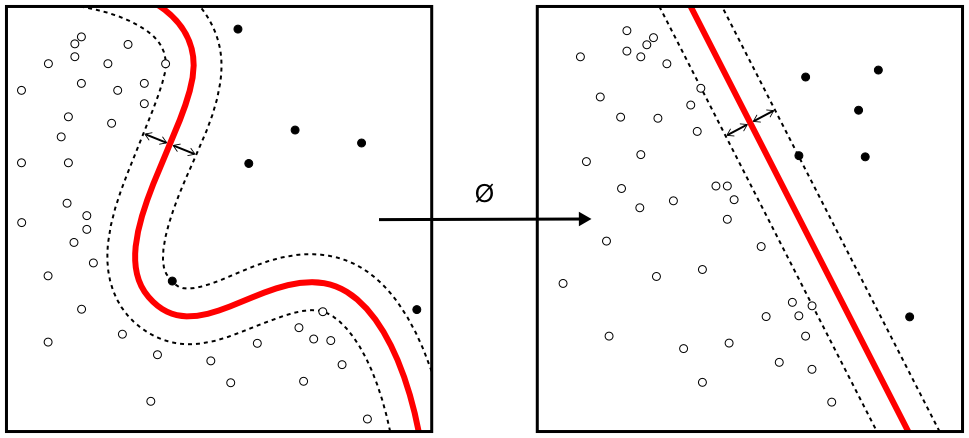}
\caption{Illustration of a kernel for a support vector machine. From Wikipedia.
\label{fig:SVMKernel}}
\end{figure}

In many cases, the separation between two classes of data points cannot be parameterized as a simple hyperplane, as illustrated to the left in \figref{fig:SVMKernel}. The solution commonly used is then to transform the data points to a feature space that admits a hyperplane as a separator. This is encoded by a kernel function $K(\mathbf{x}_i, \mathbf{x}_j)$, which then replaces the dot product in \eqref{eq:SVMLagrangian2}.

With this addition, and some further work, the maximization problem in \eqref{eq:SVMLagrangian2} can be shown to lead to the following system of linear equations:
\be
\begin{pmatrix}
0 & 1 \\
1 & K
\end{pmatrix}
\begin{pmatrix}
b \\
\mathbf{\lambda}
\end{pmatrix}
=
\begin{pmatrix}
0 \\
\mathbf{y}
\end{pmatrix},
\label{eq:SVMLinEqs}
\ee
where the ones are $1\times M$ row and column vectors ($M$ is the number of data points) and the entries in the $M\times M$ matrix $K$ are given by $K_{ij} = K(\mathbf{x}_i, \mathbf{x}_j)$.

We can now estimate the time it takes to find the support vectors on a classical computer. If the data points $\mathbf{x}_j \in \R^N$, calculating one entry in $K$ takes $\mathcal{O} (N)$ time, so calculating all of $K$ takes $\mathcal{O} (M^2 N)$ time. Solving the system of linear equations takes $\mathcal{O} (M^3)$ time, so in total, the classical computer will require $\mathcal{O} (M^2 [N+M])$ time.

\subsection{Quantum computation}

As we saw at the end of the previous subsection, the problem of SVMs boils down to two costly computations: calculating the entries in the matrix $K$ and solving the system of linear equations in \eqref{eq:SVMLinEqs}. Quantum algorithms can be applied to both these computations~\citep{Rebentrost2014, Wittek2013}.

\begin{figure}
\centering
\begin{quantikz}[row sep={0.8cm,between origins}]
    \lstick{$\dfrac{\ket{0}+\ket{1}}{\sqrt{2}}$} & \ctrl{2} & \gate{H} & \meter{}\\
    \lstick{$\ket{f}$} & \swap{1} & & \\
    \lstick{$\ket{f'}$} & \targX{} & & 
\end{quantikz}
\caption{The quantum circuit for a swap test. Figure adapted from Ref.~\citep{Wittek2013}. The resulting state of the system before the measurement is $\frac{1}{2} \mleft[ \ket{0} \mleft( \ket{f}\ket{f'} + \ket{f'} \ket{f} \mright) + \ket{1} \mleft( \ket{f}\ket{f'} - \ket{f'} \ket{f} \mright) \mright]$. This means that the probability of measuring 1 becomes zero when $f = f'$. 
\label{fig:SwapTest}}
\end{figure}
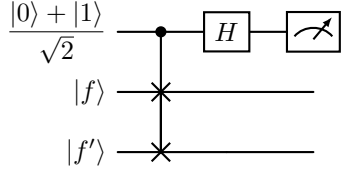

To calculate the dot product $\mathbf{x}_i \cdot \mathbf{x}_j$, we assume that $\abs{\mathbf{x}_i}$ and $\abs{\mathbf{x}_j}$ are known. We then use that
\be
\mathbf{x}_i \cdot \mathbf{x}_j = \frac{\abssq{\mathbf{x}_i} + \abssq{\mathbf{x}_j} - \abssq{\mathbf{x}_i - \mathbf{x}_j}}{2},
\label{eq:dotproduct}
\ee
which reduces our problem to finding the distance $\abssq{\mathbf{x}_i - \mathbf{x}_j}$. To find this distance, we first construct the two states
\bea
\ket{\psi} &=& \frac{1}{\sqrt{2}} \mleft( \ket{0} \ket{\mathbf{x}_i} + \ket{1} \ket{\mathbf{x}_j} \mright), \\
\ket{\phi} &=& \frac{1}{\sqrt{\abssq{\mathbf{x}_i} + \abssq{\mathbf{x}_j}}} \mleft( \abs{\mathbf{x}_i} \ket{0} - \abs{\mathbf{x}_j} \ket{1} \mright).
\eea
Note that we require QRAM to construct the state $\ket{\mathbf{x}_i}$. Next we do a ``swap test'' (see \figref{fig:SwapTest}) on $\ket{\phi}$ and the auxiliary qubit in $\ket{\psi}$. The probability of measuring 0 in the swap test is
\be
P (0) = \frac{1}{2} \mleft( 1 + \abs{\braket{\phi}{\psi}}^2 \mright) = \frac{1}{2} \mleft( 1 + \frac{1}{\sqrt{ 2 \mleft(\abssq{\mathbf{x}_i} + \abssq{\mathbf{x}_j} \mright)}} \abs{\mathbf{x}_i - \mathbf{x}_j} \mright) ,
\ee
which lets us extract $\abssq{\mathbf{x}_i - \mathbf{x}_j}$ and then calculate the dot product using \eqref{eq:dotproduct} [see also Ref.~\citep{Kopczyk2018} for an overview of this calculation]. The computational complexity for this distance calculation is $\mathcal{O} (\log N)$.

To solve the system of linear equations, we need to invert the matrix
\be
F = 
\begin{pmatrix}
0 & 1 \\
1 & K
\end{pmatrix}.
\ee
Briefly, this is done by approximating $\exp {- i F t}$ (which is not trivial, since $K$ is not sparse), and then using quantum phase estimation to extract eigenvalues and eigenvectors. These eigenvalues, together with $\mathbf{y}$ in the eigenbasis, let us construct the solution state
\be
\ket{b, \mathbf{\lambda}} \propto b \ket{0} + \sum_j \lambda_j \ket{j}.
\ee
The complexity for this part of the computation is $\mathcal{O} (\log M)$. The total complexity for the quantum SVM is thus $\mathcal{O} (\log NM)$.

\section{Quantum principal component analysis}

The methods applied in the quantum approach to SVMs can also be used in other machine-learning problems. For example, the algorithm for distance calculation can be applied to clustering. Another example is principal component analysis (PCA), where the second part of the quantum SVM algorithm can be re-used. In PCA, an unlabelled set of (high-dimensional) data points $\mathbf{x}_j$ is analyzed to find which are the axes along which the data varies the most (and which thus are most useful for classification). A nice example of PCA for physics researchers is Paperscape (\href{http://paperscape.org/}{paperscape.org}), which takes data from all papers on the arXiv and uses PCA to show the relations between the papers on a two-dimensional map.

A quantum algorithm for PCA~\citep{Lloyd2014} uses the matrix exponentiation and phase estimation from the quantum algorithm for SVMs to find the largest eigenvalues and eigenvectors of the covariance matrix $\sum_j \mathbf{x}_j \mathbf{x}_j^T$. Those eigenvectors are the principal components. We note here that there is recent work on ``quantum-inspired'' algorithms for PCA~\citep{Tang2018}, where the methods from the quantum algorithm have been adapted to find an improved (in the scaling of some parameters) classical algorithm.

\section{Quantum neural networks}

In this final section, we discuss quantum versions of neural networks. This is a quite new and rapidly developing field, so it is possible that these notes can become outdated fast. We therefore only try to give a few examples and discuss some general properties of quantum neural networks.

Although combining quantum computing and neural networks sounds interesting, given how they are both promising computing paradigms, it is not straightforward to do so. There are certainly potential upsides to quantum neural networks: they can work with quantum data, or a compact representation of classical data, and there may be quantum algorithms that can speed up training. However, there are several potential downsides or questions. For example, the input problem may be a factor here also. Furthermore, classical neural networks require nonlinear activation functions, but quantum mechanics is linear. Also, classical neural networks have a large number of neurons and many connections between them, which may be challenging for NISQ devices with few qubits and limited connectivity. Another question is whether back-propagation can be implemented in a quantum neural network, since classical back-propagation requires measuring the output at each point in the network, something that would destroy a quantum superposition.

\subsection{Quantum feedforward neural networks}

\begin{figure}
\centering
\includegraphics[width=0.9\linewidth]{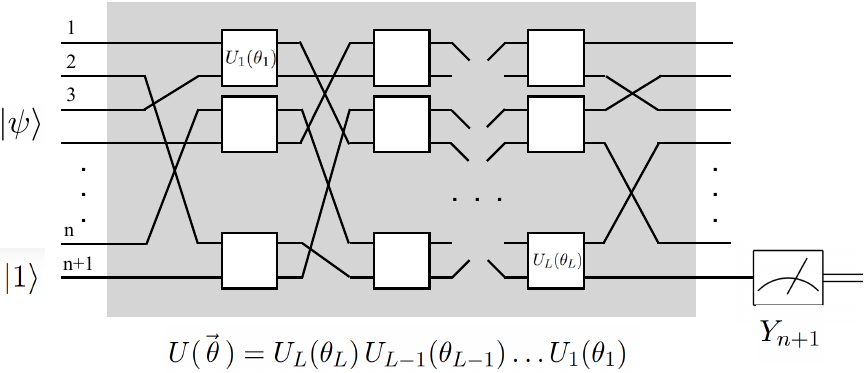}
\caption{A quantum feedforward neural network. Figure from Ref.~\citep{Farhi2018}.
\label{fig:QFFNN}}
\end{figure}

In 2018, Farhi (yes, the same Farhi that proposed the QAOA that you will meet in \secref{se:qaoa}) and Neven proposed an architecture for a quantum feedforward neural network~\citep{Farhi2018}, as depicted in \figref{fig:QFFNN}. Here, instead of layers of neurons, there are layers of quantum gates, which act on an $n$-qubit input state $\ket{\psi}$ and an auxiliary qubit prepared in $\ket{1}$. Instead of the weights and biases in a classical neural network, we now have parameters $\theta_j$ parameterizing these gates; the action of the whole quantum circuit is a unitary operation $U(\mathbf{\theta})$. At the end, the auxiliary qubit is measured (here, in the $Y$ basis). The outcome of the measurement is used to classify the input state, giving it one of two possible labels.

To train this quantum neural network for its classification task, the authors propose using a loss function
\be
C (\theta, z) = 1 - l(z) \brakket{z,1}{U^\dag (\theta) Y_{n+1} U (\theta)}{z, 1},
\ee
where $z$ is the input and $l(z)$ is a label function giving the correct label ($\pm 1$). This cost function is zero if the network spits out the correct classification, and greater than zero otherwise. The authors use stochastic gradient descent to find parameters $\mathbf{\theta}$ that minimize this cost function. However, no back-propagation is used (since this does not seem to be applicable to this architecture, as discussed above). Furthermore, to evaluate the gradient, multiple runs of the quantum circuit are required for each partial derivative, since one needs to collect enough statistics to find the expectation value of the output with sufficient precision. The authors point out that a possible advantage of the quantum network is that the form of the unitary $U$ guarantees that the gradient does not ``blow up'', which can be a problem in some classical machine-learning algorithms.

Since the quantum feedforward neural network lacks a nonlinear element, one can ask whether it has the ability to represent any label function. The authors show that the network indeed has this ability, but some label functions may require an exponential circuit depth.

\subsection{Quantum convolutional neural networks}

\begin{figure}
\centering
\includegraphics[width=0.85\linewidth]{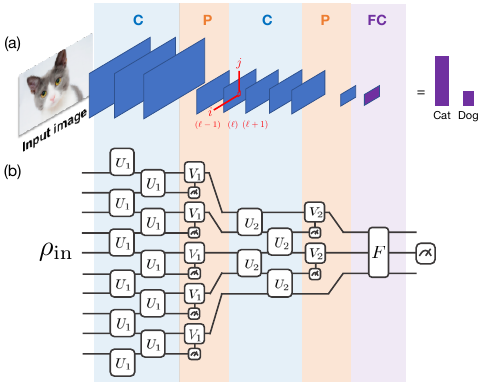}
\caption{Illustrations of (a) a classical convolutional neural network and (b) a quantum convolutional neural network. Figure from Ref.~\citep{Cong2019}.
\label{fig:QCNN}}
\end{figure}

Another recent proposal~\citep{Cong2019} for a quantum neural network is a quantum version of a convolutional neural network (CNN), illustrated in \figref{fig:QCNN}. Convolutional neural networks are often used for image recognition. As depicted in \figref{fig:QCNN}(a), a classical CNN consists of convolutional layers (C) that essentially scan a filter across the image, pooling layers (P) that reduce the size of the feature map produced by the convolution, and a final part with fully connected layers (FC) that do classification based on the features extracted in previous layers. In the quantum version [\figref{fig:QCNN}(b)], the filter in convolutional layer number $j$ is replaced by a two-qubit unitary operation $U_j$, which is applied to all pairs of neighbouring qubits. The pooling layer number $k$ is replaced by measuring half of the qubits and applying a unitary single-qubit operation $V_k$ to the remaining qubits, conditioned on the measurement outcome for the neighbouring qubit. This operation conditioned on measurement has the added benefit of adding a nonlinearity to the setup. Finally, the fully connected layer is replaced by a quantum circuit similar to that proposed by Farhi and Neven, as discussed in the preceding subsection. Similar to that proposal, the training here is also done by gradient descent without any back-propagation.

\subsection{Quantum Boltzmann machines}

\begin{figure}
\centering
\includegraphics[width=0.85\linewidth]{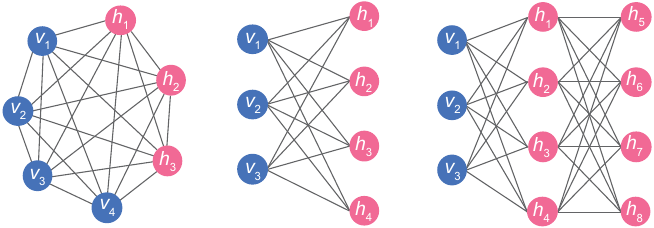}
\caption{Layouts for Boltzmann machines. Left: A fully connected Boltzmann machines, with connections both between the hidden and visible layers, and within the hidden and visible layers. Center: A restricted Boltzmann machine, with connections only going between the hidden and visible layers. Right: A deep Boltzmann machine, with multiple hidden layers. Figure from Ref.~\citep{Allcock2019}.
\label{fig:BMs}}
\end{figure}

A type of neural network that has been quite frequently studied in some quantum form is Boltzmann machines~\citep{Amin2018}, which come in a few different architectures, as shown in \figref{fig:BMs}. A Boltzmann machine has hidden ($\mathbf{h}$) and visible ($\mathbf{v}$) neurons, which take binary values (0 or 1). The aim of training a Boltzmann machine is to make the states of its visible neurons follow a probability distribution $P(\mathbf{v})$ that mimics that of the training data. This probability distribution is given by
\be
P (\mathbf{v}, \mathbf{h}) = \frac{1}{Z} e^{- E (\mathbf{v}, \mathbf{h})},
\ee
where $Z$ is a normalization factor (partition function) and the energy function is, for the case of a restricted Boltzmann machine (\figref{fig:BMs} center),
\be
E (\mathbf{v}, \mathbf{h}) = \sum_i a_i v_i + \sum_j b_j h_j + \sum_{i, j} v_i W_{ij} h_j,
\ee
with $a_i$ and $b_j$ biases, and $W$ a weight matrix. The form of the probability distribution (a Boltzmann distribution) explains the name Boltzmann machines.

To train the network, a cost function
\be
C (\mathbf{a}, \mathbf{b}, W) = - \sum_\mathbf{v} P_{\rm data}(\mathbf{v}) \log P(\mathbf{v})
\ee
is minimized using gradient descent, often aided by an efficient scheme for sampling from the training data. It is unclear if quantum algorithms can be used for speeding up training of classical Boltzmann machines.

To construct a quantum version of a Boltzmann machine, the neurons are replaced by spins and the energy function is given by a Hamiltonian for the spin system. It is believed that such quantum Boltzmann machines can represent some probability distributions, particularly ones arising in quantum systems, better than classical Boltzmann machines can (note that classical Boltzmann machines already are used for simulating some quantum systems well). However, the scaling properties and other performance metrics for quantum Boltzmann machines are still unclear. Furthermore, it seems that training of a quantum Boltzmann machine can be more complicated than the training of a classical one~\citep{Amin2018}. A potential upside is that quantum annealers could be used to implement quantum Boltzmann machines. 

\newpage

\section*{Exercises}

\begin{enumerate}

\item Explain what the ``input problem'' in quantum machine learning (and more generally in quantum computing) is.

\item Explain two possible obstacles (excluding the input problem in the preceding question) to achieving speed-ups in quantum versions of feedforward neural networks.

\end{enumerate}


\chapter{Measurement-based quantum computation}
\label{chp:MBQC}

The circuit model introduced and discussed in the preceding chapters is not the only way to perform quantum computation. In this and the next chapter, we will look at two other approaches: measurement-based quantum computation (MBQC) and adiabatic quantum computation. We will also see some additional examples later in the course; MBQC will resurface when we discuss quantum computation with continuous variables.

MBQC was first proposed by Raussendorf and Briegel in 2001~\citep{Raussendorf2001}. More details can be found in the follow-up paper~\citep{Raussendorf2003}. The first experimental demonstration of all basic components of MBQC was performed by the Zeilinger group in 2005~\citep{Walther2005}.

\section{The basic idea of MBQC}

In the circuit model of quantum computation, we prepare an initial state, apply a sequence of quantum gates to this state, and finally do some measurement on the output state to obtain the result of the computation. While this seems a natural order of operations, which agrees with our intuition for how a classical computation is performed, it is possible to mix up the order somewhat. 

In MBQC, we first prepare a certain entangled state of qubits, where a subset of these qubits represent the input state. This entangled state is the \textit{resource} for our computation. In the rest of the computation, we never need to apply any multi-qubit operations. All we do is apply single-qubit rotations and single-qubit measurements in some sequence, where later rotations are conditioned on earlier measurement results. The output state will be encoded in a subset of the qubits (different from the subset that encoded the input), and a final measurement can be performed on this output state to read out the result of the computation.

Because of the presence of measurements in earlier steps of the computation, MBQC differs from the circuit model for quantum computation in one important aspect: it is not reversible. In the circuit model, we apply a large unitary transformation (decomposed into a sequence of gates) that takes us from the input state to the output state. Before any measurement is done on the output state, this unitary transformation could be reversed to bring us back to the input state. However, since measurements are done (likely projecting some superposition states in a non-unitary fashion) before we reach the output state in MBQC, that computation cannot be reversed. For this reason, MBQC is often referred to as one-way quantum computation.

\section{The details of MBQC}
\label{sec-MBQC-dv}

We now present some details about MBQC. We first review what we need to create a universal gate set for qubits, then show how we can perform a single-qubit operation in the MBQC setting, build on this example to get to universal single-qubit operations for MBQC, discuss the concept of cluster states, and finally say how two-qubit gates can be realized in MBQC, completing a universal gate set.

\sse{Definitions of the possible operations}

Let us first recall from \chpref{chp:CircuitModel} some of the operations we can perform on single and multiple qubits.

\ssse{Single-qubit Clifford transformations}

Clifford operations $C$ are the unitary operations which map Pauli-group operators $\Sigma$ to Pauli-group operators $\Sigma'$ under conjugation, i.e., $C \Sigma C^\dagger =  \Sigma'$~\citep{horodecki2006lectures}.
\be
\lgr H, R_z(\pi/2) = Z_{\pi/2} \rgr \mbox{\bf{universal set for single-qubit Clifford operations}} \nn
\ee
%

\ssse{Multiqubit Clifford transformations}

The addition of any nontrivial two-qubit Clifford gate, e.g., $ \text{CZ} = \ketbra{0}{0} \otimes I +  \ketbra{1}{1} \otimes \hat Z $, combined with the set of single-qubit operations above, allows for any general multi-qubit Clifford operation.
\be
\lgr H, R_z(\pi/2) = Z_{\pi/2}, \text{CZ} \rgr \mbox{\bf{universal set for multi-qubit Clifford operations}} \nn
\ee
These operations are enough to perform some algorithms like error-correcting codes, but no algorithm which could not be efficiently simulated on a classical computer~\citep{horodecki2006lectures}, as we saw in the Gottesman--Knill theorem in \secref{sec:UniversalQComp}.

\ssse{Single-qubit universal transformations}

A general single-qubit transformation can be decomposed as $R_z (\gamma) R_x (\beta) R_z (\alpha)$. To implement any such arbitrary operation with arbitrary accuracy, it is sufficient to be able to perform the Clifford operations together with any non-Clifford operation provided, e.g.,
\be
\lgr H, Z_{\pi/2}, Z_{\pi/4} \rgr \mbox{\bf{universal set for single-qubit quantum computation}} \nn
\ee
%

\ssse{Multi-qubit universal transformations}

If we have universal control of single qubits, adding a nontrivial (entangling) two-qubit gate to the gate set is enough to achieve universality for multiple qubits. For example,
\be
\lgr  H, Z_{\pi/2},  Z_{\pi/4}, \text{CZ} \rgr   \mbox{   \bf{  universal set for multi-qubit quantum computation}}\nn
\ee
%
\sse{Preparing the initial state}

We first consider a setup with two qubits. One qubit contains the (arbitrary) initial state that we want to process: $\ket{\psi} = \alpha \ket{0} + \beta \ket{1}$. The other qubit is prepared in $\ket{+} = ( \ket{0} + \ket{1} ) / \sqrt{2}$, e.g., by applying a Hadamard gate to an initial state $\ket{0}$. We then apply a CZ gate between the two qubits, obtaining
\begin{align}
\text{CZ} \mleft( \ket{\psi} \otimes \ket{+} \mright) &= \mleft( \ketbra{0}{0} \otimes 1 + \ketbra{1}{1} \otimes Z \mright) \mleft[ \mleft( \alpha \ket{0} + \beta \ket{1} \mright) \otimes \frac{1}{\sqrt{2}} \mleft( \ket{0} + \ket{1} \mright) \mright] \nn \\
&= \mleft[ \alpha \ket{0} \otimes \frac{1}{\sqrt{2}} \mleft( \ket{0} + \ket{1} \mright) + \beta \ket{1} \otimes \frac{1}{\sqrt{2}} \mleft( \ket{0} - \ket{1} \mright) \mright] \nn \\
&= \alpha \ket{0} \ket{+} + \beta \ket{1} \ket{-} .
\end{align}

\sse{Measurements and their effect}

The next step is to measure the input qubit in some rotated basis. Measuring in a rotated basis with angles $(\theta,\phi)$ is like measuring $R_z(\phi + \pi/2) R_x(\theta) Z R_x(- \theta) R_z(-\phi - \pi/2)$. Here we measure the first qubit with $\theta = \pi/2$ and an arbitrary $\phi$, i.e., we measure the observable
\begin{align}
\label{eq:rotated_obs}
\hat {\sigma}_\phi &\equiv R_z(\phi + \pi/2) R_x(\pi/2) Z R_x(- \pi/2) R_z(-\phi - \pi/2) \nn \\
&= \cos \phi X + \sin \phi Y \nn \\
&= e^{-i \phi} \ketbra{0}{1} + e^{i \phi} \ketbra{1}{0} = \ketbra{\phi_+}{\phi_+} - \ketbra{\phi_-}{\phi_-} ,
\end{align}
where we have used Eqs.~(\ref{eq:PauliX})--(\ref{eq:RotX}) and introduced the rotated measurement basis $\ket{\phi_\pm} = 1/\sqrt{2} ( \ket{0} \pm e^{i \phi} \ket{1} )$. To obtain the result, note that conversely $\ket{0} = \frac{1}{\sqrt{2}} ( \ket{\phi_+} + \ket{\phi_-} )$ while $\ket{1} = \frac{1}{\sqrt{2}} e^{- i \phi} ( \ket{\phi_+} - \ket{\phi_-} )$, and rewrite the state:
\begin{align}
\alpha \ket{0} \ket{+} + \beta \ket{1} \ket{-} &= \frac{\alpha}{\sqrt{2}} \mleft( \ket{\phi_+} + \ket{\phi_-} \mright) \ket{+} + \frac{\beta}{\sqrt{2}} e^{- i \phi} \mleft( \ket{\phi_+} - \ket{\phi_-} \mright) \ket{-} \nn \\
&= \ket{\phi_+} \mleft( \frac{\alpha}{\sqrt{2}} \ket{+} + \frac{\beta}{\sqrt{2}} e^{- i \phi} \ket{-} \mright) + \ket{\phi_-} \mleft( \frac{\alpha}{\sqrt{2}} \ket{+} - \frac{\beta}{\sqrt{2}} e^{- i \phi} \ket{-} \mright) .
\end{align}
Hence a measurement of the operator in \eqref{eq:rotated_obs} projects the second qubit into:
\begin{itemize}
 \item $\ket{\psi}_{\text{out}} \propto \mleft( \alpha \ket{+} + \beta e^{- i \phi} \ket{-} \mright)$ if the outcome is $1$ ($m = 0$)
 \item $\ket{\psi}_{\text{out}} \propto \mleft( \alpha \ket{+} - \beta e^{- i \phi} \ket{-} \mright)$ if the outcome is $-1$  ($m = 1$) 
 \end{itemize}
The state of the second qubit can then be compactly written as
\be
\ket{\psi}_{\text{out}} = X^m H R_z (- 2 \phi) \ket{\psi}.
\ee
This is readily verified, since
\begin{align}
X^m H R_z (- 2 \phi) \mleft( \alpha \ket{0} + \beta \ket{1} \mright) &= X^m H \mleft( \alpha e^{i \phi} \ket{0} + \beta e^{-i \phi} \ket{1} \mright) \nn \\
&= X^m \mleft( \alpha e^{i \phi} \ket{+} + \beta e^{-i \phi} \ket{-} \mright),
\end{align}
giving the result. The extra Pauli operator $X^m$ depends on the outcome of the measurement on qubit $1$ and is said to be a ``by-product'' operator. It can be compensated for by choosing the measurement basis of the following steps in the computation (thus introducing, in general, time ordering). In the following, we rename $- 2 \phi \rightarrow \phi$.

\sse{Universal single-qubit operations}

We can repeat the preceding protocol three times in a chain of four qubits, first using qubit 1 as input and qubit 2 as output, then using qubit 2 as input and qubit 3 as output, and so on. The final output state (qubit 4) then becomes
\begin{align}
| \psi \rangle_{\text{out}} &= X^{m_3} H R_z (\phi_3) X^{m_2} H R_z (\phi_2) X^{m_1} H R_z (\phi_1) \ket{\psi} \nn \\
&= H Z^{m_3} R_z (\phi_3) H Z^{m_2} R_z (\phi_2) H Z^{m_1} R_z (\phi_1) \ket{\psi} \nn \\
&= H Z^{m_3} R_z (\phi_3) [ H Z^{m_2} H ] [ H R_z (\phi_2) H ] Z^{m_1} R_z (\phi_1) \ket{\psi} \nn \\
&= H Z^{m_3} R_z (\phi_3) X^{m_2} R_x (\phi_2) Z^{m_1} R_z (\phi_1) \ket{\psi} \nn \\
&= X^{m_3} Z^{m_2} X^{m_1} H R_z ((-1)^{m_2} \phi_3) R_x ((-1)^{m_1} \phi_2)  R_z (\phi_1) \ket{\psi},
\end{align}
where in the first step we used that $X^{m} H = H Z^{m}$, in the third that $H Z^{m} H = X^m$, and later on that $X R_z (\phi) = R_z  (-\phi) X$ and $Z R_x (\phi) = R_x (-\phi) Z$.

Since the most general rotation of a single qubit can be decomposed as $R_z (\gamma) R_x (\beta)  R_z (\alpha)$, the above steps let us implement any single-qubit operation. For the result to be deterministic, we can perform the measurements choosing the basis sequentially, depending on the preceding measurement outcomes, as $\phi_1 = \alpha$, $\phi_2 = (-1)^{m_1} \beta$, and $\phi_3 = (-1)^{m_2} \gamma$. The Pauli corrections remaining at the end of the computation are not important and never need to be physically applied; they can be accounted for in the final interpretation of the result (classical postprocessing).

If we want to implement a Clifford unitary, by definition $C \Sigma = \Sigma' C$, meaning that interchanging the order of Clifford operators and Pauli matrices will leave the Clifford operator unchanged. This means that there is no need to choose measurements adaptively.

\sse{Cluster states as a resource}

In the example above, the four qubits were entangled through CZ gates between nearest neighbours. These CZ gates could all have been done before the start of the computation (when qubit 1 was in the input state and qubits 2--4 were in the state $\ket{+}$). This entangled initial state $\text{CZ}_{\rm neighbours}\ket{\psi + + +}$ would then be a one-dimensional \textit{cluster state}, which is the resource enabling the rest of the computation being carried out by just measurements and a limited set of single-qubit operations. 

More generally, a state like this, which allows any input state and any unitary transformation on that input (using only single-qubit rotations and measurements), is said to be a universal resource. It has been shown that a square lattice graph (a cluster state) with unit weights is a universal resource for quantum computation. Despite this fact, depending on the specific kind of computation, other graphs than a square lattice could be more suitable for implementing the computation~\citep{horodecki2006lectures}.

The concept of cluster states, and MBQC, is important also for quantum computing with continuous variables. You will encounter these concepts again in that context in \secref{se:q_comp_cv}.

\sse{Two-qubit gates}

For universal quantum computation, we need some two-qubit gate in addition to the arbitrary single-qubit rotations that we constructed above. Such two-qubit gates, e.g., the CZ and CNOT gates, can be constructed in a two-dimensional cluster state where two input qubits are entangled with a few other qubits. By a series of single-qubit measurements and rotations, we can end up with two of the other qubits representing the output state corresponding to the two-qubit gate having acted on the input state. The example of the CNOT gate will be demonstrated in~\nameref{tutorial3}.

\section{Universality and efficiency}

From the previous section, it is clear that we can implement arbitrary single-qubit rotations and a two-qubit gate (e.g., CZ or CNOT), which together form a universal gate set for quantum computation (cf.~\chpref{chp:CircuitModel}). Thus, MBQC can implement universal quantum computation.

However, it is clear that MBQC, in general, requires more qubits than the circuit model does. It is also necessary to compare the number of qubit rotations and measurements needed in MBQC and in the circuit model. If MBQC would turn out to simply be an inefficient reformulation of the circuit model, it would not be useful. Fortunately, it turns out that the overhead of MBQC is polynomial in all these resources, so MBQC is an efficient paradigm for quantum computation.

The efficiency of MBQC opens up new avenues for practical implementations of quantum computation. There may be physical systems where it is hard to implement single two-qubit gates between chosen qubits at chosen times, but where it is easier to create a large entangled state in one big operation and then only perform single-qubit operations.

As an aside, we note that MBQC usually has been considered for two-dimensional cluster states. However, by working with cluster states in three dimensions, the quantum computation can be made fault-tolerant~\citep{Briegel2009}.

\newpage

\section*{Exercises}

\begin{enumerate}

\item Consider two qubits, where qubit 1 is initialized in a state $\ket{\psi}$ and qubit 2 is initialized in state $\ket{0}$. We will now implement a Hadamard gate using measurement-based quantum computing. The procedure is the following:

\begin{enumerate}
\item Apply a Hadamard gate to qubit 2.
\item Apply a CZ gate to qubits 1 and 2.
\item Measure qubit 1 in the $X$ basis.
\item Apply a feedback operation $F$, conditioned on the measurement result, on qubit 2 to obtain the desired output state.
\end{enumerate}

What should the feedback operation $F$ be in order for the final state of qubit 2 to be the result of applying a Hadamard gate to $\ket{\psi}$?

\item 
\begin{figure}[h!]
\centering
\begin{minipage}{.45\textwidth}
\centering
\includegraphics[width=0.85\linewidth]{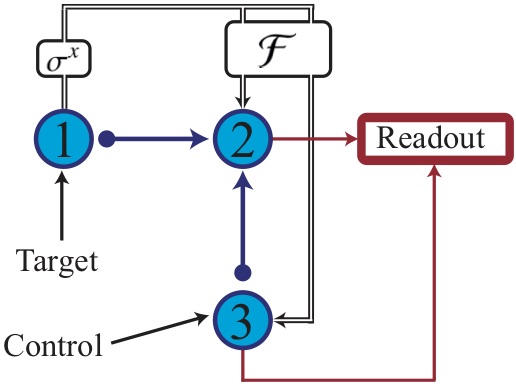}
\caption{A scheme for MBQC. \label{fig:MBQC}}
\end{minipage}%
\begin{minipage}{.45\textwidth}
\centering
\begin{quantikz}
    \lstick{Qubit 1} & \gate{Z} & \ctrl{1} & \gate{H} & \ctrl{1} & &\\
    \lstick{Qubit 2} & \gate{H} & \gate{Z} & & \gate{Z} & \gate{H} &\\
\end{quantikz}
\caption{An entangling gate $U_{1,2}$. \label{fig:U12}}
\end{minipage}
\end{figure}

The setup for one way of implementing a CNOT gate in measurement-based quantum computing is shown in \figref{fig:MBQC}. The incoming state is represented by qubits 3 (control) and 1 (target). The outgoing state will be encoded in qubits 3 (control) and 2 (target). The protocol is:
\begin{enumerate}
\item Initialize qubits 1 and 3 as the input state, and qubit 2 in $\ket{0}$.
\item Apply the entangling gate $U_{1,2}$ shown in \figref{fig:U12} to qubits 1 and 2.
\item Apply the CZ gate to qubits 2 and 3.
\item Apply H to qubit 1.
\item Measure qubit 1 in the computational basis, obtaining the result $m = 0$ or $m = 1$.
\item Apply H to qubit 2.
\item Apply a feedback operation $F$ on the state of qubits 2 and 3 to obtain the desired output state.
\end{enumerate}
What should $F$ in the last step be to achieve the CNOT gate?

\item Compute the state of a 2-qubit linear cluster state and the one of a 3-qubit linear cluster state.

\end{enumerate}


\chapter{Adiabatic quantum computation}
\label{ch-aqc}

The MBQC that we considered in \chpref{chp:MBQC} was, although distinct from the circuit model for quantum computation, still rather similar to that circuit model. The paradigm of quantum computation that we explore in this chapter, adiabatic quantum computation (AQC), is further removed from the circuit model than MBQC. The idea for AQC was evolved around the turn of the millennium. A recent extensive review of the topic is Ref.~\citep{albash2018adiabatic}, from which we quote several of the following considerations.

\section{The basic idea of AQC}

Adiabatic quantum computation is based on the \emph{adiabatic theorem} (we give a proof for this theorem in \secref{sec:ProofAdiabaticTheorem}). The theorem states that a system which starts in a nondegenerate ground state of a time-dependent Hamiltonian $\hat{\mathcal{H}}(t)$ that is \emph{slowly} changing from some initial form $\hat{\mathcal{H}}_0$ to some final form $\hat{\mathcal{H}}_1$, during time $\tau$, will remain in its instantaneous ground state throughout the evolution, provided that the Hamiltonian varies sufficiently slowly (adiabatically). Assuming a linear time dependence, the AQC Hamiltonian can be written as
\be
\label{eq:AdiabaticQuantumComputation}
\hat{\mathcal{H}}(t) = \mleft( 1 - \frac{t}{\tau} \mright) \hat{\mathcal{H}}_0 + \frac{t}{\tau} \hat{\mathcal{H}}_1, \quad (0 \leq t \leq \tau),
\ee
where the coefficient in front of $\hat{\mathcal{H}}_0$ changes from unity to zero, and the coefficient in front of $\hat{\mathcal{H}}_1$ changes from zero to unity. Moreover, it is crucial that $\hat{\mathcal{H}}_0$ and $\hat{\mathcal{H}}_1$ are two noncommuting Hamiltonians (see \secref{sec:NoncommutingHamiltonians}). 

If the initial Hamiltonian $\hat{\mathcal{H}}_0$ is such that its ground state is easy to construct, it is easy to initialize the system in this ground state. The Hamiltonian $\hat{\mathcal{H}}_1$ is chosen such that its ground state encodes the solution of a certain problem. This means that the adiabatic evolution will provide us with this solution.

Adiabatic quantum computation is closely related to quantum annealing (QA), which also encodes the solution to a problem in the ground state of some final Hamiltonian. The relation between AQC and QA is discussed later in the course.

\section{Adiabatic evolution and quantum speed-up}

What is the minimum evolution time $\tau$, such that the time evolution in \eqref{eq:AdiabaticQuantumComputation} is adiabatic? According to the adiabatic theorem, it can be shown that for a nondegenerate ground state, adiabatic evolution is assured if the evolution time $\tau$ satisfies the condition
\be
\label{eq:AdiabaticCondition}
\tau \gg 
\frac{\underset{0\leq s \leq 1}{\text{max}} \hbar \abs{\bra{\psi_1(s)} \dfrac{\partial \hat{\mathcal{H}}(s)}{\partial s}\ket{\psi_0(s)}}}
{\underset{0\leq s \leq 1}{\text{min}}\Delta^2(s)}; \qquad s\equiv\frac{t}{\tau},
\ee
where $\ket{\psi_0(s)}$ and $\ket{\psi_1(s)}$ are the instantaneous ground state and the first excited eigenstate, respectively, of the Hamiltonian in \eqref{eq:AdiabaticQuantumComputation}, and $\Delta(s) = E_1(s) - E_0(s)$ is the instantaneous energy gap between the ground-state and first-excited-state energies. If the criterion of \eqref{eq:AdiabaticCondition} is fulfilled, then we can be certain that the system will evolve into the ground state of $\hat{\mathcal{H}}_1$.

While \eqref{eq:AdiabaticQuantumComputation} is a useful sufficient condition, checking whether it is satisfied means, in general, bounding the minimum energy gap between $\ket{\psi_0(s)}$ and $\ket{\psi_1(s)}$ of a complicated many-body Hamiltonian, which is a notoriously difficult problem. On the other hand, this complication for AQC has also generated much interest among physicists, since it connects AQC to well-studied problems in condensed matter physics. For example, due to the dependence of the run time on the energy gap, the performance of AQC algorithms is related to the type of quantum phase transition that the system would undergo in the thermodynamic limit.

There are some known examples where the energy gap can be analyzed. For example, AQC can implement a process analogous to Grover's search algorithm (see \chpref{chp:Grover}), which means that it can provide a quadratic speedup over the best possible classical algorithms for search problems. Some of the other examples where the gap analysis can be performed also demonstrate that AQC can provide a speedup over classical computation; these examples include adiabatic versions of some key algorithms in the circuit model. 

However, the most common scenario is that either analyzing the energy gap shows no speedup for AQC over classical computation, or that it is not possible to clearly answer whether there is a speedup. Early motivation for the development of AQC was the hope of solving combinatorial optimization problems~\citep{Farhi2001}, but the question of speedups here is not resolved and therefore remains an area of very active research, partly due the availability of commercial QA devices, e.g., developed by D-Wave Systems Inc. We will return to quantum algorithms for combinatorial optimization in \chpref{chap:OptimizationProblems}.

\section{Universality and stoquasticity}

Adiabatic quantum computation is a universal model of quantum computation. Problems that are solvable in polynomial time with the circuit model can be solved in polynomial time with an adiabatic quantum computer, and vice versa. The fact that the circuit model can efficiently simulate AQC was proven by Edward Fahri et al.~in 2000~\citep{Farhi2000}. The fact that AQC can efficiently simulate the circuit model was proven by Dorit Aharonov et al.~in 2004~\citep{Aharonov2004}. As such, the two models are computationally equivalent. The proofs for these statements can be found in Ref.~\citep{albash2018adiabatic}.

When discussing universality and efficiency for AQC, it is important to distinguish whether the final Hamiltonian $\hat{\mathcal{H}}_1$ is \textit{stoquastic} or not. The commercial devices designed to solve optimization problems with QA are made for stoquastic Hamiltonians. A Hamiltonian $\hat{\mathcal{H}}$ is called stoquastic if it only has real-valued elements, and, in some local basis $B$, all off-diagonal elements of $\hat{\mathcal{H}}$ are either zero or negative, i.e., $\brakket{x}{\hat{\mathcal{H}}}{y} \leq 0 \:  \forall x,y \in B \land x \neq y$. Stoquastic Hamiltonians seem to exist in the borderlands between the classical and quantum worlds. It is an open question whether stoquastic AQC can be efficiently simulated by a classical computer. It has been shown that stoquastic AQC cannot be universal for quantum computation unless the polynomial hierarchy collapses (in \chpref{chp:complexity-classes}, we discuss this and other terminology for computational complexity). The proof by Aharonov et al.~of the universality for AQC thus requires non-stoquastic Hamiltonians.

\section{Reason for non-commuting Hamiltonians}
\label{sec:NoncommutingHamiltonians}

As mentioned above, it is important in AQC that the initial Hamiltonian $\hat{\mathcal{H}}_0$ and the final Hamiltonian $\hat{\mathcal{H}}_1$ do not commute, i.e., $\comm{\hat{\mathcal{H}}_0}{\hat{\mathcal{H}}_1}\neq 0$. This can be understood by considering the following trivial example, taken from Pontus Vikst\aa l's master's thesis, Ref.~\citep{Pontus-Master-Thesis}. Suppose that the initial and final Hamiltonian in an adiabatic quantum computation are given by
\be
\hat{\mathcal{H}}_0 = 
\begin{pmatrix}
1 & 0 \\
0 & -1
\end{pmatrix}
\quad\text{and}\quad
\hat{\mathcal{H}}_1 =
\begin{pmatrix}
-1 & 0 \\
0 & -\frac{1}{2}
\end{pmatrix},
\ee
which clearly commute. Since the Hamiltonians are both diagonal in the $z$ basis, we label the corresponding eigenvectors as
\be
\ket{0} =
\begin{pmatrix}
1 \\
0
\end{pmatrix}
\quad\text{and}\quad
\ket{1} =
\begin{pmatrix}
0 \\
1
\end{pmatrix}.
\ee
It is easy to see that the ground state of $\hat{\mathcal{H}}_0$ is $\ket{1}$ and that the ground state of $\hat{\mathcal{H}}_1$ is $\ket{0}$. We described above how AQC is based on the adiabatic theorem. For the theorem to hold there must always exist an energy gap between the eigenstates (see the proof below in \secref{sec:ProofAdiabaticTheorem}). Following the AQC algorithm \eqref{eq:AdiabaticQuantumComputation}, it can be seen that the energy gap between the eigenstates $\ket{1}$ and $\ket{0}$ closes at some point for the Hamiltonians considered here. Specifically, in this example, the energies becomes equal at the point $t/\tau=4/5$, and so the gap between them closes. 

More generally, if two matrices $A$ and $B$ commute, they have the same eigenvectors. This means that if $\comm{\hat{\mathcal{H}}_0}{\hat{\mathcal{H}}_1} = 0$, then either $\hat{\mathcal{H}}_0$ and $\hat{\mathcal{H}}_1$ have the same ground state, in which case there is no need to run a computation, or they have different ground states (which correspond to common eigenvectors of the two Hamiltonians) and end up in the situation described above when trying to run AQC. This is enough to see that $\comm{\hat{\mathcal{H}}_0}{\hat{\mathcal{H}}_1}\neq 0$ is a necessary condition for keeping the gap open and allowing AQC.

\section{Proof of the adiabatic theorem}
\label{sec:ProofAdiabaticTheorem}

\label{sec:AdiabaticTheorem}

In this section, we provide a simple and straightforward proof of the famous \emph{adiabatic theorem}, first published by Max Born and Vladimir Fock in 1928. This section is taken from Pontus Vikst\aa l's master's thesis, Ref.~\citep{Pontus-Master-Thesis}. \\

\noindent \textbf{Theorem:} A particle that begins from the $n$th eigenstate of a Hamiltonian that is gradually (adiabatically) changing from an initial form $\hat{\mathcal{H}}_i$ into a final form $\hat{\mathcal{H}}_f$, will remain in the $n$th eigenstate. \\ 

\noindent \textbf{Proof:} Consider an arbitrary time-independent Hamiltonian $\hat{\mathcal{H}}$, for which the Schr\"{o}dinger equation reads
\be
\label{eq:SchrodingerEquation}
i\hbar\frac{d\Psi(x,t)}{dt} = \hat{\mathcal{H}} \Psi(x,t).
\ee
Through separation of variables, the time-independent Schr\"{o}dinger equation can be obtained:
\be
\hat{\mathcal{H}} \psi_n(x) = E_n \psi_n(x).
\ee
The general solution to the Schr\"{o}dinger equation is given by a superposition of the separable solutions
\be
\Psi(x,t) = \sum_n c_n \Psi_n(x,t) = \sum_n c_n \psi_n(x) e^{-i E_n t/\hbar},
\ee
or by simply considering a specific eigenfunction
\be
\Psi_n(x,t) = \psi_n(x) e^{-i E_n t/\hbar}.
\ee
Hence, the $n$th eigenstate for a time-independent Hamiltonian remains in the $n$th eigenstate, simply picking up a phase factor $-E_nt/\hbar$. 

For a time-dependent Hamiltonian, the eigenenergies and eigenfunctions are themselves time-dependent. The instantaneous eigenstates and eigenenergies are defined as
\be
\hat{\mathcal{H}}(t) \psi_n(x,t) = E_n(t) \psi_n(x,t).
\ee
At any instant of time, the eigenfunctions form a complete orthogonal set
\be
\label{eq:CompleteSet}
\braket{\psi_m(t)}{\psi_n(t)} = \delta_{mn},
\ee
where the dependence on position is implicit. We have also introduced the bra-ket notation $\psi_n(x) = \braket{x}{\psi_n}$. The general solution to the Schr\"{o}dinger equation is now given by
\be
\label{eq:GeneralTimeDepSchrodinger}
\Psi(x,t) = \sum_n c_n(t) \Psi_n(x,t) = \sum_n c_n(t) \psi_n(x,t) e^{i \theta_n(t)},
\ee
where $\theta_n(t)$ is known as the \emph{dynamical phase} factor;
\be
\theta_n(t) = -\frac{1}{\hbar} \int^t_0 E_n(s) \dd s.
\ee

Our task is to determine the coefficients $c_n(t)$. By substituting (\ref{eq:GeneralTimeDepSchrodinger}) into the Schr\"{o}dinger equation, we obtain
\be
i\hbar \sum_n \mleft( \dot{c}_n \psi_n + c_n \dot{\psi}_n + i c_n \psi_n \dot{\theta}_n \mright) e^{i \theta_n} = \sum_n c_n E_n \psi_n e^{i \theta_n}.
\ee
The third term on the left cancels the term on the right, since $\dot{\theta}_n = - E_n / \hbar$. We are thus left with
\be
i\hbar \sum_n \mleft( \dot{c}_n \psi_n + c_n \dot{\psi}_n \mright) e^{i \theta_n} = 0.
\ee
Multiplying with an arbitrary eigenfunction $\bra{\psi_m}$ from the left and using the orthogonality condition of \eqref{eq:CompleteSet} yields
\be
\dot{c}_m = -\sum_n c_n \braket{\psi_m}{\dot{\psi}_n} e^{i(\theta_n - \theta_m)}.
\label{eq:cmdot}
\ee

To calculate the quantity $\braket{\psi_m}{\dot{\psi}_n}$, we first observe that, for $m \neq n$,
\begin{align}
\frac{d}{dt} \mleft( \bra{\psi_m} \hat{\mathcal{H}} \ket{\psi_n} \mright) = 0
&= \bra{\dot{\psi}_m} \underbrace{\hat{\mathcal{H}} \ket{\psi_n}}_{E_n \ket{\psi_n}}
+ \bra{\psi_m} \dot{\hat{\mathcal{H}}} \ket{\psi_n}
+ \underbrace{\bra{\psi_m} \hat{\mathcal{H}}}_{E_m \bra{\psi_m}} \ket{\dot{\psi}_n} \nn \\
&= E_n \braket{\dot{\psi}_m}{\psi_n}
+ \bra{\psi_m} \dot{\hat{\mathcal{H}}} \ket{\psi_n}
+ E_m \braket{\psi_m}{\dot{\psi}_n} .
\end{align}
We then note that
\be
\frac{d}{dt} ( \underbrace{\braket{\psi_m}{\psi_n}}_{\delta_{mn}} ) = 0 
= \braket{\dot{\psi}_m}{\psi_n} + \braket{\psi_m}{\dot{\psi}_n} ,
\ee
which implies the relation $\braket{\dot{\psi}_m}{\psi_n} = - \braket{\psi_m}{\dot{\psi}_n}$, so
\be
\braket{\psi_m}{\dot{\psi}_n} = \frac{\bra{\psi_m} \dot{\hat{\mathcal{H}}} \ket{\psi_n}}{E_n - E_m} , \quad (m \neq n) .
\ee
This holds as long as no transitions between eigenstates occur. 

The differential equation in \eqref{eq:cmdot} can now be written
\be
\label{eq:AdiabaticApproximation}
\dot{c}_m = - c_m \braket{\psi_m}{\dot{\psi}_m} - \sum_{m \neq n} \frac{\bra{\psi_m} \dot{\hat{\mathcal{H}}} \ket{\psi_n}}{E_n - E_m} .
\ee
Now, if the Hamiltonian is slowly changing, such that its time derivative can be considered to be very small and that the energy difference $\abs{E_n - E_m}$ is large compared to $\abs{\bra{\psi_m} \dot{\hat{\mathcal{H}}} \ket{\psi_n}}$, the second term on the right-hand side of in \eqref{eq:AdiabaticApproximation} becomes negligible. This approximation is known as the \emph{the adiabatic approximation}. It lets us conclude that
\be
\dot{c}_m \approx - c_m \braket{\psi_m}{\dot{\psi}_m} .
\ee
By solving this equation, we find
\be
c_m(t) = c_m(0) \exp \mleft( - \int^t_0 \braket{\psi_m(s)}{\dot{\psi}_m(s)} \dd s \mright) = c_m(0) e^{i \gamma_m(t)},
\ee
where
\be
\gamma_m(t) = i \int^t_0 \braket{\psi_m(s)}{\dot{\psi}_m(s)} \dd s
\ee
is the \emph{geometrical} (Berry) phase factor. Putting the obtained expression for the coefficients $c_m(t)$ back into \eqref{eq:GeneralTimeDepSchrodinger}, we obtain that the $n$th eigenstate is given by
\be
\ket{\Psi_n(t)} = e^{i \theta_n(t)} e^{i \gamma_n(t)} \ket{\psi_n(t)}.
\ee
Hence, a system that starts out in the $n$th eigenstate, will remain in the $n$th eigenstate, simply picking up some phase factors.

\newpage

\section*{Exercises}

\begin{enumerate}

\item

Why is it important that the initial and final Hamiltonians used in adiabatic quantum computation do not commute?

\end{enumerate}


\chapter{Quantum complexity theory}
\label{chp:complexity-classes}

The time it takes to solve a computational problem can scale differently with the problem size, and can depend on what type of computational machine is used. This is the basis for defining computational complexity classes, which are the topic of this section. We begin with an overview of definitions of various complexity classes, and then zoom in on the class of problems known as combinatorial optimization problems, studying their computational complexity.

\section{Complexity classes and conjectures}

In this section, we define complexity classes for both classical and quantum computers. Most of the following is based on the supplementary material of Ref.~\citep{Douce2017} and Section 2 in Ref.~\citep{Wendin2017}. A comprehensive discussion of complexity classes can be found in the book by Arora and Barak, the draft version of which is available at \url{https://theory.cs.princeton.edu/complexity/book.pdf}.

For classical computers, the scaling with problem size of the time it takes to solve a problem is usually defined either for a deteriministic Turing machine (DTM) or a probabilistic Turing machine (PTM). The DTM is the model that corresponds to ordinary classical computers; it is a finite state machine that reads and writes on a infinite tape. 

\sse{Complexity classes for a deterministic Turing machine}

The most basic and well-known complexity classes are those defined for a DTM:
\begin{itemize}
\item P (polynomial): The set of decision problems solvable in polynomial (poly) time by a DTM.
\item NP (non-deterministic polynomial): The set of decision problems whose solutions can be verified in polynomial time by a DTM.
\item NP-hard: The set of problems whose solutions allows solving all problems in NP.
\item NP-complete: The set of problems that are in NP and whose solutions allows solving all problems in NP.
\item $\#$P: The set of problems associated with counting the number of solutions of NP problems.
\item $\#$P-hard: The set of problems whose solutions allows solving all other problems in $\#$P.
\end{itemize}

It is known that P $\subseteq$ NP, but the question whether the inclusion holds strictly (and hence ultimately P $\neq$ NP) stands as one of the most important open problems in the modern age of science.

A concept that is often mentioned in connection to P and NP is the polynomial hierarchy (PH). It is a hierarchy of complexity classes that generalizes the classes P, NP to the case in which oracles are accessible. An oracle is a black box that can output the solution of a decision problem contained in a given complexity class using a single call. For example, A$^\text{B}$ is the set of decision problems that belong to class A when solved by a DTM augmented by an oracle for some complete problem in class B.

The first level of the PH is the class P; in symbols, $\Sigma_0 = \text{P}$. Successive levels are refined recursively:
\be
\Sigma_{k+1} = \text{NP}^{\Sigma_{k}}.
\ee
%
A problem is in the polynomial hierarchy if it is in some ${\Sigma_{k}}$, i.e., the polynomial hierarchy is the union of all ${\Sigma_{k}}$. 

Analogously to what was said above concerning the relation between P and NP, it is known that $\Sigma_{k} \subseteq \Sigma_{k + 1}$, i.e., higher levels of the PH contain lower levels, and it is strongly believed that the inclusion is strict, namely that $\Sigma_{k} \neq \Sigma_{k + 1}$. If there is a $k$ for which $\Sigma_{k} = \Sigma_{k + 1}$, the PH is said to collapse to level $k$. It can be shown that if a collapse occurs at level $k$ then for all $k' > k$ it would hold that $ \Sigma_{k'} = \Sigma_{k} $, which justifies the terminology ``collapse''.

\sse{Complexity classes for a probabilistic Turing machine}

A PTM is much like a DTM, but it makes random choices of the state of the finite state machine when reading from the tape, and it traverses all states in a random sequence. This randomness means that a PTM will not get stuck away from a solution, which could happen for a DTM. This leads to the definition of the following complexity classes:
\begin{itemize}
\item BPP (bounded probabilistic polynomial): The class of decision problems that a PTM solves in polynomial time with a success probability (i.e., probability of accepting a correct answer and of rejecting an incorrect answer) larger than or equal to 2/3. In other words, the error probability is bounded by (i.e., strictly less than) 1/3 for all instances. The postselected version of BPP, i.e., PostBPP, is contained in the third level of the polynomial hierarchy. 

\item PP (probabilistic polynomial): The class of decision problems that a PTM solves in polynomial time with a success probability larger than 1/2. In other words, the error probability is less than or equal to 1/2 for all instances. Toda's theorem states that PH $\subseteq \mathrm{P}^{\mathrm{PP}}$.
\end{itemize}
%

Several arguments point towards that BPP = P, which therefore stands as a widely believed conjecture. Some evidence is based on the fact that for many problems that were known to be solvable with randomized algorithms, deterministic algorithms to solve the same problems were discovered a few years later ("de-randomization"). Another argument is that the random-number generators used in our laptops to implement randomized algorithms are actually relying on pseudo-random numbers, i.e., numbers (such as the digits of $\pi$) that pass statistical tests for random numbers, but that are in fact deterministically available, e.g., in the RAM of the processor used to implement the randomized algorithm. As such, true randomness might actually not be needed to solve these problems.

\sse{Complexity classes for a quantum Turing machine}
\label{sse:complexity}

We can define a quantum Turing machine (QTM) in analogy with the classical Turing machines, using a quantum memory (tape) and a quantum processor. We then have a few new complexity classes:
\begin{itemize}
\item BQP (bounded quantum polynomial): This is the quantum analogue of BPP. Intuitively, BQP is the class of problems that can be solved using at most a polynomial number of gates, with at most $1/3$ probability of error.
\item PostBQP (Post-selected bounded quantum polynomial): PostBQP is an extension of BQP where, during a polynomial time computation, one is allowed to abort and start all over again for free whenever the result on a specific conditioning qubit (or subset of qubits) is not satisfying. Scott Aaronson has shown that PostBQP = PP, thereby relating a quantum complexity class to a classical one.
\item QMA (quantum Merlin--Arthur): The quantum analog of NP problems can be viewed as an instance of a so-called Merlin--Arthur protocol. For a given problem, Merlin, an all-powerful being, provides a candidate solution to Arthur, a polynomial-time classical algorithm (that can also use randomness). Arthur can then verify the solution and either accept or reject. For QMA, Merlin is allowed to send quantum states to Arthur and Arthur's verification procedure can now be performed on a quantum computer running in polynomial time.
\item QMA-hard: The set of decision problems whose solutions allows solving all problems in QMA.
\end{itemize}

When we talk about problems being efficiently solved by a universal quantum computer, the class we refer to is BQP. Note that we do not have to specify which gates the definition is based upon, as long as they constitute a universal set. Thanks to the Solovay--Kitaev theorem (see \secref{sec:SolovayKitaev}), using one universal set or another merely results in a polylogarithmic overhead; this cost is dominated by a polynomial function. 

Quantum computing subsumes classical computing. In terms of complexity classes, this is summarized by the statement ${\rm BPP} \subseteq {\rm BQP}$.

\ssse{More details on PostBQP}

The postselection procedure in PostBQP, which is not specific to quantum computing (one can define the classical complexity class PostBPP similarly), is highly unrealistic and brings in a lot of power to the model~\citep{aaronson05}. More formally, PostBQP is the class of problems solvable by a BQP machine such that:
\bi
\item If the answer is yes, then the second qubit has at least 2/3 probability of being measured 1, conditioned on the first qubit having been measured 1.
\item If the answer is no, then the second qubit has at most 1/3 probability of being measured 1, conditioned on the first qubit having been measured 1.
\item On any input, the first qubit has a nonzero probability of being measured 1. This condition can actually be refined to an $n$-dependent probability.
\ei
Denoting $q_o$ ($q_c$) the output (postselected) qubit, the relevant mathematical object is the conditional probability, which reads, by definition,
\be
P(q_o=1/q_c=1) = \frac{P(q_o=1\wedge q_c=1)}{P(q_c=1)}.
\ee
Intuitively, the power of PostBQP relies upon the denominator $P(q_c=1)$. Since it can be arbitrarily low, it may compensate for very unlikely events corresponding to the solution.

We now want to be more specific about the success probability $P(+_1)$. The Solovay--Kitaev theorem actually sets a lower bound on the acceptable probabilities. It lets us approximate any desired unitary within exponentially small error for only a polynomial increase in the circuit size. In other words, for an exponentially unlikely probability, the theorem still ensures that arbitrary universal gate sets can be used for polynomially long computations like BQP circuits, since a polynomial overhead remains in the BQP class. And indeed the class PostBQP is based upon BQP circuits. Thus it is well-defined only if the relevant output probabilities are at worst exponentially unlikely:
\be
P(+_1)\gtrsim\frac{1}{2^n}.
\ee 
It has been shown in Ref.~\citep{Kuperberg15} that this condition is fulfilled whenever ``reasonable'' universal gate sets are considered. 

Additionally, suppose now that there is a polynomial $p(n)$ such that $P(+_1) \geq 1/p(n)$. In that case, $P(+_1)$ is polynomially unlikely. Then running the BQP circuit $p(n)$ more times would still correspond to a polynomial-time computation and remain in BQP. On the other hand, such redundancy would enable recording enough statistics to simulate the quantum postselection through classical postprocessing. Hence, conditioning on an event whose probability scales as $1/p(n)$ does not give any power to the postselection. So $P(+_1)$ has to be worst than polynomially unlikely.

Following the discussion in Ref.~\citep{AaronsonBlog}, the definition of the class PostBQP could be refined to account for this feature: the conditioning probability $P(+_1)$ scales as the inverse of an exponential function,
\be
\label{eqExpScaling}
P(+_1)\sim\frac{1}{2^n},
\ee
up to some scaling factor irrelevant in terms of computational classes.

\sse{Summary of the complexity classes}

\begin{figure}
\centering
\includegraphics[width=0.7\linewidth]{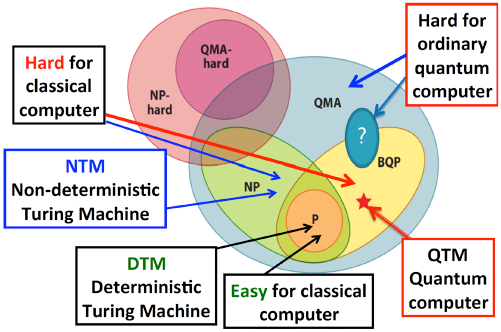}
\caption{A Venn diagram of various classical and quantum complexity classes. From Ref.~\citep{Wendin2017}.
\label{fig:ComplexityClassesVennDiagram}}
\end{figure}

\begin{figure}[!ht]
\bc
\centering
\includegraphics[width=0.3\columnwidth]{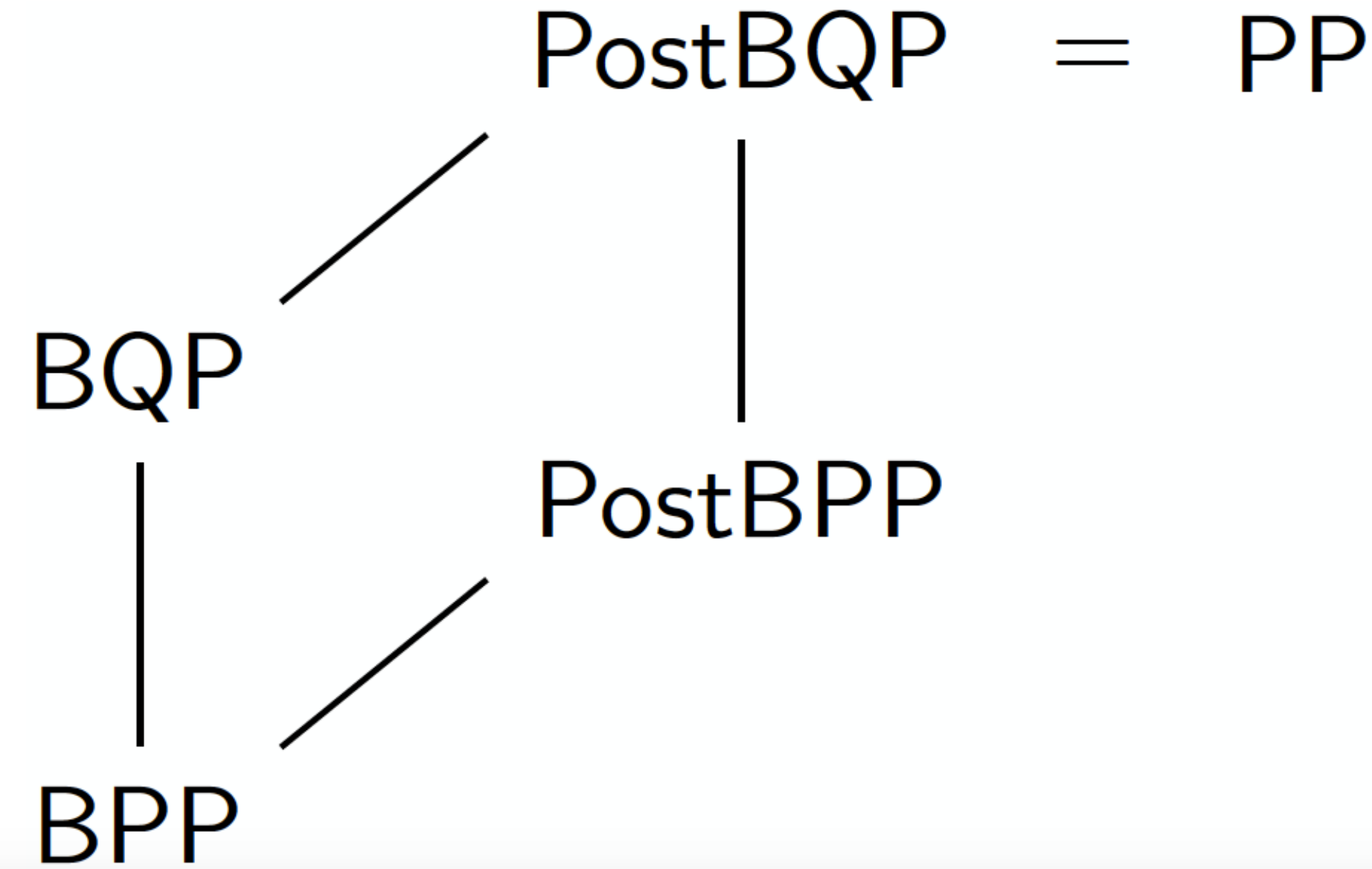}
\caption{Main complexity classes useful for our purposes and the inclusion relationships (black line, inclusion from bottom to top).\label{fig:CCTree}}
\ec
\end{figure}

The Venn diagram in \figref{fig:ComplexityClassesVennDiagram} summarizes most of the complexity classes defined above and their relations. Another plot of the relationships between a few of the complexity classes is shown in \figref{fig:CCTree}.

\begin{figure}[h!]
\begin{minipage}{\columnwidth}
\includegraphics*[width=\columnwidth]{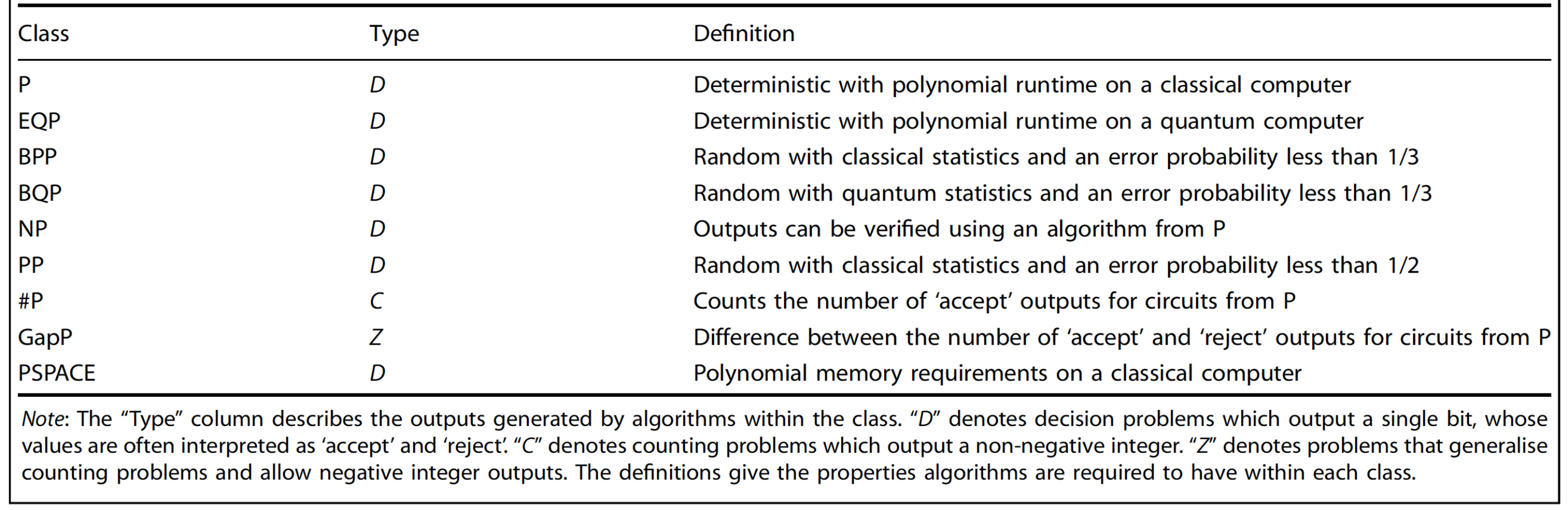}
\end{minipage}
\caption{Summary of the definitions of most of the complexity classes used in this chapter (and some others). From Ref.~\citep{lund2017quantum}.}
\label{fig:ComplexityClassesTable}
\end{figure}

Some of the definitions above, and a few more complexity classes, are given in \figref{fig:ComplexityClassesTable}. The definition of further complexity classes can be found in Refs.~\citep{Watrous09, Zoo}.

\section{Combinatorial optimization problems}

For this section, we follow Ref.~\citep{rodriguez2018building}.

A combinatorial optimization problem is about finding the best answer to a given problem from a vast collection of configurations. A typical example of a combinatorial optimization problem is the travelling salesperson problem (TSP), where a salesperson seeks to find the shortest travel distance between different locations, such that all locations are visited once. The naive method to solve this problem would be to make a list of all the different routes between locations that the salesperson could take and from that list find the best answer. This could work when the number of locations is small, but the naive method would fail if the number of locations grows large, since the number of possible routes increases exponentially with the number of locations. Thus the naive method is not efficient in practice, and we should therefore develop more clever optimization algorithms.

Typical examples of this kind of relevant combinatorial optimization problems are summarized in the list below:
\begin{itemize}
\item The TSP, as already mentioned. You are given a list of cities and the distances between them, and you have to provide the shortest route to visit all cities.
\item The knapsack problem. You are given the weights and values of a set of objects, and you have to provide the most valuable subset of them to take with you, given a certain bound on the total weight.
\item Sorting. Given $N$ numbers, return them in non-ascending order.
\item The integer factorization problem. You are given a big number $M$, and you have to provide two integer factors, $p$ and $q$, such that $M = pq$.
\item The satisfiability problem (SAT). You are given a Boolean expression of many variables $z_i \in \{0, 1\}$, for example, $P(z_1,z_2,z_3,z_4) = z_1 \vee \bar{z}_2 \wedge (z_3 \vee  \bar{z}_4)$. Then, you are asked whether there is a set of values of those variables which will make the complete expression true. For example, in the case above, making all $z_i=1$ is a valid solution.
\end{itemize}

Notice that all these problems have a similar structure: you are given certain input data (the initial numbers, the list of the cities, the list of objects or the integer to factorize) and you are asked to provide a response. All of these problems admit a formulation in terms of optimization problems, along with one formulation in terms of a decision problem. For instance, the decision version of the traveling salesperson problem reads as follows: given a length $L$, decide whether all cities can be visited with a route of length at most $L$ (technically, if the associated graph has a tour of at most $L$).

Let us focus for the moment on the optimization formulations. The first two problems in the above list are already written as optimization problems, in which a certain target function should be minimized (or maximized). The sorting problem can be restated as an optimization problem: we can design a penalty function to be minimized, by counting the misplaced consecutive numbers. The factorization problem can also be expressed in that way: find $p$ and $q$ such that $E =(M - pq)^2$ becomes minimal, and zero if possible. SAT can also be regarded as an optimization problem in which the evaluation of the Boolean formula should be maximized.

Thus, all those problems can be seen as combinatorial optimization problems. This means that, in order to solve them by brute force, a finite number of possibilities must be checked. But this number of possibilities grows very fast with the number of variables or the size of the input. The number of configurations typically involved becomes easily gigantic. It is enough to consider 100 guests at a party with 100 chairs, to obtain $100! \sim 10^{157}$ possible configurations to choose among when assigning each guest to a chair. This is roughly the square of number of particles in the universe, estimated to be about $10^{80}$. To calculate the correct configuration in this ocean of possible configurations is often hopeless for classical computers --- even for our best and future best super-computers. This is why it makes sense to ask the question: could a quantum computer help?


\subsection{Hardness of combinatorial optimization problems and promises of quantum computers for solving them}

Not all optimization problems are equally hard. Let us focus on the minimal possible time required to solve them as a function of the input size, i.e., to which time-complexity class they belong. 

We recall that a problem is said to be NP-hard if an algorithm solving it can be translated into an algorithm for solving any NP problem with a polynomial-time overhead. NP-hard problems are not necessarily decision problems, but for any optimization problem that is NP-hard, the corresponding decision problem is NP-complete. 

All the decision problems associated to the optimisation problems in the list introduced in the previous subsection are in NP: given a candidate solution, it can always be checked in polynomial time. But only one of them is known to be in P: the sorting problem, because it can always be solved in time $O(N \log[N])$, which is less than $O(N^2)$. For the factorization problem, we do not know whether it is in P or not. The other three belong to a special subset: they are NP-complete. As we have seen, this means that they belong to a special group with this property: if an algorithm to solve one of them exist, then we will have an algorithm to solve all NP problems that runs with a poly-time overhead.

\begin{figure}[!ht]
\centering
\includegraphics[width=15cm]{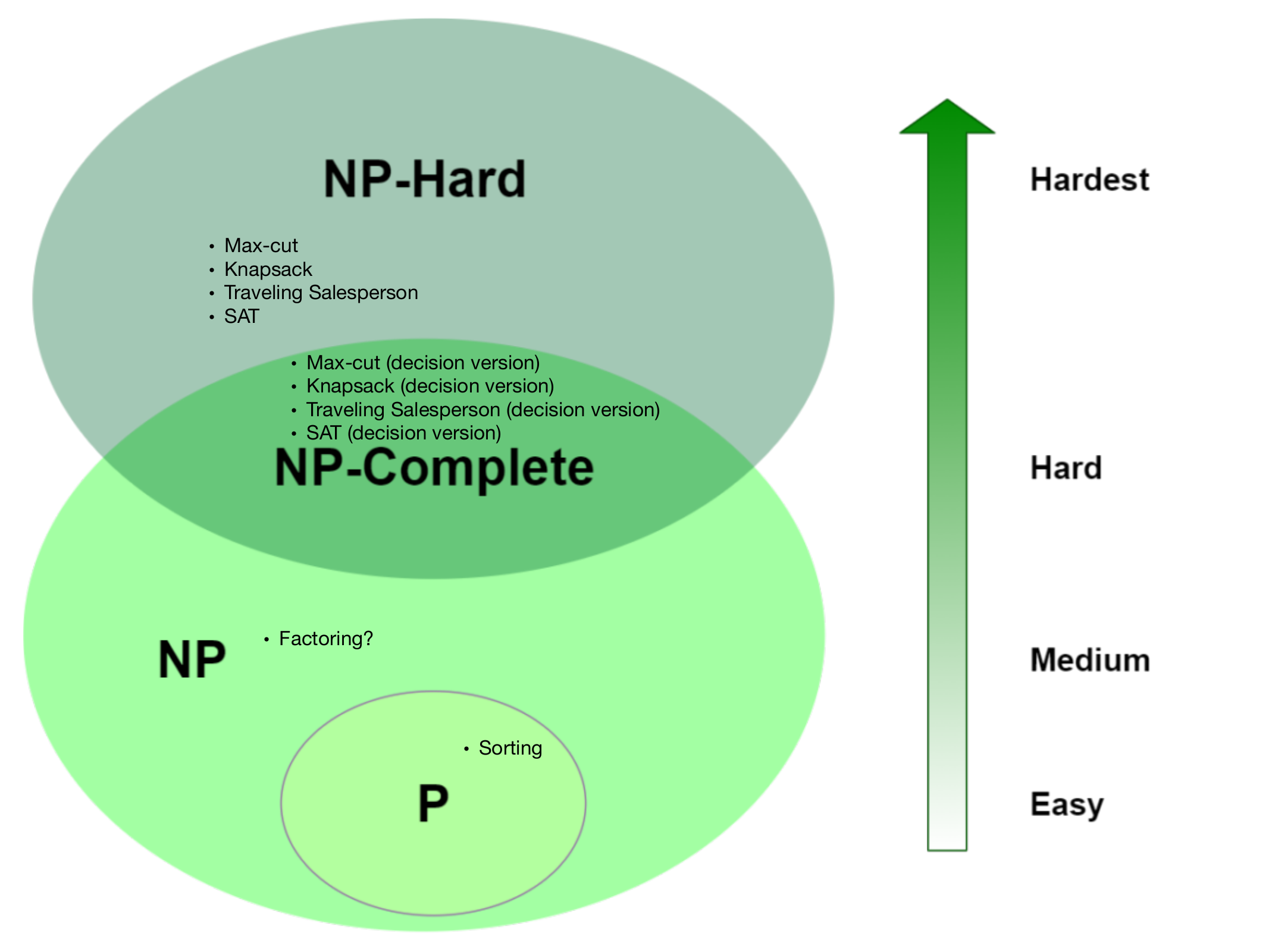}
\caption{Complexity classes relevant for optimization problems.}
\label{fig:complexity-optimisation}
\end{figure}

How can the solution to an NP-complete problem be useful to solve all NP problems? By using a strategy called reduction. An instance of the sorting problem, for example, can be converted into an instance of SAT in polynomial time. Thus, the strategy to solve any NP problem would be: (i) translate your instance to an instance of SAT, (ii) solve that instance of SAT, and (iii) translate back your solution. This reduction of all NP problems to SAT is known as the Cook--Levin theorem. A follow-up to this result by Karp shows that 21 graph problems all are NP-complete (Karp's 21 NP-complete problems). It is very relevant for physicists to know which combinatorial optimization problems are NP-complete for many reasons, and one of the most important is to avoid losing valuable time with a naive attempt to solve them in polynomial time. The complexity classes relative to the problems listed above are summarized in \figref{fig:complexity-optimisation}.

Actually, we do not expect quantum computers to efficiently solve problems that are NP-complete. That would mean that NP is contained in BQP, and we do not believe that this is the case (see, e.g., \url{https://www.scottaaronson.com/blog/?p=206}). Yet, we expect that some problems that are in NP and not in P could be efficiently solvable by a quantum computer, i.e., they belong to BQP (such as factoring). For these problems, quantum algorithms could provide an exponential advantage.

Furthermore, NP-complete problems might, in principle, be possible to solve on a quantum computer with a constant or polynomial advantage, e.g., quadratic, with respect to the classical solution, yielding practical improvement. This kind of quantum advantage is of the same kind as the one in the Grover algorithm that we studied in \chpref{chp:Grover}. The Grover search algorithm can be run on a quantum computer, either within the circuit model or using adiabatic quantum computation~\citep{albash2018adiabatic} (see \chpref{ch-aqc}), providing the same quadratic advantage, i.e., $\mathcal O (\sqrt{N})$ vs.~$\mathcal O ({N})$ running time. Some combinatorial optimisation problems can be solved by applying Grover's algorithm, yielding polynomial advantage; however, not all of them. For instance, for the TSP problem, the brute-force classical algorithm runs in $\mathcal O (N!) \sim \mathcal O (N^N) \sim \mathcal O (2^{N log N}) > \mathcal O (2^N)$ time. However, a better classical algorithm exists, which solves the TSP problem in $\mathcal O (2^N)$ time. Ambainis and collaborators~\citep{Ambainis2018} have found a quantum algorithm scaling with $\mathcal O(1.728^N)$, which is better than $\mathcal O (2^N)$. However, it is not known if Grover's type of advantage can be achieved with a better quantum algorithm, which would run in $\mathcal O(2^{N/2}) \sim \mathcal O(1.41^N)$ time.
Explicit use of the Grover algorithm for solving some combinatorial optimization problems, such as k-SAT and graph coloring, along with the characterization of the corresponding quantum advantage, can be found in Ref.~\citep{campbell2019applying}.

Finally, \emph{approximate} solutions of combinatorial optimization problems (even NP-complete ones) might be possible to obtain efficiently with the QAOA algorithm, that we will see in \secref{se:qaoa}. Not much is currently known about the time-complexity of finding approximate solutions to those problems, nor how this compares to the corresponding approximate classical algorithms.

\newpage

\section*{Exercises}

\begin{enumerate}

\item Which complexity class intuitively corresponds to the set of problems that are easy to solve for classical computers? Name at least 2 complexity classes that contain that class.

\item For which of the following complexity classes are quantum computers expected to be able to solve all problems efficiently?
\be
\text{P} \quad \text{NP} \quad \text{NP-hard} \quad \text{BPP} \quad \text{BQP} \quad \text{QMA} \quad \text{QMA-hard} \nn
\ee

\end{enumerate}


\chapter{Algorithms for solving combinatorial optimization problems}
\label{chap:OptimizationProblems}

In this chapter, we first introduce the Ising Hamiltonian, and we then dwell on two heuristic quantum algorithms that allow for approximately solving combinatorial optimization problems, namely quantum annealing (QA) and the quantum approximate optimisation algorithm (QAOA). For QA we reuse material from Refs.~\citep{rodriguez2018building, albash2018adiabatic}, while for QAOA we use material from Refs.~\citep{Vikstal-licentiate, farhi2014quantum}. 


\section{Combinatorial optimization and the Ising model}

To solve a combinatorial optimization problem using a quantum algorithm, be it quantum annealing or the QAOA, as we are going to see in the next sections, the quantum algorithm must be able to interpret the specific problem that we wish to solve. This can be done by encoding the optimization problem onto a quantum system. In this section, we will see how a combinatorial optimization can be framed as a cost Hamiltonian in the Ising form.

Let $C:\{0,1\}^n\rightarrow\mathbb{R}$ be a cost function that encodes a combinatorial optimization problem. There are in total $2^n$ possible strings; the goal is to find the bit-string $z=z_1\ldots z_n$ that minimizes the cost function $C(z)$. Note that a minimization problem can be transformed into a maximization problem by a minus sign $C(z) \rightarrow -C(z)$. 

One of the most widely used models in physics, which also is used to represent optimization problems, is the \emph{Ising model}. It is was developed in the 1920s as a way to understand phase transitions in magnetic materials. The Ising model consists of $n$ Ising spins on a lattice that can take the values $s_i=+1$ or $s_i=-1$ ($i$ labels the lattice site), which corresponds to the spin-$\uparrow$ and the spin-$\downarrow$ direction. The Ising spins are coupled together through long-range magnetic interactions, which can be either \emph{ferromagnetic} or \emph{antiferromagnetic}, corresponding to the spins being encouraged to be aligned or anti-aligned, respectively. Moreover, an external magnetic field can be applied at each individual spin site, which will give a different energy to the spin-$\uparrow$ and spin-$\downarrow$ directions. 

The energy configuration of the Ising model with $n$ Ising spins is given by the Ising Hamiltonian
\begin{equation}
	\label{eq:Ising}
    \mathcal{H}(s_1,\ldots,s_n)=-\sum_{i,j=1}^n
    J_{ij}s_js_i 
    -\sum^n_{i=1}h_is_i,
    \quad(s_i=\pm 1),
\end{equation}
where $J_{ij}$ is the coupling strength between the $i$th spin and the $j$th spin, and $h_i$ is the magnetic field at the $i$th spin site. Note that for a given set of couplings, magnetic fields, and spin configurations, the Ising Hamiltonian will just ``\emph{spit out}'' a number that represents the energy of that particular spin configuration. For a given set of couplings and magnetic fields, there always exists at least one spin configuration that minimizes the energy of the Ising Hamiltonian. 

A quantum version of this model is obtained by simply replacing the spin-variables $s_i$ with Pauli-$Z$ operators
\begin{equation}
    \label{eq:ising-hamiltonian}
    \hat H_C \equiv \hat H(\hat Z_1,\ldots,\hat Z_n) = - \sum_{1\leq i < j \leq n}J_{ij}\hat Z_i\hat Z_j - \sum_{i=1}^n h_i \hat Z_i,
\end{equation}
where $\hat Z_i$ refers to the Pauli-$Z$ matrix acting on the $i$th qubit. 
The spectral decomposition of this Hamiltonian then encodes the different candidate solutions in the computational basis:
\begin{equation}
    \label{eq:cost-hamiltonian}
    \hat H_C = \sum_{z\in\{0,1\}^n} C(z) \dyad{z},
\end{equation}
such that the eigenstate $\ket{z}$ with the lowest eigenvalue corresponds to the optimal solution. The Hamiltonian of \eqref{eq:ising-hamiltonian} is known as the Ising Hamiltonian after its inventor, but it can also be referred to as the \textit{cost Hamiltonian} in the context of optimization problems, because its eigenvalues in the computational basis correspond to the different values of the cost function, i.e., to the different costs. 

Also note that in the context of QAOA, we will refer to the Ising or cost Hamiltonian as \eqref{eq:ising-hamiltonian}, where however we will take the plus sign in front of both terms of the Hamiltonian.

\subsection{Mapping combinatorial optimization problems to spin Hamiltonians}
\label{sec:MappingIsing}

Many optimization problems, including all of Karp's 21 NP-complete problems, can be written in the form of \eqref{eq:ising-hamiltonian} and hence solved on a quantum computer, by choosing appropriate values for $J_{ij}$ and $h_i$~\citep{lucas2014ising}. 
We now turn to a couple of specific examples, and we will disregard here the minus sign in front of the terms in the Ising Hamiltonian. In this subsection, we denote the number of variables with $N$ to avoid confusion with the integer numbers appearing in the problems, and we might disregard the minus sign appearing in the definition of the Ising Hamiltonian (\ref{eq:ising-hamiltonian}).

\subsubsection{A special case of the knapsack problem: the subset sum problem}
\label{sec:SubsetSum}

The \emph{subset sum problem} is a famous combinatorial optimization problem that is known to be NP-complete. It is a special case of the decision version of the knapsack problem, where the weights $w_i$ are equal to the values $v_i$ for each object $i$, i.e., $w_i = v_i$. The subset sum problem can be formulated as a decision problem, as follows: Given an integer $m$ (the total value) and a set of positive and negative integers $n=\{n_1,n_2,\ldots,n_N\}$ of length $N$, is there a subset of those integers that sums exactly to $m$?

\textbf{Example}: Consider the case when $m=7$ and the set $n=\{-5,-3,1,4,9\}$. In this particular example, the answer is ``\emph{yes}''. The subset is $\{-3,1,9\}$.

\textbf{Example}: Consider $m=13$ and $n=\{-3,2,8,4,20\}$. This time the answer is ``\emph{no}''.

This problem can be framed as an energy minimization problem. The energy function for the subset sum problem can be formulated as 
\begin{equation}
\label{energy-subset-sum}
	\mathcal{E}(z_1,\ldots,z_N)
    = \mleft( \sum^N_{i=1} n_i z_i - m \mright)^2,
    \quad z_i\in\{0,1\},
\end{equation}
where $N$ corresponds to the size of the subset. Hence, if a configuration of $z_i$ exists such that $\mathcal{E}=0$, then the answer is ``\emph{yes}''. Likewise, if $\mathcal{E}>0$ for all possible configurations of $z_i$, then the answer is ``\emph{no}''. To map this energy function onto an Ising Hamiltonian, we introduce the Ising spins $s_i=\pm 1$ instead of $z_i$ as
\begin{equation}
\label{eq:mapping-Ising}
	z_i = \frac{1}{2}(1-s_i),
\end{equation}
such that $s_i=+1$ (spin-$\uparrow$) corresponds to $z_i=0$, and $s_i=-1$ (spin-$\downarrow$) corresponds to $z_i=1$. Then the Hamiltonian is written as
\begin{equation}
\begin{split}
	\mathcal{H}(s_1,\ldots,s_N)& = \mleft( \sum^N_{i=1}n_i\frac{1}{2}(1-s_i) - m \mright)^2
    \\
    &= \mleft( \sum^N_{i=1} -\frac{1}{2}n_is_i + \frac{1}{2} \sum^N_{i=1}n_i - m \mright)^2
    \\
    &= \frac{1}{4} \sum_{1\leq i,j\leq N} n_i n_j s_i s_j
    - \sum^N_{i=1} \mleft( \frac{1}{2} \sum^N_{j=1} n_j - m \mright) n_i s_i
    + \mleft( \frac{1}{2} \sum^N_{j=1} n_j - m \mright)^2.
\end{split}
\end{equation}

We now introduce the coupling $J_{ij}$ and the magnetic field $h_i$ as
\begin{equation}
	\label{eq:CouplingsAndMagneticFields}
	J_{ij} \equiv \frac{n_i n_j}{4}, 
    \quad\text{and}\quad
    h_i = - \mleft( \frac{1}{2} \sum^N_{j=1} n_j - m \mright) n_i,
\end{equation}
and finally obtain
\begin{equation}
	\mathcal{H}(s_1,\ldots,s_N) = \sum_{1\leq i,j\leq N} J_{ij} s_i s_j
    + \sum^N_{i=1} h_i s_i
    + \mleft( \frac{1}{2} \sum^N_{j=1} n_j - m \mright)^2.
\end{equation}
Notice that the last term is simply a constant. This Hamiltonian is an Ising Hamiltonian [cf.~\eqref{eq:Ising}]. Furthermore, by observing that $J_{ij}$ is symmetric and that the sum of the diagonal elements $i=j$ is equal to the trace of $J_{ij}$, we obtain
\begin{equation}
	\label{eq:SubsetSumHamiltonian}
\begin{split}
	\mathcal{H}(s_1,\ldots,s_N) &= 2\sum_{1\leq i<j \leq N}J_{ij} s_i s_j
    + \sum^N_{i=1} h_i s_i
    + \mleft( \frac{1}{2} \sum^N_{j=1} n_j - m \mright)^2
    + \text{Tr} \mleft[ J_{ij} \mright]
    \\
    &= \sum_{1\leq i<j \leq N} J_{ij} s_i s_j
    + \sum^N_{i=1} h_i s_i
    + \text{const},
\end{split}
\end{equation}
where we have absorbed the $1/2$ into $J_{ij}$ in the second step and made an implicit redefinition of the couplings $J_{ij} \equiv n_i n_j/2$. 
To solve the problem on a quantum computer, one can now quantize the spin variables $s_i \rightarrow \hat{Z}_i$.

\subsubsection{Number partitioning problem}
\label{sec:NumberPartitioningProblem}

The \emph{number partitioning problem} is another well known combinatorial optimization problem that is also NP-complete~\citep{albash2018adiabatic}. The number partitioning problem can be defined as a decision problem, as follows: Given a set $\mathcal{S}$ of positive integers $\{n_1,n_2,\ldots,n_N\}$ of length $N$, can this set be partitioned into two sets $\mathcal{S}_1$ and $\mathcal{S}_2$ such that the sum of the sets are equal?

\textbf{Example}: Consider the set $\{1,2,3,4,6,10\}$. Can this set be partitioned into two sets, such that the sum of both sets are equal? The answer is ``\emph{yes}''. The partitions are $\mathcal{S}_1=\{1,2,4,6\}$ and $\mathcal{S}_2=\{3,10\}$.

\textbf{Example}: Consider the set $\{1,2,3,4,6,7\}$. Can this set be partitioned into two sets, such that the sum of both sets are equal? This time the answer is ``\emph{no}'', which you can try to convince yourself that it is.

The Ising Hamiltonian for the number partitioning problem can be straightforwardly written down as
\begin{equation}
	\label{eq:NPP}
	\mathcal{H}(s_1,\ldots,s_N) = \mleft( \sum^N_{i=1} n_i s_i \mright)^2, \quad (s_i=\pm 1).
\end{equation}
It is clear that the answer to the number partitioning problem is ``\emph{yes}'' if $\mathcal{H} = 0$, because then there exists a spin configuration where the sum of the $n_i$ for the $+1$ spins is equal to the sum of the $n_i$ for the $-1$ spins~\citep{lucas2014ising}. Likewise, the answer is ``\emph{no}'' if $\mathcal{H}> 0$ for all possible spin configurations. Expanding the square of \eqref{eq:NPP}, we get
\begin{equation}
	\label{eq:NPPIsing}
	\mathcal{H}(s_1,\ldots,s_N) = \sum_{1\leq i,j \leq N} J_{ij} s_i s_j
    = 2 \sum_{1\leq i<j \leq N} J_{ij} s_i s_j + \text{Tr} \mleft[ J_{ij} \mright],
\end{equation}
where we have introduced the couplings as
\begin{equation}
	J_{ij} = n_i n_j.
\end{equation}
It should be noted that the ground state of the Ising Hamiltonian for the number partitioning problem is always at least two-fold degenerate. This has to do with the fact that changing $s_i$ to $-s_i$ does not change $\mathcal{H}$. What is also notable about the number partitioning problem compared to the subset sum problem is that the number partitioning problem does not require any additional magnetic fields on each spin site. 

Number partitioning is known as the ``easiest hard problem" ~\citep{hayes2002computing} due to the existence of efficient approximation algorithms that apply in most (although of course not all) cases, e.g., a polynomial-time
approximation algorithm known as the differencing method ~\citep{karmarkar1982differencing}. 

\subsubsection{Max-Cut}

The Max-Cut problem is one of the most extensively studied problems in the context of algorithms for combinatorial optimization. 
The objective of Max-Cut is to partition the set of vertices of a graph into two subsets, such that the sum of the edge weights going from one partition to the other is maximized. Max-Cut is NP-hard. It is not a problem with a \textit{yes} or \textit{no} answer. However, its decision version asks whether there is a cut of at least size $k$ in a graph. This version is NP-complete. 

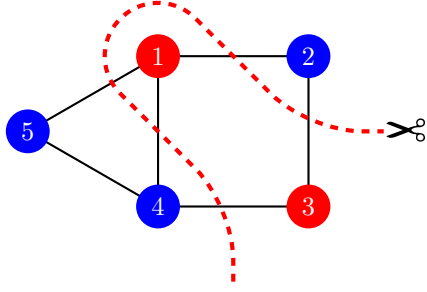
\begin{figure}
    \centering
  \begin{tikzpicture} [scale=1,
vertex/.style={circle, text=white, inner sep=3pt}]
  \centering
    \def\edge{2} 
    \draw[thick] 
    (0,0) node [vertex,fill=red] {$1$} -- 
    ++(\edge,0) node [vertex,fill=blue] {$2$} -- 
    ++(0,-\edge) node [vertex,fill=red] {$3$} -- 
    ++(-\edge,0) node [vertex,fill=blue] {$4$} -- 
    +(150:\edge) node [vertex,fill=blue] {$5$} -- 
    (0,0) -- (0,-\edge);
    \draw[ultra thick, dashed,rounded corners=20pt, red] (\edge/2,-\edge-1) -- ++(0,1) -- (0,-1) -- (-1,0) -- (0,1) -- (1,0) -- (2,-1) -- (3,-1);
    \node (scissors) at (3.3,-1) {\huge\Leftscissors};
\end{tikzpicture}
  \caption{A maximum cut of a graph with $5$ vertices. The dashed red line corresponds to the cut edges. An edge is cut if two vertices connected by an edge are assigned different colors.}
    \label{fig:example-graph}  
\end{figure}

Given an undirected graph $G=(V,E)$, where $V$ is the set of vertices, $E$ is the set of edges with nonnegative edge weights $w_{ij}=w_{ji}:(i,j)\in E$, the formulation of Max-Cut is
\begin{align}
    &\text{maximize }~\frac{1}{2}\sum_{1\leq i<j\leq N}w_{ij}(1-s_is_j), \quad
    \\
    &\text{subject to: } s_i\in\{-1,1\} \quad i\in V.
\end{align}
An example of a graph, as well as its maximum cut, is shown in Fig~\ref{fig:example-graph}.

To map this problem onto a cost Hamiltonian, all we have to do is to replace the classical variables $s_i$ with Pauli-$Z$ matrices. The corresponding Max-Cut Hamiltonian then reads
\begin{equation}
    \label{eq:max-cut-hamiltonian}
    \hat H_C = \frac{1}{2}\sum_{1\leq i<j\leq n}w_{ij}(1-\hat Z_i\hat Z_j).
\end{equation}
The eigenstate to this Hamiltonian with the highest eigenvalue corresponds to the maximum cut. This concludes this section of examples.

\section{Quantum annealing}

There is no consensus in the literature regarding the terminology for quantum annealing. We adopt the following definition~\citep{hauke2020perspectives}:
\emph{Quantum annealing} (QA) is a heuristic quantum algorithm that is based on the \textit{adiabatic quantum computation} model seen in Chapter~\ref{ch-aqc}, and that aims at solving hard \emph{combinatorial optimization problems}, by using an Ising Hamiltonian as the target Hamiltonian. In other words, the main idea of QA is to take advantage of adiabatic evolution expressed by \eqref{eq:AdiabaticQuantumComputation} to go from a simple nondegenerate ground state of an initial Hamiltonian to the ground state of an Ising Hamiltonian that encodes the solution of the desired combinatorial optimization problem. It can hence be thought of as the restriction of adiabatic quantum computation to optimization problems. As a consequence, a less general Hamiltonian is sufficient for addressing quantum annealing, as compared to adiabatic quantum computation [see also Scott Pakin's lecture on quantum annealing at the Quantum Science summer school 2017~\citep{Pakin2017}].

\subsection{Solving optimization problems on a quantum annealer}
\label{eq:quantum annealing}

To solve an Ising problem on a qubit-based quantum annealer, one defines a set of qubits $\ket{z_1z_2\ldots z_n}$ to store the answer to the problem and interpret $z_i=0$ to mean $s_i=+1$ (spin-$\uparrow$) and $z_i=1$ to mean $s_i=-1$ (spin-$\downarrow$). 
Then one chooses a pair of noncommuting Hamiltonians $\hat{\mathcal{H}}_0$ and $\hat{\mathcal{H}}_1$, such that the ground state of $\hat{\mathcal{H}}_0$ is easy to prepare and the ground state of $\hat{\mathcal{H}}_1$ encodes the solution to the Ising problem. To achieve the former, the initial Hamiltonian is typically assumed to be 
\begin{equation}
\label{eq:annealing-starting-Ham}
	\hat H_0 = - \sum_i \hat X_i, \quad \hat X = 
    \begin{pmatrix}
    0 & 1 \\
    1 & 0
    \end{pmatrix}
\end{equation}
where $(\hat{X}_i = \hat{I}\otimes\ldots\otimes\hat{X}\otimes\ldots\hat{I}$) with the Pauli $\hat{X}$ matrix in the $i$th place. The qubits or (spins) can then be regarded as pointing simultaneously in the spin-$\uparrow$ and spin-$\downarrow$ directions along the $z$-axis at the beginning. Indeed, it is easy to show, using basic quantum mechanics, that the ground state of $\hat{\mathcal{H}}_0$ for $N$ qubits is
\begin{equation}
	\ket{\psi(0)} = \ket{+}^{\otimes n} = 
    \mleft( \frac{1}{\sqrt{2}} \mright)^n \mleft( \ket{0} + \ket{1} \mright) \otimes \ldots \otimes \mleft( \ket{0} + \ket{1} \mright) = 
    \frac{1}{\sqrt{2^n}} \sum_{z\in\{0,1\}^n} \ket{z},
\end{equation}
which corresponds to a superposition of all possible spin configurations.

Constructing $\hat H_1$ can be done in a straightforward manner by replacing each $s_i$ in \eqref{eq:Ising} with a corresponding Pauli-$Z$ matrix, to yield \eqref{eq:ising-hamiltonian}, that we restate here for convenience:
\begin{equation}
\label{ham-ising}
	\hat H_1 = - \sum_{i,j=1}^n J_{ij} \hat{Z}_j \hat{Z}_i - \sum^n_{i=1} h_i \hat{Z}_i,
\end{equation}
where $\hat{Z}_i = \hat{I} \otimes \ldots \otimes \hat{Z} \otimes \ldots \hat{I}$ with $\hat{Z}$ in the $i$th place. When the time-dependent AQC Hamiltonian smoothly interpolates between $\hat H_0$ and $\hat H_1$,
\begin{equation}
\label{eq:aqc-hamiltonian}
    \hat H(t) = (1-s(t))\hat H_0 + s(t)\hat H_1,
\end{equation}
with $s(0)=0$, $s(\tau)=1$, and $\tau$ the total time of the algorithm, the qubits will gradually choose between $0$ and $1$, corresponding to spin-$\uparrow$ and spin-$\downarrow$, depending on which spin configuration minimizes the energy of the Ising Hamiltonian. At the end, $t=\tau$, the system will have evolved into $\ket{\psi(\tau)} = \ket{z_1 z_2 \ldots z_N}$ and a readout of this state in the computational basis will reveal each bit value, from which the corresponding values of the Ising spins can be obtained. If the annealing time was not slow enough, which is in practice occurring in any realistic situation, the state that is read out will only encode the optimal solution to the problem with a certain success probability $p = |\langle z_{\text{sol}}|\psi(\tau) \rangle |^2$, given by the overlap of the final state with the solution state. It is possible to estimate the success probability for each run by calculating the ratio of the number of runs where the optimal solution was found and the total number of runs.

In order to calculate the time to solution with \unit[99]{\%} certainty, we can evaluate the success probability after repeating the annealing procedure $m$ times and equate it to 0.99, i.e.,
\begin{equation}
	P_{\rm succ}^m = 1 - (1-p)^m = 0.99,
\end{equation}
from which we can extract
\begin{equation}
	m = \frac{\ln(1-0.99)}{\ln(1-p)},
\end{equation}
where we have used the change of basis of logarithms, $\log_b x = \log_a x / \log_a b$, and where $\ln$ is the natural logarithm.
Therefore we obtain in the end
\begin{equation}
\label{eq:time-to-solution}
	T_{99} = m \tau = \frac{\ln(1 - 0.99)}{\ln(1-p)} \tau.
\end{equation}

A challenge in quantum annealing is that full connectivity, which manifests in the interaction parameters $J_{ij}$ being $\neq 0$ beyond nearest-neighbour interactions, is often required in the problem Hamiltonian (for typical hard problems). This full connectivity may be difficult to achieve experimentally. In case the hardware has limited connectivity, embeddings allow one to map the fully connected problem onto a locally connected spin-system, at the price of an overhead in the total number of physical qubits to use. Examples of embedding are the Lechner, Hauke, and Zoller (LHZ) scheme or the minor embedding (we will talk about this later).

\subsection{Trotter expansion in quantum annealing}

In practical scenarios, instead of evolving the system adiabatically, it might be convenient to Trotterize the evolution stemming from the Hamiltonian in Eq.(\eqref{eq:aqc-hamiltonian}), i.e. to divide it in several small time steps:
\begin{equation}
    \label{eq:time-evolution-operator}
    \hat U(\tau) \equiv \mathcal{T} \exp \mleft[ - i \int_0^\tau \hat H(t)\,\dd t \mright] \approx 
    \prod_{k=1}^{p} \exp \mleft[ - i \hat H(k\Delta t) \Delta t \mright].
\end{equation}
Here $\hat U(\tau)$ is the evolution operator from $0$ to $\tau$, $\mathcal{T}$ is the time-ordering operator, and $p$ is a large integer such that $\Delta t = \tau / p$ is a small time segment. Next, for two non-commuting operators $A$ and $B$ and sufficiently small $\Delta t$, one can use the Trotter formula
\begin{equation}
    e^{i (A+B)\Delta t} = e^{i A\Delta t} e^{i B\Delta t} + \mathcal{O}((\Delta t)^2),
\end{equation}
and apply it to the discretized time evolution operator in \eqref{eq:time-evolution-operator}, yielding
\begin{equation}
\label{eq.trotterized-qaoa}
    \hat U(\tau) \approx
    \prod_{k=1}^{p} \exp \mleft[ -i (1-s(k\Delta t)) \hat H_0 \Delta t \mright] \exp \mleft[ -i s(k\Delta t)\hat H_1 \Delta t \mright]
\end{equation}
and the final state
\begin{equation}
\label{eq.trotterized-qaoa-final-state}
    \hat U(\tau) \ket{+}^{\otimes n} \approx
    \prod_{k=1}^{p} \exp \mleft[ -i (1-s(k\Delta t)) \hat H_0 \Delta t \mright] \exp \mleft[ -i s(k\Delta t)\hat H_1 \Delta t \mright] \ket{+}^{\otimes n}.
\end{equation}
Thus it is possible to approximate quantum annealing by applying the initial and final Hamiltonians in an alternating sequence, and the approximation is correct in the limit of very large $p$.
\subsection{Heuristic understanding of quantum annealing}
\label{sec:heuristic-annealing}

The term ``annealing'' is due to the analogy with the procedure used by metallurgists in order to make perfect crystals: the metal is melted and allowed to reduce its temperature very slowly. When the temperature becomes low enough, the system is expected to be in its global energy minimum with very high probability. Yet, mistakes are bound to take place if the temperature is reduced too fast.

Why does this procedure work? Thermal fluctuations allow the system to explore a huge number of configurations. Sometimes, a metastable state is found, but if the temperature is still large enough, fluctuations will allow the system to escape from this state. The escape probability is always related to the energy barrier separating the metastable state from the deeper energy minima. So, annealing can be considered an analog computation.

Annealing (meaning this metallurgic version) works nicely when the metastable states are separated by low energy barriers. Some target functions have barriers that are tall but very thin. Others will have different peculiarities. 

Is there any alternative analog computation suitable for problems with tall and thin barriers? Yes, there is one. Instead of escaping metastable states through thermal fluctuations, we may try to escape them through quantum tunneling, since the probability of such an event is known to decay with the product of the height and width of the energy barriers. 

This is what quantum annealing does: in QA, we engineer a quantum system so that its energy is related to the target function that we want to minimize. Thus, its ground state will provide the solution to our problem. If we start out at high temperature and cool the system down, then we are simply performing an annealing analog computation. Alternatively, we can operate always at an extremely low temperature, so that quantum effects are always important, but add an extra element to the system which promotes strong quantum fluctuations. This extra element (the starting Hamiltonian) is then slowly reduced and, when it vanishes, the ground state will give us the solution to our problem. In other terms: we make the system Hamiltonian evolve from a certain starting Hamiltonian, $H_0$, whose ground state presents strong quantum fluctuations in the basis of the final Hamiltonian eigenstates, to our target Hamiltonian, $H_1$, whose ground state is the solution to our problem.

\subsection{QUBO optimization}
\label{Qubo}

Many optimization problems can be recast in terms of QUBOs, where the acronym stands for quadratic unconstrained binary optimization, i.e., the function to be optimized is a quadratic function over binary variables without further constraints. 

The standard format for a QUBO objective function to be minimized over binary variables $z \in \{0,1\}^n$ is
\begin{equation}
\label{eq:qubo}
	q(z) = z^\top Q z = \sum_{j=1}^n Q_{jj} z_j + \sum_{\substack{j,k = 1 \\ j<k}}^n Q_{jk} z_j z_k 
\end{equation}
with an upper-triangular quadratic matrix $Q \in \mathbb{R}^{n \times n}$.

Usually these transformations produce overhead, e.g., in terms of increasing the number of variables, the required connections between them, or the value of the coefficients. Note that the expansion of the square in the cost function in \eqref{energy-subset-sum} directly yields a QUBO form.

\subsection{D-Wave quantum annealer}
\label{D-wave}

\begin{figure}[!ht]
\centering
\includegraphics[width=6cm]{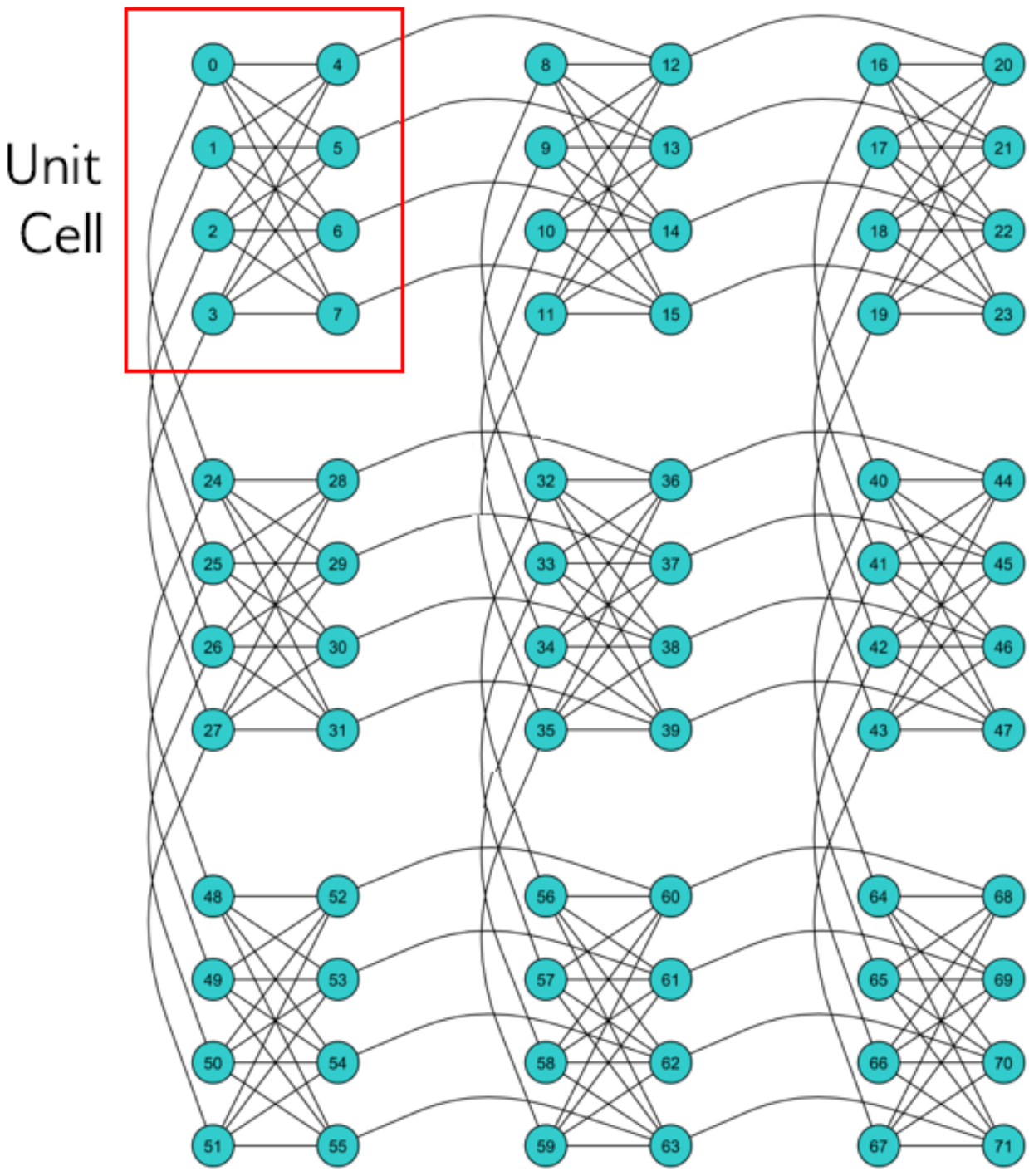}
\caption{A 3x3
 Chimera graph, denoted C3. Qubits are arranged in 9 unit cells. From the D-Wave website.}
\label{chimera.pdf}
\end{figure}

D-Wave 2000Q is the first commercially available quantum annealer, developed by the Canadian company D-Wave Systems. It is a heuristic solver using quantum annealing for solving optimization problems, and is based on rf-SQUIDs (superconducting quantum interference devices). 

In order to fit on D-Wave machines, optimization problems can be formulated either in terms of Ising Hamiltonians, or in terms of QUBOs.

The hardware layout of the D-Wave 2000Q quantum annealer restricts the connections between binary variables to the so-called Chimera graph (see \figref{chimera.pdf}).
In order to make problems with higher connectivity amenable to the machine, an embedding must be employed, for instance, minor embedding.
This means coupling various physical qubits together into one logical qubit, representing one binary variable, with a strong ferromagnetic coupling $J_F$ in order to ensure that all physical qubits have the same value after readout.

For how to program a D-Wave machine, see 
\url{https://docs.dwavesys.com/docs/latest/doc_getting_started.html}.
For a comparison of the performances between different generations of D-Wave hardware, see Ref.~\citep{Pokharel2021}, as well as the work featuring WACQT industry PhD student Marika Svensson together with Aachen-based researcher Kristel Michielsen, where different generations of D-Wave processors are compared for solving aircraft assignment~\citep{Willsch_2022}.  
A comprehensive review on using quantum annealers for solving industry problems is Ref.~\citep{Yarkoni_2022}.


\subsection{Summary of pros and cons for quantum annealing}

As a summary of this quantum-annealing section, we present a list of pros and cons for quantum annealing, taken from Ref.~\citep{Pakin2017}.

The bad:
\begin{itemize}
\item Very difficult to analyze an algorithm's computational complexity
\begin{itemize}
\item Need to know the energy gap between the ground state and the first excited state, which can be costly to compute
\item In contrast, circuit-model algorithms tend to be more straightforward to analyze
\end{itemize}
\item Unknown if quantum annealing can outperform classical computation
\begin{itemize}
\item If the gap always shrinks exponentially, then no
\end{itemize}
\end{itemize}

The good:
\begin{itemize}
\item Constants do matter
\begin{itemize}
\item If the gap is such that a correct answer is expected only once every million anneals, and an anneal takes 5 microseconds, that is still only 5 seconds to get a correct answer --- may be good enough
\item On current systems, the gap scaling may be less of a problem than the number of available qubits
\end{itemize}
\item We may be able to (classically) patch the output to get to the ground state
\begin{itemize}
\item Hill climbing or other such approaches may help to quickly get from a near-groundstate solution into the ground state
\item We may not even need the exact ground state
\item For many optimization problems, ``good and fast'' may be preferable to ``perfect but slow''
\end{itemize}
\end{itemize}

\section{Quantum approximate optimization algorithm (QAOA)}
\label{se:qaoa}

The quantum approximate optimization algorithm is a hybrid quantum-classical algorithm that allows for optimizing a cost function and finding approximate solutions. The role of the quantum processor in QAOA is to prepare a trial state depending on some parameters. Then, the state is measured several times (upon repeated preparation of the same state), the expectation value of the cost function is computed, and the result is fed into a classical optimiser, which determines how to change the parameters for the next state-preparation stage, so that the expectation value of the cost function will be lowered. In the next chapter, you will encounter another hybrid algorithm with a similar structure, the variational quantum eigensolver (VQE). Both algorithms are variational quantum algorithms (VQAs). An extensive review on VQAs can be found in Ref.~\citep{cerezo2021variational}.

\subsection{Introduction to QAOA: From the quantum adiabatic algorithm to QAOA}
\label{sec:aqc}

This section and the following one are taken from the licentiate thesis of Pontus Vikst{\aa}l~\citep{Vikstal-licentiate}.

The QAOA~\citep{farhi2014quantum} is inspired by adiabatic quantum computing (AQC), and in particular its application to optimization problems in the context of quantum annealing, that you have seen in Chapters~\ref{ch-aqc} and~\ref{eq:quantum annealing}. We restate here, for convenience, the Hamiltonian of \eqref{eq:aqc-hamiltonian}:
\begin{equation}
    \label{eq:aqc-hamiltonian2}
    \hat H(t) = (1-s(t))\hat H_M + s(t)\hat H_C. 
\end{equation}
Here $s(0)=0$ and $s(\tau)=1$, $\tau$ is the total time of the algorithm, $\hat H_M$ is the initial Hamiltonian, whose ground state (or maximally excited state) is easy to prepare, and $\hat H_C$ is the cost Hamiltonian, whose ground state (or maximally excited state) encodes the solution to an optimization problem.

The QAOA is based on the observation that the easiest way to practically implement the quantum-annealing Hamiltonian evolution expressed by \eqref{eq:aqc-hamiltonian2} is to Trotterize it, i.e., to decompose it in small time steps, as we have seen in Eq.(\ref{eq.trotterized-qaoa}).
However, the key idea underlying QAOA, which yields a key difference with quantum annealing,  is to truncate this product to an arbitrary (small) positive integer and redefine the time dependence in each exponent $(1-s(k\Delta t))\Delta t \rightarrow \beta_k$ and $s(k\Delta t) \Delta t \rightarrow \gamma_k$. In this way, and as a crucial difference with quantum annealing, {\it the fixed time segments instead become variational parameters to be optimized}. Finally, letting the product act on the initial state of quantum annealing, the plus state, one obtains the variational state
\begin{equation}
\label{eq.variational-state}
    \ket{\vec{\gamma},\vec{\beta}} \equiv \prod_{k=1}^{p} e^{-i \beta_k\hat H_M} e^{-i \gamma_k\hat H_C}\ket{+}^{\otimes n} = \sum_z d^{(\vec{\gamma},\vec{\beta})}_z  |z \rangle,
\end{equation}
where $\vec{\gamma} = (\gamma_1,\gamma_2,\ldots,\gamma_p)$ and $\vec{\beta}=(\beta_1,\beta_2,\ldots,\beta_p)$, and where in the last step we have just made explicit that the variational state is a given superposition of $Z$ eigenstates. Eq.(\ref{eq.variational-state}) can be compared to the final state of the anneling evolution  Eq.(\ref{eq.trotterized-qaoa-final-state}). If the eigenvalues of the cost Hamiltonian are all integers, then $\gamma$ is $2\pi$-periodic, and can be restricted to $\gamma_k \in [0, 2\pi)$, and the mixer is $\pi$-periodic such that $\beta_k \in [0,\pi)$.
However, the task of choosing the time dependence of the angular variational parameters $(\vec{\gamma},\vec{\beta})$ remains. 
The possibility of optimizing over variational parameters makes it irrelevant whether the initial state is a ground state or the highest excited state.  
In the context of QAOA, we might hence redefine the mixing Hamiltonian without the minus sign, i.e., we set $\hat H_M$ to be
\begin{equation}
    \label{eq:mixing-hamiltonian}
    \hat H_M = \sum_i \hat X_i.
\end{equation}
Also, different authors use QAOA for minimization or maximization of a given cost function. What you should pay attention to is the following: once the problem is given, if your cost function $\hat H_C$ encodes the solution in its ground state, then you should proceed to minimization of the cost function. 

\begin{figure}[ht]
    \centering
   \includegraphics[width=0.9\linewidth]{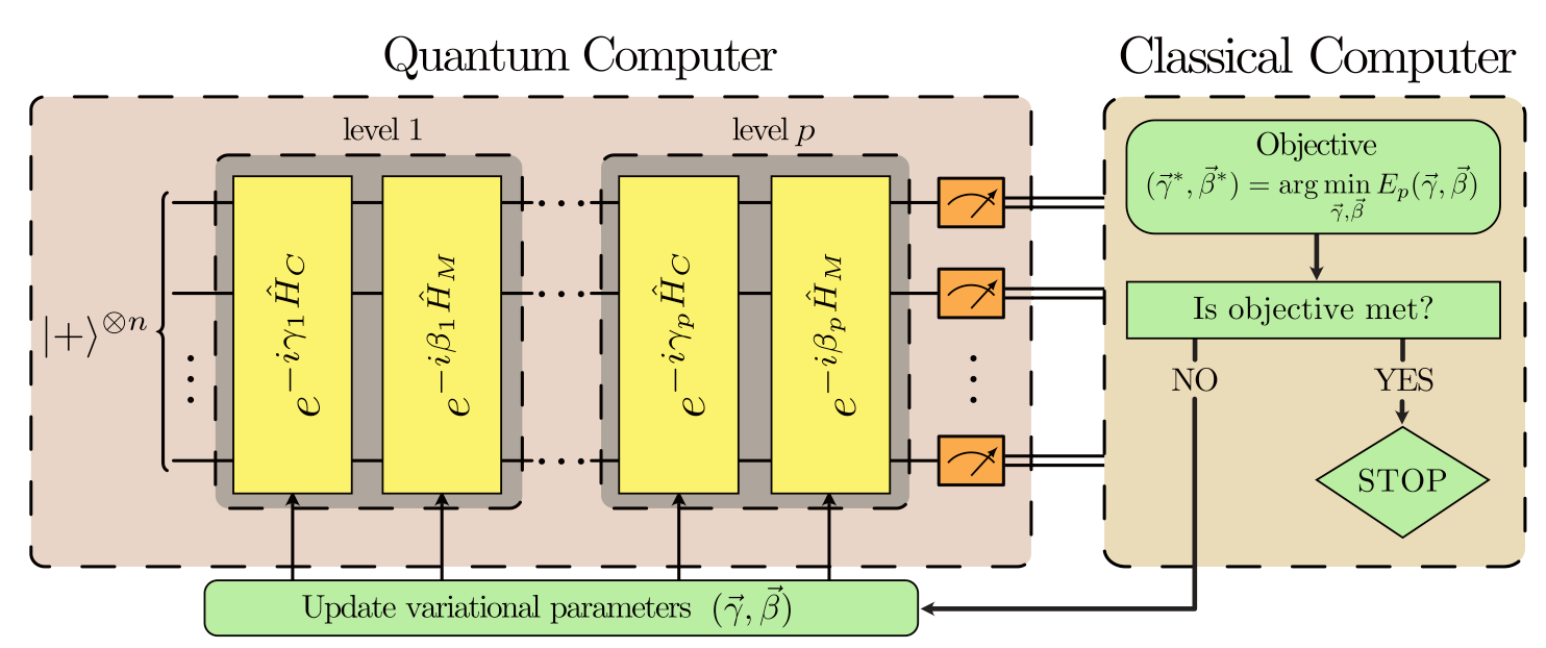}
    \caption{Schematic representation of the QAOA. The quantum processor prepares the variational state, depending on variational parameters. The variational parameters $(\vec{\gamma},\vec{\beta})$ are optimized in a closed loop using a classical optimizer. From Ref.~\citep{vikstaal2020applying}.}
    \label{fig:qaoa}
\end{figure}

Let $E_p(\vec{\gamma},\vec{\beta})$ be the expectation value of $\hat H_C$ in the variational state of \eqref{eq.variational-state}. Then
\begin{equation}
    \label{eq:expectation-value}
    E_p(\vec{\gamma},\vec{\beta})\equiv\matrixelement*{\vec{\gamma},\vec{\beta}}{\hat H_C}{\vec{\gamma},\vec{\beta}} = \sum_z \text{Prob}^{(\vec{\gamma},\vec{\beta})}_z C(z),
\end{equation}
where, by looking at the last term in \eqref{eq.variational-state}, we see that $\text{Prob}^{(\vec{\gamma},\vec{\beta})}_z = | d^{(\vec{\gamma},\vec{\beta})}_z|^2$ and $C(z) = \langle z | H_C | z\rangle$.
By finding good angles $\vec{\gamma}$ and $\vec{\beta}$ that minimize the expectation value above, the probability of finding the qubits in their lowest energy configuration when measuring is increased, because the candidate variational states become closer to the ground state of the cost Hamiltonian. Therefore the angles are chosen such that the expectation value is minimized:
\begin{equation}
    \label{eq:BestAngles}
    (\vec{\gamma}^*,\vec{\beta}^*)=\mathrm{arg}\min_{\vec{\gamma},\vec{\beta}}E_p(\vec{\gamma},\vec{\beta}).
\end{equation}
In general, this requires the quantum computer to query a classical optimizer, to tell the quantum computer how it should update the variational state by slightly changing the angles in order to minimize the expectation value, see \figref{fig:qaoa}. This has to be repeated until some convergence criteria are met or if an optimal or a good enough solution is found. In other words, QAOA is converting the search in a space of a combinatorial number of discrete configurations to a search for $2p$ optimal angles in a continuous energy landscape.


We end this section by giving a summary of the QAOA:
\begin{enumerate}
  \item Pick a $p$ and start with an initial set of $2p$ angles $(\vec{\gamma},\vec{\beta})$.
  \item Construct the state $\ket{\vec{\gamma},\vec{\beta}}$ using a quantum computer and measure this state in the computational basis. The output is a string $z$ with a probability given by the distribution of states $\ket{z}$. 
  \item Calculate $C(z) = \langle z | H_C | z \rangle$ using a classical computer. This step is classically efficient.
  \item Repeat steps 2 and 3 $m$ times. Record the best observed string $z_\text{best}$ and the sample mean $1/m\sum_{i=1}^m C(z_i)$, where $z_i$ is the $i$th measurement outcome. Note that when $m \rightarrow \infty$, the sample mean approaches the expectation value \eqref{eq:expectation-value} by the law of large numbers, as is clear in particular from the right-most term.
  \item If the optimal or a ``good enough'' solution is found, output $C(z_\text{best})$ together with the string $z_\text{best}$. Else, query a classical optimizer that updates the angles $(\vec{\gamma},\vec{\beta})$ based on the minimization of the expectation value and repeat from step 2.
\end{enumerate}

Since QAOA is an algorithm that provides approximate solutions, a relevant metric in order to compare its performance with respect to efficient classical algorithms also producing approximate solutions, is the approximation ratio. The approximation ratio is $C(z)$, where $z$ is the output of the quantum algorithm, divided by the maximum of $C$. 

Another relevant quantity in the context of QAOA is its success probability. Analogously to quantum annealing, it is defined as the square of the overlap between the variational states obtained for the optimal angles, with the actual solution of the problem. 

\subsection{\label{sec:implementation}Implementation of the QAOA}
The QAOA has already been implemented on several different hardware systems, including photonic~\citep{Qiang2018}, trapped-ion~\citep{Pagano2019}, and superconducting quantum processors~\citep{Bengtsson2020, Arute2020, Lacroix2020}. The unitary matrices that create the variational state $\ket{\vec{\gamma},\vec{\beta}}$ need to be decomposed into $1$-qubit and $2$-qubit gates in order to run on an actual quantum computer. In general, the decomposition will depend on the primitive quantum gates for the specific hardware. Here we will give an example of how the variational state $\ket{\vec{\gamma},\vec{\beta}}$ can be constructed in terms of single- and two-qubit gates, starting from the $\ket{0}^{\otimes n}$ state and using the gate set $\{H, R_x(\theta),R_z(\theta),\text{CZ}\}$.

The starting state $\ket{+}^{\otimes n}$ of the QAOA can be constructed by applying the Hadamard gate to each individual qubit in the $\ket{0}^{\otimes n}$ state. 

Next, the unitary that involves the sum of Pauli-$X$ matrices can be written as a product because all terms in the Hamiltonian commute:
\begin{equation}
    e^{-i\beta\hat H_B}= e^{-i\beta \sum_{i=1}^n\hat X_i}=\prod_{i=1}^n e^{-i \beta \hat X_i}.
\end{equation}
This unitary can be implemented as $n$ parallel single-qubit rotations around the $x$ axis of the Bloch sphere with the same angle. The rotation on qubit $i$ is
\begin{figure}[H]
    \centering
   \includegraphics[width=0.3\linewidth]{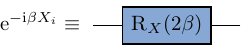}
\end{figure}

The cost Hamiltonian consists of two parts, a two-body Hamiltonian with $\hat Z_i\hat Z_j$ terms and a single-body Hamiltonian with $\hat Z_i$ terms:
\begin{equation}
  e^{-i\gamma\hat H_C} = \prod_{1\leq i<j\leq n} e^{-i\gamma J_{ij}\hat{Z}_i\hat{Z}_j}\prod_{i=1}^n e^{-i\gamma h_i\hat{Z}_i}.
\end{equation}
All terms in this product commute, so the order in which they are applied does not matter. Starting with the single-qubit term, it can be implemented as $n$ rotations around the $z$ axis of the Bloch sphere, where the rotation for the $i$th qubit is
\begin{figure}[H]
    \centering
   \includegraphics[width=0.3\linewidth]{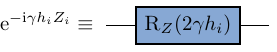}
\end{figure}

The two-qubit interaction terms $\hat Z_i\hat Z_j$ in the cost Hamiltonian can be implemented using a local single-qubit gate between two CNOT gates~\citep{Crooks2018}:
\begin{figure}[H]
    \centering
   \includegraphics[width=0.8\linewidth]{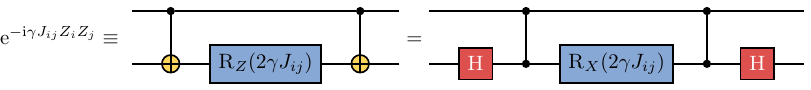}
\end{figure}
\noindent Since CNOT is not in our available gate set, we have used the following identity to express the CNOT gates in terms of CZ gates:
\begin{figure}[H]
    \centering
   \includegraphics[width=0.4\linewidth]{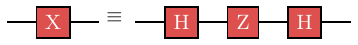}
\end{figure}
\noindent where X and Z are the Pauli-$x$ and -$z$ matrices. 

A problem where every element of $J_{ij}$ is nonzero would require a total of $n(n-1)$ CZ gates for each level $p$. This is of course under the assumption that it is possible to apply the two-qubit gates directly between any two qubits. In reality, many quantum processors have limited hardware connectivity. Still, SWAP gates, such as the one below, can be used to move distant qubits:
\begin{figure}[H]
    \centering
    \begin{quantikz}[row sep = 0.7cm]
        \lstick{$\ket{\psi}$} & \swap{1} & \rstick{$\ket{\phi}$}\\
        \lstick{$\ket{\phi}$} & \targX{} & \rstick{$\ket{\psi}$}
    \end{quantikz}
\end{figure}
\noindent The number of SWAP gates needed to make all the qubits interact with each other depends on the connectivity of the hardware. For example, in a linear array of qubits, a swap network can be implemented using $\mathcal{O}(n)$ number of SWAP gates~\citep{Crooks2018}.

\subsection{Other interesting remarks and extensions of QAOA}\label{sec:extensions}

\begin{itemize}

\item It is still an open area of research to establish whether QAOA with a fixed or logarithmic $p$ allows for a better performance in solving approximately NP-complete problems, compared to classical algorithms. It is also an open question how QAOA performs compared to quantum annealing. An interesting comparative study for Max-Cut on unweighted graphs and 2-SAT can be found in Ref.~\citep{willsch2019benchmarking}.

\item Originally (see version 1 of Ref.~\citep{Fahri-E3LIN}) the inventors of QAOA also considered the problem Max E3LIN2, where they showed that level-one QAOA achieves a better approximation ratio than, at that time, the best classical approximation algorithm. The computer science community took on the challenge and soon came up with a better classical algorithm~\citep{Barak2015}. 

\item It has been shown that QAOA is universal, meaning that for a problem of size $n$, and a choice of $\hat H_C$ and $\hat H_M$, QAOA can approximate any unitary $U$ of dimension $2^n \times 2^n$ to arbitrary precision when using a sufficiently large number of iterations $p$, which in general depends on $n$~\citep{farhi2017quantum, Lloyd2018, Morales2019}. 

\item Farhi et al.~showed that level $p=1$ QAOA is computationally hard to simulate on a classical computer without collapsing the polynomial hierarchy to the third level~\citep{Farhi2019}. However, this does not imply that QAOA for $p = 1$ is able to solve problems more efficiently than classical computers.

\item In the absence of a fully connected set-up, one can split the role of the cost function used for the classical optimization within QAOA and the Hamiltonian implementing the evolution, and still retain nontrivial approximation ratios, e.g., for Max-Cut~\citep{farhi2017quantum}. Furthermore, one can explore the advantage stemming from rotating the qubits with different angles when acting with the mixer Hamiltonian, and more generally introducing more free parameters.  

\item Brandao et al.~\citep{Brandao2018} demonstrated that if the problem instances come from a reasonable distribution, then the expectation value of the cost function concentrates. This suggests that it is possible to train a classical optimizer to find good angles on small instances and reuse those angles on larger instances, as long as they come from the same distribution. 

\item Furthermore, it was shown that the QAOA is able to realize Grover's search algorithm~\citep{Jiang2017, Niu2019}. 

\item In 2019, Hadfield et al.~put forward the quantum alternating operator ansatz, which generalizes the original QAOA ansatz to allow for more general types of Hamiltonians and initial states~\citep{Hadfield2019}.

\item A vast literature also tries to assess the limitations of QAOA as well as of its trainability, i.e., the possibility to carry the classical optimisation in an efficient way. The trainability is, in particular, hindered for certain cost functions by the presence of barren plateaus, i.e., regions of the cost-function landscape that are flat with respect to the variational parameters, or where the derivatives vanish exponentially with the system size, requiring exponential precision to minimise the cost function~\citep{cerezo2021cost}.   

\item In general, QAOA is nowadays an active field of research, and researchers are fervidly trying to apply it to solve real-world problems, understand its potential for quantum advantage, and characterise the bounds on its performance, as well as finding what is the best classical optimization method to use. For a comprehensive review updated to 2021, see Ref.~\citep{cerezo2021variational}.

\end{itemize}

\newpage
\section*{Exercises}

\begin{enumerate}

\item Derive explicitly the expression of the cost function $C(z)$ in \eqref{eq:cost-hamiltonian} in terms of the coefficients $J_{ij}$ and $h_{i}$ from the equivalent expression \eqref{eq:ising-hamiltonian} for the case of two qubits.

\item \label{ex:Ising-Hamiltonian} Compute explicitly the Ising Hamiltonian for the examples of instances for the subset sum problem and the set-partitioning problem given in the text.

\item \label{ex:Guzik} Write down the Ising Hamiltonian for solving factoring.

\item An exercise on QAOA: \\
\url{https://drive.google.com/file/d/11fnCUddtr1pWMRxnIMPdtPmfBpBYC28l/view?usp=share_link}

\item For implementing QAOA, we have used $R_X$ rotations, $R_Z$ rotations, CZ gates, and $H$ as the native gate set. However, in the first part of the course, we had provided another universal gate set. How would you express these operations in terms of $\{Z, H, T, \text{CZ}\}$? 

\end{enumerate}


\chapter{The variational quantum eigensolver}

In this chapter, we discuss the variational quantum eigensolver (VQE). The VQE is a heuristic approach to solving various problems with a combination of quantum and classical computation. As we will see later, the QAOA of the preceding chapter can be considered a special case of the VQE. 

The content of this chapter is mostly based on the review in Ref.~\citep{Moll2018}. We first outline how the VQE works and then discuss details of some of the steps in the algorithm.


\section{Outline of the algorithm}

The VQE is designed to solve problems that can be cast in the form of finding the ground-state energy $E_{\rm GS}$ of a Hamiltonian $H$. The ground-state energy is the lowest eigenvalue of the Hamiltonian,
\be
H \ket{\Psi_{\rm GS}} = E_{\rm GS} \ket{\Psi_{\rm GS}}.
\ee
How hard is this problem in general? If the Hamiltonian is $k$-local, i.e., if terms in $H$ act on at most $k$ qubits, the problem is known to be QMA-complete for $k \geq 2$. The general problem would thus be hard even for an ideal quantum computer. However, it is believed that physical systems have Hamiltonians that do not correspond to hard instances of this problem, and a heuristic quantum algorithm could still outperform a classical one.

A general Hamiltonian for $N$ qubits can be written
\be
H = \sum_\alpha h_\alpha P_\alpha = \sum_\alpha h_\alpha \bigotimes_{j = 1}^N \sigma_{\alpha_j}^{(j)},
\label{eq:HVQE}
\ee
where the $h_\alpha$ are coefficients and the $P_\alpha$ are called Pauli strings. The latter are products of single-qubit Pauli matrices (including the identity matrix).

\begin{figure}
\centering
\includegraphics[width=\linewidth]{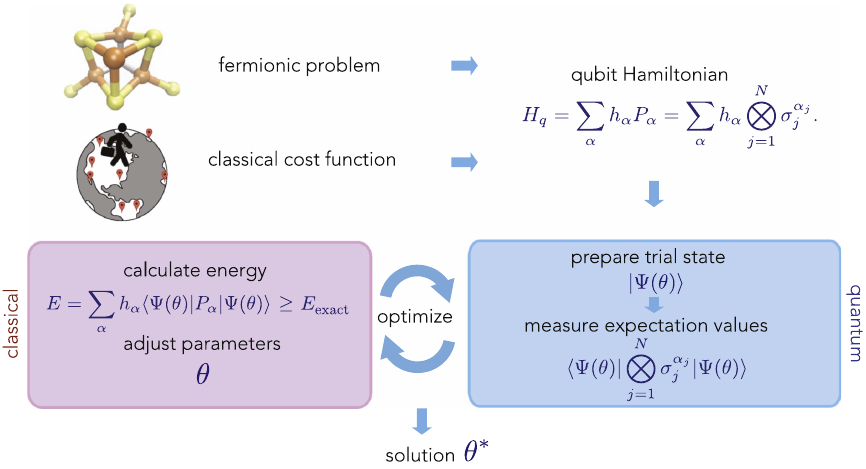}
\caption{The steps of the VQE algorithm. From Ref.~\citep{Moll2018}.
\label{fig:VQE}}
\end{figure}

The steps of the VQE algorithm are the following (see also \figref{fig:VQE}):
\begin{enumerate}
\setcounter{enumi}{-1}
\item Map the problem that you wish to solve to finding the ground-state energy of a Hamiltonian on the form in \eqref{eq:HVQE}.
\item Prepare a trial state $\ket{\Psi(\theta)}$ set by a collection of parameters $\theta$.
\item Measure expectation values of the Pauli strings in the Hamiltonian, i.e., measure $\brakket{\Psi(\theta)}{P_\alpha}{\Psi(\theta)}$.
\item Calculate the energy $E$ corresponding to the trial state, $E = \sum_\alpha h_\alpha \brakket{\Psi(\theta)}{P_\alpha}{\Psi(\theta)}$, by summing up the results of the measurements in the preceding step.
\item Update the parameters $\theta$ based on the result (and results in previous iterations).
\end{enumerate}
Steps 1 and 2 are run on a quantum computer, which can handle the quantum states more efficiently than a classical computer. Steps 3 and 4 are done on a classical computer. The algorithm is iterative, i.e., it starts over from step 1 after step 4, and continues to iterate until some convergence criterion is met, indicating that the ground-state energy has been found. In the following sections, we discuss steps 0, 1, and 4 in more detail.


\section{More on step 0 -- mapping to a Hamiltonian}

Broadly speaking, the VQE is currently mostly being considered for two types of problems: optimization problems, where $H$ is a cost function for the problem, and many-body fermionic quantum systems, e.g. molecules (quantum chemistry). The first type of problems was discussed extensively in \chpref{chap:OptimizationProblems}, where we saw several examples of how optimization problems can be mapped to a Hamiltonian. In this chapter, we therefore focus on quantum-chemistry problems.

Even though a Hamiltonian can be written down for a molecular system, that is not the Hamiltonian that is used in VQE. In classical simulations of molecular systems, there are many different methods, e.g., density functional theory (DFT), where the actual system of interacting electrons is described as non-interacting electrons moving in a modified external potential. An approach more suited to VQE is to describe the system in second quantization. This requires calculating a number of spatial integrals on a classical computer, but that task can be accomplished efficiently. The Hilbert space consists of electron orbitals. The Hamiltonian is
\be
H_F = \sum_{i,j} t_{ij} a^\dag_i a_j + \sum_{i,j,k,l} u_{ijkl} a^\dag_i a^\dag_k a_l a_j,
\label{eq:HMolecule}
\ee
where $a_i$ ($a^\dag_i$) annihilates (creates) an electron in the $i$th orbital. The coefficients $t_{ij}$ and $u_{ijkl}$ describing one- and two-electron interactions are calculated from the spatial integrals mentioned above.

The operators in \eqref{eq:HMolecule} are fermionic. They thus obey the fermionic anti-commutation relations, e.g., $\mleft\{ a_i, a^\dag_j \mright\} = \delta_{ij}$. These are not the relations that the qubit Pauli operators obey. We thus need to translate the Hamiltonian in \eqref{eq:HMolecule} to a form that can be implemented on the quantum computer. One well-known mapping from fermionic operators to qubit operators is the Jordan-Wigner transformation:
\be
a^\dag_i \rightarrow \mathbf{1}^{\otimes i -1 } \otimes \sigma_- \otimes \sigma_z^{\otimes N-i},
\ee
where $N$ is the number of orbitals and qubits. This mapping is not well suited to the VQE, because it creates highly non-local terms in the qubit Hamiltonian. In actual applications of VQE to quantum chemistry, other mappings are used (Bravyi-Kitaev, parity, ...). There is ongoing research on finding more suitable mappings.


\section{More on step 1 -- the trial state}

The trial state $\ket{\Psi(\theta)}$ can essentially be parameterized by $\theta$ in two ways: to form states that have a form that is suggested by the problem Hamiltonian, or to form states that are easy to create with the available quantum-computing hardware.

\subsection{Problem-specific trial states}

In quantum chemistry, a common class of trial states are created using a so-called coupled-cluster approach, often the unitary coupled-cluster (UCC) one. Here, the unitary operator $U(\theta)$ creates the trial state:
\be
\ket{\Psi(\theta)} = U(\theta) \ket{\Phi} = \exp \mleft[ T(\theta) - T^\dag(\theta) \mright] \ket{\Phi},
\ee
where $\ket{\Phi}$ is a simple state formed by the Slater determinant for low-energy orbitals. The operator $T(\theta)$ is known as a cluster operator. It is given by
\bea
T (\theta) &=& \sum_k T_k (\theta), \\
T_1 (\theta) &=& \sum_{i \in \text{occ}, j \in \text{unocc}} \theta_i^j a^\dag_j a_i, \\
T_2 (\theta) &=& \sum_{i,j \in \text{occ}, k,l \in \text{unocc}} \theta_{ij}^{kl} a^\dag_l a^\dag_k a_j a_i,
\eea
where the sums go over occupied and unoccupied orbitals. The coefficients of the higher-order cluster operators decrease rapidly as more orders are included. For this reason, the expansion is usually truncated at the second ,  ``double'', order (UCCSD) or the third, ``triple'', order (UCCSDT).

\subsection{Hardware-efficient trial states}

On an actual quantum computer, particularly a NISQ one, implementing the cluster operators can be hard, especially since the fermionic operators in the cluster operators must be mapped to qubit operators first. Therefore, hardware-efficient trial states are preferred. In the work of Ref.~\citep{Kandala2017}, where the H$_2$, LiH, and BeH$_2$ molecules were simulated using 2, 4, and 6 qubits, respectively, the trial states were of the form
\be
\ket{\Psi(\theta)} = \underbrace{U_{\rm single} (\theta) U_{\rm ent} (\theta) U_{\rm single} (\theta) U_{\rm ent} (\theta) \ldots U_{\rm single} (\theta) U_{\rm ent} (\theta)}_{d \: \rm repetitions} U_{\rm single} (\theta) \ket{0 0 \ldots 0}.
\label{eq:HWEfficientTrial}
\ee
Here, $U_{\rm single} (\theta)$ represent arbitrary-single qubit rotations on each of the $N$ qubits (different rotations in each of the $d+1$ steps) and $U_{\rm ent} (\theta)$ represent two-qubit entangling operations (same in each step) that were easy to implement in the available hardware. For the single-qubit operations alone, there are $N(3d + 2)$ independent rotation angles in the parameter vector $\theta$ (an arbitrary single-qubit rotation can be characterized by three Euler angles).

Already for relatively small molecules, $d$ needs to be more than just a few repetitions to reach accuracy that can compete with classical methods. However, a larger $d$ means that the quantum circuit takes longer to run, and thus decoherence will limit the achievable $d$. Recently, researchers are exploring ``error mitigation'' to get around this problem. In one type of error mitigation, the experiment is rerun several times with varying levels of added noise. From this, one can extrapolate the answer towards what it would have been for zero noise.

Note that the form of \eqref{eq:HWEfficientTrial} is that of the QAOA in \eqref{eq.variational-state}. This shows that both the QAOA and VQE are examples of the broader class of variational quantum algorithms (VQAs).


\section{More on step 4 -- updating the parameters}

Just like the other steps in the VQE that we have discussed so far, step 4 is also the subject of ongoing research. When searching for the ground-state energy of the problem Hamiltonian, there are several pitfalls that the update step must deal with. For example, the parameter landscape may have local minima. Furthermore, there is evidence that the landscape for larger problems can contain ``barren plateaus''. Both these problems are hard to deal with if one uses a standard gradient-descent-based search for the optimal parameters. Also, the value of $E$ obtained in step 3 is noisy, since it is based on limited sampling of the expectation values for the Pauli strings making up the Hamiltonian (at some point, it becomes too costly to run the quantum computer enough times to sample all strings enough time eliminate the noise). The search method used needs to be robust against this noise. Another issue is that the number of parameters will be large for a larger problem. One possibility is to use gradient-free algorithms like Nelder-Mead.

There are many considerations that go into choosing the right method for updating the parameters. Yet another is that it can take non-negligible time to change all parameters and set up the instructions (pulse shapes, etc.) needed to implement step 1 on the quantum computer again.

Although VQE is an interesting heuristic hybrid quantum-classical algorithm for NISQ devices, it is clear that there is still much to be understood about the different steps of the algorithm. It is still unclear how well the VQE will scale with the size of the problems it is applied to. A nice recent introductory overview discussing how to apply VQE or quantum phase estimation to problems in quantum chemistry can be found in Ref.~\citep{Arrazola2021}. That overview also contains some statements on how these algorithms can be expected to scale with problem size and required precision.

\chapter*{Tutorial 3: MBQC, QAOA, and SWAP network\label{tutorial3}}
\addcontentsline{toc}{chapter}{Tutorial 3} 
\renewcommand{\thesection}{\arabic{section}} 
\setcounter{section}{0} 
\setcounter{tutorial}{3}
\setcounter{ex}{0}
\setlength{\parindent}{0pt} 

\setcounter{equation}{0} 
\renewcommand{\theequation}{T3.\arabic{equation}} 

\section{MBQC}

In class, you have seen how to implement a Hadamard gate with measurement-based quantum computation (MBQC). The Hadamard gate is a single-qubit gate. For universal quantum computation, we need some two-qubit gate in addition to the arbitrary single-qubit rotations that we constructed above. Such two-qubit gates, e.g., the CZ and CNOT gates, can be constructed in a two-dimensional cluster state, where two input qubits are entangled with a few other qubits. By a series of single-qubit measurements and rotations, we can end up with two of the other qubits representing the output state corresponding to the two-qubit gate having acted on the input state. We will now see the example of the CNOT gate.

\begin{ex}
Show that it is possible to simulate the $\CNOT$ gate with an initial state given by
\be
\ket{\psi}_{1234} = \CZ_{23}\CZ_{13}\CZ_{34}\ket{\psi}_{12}\ket{+}_3\ket{+}_4,
\ee
and measuring qubits 2 and 3 in the $X$ basis. Note that the input state is given by qubits 1 and 2, and the output state by qubits 1 and 4, i.e.,
\be
\ket{\psi_{\rm in}}_{12} \to \ket{\psi_{\rm out}}_{14} \equiv \CNOT_{1\to 4} \ket{\psi_{\rm in}}_{14}.
\ee
\end{ex}

\begin{proof}[Solution]
Let us start by preparing the initial state. From the input state $\ket{\psi_{\rm in}}_{12}\ket{+}_3\ket{+}_4$, where $\ket{+}$ can be prepared by applying a Hadamard gate on $\ket{0}$, we create a cluster state. That is,
\begin{align}
    \ket{\psi_{\rm in}}_{12}\ket{+}_3\ket{+}_4 &=
    \mleft( a\ket{0} + b\ket{1} \mright)_1 \mleft( c\ket{0} + d\ket{1} \mright)_2 \ket{+}_3 \ket{+}_4 \nn \\ 
    &\stackrel{\CZ_{34}}{\longrightarrow} \mleft( a\ket{0} + b\ket{1} \mright)_1 \mleft( c\ket{0} + d\ket{1} \mright)_2 \frac{1}{\sqrt{2}} \mleft( \ket{0}_3 \ket{+}_4 + \ket{1}_3 \ket{-}_4 \mright) \nn \\
    &\stackrel{\CZ_{13}}{\longrightarrow} a\ket{0}_1 \mleft( c\ket{0} + d\ket{1} \mright)_2 \frac{1}{\sqrt{2}} \mleft( \ket{0}_3 \ket{+}_4 + \ket{1}_3 \ket{-}_4 \mright) \nn \\
    &\quad\; + b\ket{1}_1 \mleft( c\ket{0} + d\ket{1} \mright)_2 \frac{1}{\sqrt{2}} \mleft( \ket{0}_3 \ket{+}_4 - \ket{1}_3 \ket{-}_4 \mright) \nn \\
    &\stackrel{\CZ_{23}}{\longrightarrow} a\ket{0}_1 c\ket{0}_2 \frac{1}{\sqrt{2}} \mleft( \ket{0}_3 \ket{+}_4 + \ket{1}_3 \ket{-}_4 \mright) \nn \\
    &\quad\; + a\ket{0}_1 d\ket{1}_2 \frac{1}{\sqrt{2}} \mleft( \ket{0}_3 \ket{+}_4 - \ket{1}_3 \ket{-}_4 \mright) \nn \\
    &\quad\; + b\ket{1}_1 c\ket{0}_2 \frac{1}{\sqrt{2}} \mleft( \ket{0}_3 \ket{+}_4 - \ket{1}_3 \ket{-}_4 \mright) \nn \\
    &\quad\; + b\ket{1}_1 d\ket{1}_2 \frac{1}{\sqrt{2}} \mleft( \ket{0}_3 \ket{+}_4 + \ket{1}_3 \ket{-}_4 \mright) = \ket{\psi}_{1234}.
\end{align}
Since we want to measure qubits 2 and 3 in the $X$ basis, it is convenient to rewrite $\ket{\psi}_{1234}$ as follows:
\begin{align}
    \ket{\psi}_{1234} =& \frac{1}{2} \mleft( \ket{+}_3 + \ket{-}_3 \mright) \frac{1}{\sqrt{2}} \mleft( \ket{0}_4 + \ket{1}_4 \mright) \mleft( a\ket{0}_1 + b\ket{1}_1 \mright) \mleft( \frac{c+d}{\sqrt{2}} \ket{+}_2 + \frac{c-d}{\sqrt{2}} \ket{-}_2 \mright) + \nn \\
    &+ \frac{1}{2} \mleft( \ket{+}_3 - \ket{-}_3 \mright) \frac{1}{\sqrt{2}} \mleft( \ket{0}_4 - \ket{1}_4 \mright) \mleft( a\ket{0}_1 - b\ket{1}_1 \mright) \mleft( \frac{c-d}{\sqrt{2}} \ket{+}_2 + \frac{c+d}{\sqrt{2}} \ket{-}_2 \mright).
\end{align}
Now, if we measure qubits 2 and 3 in the $X$ basis, the possible outcomes are the following:
\begin{itemize}
    \item If we measure $+1$ in both qubits, this is equivalent to projecting $\ket{\psi}_{1234}$ onto $\ket{++}_{23}$. That is, omitting normalization,
    \begin{align}
        \prescript{}{23}{\braket{++}{\psi}}_{1234} =&\, (c+d) \mleft( a\ket{0}_1 + b\ket{1}_1 \mright) \mleft( \ket{0}_4 + \ket{1}_4 \mright) + (c-d) \mleft( a\ket{0}_1 - b\ket{1}_1 \mright) \mleft( \ket{0}_4 - \ket{1}_4 \mright) \nn \\
        =&\, a\ket{0}_1 \mleft( c\ket{0}_4 + d\ket{1}_4 \mright) + b\ket{1}_1 \mleft( d\ket{0}_4 + c\ket{1}_4 \mright) \nn \\
        =&\, {\color{NavyBlue}\CNOT_{1\to4}} \mleft( a\ket{0} + b\ket{1} \mright)_1 \mleft( c\ket{0} + d\ket{1} \mright)_4 
    \end{align}
    So if we measure $+1$ in qubits 2 and 3, it means that we ``perform'' a CNOT gate on our input state. The only thing to pay attention to is that the input state from qubits 1 and 2 was transferred to qubits 1 and 4, and then the CNOT is ``performed'' on those.
    \item Similarly, if we measure $+1$ in qubit 2 and {\color{ForestGreen}$-1$ in qubit 3},
    \begin{align}
        \prescript{}{23}{\braket{+-}{\psi}}_{1234} =&\, (c+d) \mleft( a\ket{0}_1 + b\ket{1}_1 \mright) \mleft( \ket{0}_4 + \ket{1}_4 \mright) {\color{ForestGreen}\boldsymbol{-}} (c-d) \mleft( a\ket{0}_1 - b\ket{1}_1 \mright) \mleft( \ket{0}_4 - \ket{1}_4 \mright) \nn \\
        =&\, a\ket{0}_1 \mleft( d\ket{0}_4 + c\ket{1}_4 \mright) + b\ket{1}_1 \mleft( c\ket{0}_4 + d\ket{1}_4 \mright) \nn \\
        =&\, {\color{NavyBlue}X_4 \CNOT_{1\to4}} \mleft( a\ket{0} + b\ket{1} \mright)_1 \mleft( c\ket{0} + d\ket{1} \mright)_4 
    \end{align}
    \item If we measure instead {\color{red}$-1$ in qubit 2} and $+1$ in qubit 3,
    \begin{align}
        \prescript{}{23}{\braket{-+}{\psi}}_{1234} =&\, (c{\color{red}-}d) \mleft( a\ket{0}_1 + b\ket{1}_1 \mright) \mleft( \ket{0}_4 + \ket{1}_4 \mright) + (c{\color{red}+}d) \mleft( a\ket{0}_1 - b\ket{1}_1 \mright) \mleft( \ket{0}_4 - \ket{1}_4 \mright) \nn \\
        =&\, a\ket{0}_1 \mleft( c\ket{0}_4 - d\ket{1}_4 \mright) - b\ket{1}_1 \mleft( d\ket{0}_4 - c\ket{1}_4 \mright) \nn \\
        =&\, {\color{NavyBlue}Z_1Z_4 \CNOT_{1\to4}} \mleft( a\ket{0} + b\ket{1} \mright)_1 \mleft( c\ket{0} + d\ket{1} \mright)_4 
    \end{align}
    \item Finally, if we measure {\color{red}$-1$ in qubit 2} and {\color{ForestGreen}$-1$ in qubit 3},
    \begin{align}
        \prescript{}{23}{\braket{--}{\psi}}_{1234} =&\, (c{\color{red}-}d) \mleft( a\ket{0}_1 + b\ket{1}_1 \mright) \mleft( \ket{0}_4 + \ket{1}_4 \mright) {\color{ForestGreen}\boldsymbol{-}} (c{\color{red}+}d) \mleft( a\ket{0}_1 - b\ket{1}_1 \mright) \mleft( \ket{0}_4 - \ket{1}_4 \mright) \nn \\
        =&\, a\ket{0}_1 \mleft( -d\ket{0}_4 + c\ket{1}_4 \mright) + b\ket{1}_1 \mleft( c\ket{0}_4 - d\ket{1}_4 \mright) \nn \\
        =&\, {\color{NavyBlue} X_4 Z_1 Z_4 \CNOT_{1\to4}} \mleft( a\ket{0} + b\ket{1} \mright)_1 \mleft( c\ket{0} + d\ket{1} \mright)_4 
    \end{align}
\end{itemize}
We can summarize these four cases as
\begin{equation}
\ket{\psi_{out}}_{14} = X_4^{m_3}(Z_1Z_4)^{m_2}\CNOT_{1\to4}\ket{\psi_{in}}_{14}, \quad m_{2,3} = 
\begin{cases} 
0 & \text{if we measure } +1 \\ 
1 & \text{if we measure } -1 
\end{cases}
\label{eq:CNOT}
\end{equation}
where $m_2$ and $m_3$ are given by the measurement results for qubits 2 and 3, respectively.\\

The circuit implementation of this MBQC CNOT gate is displayed below, where the first slice denotes the preparation of qubits 3 and 4 into states $\ket{++}_{34}$ and the second shows the entanglement of the four qubits that creates the initialized cluster state $\ket{\psi}_{1234}$.
Note that we implement X and Z gates in the inverse order of how they appear in \eqref{eq:CNOT} to correct them and obtain $\ket{\psi_{out}}_{14} \equiv \CNOT_{1\to4}\ket{\psi_{in}}_{14}$.

\begin{figure}[H]
\centering
\begin{quantikz}[column sep = 0.6cm, row sep = 0.8cm]
\lstick[wires=2]{$\ket{\psi_{in}}_{12}$} &  &  & \ctrl{2} &  &  &\gate{Z}  & \rstick[wires=4]{$\ket{\psi_{out}}_{14}\equiv\CNOT_{1\to4}\ket{\psi_{in}}_{14}$}\\
& &  &  & \ctrl{1} &  &\meter{m_2, \;\ket{\pm}} \wire[u][1]{c}\wire[d][2]{c} \\
\lstick{$\ket{0}_3$} & \gate{H}\slice{$\underbrace{\ket{\psi_{in}}_{12}\ket{++}_{34}}$} & \ctrl{1} & \control{} & \control{}\slice{$\underbrace{\ket{\psi}_{1234}}$}  & \meter{m_3,\ket{\pm}}\wire[d][1]{c} \\
\lstick{$\ket{0}_4$} & \gate{H} & \control{} &  &  & \gate{X} & \gate{Z} & 
\end{quantikz}
\end{figure}

We can also represent this scheme in a compact form as follows, where the black lines connecting the qubits denote CZ gates.

\begin{figure}[H]
\centering
\begin{tikzpicture}
\draw (0,0) circle (12pt) node{Q1};
\draw[red, very thick] (0,12pt) arc [start angle=90, end angle=270, x radius=12pt, y radius =12pt];
\draw[NavyBlue, very thick] (0,12pt) arc [start angle=90, end angle=270, x radius=-12pt, y radius =12pt];
\draw[thick] (0,2) circle (12pt) node{Q3};
\node[] at (0,2.7) {X};
\draw[red, very thick] (-2,2) circle (12pt) node{Q2};
\node[] at (-2,2.7) {X};
\draw[NavyBlue, very thick] (2,2) circle (12pt) node{Q4};
\draw[very thick] (0, 0.4) -- (0, 1.6);
\draw[very thick] (0.4, 2) -- (1.6, 2);
\draw[very thick] (-0.4, 2) -- (-1.6, 2);
\node[red] at (-1.5,1) {input, $\ket{\psi_{in}}_{12}$};
\node[NavyBlue] at (1.5,1) {output, $\ket{\psi_{out}}_{14}$};
\end{tikzpicture}
\end{figure}

Because we need to draw this qubit arrangement in a lattice and not in a single line, we call this state a 2D entangled cluster state. Two-dimensional cluster states are needed to construct two-qubit gates, and therefore for universal quantum computation.
\end{proof}

\subsection{Extra material}
For more on MBQC, check out these references:
\begin{itemize}
    \item~\citep{Browne2011} 
    on the potential of MBQC compared to circuit model.
    \item~\citep{Yoshikawa2016} 
    on experimentally realizing giant cluster states.
\end{itemize}

\section{QA and QAOA}

\textit{Note: This section is no longer taught in the tutorial, but it is left here for completeness.}\\

Quantum annealing (QA) and the quantum approximate optimization algorithm (QAOA) are two special cases of the following control problem: apply a combination of two Hamiltonians to minimize the energy of a quantum state.
QA smoothly interpolates between the two Hamiltonians, whereas QAOA applies one or the other in sequence.
It has previously been unclear which method, if either, is the most efficient. A recent paper~\citep{Brady2021} 
shows that hybrid protocols are, in theory, the best solution, although the practicality of implementing such protocols remains unknown.\\

It is useful to map QA to QAOA for a given number of steps, for a few reasons:
\begin{itemize}
    \item Inspiration for QAOA came from QA, by wanting to run the QA algorithm in gate-based quantum computing.
    \item In Ref.~\citep{willsch2019benchmarking}, 
    they found that the overall performance of the QAOA strongly depends on the problem instance, so mapping QA and QAOA allows us to compare the two methods.
    \item Initializing QAOA with parameters given from Trotterized QA gives comparable performance to the search over an exponentially scaling number of random initializations~\citep{Sack2021}. 
\end{itemize}

\begin{ex}
Given the following Hamiltonian describing a quantum annealing scheme
\begin{align}
  H(s) = A(s)(-H_0) + B(s)H_C, \quad s=t/t_a\in[0,1],
\end{align}
with $t_a$ denoting the annealing time, and the initial and final Hamiltonians
\begin{align}
   H_0 &= \sum_i \sigma_i^x, \\
   H_C &= \sum_i h_i\sigma_i^z +\sum_{ij}J_{ij}\sigma_i^z\sigma_j^z.
\end{align}

\begin{enumerate}
\item Find the mapping of the quantum annealing scheme to the QAOA for a given number of steps.

\item Give the value of the QAOA angles $\beta_n$ and $\gamma_n$ for the particular example with $A(s) = 1-s$ and $B(s) = s$.
\end{enumerate}
\end{ex}

\begin{proof}[Solution]
\hfill
\begin{enumerate}
\item QA is a smooth evolution from one Hamiltonian to another, whereas QAOA is a sequence of Hamiltonians. Therefore, the first thing to do is to discretize the QA evolution. To do so, we consider the Lie--Trotter--Suzuki decomposition for an evolution operator,
\begin{align}
  U(0,t) \approx \prod_{j=1}^N \prod_k \exp \mleft[ -i H_k(t_j) \frac{t}{N} \mright],
\end{align}
where $N$ is the number of time steps and the Hamiltonian has the form
\begin{align}
  H(t) = \sum_{k \subset \{1,2,\ldots, X\}} H_k(t).
\end{align}
This Trotterization is just an approximation, as in general $e^{A+B} \neq e^A e^B$. In particular, we will consider a symmetrized second-order decomposition of the form
\begin{align}
  \exp \mleft[ (A+B) \Delta t \mright] = \exp \mleft( \frac{A}{2} \Delta t \mright) \exp \mleft( B \Delta t \mright) \exp \mleft( \frac{A}{2} \Delta t \mright) + {\cal O} \mleft( (\Delta t)^3 \mright)
\end{align}
for each time step of approximated time-evolution operator of the annealing process. Thus, the sequence of operations for $N$ time steps of size $\tau=t_a/N$ yields
\begin{align}
   U_\text{QA}(t_a,0) &\approx e^{+i \tau A(s_N) H_0/2} e^{-i \tau B(s_N) H_C} e^{+i \tau (A(s_N) + A(s_{N-1})) H_0/2} \dots \nn \\
   &\quad \times e^{-i \tau B(s_2) H_C} e^{+i \tau (A(s_2) + A(s_1)) H_0/2} e^{-i \tau B(s_1) H_C} e^{+i \tau A(s_1) H_0/2},
\end{align}
where $s_n = (n-1/2)/N$, and $n = 1, \dots, N$, such that we use the middle-point Hamiltonian values.

To map the sequence $U$ to the QAOA sequence of gates, we recall that the QAOA evolution for $p$ steps reads
\begin{align}
  U_{\rm{QAOA}} = e^{-i \beta_p H_0} e^{-i \gamma_p H_C} \cdots e^{-i \beta_1 H_0} e^{-i \gamma_1 H_C},
\end{align}
which yields the variational state
\be
\ket{\vec{\gamma}, \vec{\beta}} = U_\text{QAOA} \ket{+}^{\otimes M},
\ee
where $M$ is the number of spins.

Now, we can neglect $e^{+i \tau A(s_1) H_0/2}$ in $U_\text{QA}$ because its action on $\ket{+}^{\otimes M}$ only yields a global phase factor (this term is the evolution of $\sigma_x$ on $\ket{+}$). Thus, we can choose
\begin{align}
   \gamma_n &= \tau B(s_n),\qquad n = 1,\dots ,N \\
   \beta_n &= -\tau \mleft( A(s_{n+1}) + A(s_n) \mright)/2, \qquad n = 1, \dots, N-1 \\
   \beta_{N} &= -\tau A(s_N)/2.
\end{align}
So $N$ time steps for the second-order annealing scheme correspond to $p=N$ steps for the QAOA.

\item We take as a particular example of a quantum annealing process
\begin{align}
   A(s) = 1-s, \qquad B(s) = s.
\end{align}
Using our previous results and $s_n=(n-1/2)/N$, we obtain
\begin{align}
   \gamma_n &= \frac{\tau(n-1/2)}{N} \\
   \beta_n &= -\tau \mleft( 1 - \frac{n}{N} \mright) \\
   \beta_N &= -\frac{\tau}{4N}.
\end{align}
\end{enumerate}
\end{proof}

\section{SWAP Network}

For hard problems, the cost function typically involves interactions beyond nearest-neighbour spins. However, experimentally, full connectivity is very difficult to achieve; we are usually restricted to linear chains or planar architectures. This means that it is not always trivial to find how to implement gates between far qubits with the minimum circuit depth. For instance, take four superconducting qubits laid out in a linear chain. If we want to entangle qubits 1 and 4, we have to do something like this:

\begin{figure}[H]
\centering
\begin{quantikz}
 & \ctrl{3} &\\
 & & \\
 & &\\
& \control{} &
\end{quantikz}
\quad = \quad 
\begin{quantikz}
 &  &  & \ctrl{1} &  &  &  \\
 &  & \swap{1} & \control{} & \swap{1} & & \\
 & \swap{1} & \targX{} &  & \targX{} & \swap{1}& \\
 & \targX{} &  &  &  & \targX{}& 
\end{quantikz}
\end{figure}
which substantially increases the circuit depth, thus increasing runtime and leaving room for errors to occur. This is why there is a very active research field on finding smart ways to implement gates when lacking all-to-all connectivity. An example would be to use CNOT-SWAP gates instead of SWAPs~\citep{Cruz2024}, and another would be to use a SWAP network~\citep{Babbush2018}. 
The latter, which we will see now more in detail, can be applied at linear cost and is very useful for variational algorithms such as QAOA.

\begin{ex}
We consider a $4\times 4$ planar lattice of qubits in an architecture that can implement nearest-neighbour gates. If we want to have interactions of the kind $\sigma_i^z\sigma_j^z$, like in QAOA, what is the optimal way to do that? 
\end{ex}

\begin{proof}[Solution]
We build the cost Hamiltonian, $H_c = \sum_{ij} J_{ij} \sigma_i^z \sigma_j^z = \sum_k H_k$, where $H_k$ are Hamiltonians containing only nearest-neighbour interactions, as follows:

\begin{figure}[H]
\centering
\begin{quantikz}
 & \gate[wires=4]{H_1}\gategroup[wires=4, steps=4, style={dashed, rounded corners}]{{$H_c$}} & \gate[wires=4, disable auto height]{\begin{array}{c} \text{SWAP}\\ \text{network}\end{array}} & \gate[wires=4]{H_2} & \ldots\\
\lstick[]{$\vdots$} &  & & & \ldots\\
 & & & & \ldots\\
 & & & & \ldots
\end{quantikz}
\end{figure}

The SWAP network is implemented in 4 steps:\\

\underline{Step 1.} Define a closed-loop 1D path through the qubits. Then, decompose this path into two different, disconnected graphs which we will call the ``left stagger'' ($U_L$) and ``right stagger'' ($U_R$). These two are made from the 1D closed-loop path taking alternating connections.

\begin{figure}[H]
\centering
\subfloat[1D closed-loop path]{\label{fig:closed_loop}
\begin{tikzpicture}
\foreach \x in {0,2}
\foreach \y in {0,2}
{\pgfmathtruncatemacro{\label}{12 + \x - 4 * \y}
\node [circle,draw,inner sep=0pt, text width=6mm, align=center,text opacity=0] (\x\y) at (\x,\y) {\label};}
\foreach \x in {0,2}
\foreach \y in {1,3}
{\pgfmathtruncatemacro{\label}{12 + \x - 4 * \y}
\node [circle,draw,inner sep=0pt, text width=6mm, align=center,text opacity=0] (\x\y) at (\x,\y) {\label};}
\foreach \x in {1,3}
\foreach \y in {0,2}
{\pgfmathtruncatemacro{\label}{12 + \x - 4 * \y}
\node [circle,draw,inner sep=0pt, text width=6mm, align=center,text opacity=0] (\x\y) at (\x,\y) {\label};}
\foreach \x in {1,3}
\foreach \y in {1,3}
{\pgfmathtruncatemacro{\label}{12 + \x - 4 * \y}
\node [circle,draw,inner sep=0pt, text width=6mm, align=center,text opacity=0] (\x\y) at (\x,\y) {\label};}
\draw (13) -- (23);
\draw (23) -- (33);
\draw (33) -- (32);
\draw (32) -- (22);
\draw (22) -- (12);
\draw (12) -- (11);
\draw (11) -- (21);
\draw (21) -- (31);
\draw (31) -- (30);
\draw (30) -- (20);
\draw (20) -- (10);
\draw (10) -- (00);
\draw (00) -- (01);
\draw (01) -- (02);
\draw (02) -- (03);
\draw (03) -- (13);
\end{tikzpicture}}
\hfill
\subfloat[``Left stagger'' $(U_L)$]{\label{fig:left_stagger}
\begin{tikzpicture}
\foreach \x in {0,2}
\foreach \y in {0,2}
{\pgfmathtruncatemacro{\label}{12 + \x - 4 * \y}
\node [circle,draw,inner sep=0pt, text width=6mm, align=center,text opacity=0] (\x\y) at (\x,\y) {\label};}
\foreach \x in {0,2}
\foreach \y in {1,3}
{\pgfmathtruncatemacro{\label}{12 + \x - 4 * \y}
\node [circle,draw,inner sep=0pt, text width=6mm, align=center,text opacity=0] (\x\y) at (\x,\y) {\label};}
\foreach \x in {1,3}
\foreach \y in {0,2}
{\pgfmathtruncatemacro{\label}{12 + \x - 4 * \y}
\node [circle,draw,inner sep=0pt, text width=6mm, align=center,text opacity=0] (\x\y) at (\x,\y) {\label};}
\foreach \x in {1,3}
\foreach \y in {1,3}
{\pgfmathtruncatemacro{\label}{12 + \x - 4 * \y}
\node [circle,draw,inner sep=0pt, text width=6mm, align=center,text opacity=0] (\x\y) at (\x,\y) {\label};}
\draw (23) -- (33);
\draw (32) -- (22);
\draw (12) -- (11);
\draw (21) -- (31);
\draw (30) -- (20);
\draw (10) -- (00);
\draw (01) -- (02);
\draw (03) -- (13);
\end{tikzpicture}}
\hfill
\subfloat[``Right stagger'' $(U_R)$]{\label{fig:right_stagger}
\begin{tikzpicture}
\foreach \x in {0,2}
\foreach \y in {0,2}
{\pgfmathtruncatemacro{\label}{12 + \x - 4 * \y}
\node [circle,draw,inner sep=0pt, text width=6mm, align=center,text opacity=0] (\x\y) at (\x,\y) {\label};}
\foreach \x in {0,2}
\foreach \y in {1,3}
{\pgfmathtruncatemacro{\label}{12 + \x - 4 * \y}
\node [circle,draw,inner sep=0pt, text width=6mm, align=center,text opacity=0] (\x\y) at (\x,\y) {\label};}
\foreach \x in {1,3}
\foreach \y in {0,2}
{\pgfmathtruncatemacro{\label}{12 + \x - 4 * \y}
\node [circle,draw,inner sep=0pt, text width=6mm, align=center,text opacity=0] (\x\y) at (\x,\y) {\label};}
\foreach \x in {1,3}
\foreach \y in {1,3}
{\pgfmathtruncatemacro{\label}{12 + \x - 4 * \y}
\node [circle,draw,inner sep=0pt, text width=6mm, align=center,text opacity=0] (\x\y) at (\x,\y) {\label};}
\draw (13) -- (23);
\draw (33) -- (32);
\draw (22) -- (12);
\draw (11) -- (21);
\draw (31) -- (30);
\draw (20) -- (10);
\draw (00) -- (01);
\draw (02) -- (03);
\end{tikzpicture}}
\end{figure}


\underline{Step 2.} Alternate layers of SWAP gates on the ``left stagger'' and ``right stagger'' layouts until all of the qubits return to their original positions. If we color the qubits in a checkerboard fashion, then all qubits of one color (teal) will move along the 1D path in a clockwise fashion whereas all qubits of the other color (lime green) will move along the 1D path in a counterclockwise fashion. We show the first four out of sixteen layers required to circulate these qubits all the way through the 1D path:

\begin{figure}[H]
\centering
\subfloat[Cycle 1]{
\begin{tikzpicture}[>=stealth, >-<]
\foreach \x in {0,2}
\foreach \y in {0,2}
{\pgfmathtruncatemacro{\label}{12 + \x - 4 * \y}
\node [circle,draw,fill=SpringGreen	,inner sep=0pt, text width=6mm, align=center] (\x\y) at (\x,\y) {\label};}
\foreach \x in {0,2}
\foreach \y in {1,3}
{\pgfmathtruncatemacro{\label}{12 + \x - 4 * \y}
\node [circle,draw,fill=TealBlue,inner sep=0pt, text width=6mm, align=center] (\x\y) at (\x,\y) {\label};}
\foreach \x in {1,3}
\foreach \y in {0,2}
{\pgfmathtruncatemacro{\label}{12 + \x - 4 * \y}
\node [circle,draw,fill=TealBlue,inner sep=0pt, text width=6mm, align=center] (\x\y) at (\x,\y) {\label};}
\foreach \x in {1,3}
\foreach \y in {1,3}
{\pgfmathtruncatemacro{\label}{12 + \x - 4 * \y}
\node [circle,draw,fill=SpringGreen	,inner sep=0pt, text width=6mm, align=center] (\x\y) at (\x,\y) {\label};}
\draw [thick, >-<] (23) -- (33);
\draw [thick, >-<] (32) -- (22);
\draw [thick, >-<] (12) -- (11);
\draw [thick, >-<] (21) -- (31);
\draw [thick, >-<] (30) -- (20);
\draw [thick, >-<] (10) -- (00);
\draw [thick, >-<] (01) -- (02);
\draw [thick, >-<] (03) -- (13);
\end{tikzpicture}}
\qquad \qquad
\subfloat[Cycle 2]{
\begin{tikzpicture}[>=stealth, >-<]
\node [circle,draw,fill=SpringGreen	,inner sep=0pt, text width=6mm, align=center] (03) at (0,3) {1};
\node [circle,draw,fill=TealBlue,inner sep=0pt, text width=6mm, align=center] (13) at (1,3) {0};
\node [circle,draw,fill=SpringGreen	,inner sep=0pt, text width=6mm, align=center] (23) at (2,3) {3};
\node [circle,draw,fill=TealBlue,inner sep=0pt, text width=6mm, align=center] (33) at (3,3) {2};
\node [circle,draw,fill=TealBlue,inner sep=0pt, text width=6mm, align=center] (02) at (0,2) {8};
\node [circle,draw,fill=SpringGreen	,inner sep=0pt, text width=6mm, align=center] (12) at (1,2) {9};
\node [circle,draw,fill=TealBlue,inner sep=0pt, text width=6mm, align=center] (22) at (2,2) {7};
\node [circle,draw,fill=SpringGreen	,inner sep=0pt, text width=6mm, align=center] (32) at (3,2) {6};
\node [circle,draw,fill=SpringGreen	,inner sep=0pt, text width=6mm, align=center] (01) at (0,1) {4};
\node [circle,draw,fill=TealBlue,inner sep=0pt, text width=6mm, align=center] (11) at (1,1) {5};
\node [circle,draw,fill=SpringGreen	,inner sep=0pt, text width=6mm, align=center] (21) at (2,1) {11};
\node [circle,draw,fill=TealBlue,inner sep=0pt, text width=6mm, align=center] (31) at (3,1) {10};
\node [circle,draw,fill=TealBlue,inner sep=0pt, text width=6mm, align=center] (00) at (0,0) {13};
\node [circle,draw,fill=SpringGreen	,inner sep=0pt, text width=6mm, align=center] (10) at (1,0) {12};
\node [circle,draw,fill=TealBlue,inner sep=0pt, text width=6mm, align=center] (20) at (2,0) {15};
\node [circle,draw,fill=SpringGreen	,inner sep=0pt, text width=6mm, align=center] (30) at (3,0) {14};
\draw [thick, >-<] (13) -- (23);
\draw [thick, >-<] (33) -- (32);
\draw [thick, >-<] (22) -- (12);
\draw [thick, >-<] (11) -- (21);
\draw [thick, >-<] (31) -- (30);
\draw [thick, >-<] (20) -- (10);
\draw [thick, >-<] (00) -- (01);
\draw [thick, >-<] (02) -- (03);
\end{tikzpicture}}
\\[12pt]
\subfloat[Cycle 3]{
\begin{tikzpicture}[>=stealth, >-<]
\node [circle,draw,fill=TealBlue,inner sep=0pt, text width=6mm, align=center] (03) at (0,3) {8};
\node [circle,draw,fill=SpringGreen	,inner sep=0pt, text width=6mm, align=center] (13) at (1,3) {3};
\node [circle,draw,fill=TealBlue,inner sep=0pt, text width=6mm, align=center] (23) at (2,3) {0};
\node [circle,draw,fill=SpringGreen	,inner sep=0pt, text width=6mm, align=center] (33) at (3,3) {6};
\node [circle,draw,fill=SpringGreen	,inner sep=0pt, text width=6mm, align=center] (02) at (0,2) {1};
\node [circle,draw,fill=TealBlue,inner sep=0pt, text width=6mm, align=center] (12) at (1,2) {7};
\node [circle,draw,fill=SpringGreen	,inner sep=0pt, text width=6mm, align=center] (22) at (2,2) {9};
\node [circle,draw,fill=TealBlue,inner sep=0pt, text width=6mm, align=center] (32) at (3,2) {2};
\node [circle,draw,fill=TealBlue,inner sep=0pt, text width=6mm, align=center] (01) at (0,1) {13};
\node [circle,draw,fill=SpringGreen	,inner sep=0pt, text width=6mm, align=center] (11) at (1,1) {11};
\node [circle,draw,fill=TealBlue,inner sep=0pt, text width=6mm, align=center] (21) at (2,1) {5};
\node [circle,draw,fill=SpringGreen	,inner sep=0pt, text width=6mm, align=center] (31) at (3,1) {14};
\node [circle,draw,fill=SpringGreen	,inner sep=0pt, text width=6mm, align=center] (00) at (0,0) {4};
\node [circle,draw,fill=TealBlue,inner sep=0pt, text width=6mm, align=center] (10) at (1,0) {15};
\node [circle,draw,fill=SpringGreen	,inner sep=0pt, text width=6mm, align=center] (20) at (2,0) {12};
\node [circle,draw,fill=TealBlue,inner sep=0pt, text width=6mm, align=center] (30) at (3,0) {10};
\draw [thick, >-<] (23) -- (33);
\draw [thick, >-<] (32) -- (22);
\draw [thick, >-<] (12) -- (11);
\draw [thick, >-<] (21) -- (31);
\draw [thick, >-<] (30) -- (20);
\draw [thick, >-<] (10) -- (00);
\draw [thick, >-<] (01) -- (02);
\draw [thick, >-<] (03) -- (13);
\end{tikzpicture}}
\qquad\qquad
\subfloat[Cycle 4]{
\begin{tikzpicture}[>=stealth, >-<]
\node [circle,draw,fill=SpringGreen	,inner sep=0pt, text width=6mm, align=center] (03) at (0,3) {3};
\node [circle,draw,fill=TealBlue,inner sep=0pt, text width=6mm, align=center] (13) at (1,3) {8};
\node [circle,draw,fill=SpringGreen	,inner sep=0pt, text width=6mm, align=center] (23) at (2,3) {6};
\node [circle,draw,fill=TealBlue,inner sep=0pt, text width=6mm, align=center] (33) at (3,3) {0};
\node [circle,draw,fill=TealBlue,inner sep=0pt, text width=6mm, align=center] (02) at (0,2) {13};
\node [circle,draw,fill=SpringGreen	,inner sep=0pt, text width=6mm, align=center] (12) at (1,2) {11};
\node [circle,draw,fill=TealBlue,inner sep=0pt, text width=6mm, align=center] (22) at (2,2) {2};
\node [circle,draw,fill=SpringGreen	,inner sep=0pt, text width=6mm, align=center] (32) at (3,2) {9};
\node [circle,draw,fill=SpringGreen	,inner sep=0pt, text width=6mm, align=center] (01) at (0,1) {1};
\node [circle,draw,fill=TealBlue,inner sep=0pt, text width=6mm, align=center] (11) at (1,1) {7};
\node [circle,draw,fill=SpringGreen	,inner sep=0pt, text width=6mm, align=center] (21) at (2,1) {14};
\node [circle,draw,fill=TealBlue,inner sep=0pt, text width=6mm, align=center] (31) at (3,1) {5};
\node [circle,draw,fill=TealBlue,inner sep=0pt, text width=6mm, align=center] (00) at (0,0) {15};
\node [circle,draw,fill=SpringGreen	,inner sep=0pt, text width=6mm, align=center] (10) at (1,0) {4};
\node [circle,draw,fill=TealBlue,inner sep=0pt, text width=6mm, align=center] (20) at (2,0) {10};
\node [circle,draw,fill=SpringGreen	,inner sep=0pt, text width=6mm, align=center] (30) at (3,0) {12};
\draw [thick, >-<] (13) -- (23);
\draw [thick, >-<] (33) -- (32);
\draw [thick, >-<] (22) -- (12);
\draw [thick, >-<] (11) -- (21);
\draw [thick, >-<] (31) -- (30);
\draw [thick, >-<] (20) -- (10);
\draw [thick, >-<] (00) -- (01);
\draw [thick, >-<] (02) -- (03);
\end{tikzpicture}}
\end{figure}

Up until this point, not all qubits have been next to all other qubits. This is clear to see, since all cycles are colored as a checkerboard.\\

\underline{Step 3.} Alternate between two staggered layers of parallel SWAP gates to move all the ``colors'' of the checkerboard pattern to separate sides of the qubit array.

\begin{figure}[H]
\centering
\subfloat[Color division layer 1]{\label{fig:color_1}
\begin{tikzpicture}[>=stealth, >-<]
\foreach \x in {0,2}
\foreach \y in {0,2}
{\pgfmathtruncatemacro{\label}{12 + \x - 4 * \y}
\node [circle,draw,fill=SpringGreen	,inner sep=0pt, text width=6mm, align=center] (\x\y) at (\x,\y) {\label};}
\foreach \x in {0,2}
\foreach \y in {1,3}
{\pgfmathtruncatemacro{\label}{12 + \x - 4 * \y}
\node [circle,draw,fill=TealBlue,inner sep=0pt, text width=6mm, align=center] (\x\y) at (\x,\y) {\label};}
\foreach \x in {1,3}
\foreach \y in {0,2}
{\pgfmathtruncatemacro{\label}{12 + \x - 4 * \y}
\node [circle,draw,fill=TealBlue,inner sep=0pt, text width=6mm, align=center] (\x\y) at (\x,\y) {\label};}
\foreach \x in {1,3}
\foreach \y in {1,3}
{\pgfmathtruncatemacro{\label}{12 + \x - 4 * \y}
\node [circle,draw,fill=SpringGreen	,inner sep=0pt, text width=6mm, align=center] (\x\y) at (\x,\y) {\label};}
\draw [thick, >-<] (02) -- (12);
\draw [thick, >-<] (22) -- (32);
\draw [thick, >-<] (00) -- (10);
\draw [thick, >-<] (20) -- (30);
\draw [thick, >-<] (13) -- (23);
\draw [thick, >-<] (11) -- (21);
\end{tikzpicture}}
\hfill
\subfloat[Color division layer 2]{\label{fig:color_2}
\begin{tikzpicture}[>=stealth, >-<]
\node [circle,draw,fill=TealBlue,inner sep=0pt, text width=6mm, align=center] (03) at (0,3) {0};
\node [circle,draw,fill=TealBlue,inner sep=0pt, text width=6mm, align=center] (13) at (1,3) {2};
\node [circle,draw,fill=SpringGreen	,inner sep=0pt, text width=6mm, align=center] (23) at (2,3) {1};
\node [circle,draw,fill=SpringGreen	,inner sep=0pt, text width=6mm, align=center] (33) at (3,3) {3};
\node [circle,draw,fill=TealBlue,inner sep=0pt, text width=6mm, align=center] (02) at (0,2) {5};
\node [circle,draw,fill=SpringGreen	,inner sep=0pt, text width=6mm, align=center] (12) at (1,2) {4};
\node [circle,draw,fill=TealBlue,inner sep=0pt, text width=6mm, align=center] (22) at (2,2) {7};
\node [circle,draw,fill=SpringGreen	,inner sep=0pt, text width=6mm, align=center] (32) at (3,2) {6};
\node [circle,draw,fill=TealBlue,inner sep=0pt, text width=6mm, align=center] (01) at (0,1) {8};
\node [circle,draw,fill=TealBlue,inner sep=0pt, text width=6mm, align=center] (11) at (1,1) {10};
\node [circle,draw,fill=SpringGreen	,inner sep=0pt, text width=6mm, align=center] (21) at (2,1) {9};
\node [circle,draw,fill=SpringGreen	,inner sep=0pt, text width=6mm, align=center] (31) at (3,1) {11};
\node [circle,draw,fill=TealBlue,inner sep=0pt, text width=6mm, align=center] (00) at (0,0) {13};
\node [circle,draw,fill=SpringGreen	,inner sep=0pt, text width=6mm, align=center] (10) at (1,0) {12};
\node [circle,draw,fill=TealBlue,inner sep=0pt, text width=6mm, align=center] (20) at (2,0) {15};
\node [circle,draw,fill=SpringGreen	,inner sep=0pt, text width=6mm, align=center] (30) at (3,0) {14};
\draw [thick, >-<] (12) -- (22);
\draw [thick, >-<] (10) -- (20);
\end{tikzpicture}}
\hfill
\subfloat[Loops in subdivisions]{\label{fig:loops}
\begin{tikzpicture}
\node [circle,draw,fill=TealBlue,inner sep=0pt, text width=6mm, align=center] (03) at (0,3) {0};
\node [circle,draw,fill=TealBlue,inner sep=0pt, text width=6mm, align=center] (13) at (1,3) {2};
\node [circle,draw,fill=SpringGreen	,inner sep=0pt, text width=6mm, align=center] (23) at (2,3) {1};
\node [circle,draw,fill=SpringGreen	,inner sep=0pt, text width=6mm, align=center] (33) at (3,3) {3};
\node [circle,draw,fill=TealBlue,inner sep=0pt, text width=6mm, align=center] (02) at (0,2) {5};
\node [circle,draw,fill=TealBlue,inner sep=0pt, text width=6mm, align=center] (12) at (1,2) {7};
\node [circle,draw,fill=SpringGreen	,inner sep=0pt, text width=6mm, align=center] (22) at (2,2) {4};
\node [circle,draw,fill=SpringGreen	,inner sep=0pt, text width=6mm, align=center] (32) at (3,2) {6};
\node [circle,draw,fill=TealBlue,inner sep=0pt, text width=6mm, align=center] (01) at (0,1) {8};
\node [circle,draw,fill=TealBlue,inner sep=0pt, text width=6mm, align=center] (11) at (1,1) {10};
\node [circle,draw,fill=SpringGreen	,inner sep=0pt, text width=6mm, align=center] (21) at (2,1) {9};
\node [circle,draw,fill=SpringGreen	,inner sep=0pt, text width=6mm, align=center] (31) at (3,1) {11};
\node [circle,draw,fill=TealBlue,inner sep=0pt, text width=6mm, align=center] (00) at (0,0) {13};
\node [circle,draw,fill=TealBlue,inner sep=0pt, text width=6mm, align=center] (10) at (1,0) {15};
\node [circle,draw,fill=SpringGreen	,inner sep=0pt, text width=6mm, align=center] (20) at (2,0) {12};
\node [circle,draw,fill=SpringGreen	,inner sep=0pt, text width=6mm, align=center] (30) at (3,0) {14};
\draw (03) -- (13);
\draw (13) -- (12);
\draw (12) -- (11);
\draw (11) -- (10);
\draw (10) -- (00);
\draw (00) -- (01);
\draw (01) -- (02);
\draw (02) -- (03);
\draw (23) -- (33);
\draw (33) -- (32);
\draw (32) -- (31);
\draw (31) -- (30);
\draw (30) -- (20);
\draw (20) -- (21);
\draw (21) -- (22);
\draw (22) -- (23);
\end{tikzpicture}}
\hfill 
\end{figure}

\underline{Step 4.} Repeat steps 1 through 3, in parallel, for the divided sectors of the array. One should alternate between horizontal and vertical color divisions for step 3. Once the divided sector size has reached four, we reach the last layer of SWAPs to ensure every qubit has neighboured at least once with all others.
\end{proof}

\newpage
\vspace{1cm}
\begin{ex}
Show that, for systems of $N$ qubits, all of them are adjacent at least once with circuit depth ${\cal O}(N)$.
\end{ex}
\begin{proof}[Solution]
In \emph{step 2}, if $U_L$ is a layer of SWAP gates associated with the ``left stagger'' and $U_R$ is a layer of SWAP gates associated with the ``right stagger'', then one should implement $(U_R U_L)^{N / 2}$ where $N$ is the number of qubits. This circuit has a depth of exactly $N$ cycles and returns all of the qubits to their original positions.\\

In the worst case, \emph{step 3} will require $\sqrt{N} / 2$ cycles. One can see this by considering half the length of one side of the grid.\\

Steps 1--3 require $N + \sqrt{N}/2$ layers of gates. After every repetition of steps 1--3, the circuit is divided into sectors of half the number of qubits, which yields $\sum_{k=0}^m (N/2^k + \sqrt{N/2^k}/2)$ . Accordingly, one will need to repeat steps 1--3 a total of $m \approx \log N$ times, since we stop when $4 = N/2^m$ (i.e., when the divided sector size has reached 4).\\

Thus, the total gate depth (not total number of gates) required is
\begin{align}
\sum_{k=0}^{\log N} \mleft( \frac{N}{2^k} + \frac{1}{2} \sqrt{\frac{N}{2^k}} \mright) \in {\cal{O}} \mleft( N \mright).
\end{align}
\end{proof}

\section{Algorithm classification}

At this point of the course, we have studied several algorithms and their speedup potential with respect to their classical counterparts. 
However, it may not be so clear yet which algorithms show the most potential and why, or how many other resources they need to run.
Thus, it can be useful to classify the algorithms learned in the map below, where in the x-axis, we draw the speedup potential with respect to their classical counterparts, from polynomial to exponential.
In the y-axis, we classify according to circuit depth, which is particularly relevant in noisy systems, since a higher number of gates makes an algorithm more prone to errors.
Ideally, we would want a quantum algorithm that runs exponentially faster with very low circuit depth, but none are known to date.

\begin{figure}[H]
    \centering
    \begin{tikzpicture}
    \tikzset{every node}=[font=\small\sffamily]
        \draw[-Stealth, very thick] (0,0) -- (0,4.2);
        \node[rotate=90] at (-0.4, 2) {Circuit depth};
        \draw[-Stealth, very thick] (0,0) -- (8.2,0);
        \node at (4, -0.4) {Speedup potential};
        \node[gray] at (0.5, -0.4) {poly};
        \node[gray] at (7.5, -0.4) {exp};
        \draw[dashed, gray] (0,2) -- (8,2);
        \draw[dashed, gray] (4,0) -- (4,4);
        \node at (2, 1) {QAOA*};
        \node at (2, 3) {Grover};
        \node at (5 , 3.5) {HHL};
        \node at (7 , 3.5) {Shor};
        \node at (6 , 3) {Quantum phase estimation};
        \node at (6 , 2.5) {Quantum Fourier transform};
     \end{tikzpicture}
\end{figure}

Let us now discuss the classification we made in the map above, which is a rough approximation to the truth, but could be argued in many different ways.
The point of this exercise is to reflect on those arguments, discuss them with your peers, and note that there is not one unequivocal way of distributing the algorithms in the map.\\

Let us start with QAOA. As a variational algorithm, it requires much fewer gates than other algorithms listed (which are gate-based), because a big part of the algorithm is a classical optimization process. 
The speedup potential of QAOA depends strongly on the problem instance and the hardware it runs on (hence the asterisk), but we do not expect it to run exponentially faster than the known classical combinatorial optimization algorithms.\\

Among the gate-based algorithms in the map, Grover's is the only one that promises only a quadratic speedup: from $O(N)$ in the classical search problem, to $O(\sqrt{N})$ in the quantum version.
Although one should maybe add an asterisk here too. Traditional complexity analysis of Grover's algorithm doesn't consider the complexity of processing the so-called oracle queries ---that is, certain questions with yes or no answers.
The query process is simply treated as a black box, thus making Grover's algorithm appealing because it needs fewer queries than classical search. However, with sufficiently time-consuming query processing, Grover's algorithm can become nearly as slow as a simple (exhaustive) classical search~\citep{Hayes2004}.\\

For the rest of the algorithms, while they all offer an exponential speedup, the circuit depth increases with the complexity of the algorithm: both Shor's and the HHL algorithms contain a quantum phase estimation subroutine, which, in turn, contains a quantum Fourier transform.
It is also relevant to point out that the exponential speedup of HHL is contingent on certain problem instances, namely that the linear system of equations that one is trying to solve is sparse and well-conditioned. Furthermore, the HHL algorithm and other qBLAS algorithms often suffer from input and output problems, as discussed in \secref{sec:QML-using-qBLAS}.

\renewcommand{\thesection}{\thechapter.\arabic{section}} 
\setlength{\parindent}{15pt} 
\renewcommand{\theequation}{\thechapter.\arabic{equation}} 

\chapter{Sampling models for quantum primacy}
\label{ch-sub-univ}

These parts of the notes follow Refs.~\citep{lund2017quantum, Aaronson2013}, as well as notes by Sevag Gharibian on quantum complexity theory at University of Paderborn~\citep{Gharibian2019}.
We also highly recommend to read the short review in Ref.~\citep{harrow2017quantum}. 

\section{Introduction: motivation for sampling models}

One of the difficulties in rigorously proving quantum advantage for computation is linked to the difficulty of bounding the power of classical computers. For instance, Shor's celebrated polynomial-time quantum algorithm for factorization is an important discovery for the utility of quantum computing, but it is not satisfying in addressing the issue of quantum advantage due to the unknown nature of the complexity of factoring. The best known classical factoring algorithm, the general number field sieve, is exponential time [growing as $\exp \mleft (c n^{1/3} \ln^{2/3}n \mright)$, where $n$ is the number of bits of the input number]. However, in order to prove quantum advantage, or really any separation between classical and quantum computational models, it must be proven for all possible algorithms and not just those that are known.

The challenge of bounding the power of classical computation can be exemplified by Shor's trilemma. The extended Church--Turing thesis asserts that any ``reasonable'' model of computation can be efficiently simulated on a standard classical computational model, such as a Turing machine, a random-access machine, or a cellular automaton.
Since Shor's algorithm allows for solving factoring in polynomial time, it is clear that at least one of the following statements must be false:
\begin{enumerate}
\item The extended Church--Turing thesis is true.
\item Factoring cannot be solved efficiently on a classical computer. 
\item Large-scale universal quantum computers can be built.
\end{enumerate}
If the third statement is false, something is wrong in quantum mechanics textbooks.

One of the main motivations of studying (selected) sampling problems is that it is possible to prove that they are indeed problems that can be solved efficiently in polynomial time by a quantum computer, while they cannot be solved in polynomial time by a classical computer (up to widely believed conjectures in computer science). Experimentally validating a sampling model would hence disprove the extended Church--Turing thesis and thereby solve Shor's trilemma. 

Sampling problems consist of generating random numbers as output according to a particular probability distribution (see, e.g., \figref{fig:IQP}). All quantum computations on $n$ qubits can be expressed as the preparation of an $n$-qubit initial state $| 0 \rangle^{\otimes n}$, a unitary evolution corresponding to a uniformly generated quantum circuit $C$, followed by a measurement in the computational basis on this system. In this picture, the computation outputs a length-$n$ bitstring $x\in \{0, 1 \}^n$ with probability
\be
P_x = \mleft| \langle x|C |  0 \rangle^{\otimes n} \mright|^2.
\ee
In this way, quantum computers produce probabilistic samples from a distribution determined by the circuit $C$.

One of the appealing features of these models is that some of them are restricted models of quantum computation, or sub-universal, and hence they do not necessarily require a universal quantum computer; they might be implementable on a large scale via near-term ``noisy intermediate scale quantum (NISQ) devices''. This quest for an experimental demonstration of quantum computational speedup has fallen under the moniker of ``quantum primacy''~\citep{sanders2021quantum}, with multiple possible models: the instantaneous quantum polynomial-time (IQP) model, random circuit sampling (RCS), boson sampling and its relative Gaussian boson sampling, and the deterministic quantum computation with one qubit (DQC1) model, among others.

In 2019, the Google Quantum AI group released a demonstration of RCS with 53 qubits~\citep{Arute:2019aa}, which is at the edge of current simulation capability of classical computers~\citep{pednault2019leveraging}, followed by a demonstration with 60 qubits from the Pan group~\citep{zhu2021quantum, wu2021}. The Pan group also achieved an experimental demonstration of boson sampling, and one of Gaussian boson sampling with 100 optical modes and 50 input squeezed states~\citep{Zhongeabe8770}. Despite that disproving the extended Church--Turing thesis is inherently challenging because it involves a scaling statement, these experiments constitute first proofs of quantum primacy, in that classical computers cannot currently simulate the experiments themselves. Some comments on the Google experiments can be found in \secref{random-sampling-google}. 

Here, however, we shall focus on two of the first and historically more important models for quantum primacy: IQP and boson sampling. Note that the historically very first leap towards exhibiting a candidate model for quantum primacy was taken by Terhal and DiVincenzo in their 2004 paper ``Adaptive quantum computation, constant depth quantum circuits and Arthur-Merlin games"~\citep{terhal2002adaptive}. Their approach was already to appeal to a complexity-theoretic argument: they gave evidence that there exists a certain class of quantum circuits that cannot be simulated classically by proving that if a classical simulation existed, certain complexity classes strongly believed to be distinct would collapse to the same class.

Beyond the conceptual interest in unveiling quantum advantage with sampling models, we also underline that a convincing demonstration of the impossibility for a classical computer to produce the same output probability distribution as a quantum architecture is also a crucial milestone towards demonstrating the controllability of quantum computers, with the scope of using them to perform useful algorithms. Finally, note that classical simulation methods, too, improve constantly. So even if in the end, for very large numbers of sufficiently clean qubits, we expect the classical computer not to be able to sample from the probability distributions characterising sampling problems displaying quantum advantage, for the intermediate-size quantum processors currently available the onset of the quantum primacy is being constantly pushed forward~\citep{pan2021solving}. 


\section{Instantaneous quantum polytime}
\label{IQP-DV}

IQP circuits are an intermediate model of quantum computation where every circuit has the form $C = H^{\otimes n} DH^{\otimes n}$, where $H$ is a Hadamard gate and $D$ is an efficiently generated quantum circuit that is diagonal in the computational basis. Sampling then simply corresponds to performing measurements in the computational basis on the state $H^{\otimes n} DH^{\otimes n} |0 \rangle^{\otimes n}$, yielding outcomes distributed according to the probability distribution  $P_x = \mleft| \langle x| H^{\otimes n} DH^{\otimes n} |0  \rangle^{\otimes n} \mright|^2$. In Ref.~\citep{Bremmer2010}, it was argued that classical computers could not efficiently sample from IQP circuits (exact sampling case) where $D$ is chosen uniformly at random from circuits composed of: 
\begin{enumerate}
\item $\sqrt{\text{CZ}}$ (square root of controlled-Z) and
$T = 
\begin{pmatrix}
1 & 0 \\
0 & e^{i \pi/4}
\end{pmatrix}$
gates; or

\item $Z$, CZ, and CCZ (doubly controlled-Z) gates.
\end{enumerate}
Although the diagonal gates inbetween the two layers of Hadamard gates may act on the same qubit many times, they all commute, so in principle they could be applied simultaneously, hence the name ``instantaneous''. In case 1, the circuits correspond to random instances of the Ising model drawn from the complete graph~\citep{Bremner2015}.

Reference~\citep{Bremner2015} extends this result to the approximate sampling case relying on conjectures, similarily to the boson sampling model. In contrast to boson sampling, though, these conjectures have later been proven by Jens Eisert and collaborators.

\begin{figure}[h!]
\centering
\begin{quantikz}
    \lstick{$\ket{0}$} & \gate{H} & \ctrl{1} & \ctrl{2} & & \ctrl{3} & & \gate{T^7} & \gate{H} & \meter{}\\
    \lstick{$\ket{0}$} & \gate{H} & \gate{Z^{\frac{1}{2}}} & & \gate{Z^{\frac{3}{2}}} &  & \ctrl{1} & \gate{T^4} & \gate{H} & \meter{}\\
    \lstick{$\ket{0}$} & \gate{H} & \gate{Z^{\frac{1}{2}}} & \gate{Z} & & & \gate{Z} & & \gate{H} & \meter{}\\
    \lstick{$\ket{0}$} & \gate{H} & \ctrl{-1} & & \ctrl{-2} & \gate{Z} & & \gate{T} & \gate{H} & \meter{}
\end{quantikz}
\caption{Example of an IQP circuit, taken from Ref.~\citep{harrow2017quantum}.}
\label{fig:IQP}
\end{figure}

The worst-case complexity of the problems in both case 1 and 2 can be seen to be hard to classically sample in two steps. 
First, one proves that 
these families of circuits are examples of sets that become universal under post-selection, and as a result their output probabilities are hard to classically sample. 
Then, complexity-theoretical arguments can be applied, that allow us to draw conclusions about the hardness of the model. We are going to review both aspects briefly here.

\subsection{Hadamard gadget}
\label{appHgadget}

This section is taken from the supplementary material of Ref.~\citep{Douce2017}. For either of the gate sets 1 or 2 above, the only missing ingredient for universality is the ability to perform Hadamard gates at any point within the circuit. In Ref.~\citep{Bremmer2010}, it was shown that such gates can be replaced with a ``Hadamard gadget'', which requires one post-selected qubit and controlled-phase gate per Hadamard gate. The Hadamard gadget~\citep{Bremmer2010} is the very essence of the difficulty to simulate IQP circuits on classical computers. It shows that under postselection, an IQP circuit can implement a Hadamard gate. 

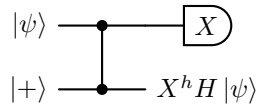
\begin{figure}[H]
\centering
\begin{quantikz}
\lstick{$\ket{\psi}$}  & \ctrl{1} & & \meterD{X} \\
\lstick{$\ket{+}$}  & \control{} & \rstick{$X^h H \ket{\psi}$} 
\end{quantikz}
\caption{\label{Hadamard-gadget-qubits} Hadamard gadget in a postselected IQP circuit, where $h$ takes value $0$ if $+1$ is measured, while $h= 1$ if the result is $-1$.}
\end{figure}

\subsubsection{Output state}

Suppose one wants to implement a Hadamard gate on an arbitrary qubit $\ket\psi=\alpha\ket0+\beta\ket1$. Following the circuit structure depicted in \figref{Hadamard-gadget-qubits}, we add an auxiliary qubit initialized in $\ket{+}$ so that we start from (omitting normalization)
\be
\ket{\psi} \ket{+} = \alpha\ket{00} + \alpha\ket{01} + \beta\ket{10} + \beta\ket{11}.
\ee
Then we apply the CZ gate and the measurement in the $X$ basis. Conditioned on getting the outcome corresponding to the state $\ket{+}$ when measuring the first qubit, we have
\begin{align}
\alpha\ket{00} + \alpha\ket{01} + \beta\ket{10} + \beta\ket{11} &\overset{\hat{C}_Z}{\longmapsto} \alpha\ket{00} + \alpha\ket{01} + \beta\ket{10} - \beta\ket{11} \nn \\
&\overset{\bra{+}}{\longmapsto} \alpha \mleft( \ket{0} + \ket{1} \mright) + \beta \mleft( \ket{0} - \ket{1} \mright) = H \ket{\psi}.
\end{align}
If instead we get the outcome corresponding to the state $\ket{-}$ when we measure the first qubit, the same kind of calculations give
\begin{align}
\alpha\ket{00} + \alpha\ket{01} + \beta\ket{10} + \beta\ket{11} &\overset{\hat{C}_Z}{\longmapsto} \alpha\ket{00} + \alpha\ket{01} + \beta\ket{10} - \beta\ket{11} \nn \\
&\overset{\bra{-}}{\longmapsto} \alpha \mleft( \ket{0} + \ket{1} \mright) - \beta \mleft( \ket{0} - \ket{1} \mright) = X H \ket{\psi}.
\end{align}
Defining $h$ as the outcome of the measurement, such that $h=0$ ($h=1$) corresponds to measuring the state $\ket{+}$ ($\ket{-}$), the result of the computation is, in the general case,
\be
X^h H \ket{\psi}.
\ee
So the point of postselecting is to ensure that it is indeed $H$, and not $-H$, that has been implemented. 

\subsubsection{Probability of measuring $\ket{+}$}

A subtlety with postselection that is worth mentioning concerns the probability of the conditioning result. Specifically, if one wants to postselect on a qubit measured in a given state, then the probability associated with this measurement must be nonzero, ensuring that the conditional probability describing the postselection is well-defined. In the case of the Hadamard gadget, we can compute the relevant success probability explicitly. We have after the $\hat{C}_Z$ gate (actually, $1/2$ times the following equation for normalization purposes)
\be
\alpha\ket{00} + \alpha\ket{01} + \beta\ket{10} - \beta\ket{11} = (\alpha + \beta) \ket{+0} + (\alpha - \beta) \ket{+1} + (\alpha - \beta) \ket{-0} + (\alpha + \beta) \ket{-1}.
\ee
It is then straightforward to show that the probability to measure $\ket+$ is
\be
\frac{1}{4} \mleft( \module{\alpha + \beta}^2 + \module{\alpha - \beta}^2 \mright) = \frac{1}{2}.
\ee

An interesting feature of this result is that it does not depend on the input state $\ket{\psi}$. So even if initialized in $\ket{-}$, the entangling CZ gate sort of smoothes the global state in such a way that the probability of measuring the first qubit in $\ket{+}$ becomes $1/2$. 
Given that the number of post-selected lines $l$ in a discrete-variable IQP circuit is on the order of the total number of lines in the circuit $n$, $l \sim O(n)$, the overall success probability distribution $1/2^l$ is exponentially low in the circuit size. However, we stress that this postselection should be regarded as a mathematical tool for the hardness proof, and its actual implementation is not required in practice.

\subsubsection{Complexity-theoretical arguments and proof of computational hardness}

In \secref{appHgadget}, we have shown that IQP with \textit{postselected} measurements is universal for PostBQP (that is, quantum polynomial-time with postselection on possibly exponentially unlikely measurement outcomes). In other words, to any computation in PostBQP corresponds to a postselected IQP circuit. 

Furthermore, Aaronson previously showed that $\text{PostBQP} = \text{PP}$. On the other hand, if a classical algorithm existed for simulating IQP, then we will show that we could simulate postselected IQP in PostBPP (that is, \textit{classical} polynomial-time with postselection, also called PostBPP). This would lead to the following chain of inclusions of complexity classes:
\begin{equation}
\label{chain-of-inclusions}
\text{PostBPP} \supseteq \text{PostIQP} \supseteq \text{PostBQP} = \text{PP}.
\end{equation}
This is known to imply a collapse of the polynomial hierarchy. The final argument why this happens is that on the one hand, PostBPP is contained in the third level of the polynomial hierarchy, PostBPP $\subseteq \Sigma_3$. On the other hand, due to Toda's theorem, $\text{PH} \subseteq \text{P}^{\text{PP}}$.

Hence, if \eqref{chain-of-inclusions} was true, then we would have
\begin{equation}
\label{chain-of-inclusions2}
\text{PP} \subseteq \text{PostBPP} \subseteq \Sigma_3, 
\end{equation}
also implying
\begin{equation}
\label{chain-of-inclusions3}
\text{PH} \subseteq \text{P}^{\text{PP}} \subseteq \text{P}^{\Sigma_3} \subseteq \Sigma_4, 
\end{equation}
implying a collapse of the polynomial hierarchy to the fourth level. A more stringent argument can be given to show that the actual implication regards a collapse to the third level. That is, to summarize, if exact IQP was efficiently classically simulatable, the full polynomial hierarchy would be contained in the third level, which implies the collapse.

\section{Random circuit sampling}
\label{random-sampling-google}

In 2019, Google published a paper demonstrating quantum primacy using a 53-qubit quantum computer~\citep{Arute:2019aa}. From an information-theoretic perspective, the classical computational hardness of sampling from this circuit family has been demonstrated in Ref.~\citep{bouland2018quantum}. 

What do these results mean and what are the implications for WACQT and other efforts to build a quantum computer around the world? This section is the WACQT statement following the release of the Google experiment, and has been written by G.~Johansson, with further edits by G.~Ferrini. For more information, and for a very clear explanation of the implications of this experiment for quantum advantage, we recommend to read the relevant blog posts on Scott Aaronson's blog \url{https://www.scottaaronson.com/blog/}. For a very nice discussion on the terminology regarding {\it quantum primacy}, take a look at Ref.~\citep{QuantumPrimacy} and \url{https://phys.cam/2019/10/quantum-supremacy/}, where you will also find a pedagogical introduction to the Google quantum primacy experiment. Also note that this experiment has been followed by a demonstration with 60 qubits from the Pan group~\citep{zhu2021quantum, wu2021}.

\begin{figure}[h!]
\centering
\includegraphics*[trim = {5cm, 0, 5cm, 0}, clip, width = 0.35 \columnwidth, angle = 270]{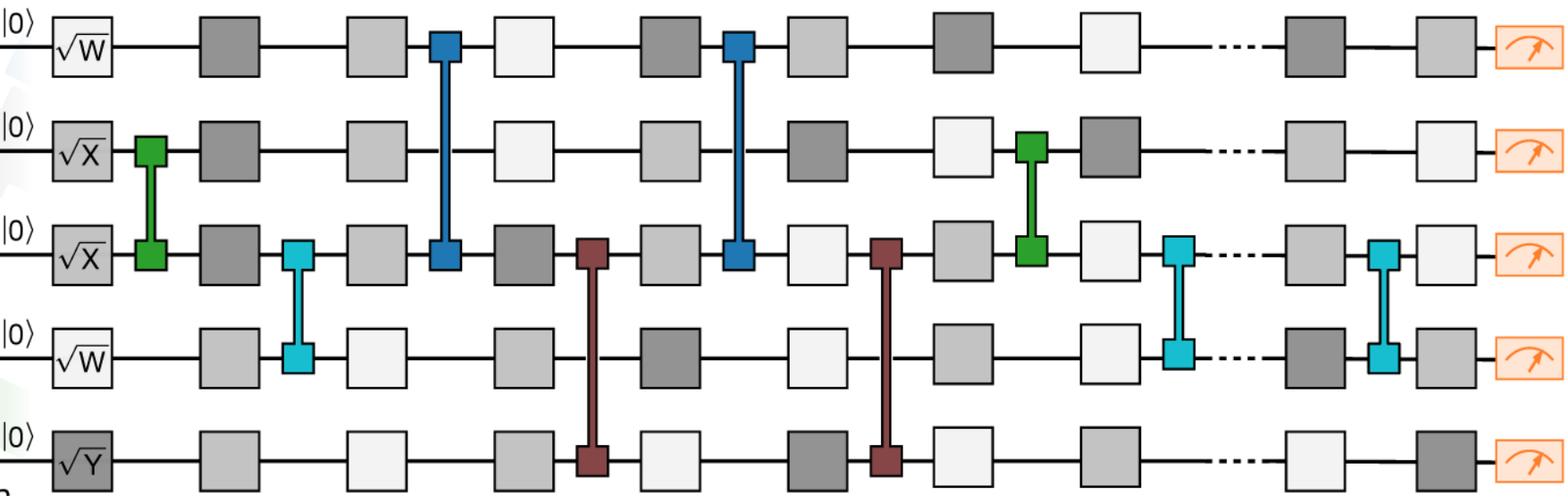}
\caption{Sketch of a Random Circuit Sampling, taken from Ref.~\citep{Arute:2019aa}. }
\label{random circuit sampling-figure}
\end{figure}

\begin{enumerate}

\item The Google experiment is a major milestone, demonstrating that a gate-based quantum computer can indeed perform a computational task in 3 minutes, which would take, at the time when the experiment result was first released, 2.5 days to solve on the most powerful supercomputer on earth. Importantly, if Google would now add a few more qubits to their chip, simulating the outcome would become impossible even for that supercomputer, in decent times. For example, for 60 qubits, you would need 33 such supercomputers for just storing the quantum state of the Google chip.

\item The computation is not claimed to be useful in any way. The task is to sample from a particular probability distribution. This task was chosen carefully, because it is easy to perform on Google's quantum computer, but hard on any classical computer.

\item This does not imply that quantum computers from now on outperform classical computers in general. However, as quantum computers evolve, the class of computational tasks where the quantum computer performs better than classical computers will grow. The hope is of course that at some point this class will also contain useful computations.

\item The computation was performed without using error correction. The error rate was low enough to give the right answer for this comparatively short quantum algorithm. This is a so-called noisy intermediate-scale quantum (NISQ) device.

\item The basic architecture of the 53-qubit device is similar to previously published devices from Google. The breakthrough consists of careful engineering of control hardware and software, as well as a thorough analysis of which computational tasks are easy for a quantum computer and hard for classical computers.

\item The algorithm creates a random entangled state by repeating layers of eight sets of gates, where most qubits are taking part in eight entangling gates, two with each nearest neighbour, interlaced with eight single-qubit gates. For the longest algorithm, each layer is repeated twenty times, giving on the order of $53 \times (8/2) \times 20 \sim 4000$ two-qubit gates. The algorithm is then repeated one million times, to give appropriate statistics. The full run-time is 200 seconds. To perform the similar sampling on a supercomputer with 1 million cores was estimated to take, when the experiment was released, 2.5 days.

\item The average lifetime ($T_1$) of the Google qubits in this experiment was 16 microseconds. Google's design gives them very fast two-qubit gates, taking less than \unit[20]{ns}. We also note the importance of automated calibration and control software.

\item In contrast to the other circuits in this chapter that are candidate to yield quantum advantage, in Ref.~\citep{Arute:2019aa}, the gates that are implemented are in principle drawn from a universal gate set. More in detail, regarding single-qubit gates, they implement $\sqrt{X}$, $\sqrt{Y}$, $\sqrt{W}$, with $W = (X + Y)/\sqrt{2}$ (see \figref{random circuit sampling-figure}). They generate random quantum circuits using the two-qubit unitaries measured for each pair during simultaneous operation, rather than a standard gate for all pairs. The typical two-qubit gate is a full iSWAP with 1/6th of a full CZ. Using individually calibrated gates in no way limits the universality of the demonstration.

\end{enumerate}

\begin{figure}[h!]
\centering
\includegraphics*[width=0.9\columnwidth]{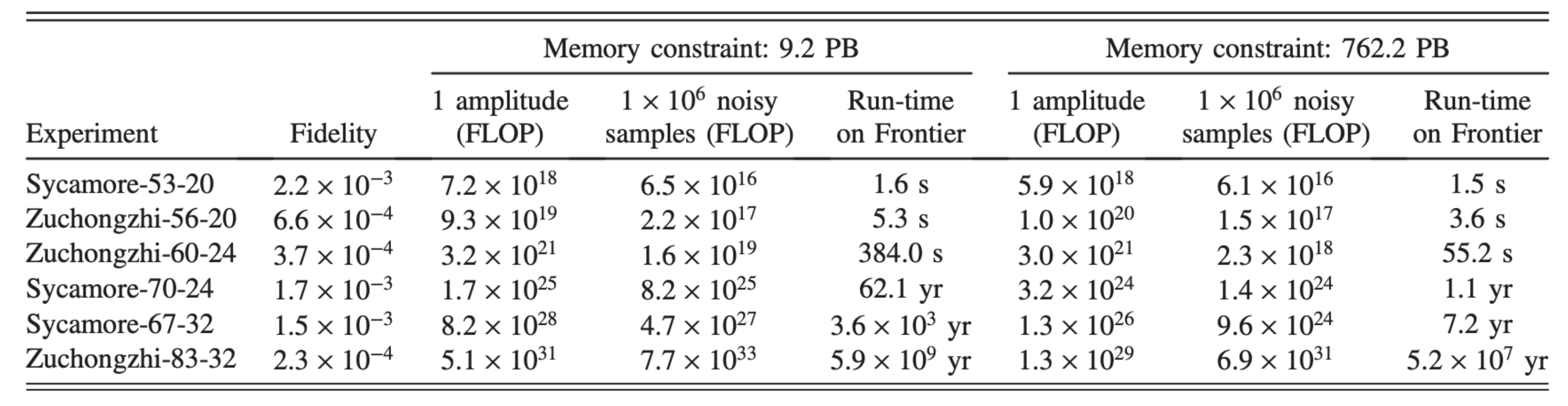}
\caption{Table 1 from Ref.~\citep{Gao2025}. Estimated classical computational cost for different experiments. The classical cost are estimated with respect to the Frontier supercomputer in two scenarios: one with 9.2 PB of memory (the actual memory of Frontier) and another with 762.2 PB (combining Frontier's actual memory with all storage, which is an impractical situation).}
\label{fig:quantum-advantage-chinese}
\end{figure}

Since the first Google experiment was performed in 2019, the debate whether this has yielded a conclusive demonstration of quantum advantage has not stopped. On the one hand, improved classical simulation algorithms based on tensor networks were able to simulate the Google 53-qubit experiment in 15 hours, using 512 GPUs~\citep{pan2021solving}. See \url{https://scottaaronson.blog/?p=6871} for a discussion on the energy cost of such simulation, which also points to the concept of energy advantage, that quantum computers might be able to provide.  

On the other hand, in 2022, it has been shown that random circuit sampling, with a constant rate of noise per gate and no error correction, cannot provide a scalable approach to quantum primacy. This is because as the number of qubits $n$ goes to infinity, and assuming that the depth of the quantum circuit is at least $\log(n)$, there is a classical algorithm to approximately sample the quantum circuit's output distribution in $\text{poly}(n)$ time (albeit, not yet a practical algorithm)~\citep{Aharonov-2022}. Further discussions and a review of results attempting to corroborate or refute Google's quantum primacy claim can be found in Ref.~\citep{Kalai-2022}. Further recent experiments on RCS were performed by the Google group with 67 qubits~\cite{morvan2024phase}, and at university of China with 83 qubits~\cite{Gao2025}, operated  in the low-noise regime, and claimed to be non-simulatable by the best available classical computers. 

\section{Boson sampling}
\label{sec-Boson-sampling}

\subsection{Definition of the boson sampling model}

Aaronson and Arkipov~\citep{Aaronson2013} describe a simple model for producing output probabilities that are hard to classically sample. Their model uses bosons that interact only by linear scattering\footnote{In this sense, therefore, the boson sampling model already lives in the bosonic space associated with an infinite dimensional Hilbert space and with continuous-variable operators, such as the quadratures of the field. However, we present this model in the discrete-variable section of the notes, to highlight the contrast with models making use of squeezed states and homodyne detection. The latter are more traditionally associated with the continuous-variable approach, and they will be presented in Chapter~\ref{ch-CV}.}. The bosons must be prepared in a Fock state and measured in the Fock basis.

\begin{figure}[h!]
\centering
\includegraphics*[trim = {0cm, 5cm, 0cm, 5cm}, clip, width=0.4\columnwidth]{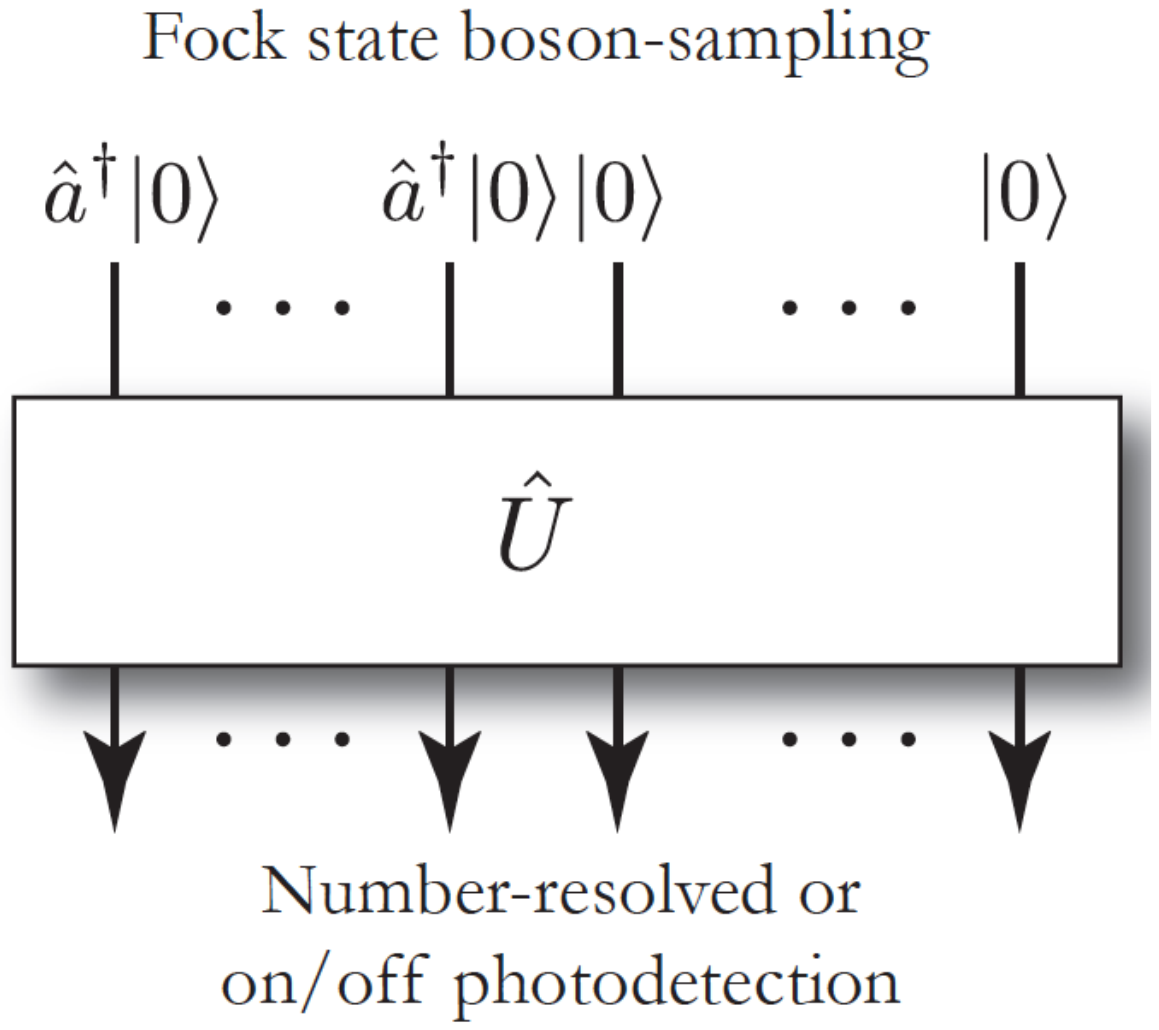}
\caption{The setup for boson sampling. The picture is taken from Ref.~\citep{Olson2015}.}
\label{fig:bs-fock}
\end{figure}

Consider for this purpose $M$ input ports of a multiport splitter, which we feed with $N$ photons. We assume that the input state only has at most one input photon per mode (see \figref{fig:bs-fock}). Without loss of generality, we order the modes such that the first $N$ input modes contain a photon, while the others are empty, i.e.,
\be
\label{eq:input state}
|\psi_{\text{in}} \rangle = |1_1, ... 1_N, ..., 0_M \rangle = \hat{a}^\dag_1  ... \hat{a}^\dag_N |0_1, ... , 0_M \rangle \equiv |T \rangle .
\ee
Now a linear optics network, described by the $M \times M$ matrix $U$, is applied to the input state. Linear bosonic interactions, or linear scattering networks, are defined by dynamics in the Heisenberg picture that generate a linear relationship between the annihilation operators of each mode, i.e.,
\begin{align}
\label{eq:linear-optics-network}
\hat{b}_j &= \mathcal{U}^\dag \hat{a}_j \mathcal{U} = \sum_{k=1}^M U_{j,k} \hat{a}_k, \mbox{ i.e.,  } \vec{b} = U \vec{a}; \\
\hat{b}^{\dagger}_j &= \mathcal{U}^\dag \hat{a}^\dag_j \mathcal{U} = \sum_{k=1}^M U^\dag_{j,k} \hat{a}^\dag_k, \mbox{ i.e.,  } \vec{b^\dag} = U^\dag \vec{a^\dag},
\end{align}
which also implies that $\hat{a}^\dag_j = \sum_{k=1}^M  U_{j,k} \hat{b}^\dag_k$. It is important to make a distinction between the unitary operator $\mathcal{U}$, which acts upon the Fock basis, and the unitary matrix defined by $U$, which describes the linear mixing of modes. For optical systems, the matrix $U$ is determined by how linear optical elements, such as beam splitters and phase shifters, are laid out. In fact, all unitary networks can be constructed using just beam splitters and phase shifters.

The set of events which are then output by the algorithm is a tuple of $M$ non-negative integers whose sum is $N$. This set is denoted $\Phi_{M,N}$. As we will explicitly show in \secref{proof-permanents}, the probability distribution of output events is related to the matrix permanent of submatrices of $U$. The matrix permanent is defined in a recursive way like the common matrix determinant, but without the alternation of addition and subtraction. For example,
\be
\text{Per}
\begin{pmatrix}
a & b \\
c & d
\end{pmatrix}
= ad + cb;
\ee
\be
\text{Per}
\begin{pmatrix}
a & b & c \\
d & e & f \\
g & h & i \\
\end{pmatrix} 
= aei + ahf + bdi + bgf + cdh + cge.
\ee
In a more general form,
\be
\text{Per} (A) = \sum_{\sigma\in \mathcal{S}_n} \prod_{i = 1}^{n} a_{i,\sigma(i)} ,
\label{eq:Permanent}
\ee
where $\mathcal{S}_n$ represents the elements of the symmetric group of permutations of $n$ elements. 

With this, we can now define the output distribution of the linear network with the input state from \eqref{eq:input state}. For an output event $S = (S_1,S_2,...,S_M) \in \Phi_{m,n}$, the probability of $S$,
\be
\label{eq:permanent-boson-sampling}
P_S = \frac{|\text{Per}(U_{S})|^2}{S_1!S_2!...S_N!} ,
\ee
where the matrix $U_S$ is an $N \times N$ submatrix of $U$ where column $i$ is repeated $S_i$ times and only the first $N$ rows are used. One critical observation of this distribution is that all events are proportional to the square of a matrix permanent derived from the original network matrix $U$. The resulting photon distribution in the output is hard sample classically, as we will show later.

\subsection{Proof that the boson sampling probability distribution is proportional to permanents}
\label{proof-permanents}

We now explicitly show that the output probability distribution of the boson sampling circuit is proportional to the permanent of the relevant submatrix. We follow a derivation similar to the one in the supplementary information of Ref.~\citep{Spring2013}. In a more general formulation, the input population is $ |T \rangle =  |T_1, ... T_M \rangle$. The general case of an arbitrary input state is treated in Ref.~\citep{scheel2004permanents}. 

In the Heisenberg representation, the input state does not evolve. However, using Eqs.~(\ref{eq:linear-optics-network}) and (\ref{eq:input state}), we can re-express it as
\be
\label{eq:sum-terms}
|\psi_{\text{out}} \rangle = |\psi_{\text{in}} \rangle 
= \mleft( \prod_{j = 1}^N \hat{a}^\dag_j \mright) |0_1, ... , 0_M \rangle  = \mleft( \prod_{j = 1}^N \sum_{k=1}^M U_{j,k} \hat{b}_k^\dag \mright) |0_1, ... , 0_M \rangle .
\ee
The sum in \eqref{eq:sum-terms} is composed of $M^N$ terms, which is the number of ways in which we can put $N$ objects in $M$ modes, allowing for repetitions. We can refer to this set of permutations as $\tilde{V}$, and then rename the terms of the sum in \eqref{eq:sum-terms}, i.e.,
\be
\prod_{j = 1}^N \sum_{k=1}^M U_{j,k} \hat{b}_k^\dag = \sum_{j=1}^{M^N} \prod_{k=1}^N U_{k,\tilde{V}_k^j} \hat{b}_{\tilde{V}_k^j}^\dag ,
\ee
where $k$ is the $k$th boson and $\tilde{V}_k^j$ is in which mode that photon is found in the permutation $j$. In other words, the ensemble $\tilde{V}$ is the set of $M^N$ permutations of N photons in M modes, with repetitions allowed. Let us consider, as an example, the case in which we have as input $N = 2$ photons in $M = 3$ modes.
Then we have to consider $\prod_{j=1}^2 \hat{a}^\dag_j$, i.e., the first two rows of the vector
\be
\label{eq:matrix}
\begin{pmatrix}
\hat{a}_1^\dag \\
\hat{a}_2^\dag \\
\hat{a}_3^\dag
\end{pmatrix}
=
\begin{pmatrix}
U_{11} & U_{12} & U_{13} \\
U_{21} & U_{22} & U_{23} \\
U_{31} & U_{32} & U_{33}
\end{pmatrix}
\begin{pmatrix}
\hat{b}_1^\dag \\
\hat{b}_2^\dag \\
\hat{b}_3^\dag
\end{pmatrix}
=
\begin{pmatrix}
U_{11} \hat{b}_1^\dag + U_{12} \hat{b}_2^\dag + U_{13} \hat{b}_3^\dag \\
U_{21} \hat{b}_1^\dag + U_{22} \hat{b}_2^\dag + U_{23} \hat{b}_3^\dag \\
U_{31} \hat{b}_1^\dag + U_{32} \hat{b}_2^\dag + U_{33} \hat{b}_3^\dag
\end{pmatrix}
,
\ee
which gives
\ba
\label{eq:sum-long}
\prod_{j = 1}^2 \hat{a}^\dag_j &=& \mleft( U_{11} \hat{b}_1^\dag + U_{12} \hat{b}_2^\dag + U_{13} \hat{b}_3^\dag \mright) \mleft( U_{21} \hat{b}_1^\dag + U_{22} \hat{b}_2^\dag + U_{23} \hat{b}_3^\dag \mright) \nn \\
&=& U_{11} U_{21} \hat{b}_1^{\dag 2} + U_{11} U_{22} \hat{b}_1^\dag \hat{b}_2^\dag + U_{11} U_{23} \hat{b}_1^\dag \hat{b}_3^\dag \nn \\
&+& U_{12} U_{21} \hat{b}_2^\dag \hat{b}_1^\dag + U_{12} U_{22} \hat{b}_2^{\dag 2} + U_{12} U_{23} \hat{b}_2^\dag \hat{b}_3^\dag \nn \\
&+& U_{13} U_{21} \hat{b}_3^\dag \hat{b}_1^\dag + U_{13} U_{22} \hat{b}_2^\dag \hat{b}_3^\dag + U_{13} U_{23} \hat{b}_3^{\dag 2}.
\ea
Hence here in the sum of \eqref{eq:sum-long} we have
\ba
j = 1:  \hspace{0.25cm} k = 1 \rightarrow \tilde{V}_k^j = 1;   \hspace{0.25cm} k = 2 \rightarrow \tilde{V}_k^j = 1; \nn \\
j = 2:  \hspace{0.25cm} k = 1 \rightarrow \tilde{V}_k^j = 1;   \hspace{0.25cm} k = 2 \rightarrow \tilde{V}_k^j = 2; \nn \\
j = 3:  \hspace{0.25cm} k = 1 \rightarrow \tilde{V}_k^j = 1;   \hspace{0.25cm} k = 2 \rightarrow \tilde{V}_k^j = 3; \nn \\
j = 4:  \hspace{0.25cm} k = 1 \rightarrow \tilde{V}_k^j = 2;   \hspace{0.25cm} k = 2 \rightarrow \tilde{V}_k^j = 1; \nn \\
...
\ea
%

Let us indicate with $S_i$ the number of photons in mode $i$ in the configuration $S$; for each configuration we have $\sum_{i=1}^M {S_i} = N$. 
The total number of configurations is the number of ways of arranging $N$ bosons in $M$ modes, i.e.,
\be
N_\text{config} = \binom{N + M - 1}{N}
\ee
(repetitions not allowed).
$S^k$ indicates the mode in which the photon $k$ is found in the configuration $S$.  
If the total number of input photons is small compared to the number of modes, such that $N \sim \sqrt{M}$, then the probability that two photons are found in the same output mode is rather small ({\it birthday paradox}). Nevertheless, we will consider here the general case of an arbitrary output distribution. The probability distribution $S$ is evaluated by projection of the output state \eqref{eq:sum-terms} onto the configuration state $|S \rangle \equiv |S_1, ... , S_M \rangle$, i.e.,
\ba
\label{eq:output-prob}
P_{S} &=& \abs{\langle S | \psi_{\text{out}} \rangle }^2 = \frac{1}{S_1! ... S_M!} \abs{ \langle  {0}_1, ... , {0}_M | \mleft( \hat{b}_1^{ S_1} ...  \hat{b}_M^{S_M} \mright) \sum_{j=1}^{M^N} \prod_{k= 1}^N U_{k,\tilde{V}_k^j} \hat{b}_{\tilde{V}_k^j}^\dag |0_1, ... , 0_M  \rangle }^2 \nn \\
&=& \frac{1}{(S_1! ... S_M!)} \abs{ \langle {0}_1, ... , {0}_M | \prod_{k = 1}^N \hat{b}_{S^k} \mleft( \sum_{j=1}^{M^N} \prod_{k=1}^N U_{k,\tilde{V}_k^j} \hat{b}_{\tilde{V}_k^j}^\dag \mright) |0_1, ... , 0_M \rangle}^2.
\ea

Before going to the general case of arbitrary output configuration, we want to fix the ideas with an example. Consider the case where we project onto the state $|\lgr S \rgr \rangle = |020 \rangle = \hat{b}_2^{\dagger 2} /\sqrt{2} |000 \rangle$.
Then the only contributing term in \eqref{eq:sum-long} is $U_{12} U_{22} \hat{b}_2^{\dag 2}$, and we obtain
\ba
\label{eq:output-prob-example}
P_S &=& \abs{\langle S | \psi_{\text{out}} \rangle }^2 = \frac{1}{2!} \abs{ \langle  000 | \hat{b}_2^{2} U \hat{b}_1^\dag \hat{b}_2^\dag |000 \rangle }^2  \nn \\
&=& \frac{1}{2} \abs{ \langle 000 | \hat{b}_2^2 U_{12} U_{22} \hat{b}_2^{\dag 2} |000 \rangle}^2 \nn \\
&=& \frac{1}{2} 4 \abs{U_{12} U_{22} }^2 = 2 \abs{U_{12} U_{22} }^2,
\ea
where we have used that $\langle 0_2 | \hat b_2^{2} \hat b_2^{\dag S_2} | 0_i\rangle = 2$.
We can now verify that this expression is equivalent to the permanent of the submatrix of $U$ identified by the rows corresponding to the input photons, and the columns identified by the output configuration of interest, where the columns are repeated as many times as the occupation of the output mode. In practice, we have
\be
U =
\begin{pmatrix}
 U_{11} &  U_{12} &  U_{13} \\
 U_{21} &  U_{22} &  U_{23} \\
 U_{31} &  U_{32} &  U_{33}
\end{pmatrix}
,
\ee
so that
\be 
U_S =
\begin{pmatrix}
 U_{12} &  U_{12} \\
 U_{22} &  U_{22}
\end{pmatrix}
,
\ee
and
\be
\text{Per}(U_{S}) = U_{12} U_{22} + U_{12} U_{22} = 2 U_{12} U_{22},
\ee
yielding immediately \eqref{eq:output-prob-example} according to \eqref{eq:permanent-boson-sampling}.

In the general case, let us now indicate with $\tilde{S}^j_k$ the $j$th permutations of the $N$ photons in the output state, where $k$ is the photon index. At maximum, there will be $N!$ permutations, corresponding to the ways of arranging $N$ photons in $N$ modes, repetitions not allowed. (Note that if in the output distribution there are more photons per mode, then the number of ways we can put $N$ photons in $M$ modes with $S_1$ in the first, $S_2$ in the second... is $N!/(S_1! S_2 ! ... S_M !)$).
For example, it is clear that if we project onto the state $|\lgr S \rgr \rangle = |011 \rangle$, then we have 
\ba
&j = 1: \hspace{0.25cm} k = 1 \rightarrow \tilde{S}^j_k = 2; \hspace{0.25cm} k = 2 \rightarrow \tilde{S}^j_k = 3; \nn \\
&j = 2: \hspace{0.25cm} k = 1 \rightarrow \tilde{S}^j_k = 3; \hspace{0.225cm} k = 2 \rightarrow \tilde{S}^j_k = 2.
\ea
It is clear that the only nonzero terms in the sum of \eqref{eq:output-prob} will be those for which $\tilde{V}_k^S = \tilde{S}_k^j$, hence\footnote{We have used that
$$
\label{step-to-check}
\langle {0}_1, ... , {0}_M | \prod_{k=1}^N \hat{b}_{S^k} \mleft( \sum_{j=1}^{M^N} \prod_{k=1}^N U_{k,\tilde{V}_k^j} \hat{b}_{\tilde{V}_k^j}^\dag \mright) |0_1, ... , 0_M \rangle 
= S_1!...S_M! \sum_{j=1}^{M^N} \prod_{k=1}^N U_{k,\tilde{S}_k^j} = \sum_{j=1}^{N!} \prod_{k=1}^N U_{k,\tilde{S}_k^j}.
$$
}
\be
\label{eq:output-prob2}
P_{S} = \frac{1}{S_1! ... S_M!} \abs{ \sum_{j=1}^{N!} \prod_{k=1}^N U_{k, \tilde{S}_k^j}}^2,
\ee
where we have used that $\langle 0_i | \hat a_i^{S_i} \hat a_i^{\dag S_i} | 0_i\rangle = S_i !$. For instance, we obtain for the state $|\lgr S \rgr \rangle = |011 \rangle$ that $P_S = \abs{U_{12} U_{23} + U_{13} U_{22}}^2$, while for the state $ |020 \rangle$ we have seen that $P_S = 2 \abs{U_{12} U_{22}}^2$.
We can now compare Eq.~(\ref{eq:output-prob2}) with the formula for the permanent of an $L \times L$ matrix $A$ in \eqref{eq:Permanent}, stated here again for convenience:
\be
\label{eq:permanentA}
\text{Per}(A) = \sum_{j=1}^{L!} \prod_{k= 1}^L a_{k, \tilde{\sigma}_k^j}.
\ee
where $\tilde{\sigma}_k^j$ is the Pauli matrix corresponding to the $k$th element of the $j$th permutation of the numbers $1, ... , L$.
It is easy to compare \eqref{eq:output-prob2} to \eqref{eq:permanentA}; note however that our original matrix was $M \times M$, while we are computing here only the permanent of the submatrix involving the first $N$ rows, and columns which correspond to configuration $S$. Hence we obtain \eqref{eq:permanent-boson-sampling}. 

\subsection{Sketch of the proof of computational hardness of the boson sampling probability distribution}

\subsubsection{Exact boson sampling}
\label{sec-boson-sampling-hardness}

The first main result of the original Aaronson paper states the following:

\emph{Theorem 1: hardness of exact boson sampling} The exact boson sampling problem is not efficiently solvable by a classical computer, unless $\text{P}^{\#\text{P}} = \text{BPP}^{\text{NP}}$ and the polynomial hierarchy collapses to the third level. 

At least for a computer scientist, it is tempting to interpret Theorem 1 as saying that ``the exact \textsc{BosonSampling} problem is ${\#\text{P}}$-hard under $\text{BPP}^{\text{NP}}$-reductions.'' Notice that this would have a shocking implication: that quantum computers (indeed, quantum computers of a particularly simple kind) could efficiently solve a $\#$P-hard\ problem!
There is a catch, though, that has to do with the fact that \textsc{BosonSampling} is a sampling problem rather than a decision problem. 
In other words, the ``reduction'' from ${\#\text{P}}$-complete problems to \textsc{BosonSampling} makes essential use of the hypothesis that we have a \textit{classical} \textsc{BosonSampling} algorithm. Details can be found in the original article. 

Two proofs of Theorem 1 can be given. In the first proof, we consider the probability $p$ of some particular basis state when a boson computer is measured. We then prove two facts:

\begin{enumerate}
\item[(1)] Even \textit{approximating} $p$ to within a multiplicative constant is a ${\#\text{P}}$-hard problem.

\item[(2)] \textit{If} we had a polynomial-time classical algorithm for exact \textsc{BosonSampling}, \textit{then} we could approximate $p$ to within a multiplicative constant in the class $\text{BPP}^{\text{NP}}$, by using a standard technique called \textit{universal hashing}.
\end{enumerate}

Combining facts (1) and (2), we find that, if the classical \textsc{BosonSampling} algorithm exists, then $\text{P}^{\#\text{P}} = \text{BPP}^{\text{NP}}$, and therefore the polynomial hierarchy collapses.

The second proof is inspired by independent work of Bremner, Jozsa, and Shepherd~\citep{Bremmer2010}, and by the proof of computational hardness for the IQP model that we have seen in \secref{IQP-DV}. In this proof, one starts with a result of Knill, Laflamme, and Milburn, which says that linear optics with \textit{adaptive measurements} is universal for BQP, giving name to the respective KLM model. A straightforward modification of their construction shows that linear optics with \textit{postselected} measurements is universal for PostBQP (that is, quantum polynomial-time with postselection on possibly exponentially unlikely measurement outcomes). Furthermore, Aaronson previously showed that $\text{PostBQP} = \text{PP}$. On the other hand, if a classical \textsc{BosonSampling} algorithm existed, then we will show that we could simulate postselected linear optics in PostBPP (that is, \textit{classical} polynomial-time with postselection, also called Path BPP). We would therefore get exactly the same chain of inclusion as in \eqref{chain-of-inclusions}, and the same conclusions on the hardness of the model apply. 

\subsubsection{Approximate boson sampling}

While theoretically interesting, Theorem 1 unfortunately does not suffice to rule out the extended Church--Turing thesis, as even an optical setup realistically cannot perform exact boson sampling due to experimental noise. Thus, one must consider approximate boson sampling, and show that it is also hard to sample. More precisely, what Aaronson has demonstrated is that even sampling from a probability distribution $\mathcal{D'}_U$ that is away from the exact boson sampling one $\mathcal{D}_U$ by a certain error bound $\epsilon$, i.e., $\sum_i \mathcal{D}_U^i - \mathcal{D'}_U^{i} < \epsilon$, it is classically hard.

The proof of computational hardness of approximate boson sampling relies on two extra conjectures, beyond the fact that the polynomial hierarchy does not collapse, namely the Permanent of Gaussians conjecture, and the Permanent anti-concentration conjecture. We will not go into the details of the proof of hardness of approximate boson sampling here. 

\subsubsection{Experimental realisations and classical simulatability of boson sampling}

Several proof-of-principle experiments have been achieved, some earlier ones with a handful of modes and single photons, see among others Ref.~\citep{Spring2013}, and later ones with 20 input photons injected in 60 modes~\citep{wang2019boson}. Due to the probabilistic nature of single-photon sources, experimentalists have now moved towards Gaussian boson sampling to seek large-scale demonstrations of quantum advantage, which was achieved shortly after the Google experiment on RCS using 50 input squeezed states, and 100 optical modes~\citep{Zhongeabe8770}, and later by Xanadu with 216 squeezed states~\citep{madsen2022quantum}. The input squeezed states can be deterministically produced, and the results on the impossibility of simulating the outcome probability distributions still stand, although with a different proof technique~\citep{hamilton2017gaussian}. Also note that, in contrast to random circuit sampling, at least one useful application of Gaussian boson sampling has been outlined, namely the calculation of vibronic molecular spectra~\citep{huh2015boson}. 
For an extensive discussion, see~\url{https://www.scottaaronson.com/blog/?p=5122}.
We will briefly review more sampling models with squeezed states and continuous variables in \secref{sse:sampling-in-CV}.

Finally, note that classical algorithms are also constantly improving, which allow the simulation and benchmarking of larger and larger boson sampling~\citep{neville2017classical, oh2022classical} and Gaussian boson sampling devices~\citep{ li2020benchmarking,mccormick2022race}.
Also, in analogy to the qubit case for RCS, imperfections such as noise (photon losses, partial distinguishability of the photons) might render boson sampling~\citep{moylett2019classically, PhysRevLett.124.100502, liu2023simulating} and Gaussian boson sampling circuits~\citep{oh2023tensor} classically efficiently simulatable.

\chapter{Continuous-variable approach to quantum information}
\label{ch-CV}

In the continuous-variable (CV) approach to quantum information processing, some of the relevant observables are characterized by a continuous spectrum, such as the amplitude $\hat q = (\hat a + \hat a^\dag)/(\sqrt{2})$ and phase $\hat p = (\hat a - \hat a^\dag)/(i\sqrt{2})$ quadratures of the electromagnetic field, satisfying $[\hat q, \hat p] = i $. The associated Hilbert space is infinite-dimensional, and the (infinite) energy levels are eigenstates of the number operator $\hat n = \hat a^\dag \hat a$. This is opposed to the  traditional discrete-variable (DV) approach, where observables have a discrete spectrum, and the Hilbert space is finite-dimensional. Generally speaking, CV systems offer the advantage that the resource states (e.g., squeezed states or large cluster states, that we will introduce in the following) can be deterministically produced.

Moreover, new methods for experimental implementations, offering solutions to scalability, have emerged in the context of CV. For instance, the Furusawa (Tokyo, Japan), Andersen (Lyngdby, Denmark), and Pfister (Charlottesville, USA) groups have been able to produce CV entangled states of up to one million optical modes; in the experiments of N.~Treps and V.~Parigi (Paris, France) several squeezed states are simultaneously available in the same optical cavity.
With microwave cavities coupled to superconducting circuits, the Yale group has demonstrated that it is possible to store quantum information for a longer time in a radiation state than if the corresponding quantum information is encoded in a qubit, using bosonic codes such as the GKP code. The Chalmers group has demonstrated the capability of generating highly non-linear states of radiation such as the cubic phase state.  Trapped ions platforms (Home, Zurich) have allow allowed for implementing and stabilizing bosonic codes.
Continuous variables are therefore promising for the implementation of scalable and robust architectures for quantum computing. 

\section{Quantum computing with continuous variables}

\sse{Basis states and quadrature operators}
\label{quantized-harmonic oscillator}

Here we provide the basic tools that we need and use in the following, starting from the basis states.
A review of the formalism underlying CV quantum information, namely the quantization of the harmonic oscillator, is provided in Refs.~\citep{gerry2005introductory,meystre2007elements}. Note that that those references use a different systems of units than the one we use in the rest of these lecture notes. 

\ssse{Fock states}

We have seen that qubit systems are characterised by two addressable states, or ``levels'', forming the basis for the two-dimensional Hilbert space of a single qubit. Continuous-variable systems are instead characterised by \emph{infinitely many} levels, yielding an infinite-dimensional Hilbert space associated to even a single bosonic mode. The corresponding quantum states are called \emph{Fock states}, or number states $\ket{n}$, where physically $n$ represents the number of quanta (e.g., photons, phonons, ...) in a single-mode bosonic field. The Fock states are eigenstates of the \emph{number operator} $\hat{n}$, satisfying
\begin{equation}
	\hat{n} \ket{n} = n \ket{n}.
\end{equation}
In particular, the Fock state corresponding to $n = 0$ is called the vacuum state $\ket{0}$.
Fock states are orthogonal
\begin{equation}
	\langle m | n \rangle = \delta_{mn}
\end{equation}
and form a complete set
\begin{equation}
	\sum^\infty_{n=0}\ketbra{n}{n} = 1.
\end{equation}

The number operator can also be expressed in terms of the non-Hermitian annihilation and creation operators $\hat{a}$ and $\hat{a}^\dag$, which satisfy the commutation relation
\begin{equation}
\label{eq:commutation-a-ad}
	[\hat{a}, \hat{a}^\dag] = 1,
\end{equation}
namely, $\hat{n} = \hat{a}^\dag \hat{a}$.
Acting with the creation and annihilation operators on the Fock states yields
\begin{align}
	\label{eq:AnnihilateFock}
	\hat{a}\ket{n} &= \sqrt{n}\ket{n-1}, \\
    \label{eq:CreateFock}
    \hat{a}^\dag\ket{n} &= \sqrt{n+1}\ket{n+1}.
\end{align}
Hence it is clear that the creation operator $\hat{a}^\dag$ creates an energy quantum and the annihilation operator $\hat{a}$ destroys an energy quantum in the single-mode bosonic field. Any Fock state can be generated by acting on the vacuum state multiple times with the creation operator:
\begin{equation}
	\label{eq:FockFromVacuum}
	\frac{\hat{a}^{\dag n}}{\sqrt{n!}} \ket{0} = \ket{n}.
\end{equation}
The action of the annihilation operator on the vacuum yields
\begin{equation}
	\label{eq:AnnihilationVaccum}
	\hat{a} \ket{0} = 0.
\end{equation}

\subsubsection{Quadrature operators}

It is convenient to introduce the two Hermitian operators
\begin{align}
    \hat{q} &= \frac{1}{\sqrt{2}} \mleft( \hat{a} + \hat{a}^\dag \mright), 
    \label{eq:QuadratureOperatorQ} \\
    \hat{p} &= \frac{1}{\sqrt{2} i} \mleft( \hat{a} - \hat{a}^\dag \mright),
    \label{eq:QuadratureOperatorP}
\end{align}
which satisfy the commutation relation
\begin{equation}
\label{eq:commutation-q-p}
	[\hat{q}, \hat{p}] = i.
\end{equation}
Note that other choices of conventions for the numerical constant in front of the linear combinations of the annihilation and creation operator in Eqs.(\ref{eq:QuadratureOperatorQ}) and (\ref{eq:QuadratureOperatorP}) are possible. Our choice corresponds to setting $\hbar = 1$.
The operators in Eqs.(\ref{eq:QuadratureOperatorQ}) and (\ref{eq:QuadratureOperatorP})  are called the {\it quadratures} of the field, and they are observables.
Measuring the quadratures, i.e. projecting into an eigenstate of $\hat q$, $\hat p$, or one of their linear combinations  is called a homodyne measurement. These observables have a continuous spectrum, hence the name ``continuous variables'':
\begin{align}
\label{eq:eigenstates-quadratures}
\hat q | s \rangle_q &= s | s \rangle_q , \\
\hat p | s \rangle_p &= s | s \rangle_p.
\end{align}
As such, the quadratures possess the spectral decomposition
\begin{align}
\hat q &= \int \dd s | s \rangle_q {_q\langle} s | s , \\
\hat p &= \int \dd s | s \rangle_p {_p\langle} s | s.
\end{align}
Quadrature eigenstates form a basis, i.e.,
\be
\mathcal{I} =  \int \dd s | s \rangle_q  {_q\langle} s |  = \int ds | s \rangle_p  {_p\langle} s | ,
\ee
which allows one to write arbitrary bosonic states in either the $q$ or $p$ representations, respectively:
\begin{equation}
	\ket{\psi} = \int \dd s \, \psi_q(s) | s \rangle_p =  \int ds \, \psi_p(s) | s \rangle_p,
\end{equation}
with $\psi_q(s) =  {_q\langle} s \ket{\psi}$ and $\psi_p(s) = {_p\langle} s \ket{\psi}$.

The variance of an arbitrary operator is defined by
\begin{equation}
	\label{eq:Variance}
    \langle(\Delta\hat{A})^2\rangle =
    \langle\hat{A}^2\rangle
    -\langle\hat{A}\rangle^2
\end{equation}
and can interpreted as the uncertainty of an observable. The expectation value of the quadratures on the Fock states is given by 
\begin{align}
	\langle\hat{q}\rangle 
    &= \frac{1}{\sqrt{2}}\bra{n}(\hat{a}+\hat{a}^\dag)\ket{n} 
    = 0 , \\
    \langle\hat{p} \rangle 
    &= \frac{1}{\sqrt{2} i}\bra{n}(\hat{a}-\hat{a}^\dag)\ket{n}
    = 0 ,
\end{align}
evaluating to zero, which means that the expectation value of the electric field is also zero. On the other hand, the expectation value of the square is nonzero:
\begin{align}
	\langle\hat{q}^2\rangle 
    &= \frac{1}{2}\bra{n}(
    \hat{a}^2
    +\hat{a}\hat{a}^\dag
    +\hat{a}^\dag\hat{a}
    +\hat{a}^{\dag 2}
    )\ket{n} 
    = \frac{1}{2}(1+2n), 
    \\
    \langle\hat{p}^2 \rangle 
    &= \frac{1}{2}\bra{n}(
    \hat{a}^2
    +\hat{a}\hat{a}^\dag
    +\hat{a}^\dag\hat{a}
    +\hat{a}^{\dag 2})\ket{n}
    = \frac{1}{2}(1+2n).
\end{align}
Thus it follows from \eqref{eq:Variance} that the uncertainty in both quadratures is equal, and when $n=0$ (corresponding to the vacuum state), the uncertainty is minimum:
\begin{equation}
	\label{eq:VarianceVacuum}
	\langle(\Delta\hat{q})^2\rangle_\text{vac} = \frac{1}{2} 
    = \langle(\Delta\hat{p})^2\rangle_\text{vac},
\end{equation}
also implying the saturation of the Heisenberg uncertainty principle,
\begin{equation}
	\label{eq:VarianceVacuum-heisenberg}
	\langle(\Delta\hat{q})^2\rangle_\text{vac} \langle(\Delta\hat{p})^2\rangle_\text{vac} = \frac{1}{4}. 
\end{equation}
This is known as vacuum fluctuations. 

\ssse{Coherent states}

In quantum optics, the \emph{coherent states} are the states with most resemblance to classical states, in the sense that they give rise to expectation values that look like the classical electric field. These states are the eigenstates of the annihilation operator:
\begin{equation}
	\label{eq:CoherentState}
	\hat{a}\ket{\alpha} = \alpha\ket{\alpha}.
\end{equation}
Since $\hat{a}$ is non-Hermitian, $\alpha$ is usually complex. For the creation operator $\hat{a}^\dag$, we have for obvious reasons
\begin{equation}
	\label{eq:CoherentStateDagger}
    \bra{\alpha} \hat{a}^\dag = \alpha^* \bra{\alpha}.
\end{equation}
It is possible to show (see Exercises) that
\begin{equation}
	\label{eq:CoherentStateFock}
    \ket{\alpha} = e^{-\abs{\alpha}^2/2} \sum^\infty_{n=0} \frac{\alpha^n}{\sqrt{n!}}\ket{n},
\end{equation}
which is a superposition of infinitely many Fock states. Furthermore, calculating the expectation value of the number operator $\hat{n}$,
\begin{equation}
	\bar{n} = \bra{\alpha}\hat{n}\ket{\alpha} = \abs{\alpha}^2,
\end{equation}
we see that $\abs{\alpha}^2$ is related to the mean number of photons in the field. Using this we can compute the probability of finding $n$ photons in the field:
\begin{equation}
	\abs{\langle n| \alpha \rangle}^2
    = e^{-\abs{\alpha}^2} \frac{\abs{\alpha}^{2n}}{n!}
    = e^{-\bar{n}} \frac{\bar{n}^n}{n!},
\end{equation}
which we recognize as a \emph{Poisson distribution} with a mean of $\bar{n}$. This distribution arises when the probability that an event occurs is independent of earlier events.

Coherent states are known to be \emph{non-orthogonal}. For example, consider the scalar product $\langle \beta | \alpha \rangle$, where $\ket{\alpha}$ and $\ket{\beta}$ are two different coherent states:
\begin{align}
	\langle \beta|\alpha \rangle &=
    e^{-|\beta|^2/2}e^{-|\alpha|^2/2} \sum_m \sum_n \frac{(\beta^m)^*\alpha^n}{\sqrt{m!}\sqrt{n!}} \langle m | n \rangle \nn \\
    &= e^{-(|\beta|^2+|\alpha|^2)/2} \sum_n \frac{(\beta^*\alpha)^n}{n!} \nn \\
    &= e^{-(|\beta|^2+|\alpha|^2 - 2 \beta^*\alpha)/2} \nn \\
    &= e^{-\abs{\alpha - \beta}^2/2} e^{(\alpha \beta^* - \beta \alpha^*)/2}.
\end{align}
Taking the modulus square, we obtain
\begin{equation}
	\label{eq:QuasiOrthogonality}
	\abs{\langle \beta | \alpha \rangle}^2 = e^{-\abs{\alpha - \beta}^2}.
\end{equation}
From \eqref{eq:QuasiOrthogonality} it is evident that two coherent states are non-orthogonal. Only when $\abs{\alpha-\beta}^2$ is large, so that $\abs{\langle \beta | \alpha \rangle}^2\sim 0$, do they become quasi-orthogonal. However, they can still be used as an (overcomplete) basis, to represent arbitrary quantum states, due to the identity resolution $1/{\sqrt{\pi}} \int \dd^2 \ketbra{\alpha}{\alpha} = 1$.

Since coherent states are nothing else than vacuum displaced in phase space, it can be easily verified that the quantum uncertainty for both quadratures on coherent states is the same as for vacuum, for all amplitudes $\alpha$:
\begin{equation}
	\langle(\Delta \hat{q})^2\rangle_\alpha 
    = \frac{1}{2} 
    = \langle(\Delta\hat{p})^2\rangle_\alpha ,
\end{equation}
also resulting in minimum-uncertainly states with
\begin{equation}
	\label{eq:Coherent-heisenberg}
	\langle(\Delta \hat{q})^2\rangle_\alpha \langle(\Delta\hat{p})^2\rangle_\alpha = \frac{1}{4}. 
\end{equation}

\subsubsection{Squeezed states}

\begin{figure}
\centering
\includegraphics*[angle = 270, width=0.6\columnwidth]{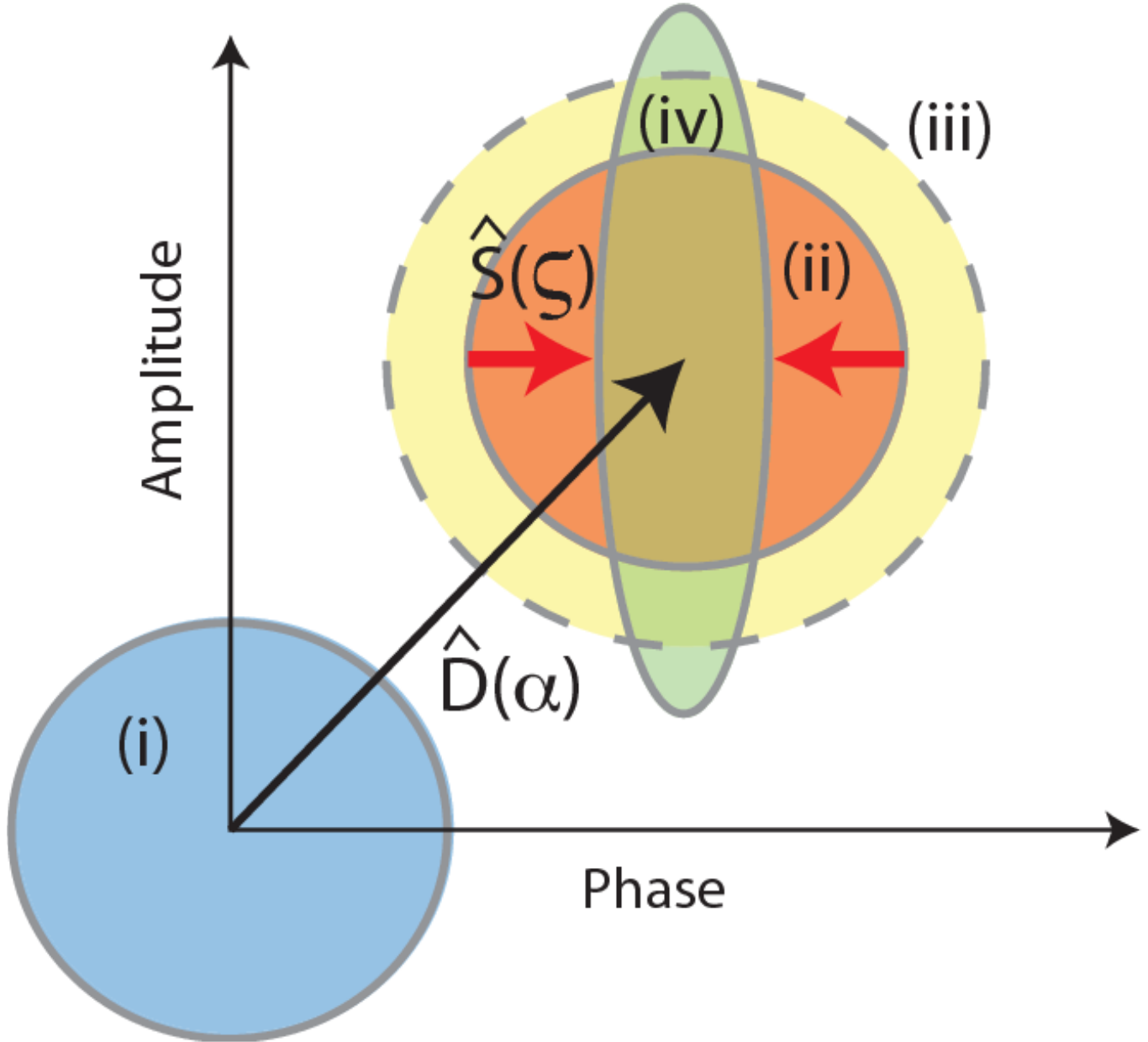}
\caption{Coherent, squeezed and thermal states. Wigner function ball-on-stick representations of (i) vacuum state (blue), (ii) coherent state (red), (iii) thermal state (yellow, dashed line), and (iv) squeezed state (green). This picture is taken from Ref.~\citep{sparkes2013storage}.}
\label{phase-space-representation}
\end{figure}

In contrast to coherent states, squeezed states are characterized by asymmetric fluctuations in $q$ and $p$ (see \figref{phase-space-representation}), i.e., $\langle(\Delta \hat{q})^2\rangle_\xi < \langle(\Delta \hat{p})^2\rangle_\xi $ for a $q$-squeezed state, and $\langle(\Delta \hat{q})^2\rangle_\xi > \langle(\Delta \hat{p})^2\rangle_\xi $ for a $p$-squeezed state, yet satisfying in both cases the Heisenberg principle with equality sign as in \eqref{eq:Coherent-heisenberg}.
Squeezed states are obtained applying the squeezing operator to the vacuum, i.e.,
\begin{equation}
\label{eq:squeezed-states}
\ket{\xi} = S(\xi) \ket{0} = e^{- \frac{i \xi (\hat q \hat p + \hat p \hat q)}{2} } \ket{0}.
\end{equation}
In the limit of infinite squeezing, which corresponds to infinite energy, one obtains the infinitely squeezed states, which are the eigenstates of position and momentum with zero eigenvalue:
\begin{align}
& \ket{0}_q \mbox{ infinitely $q$-squeezed state, such that } \hat q \ket{0}_q = 0 \ket{0}_q \\
& \ket{0}_p \mbox{ infinitely $p$-squeezed state, such that } \hat p \ket{0}_p = 0 \ket{0}_p.
\end{align}
Generalizations of these states exist, where the state that is squeezed in \eqref{eq:squeezed-states} is an arbitrary coherent state.

\ssse{Phase-space representation}

In quantum mechanics, a system can be fully described by its density operator $\hat{\rho}$. However, the density operator can be a rather abstract object and it can be hard to read off its properties. Therefore we employ the \emph{phase-space} representation. One can think of phase space as a mathematical abstract space, where the state of a harmonic oscillator is represented in terms of its quadratures. For a general quantum system, since position $\hat q$ and momentum $\hat p$ are non-commutative operators in quantum physics, the state cannot be represented as a point in phase space as it would in classical physics, because position and momentum are not allowed to both have precise values at the same time. The same holds true for the quadratures of the electromagnetic field, $\hat q$ and $\hat p$.

 Among the possible representations of states in phase space, the Wigner function one that is often used. For simplicity, we define it and look at its properties for the case of a single optical mode, however the generalization to the multimode case is straightforward.  It is defined by
\begin{equation}
\label{eq:Wigner-function}
	W_{\hat \rho(q,p)} \equiv \frac{1}{2\pi} \int^\infty_{-\infty}
    \bra{q + \tfrac{1}{2} x} \hat{\rho} \ket{q - \tfrac{1}{2} x} e^{ixp} \dd x ,
\end{equation}
where $x$, $q$, and $p$ are now interpreted as quadratures. $W\rho(q,p)$ is known as a \emph{quasi-probability distribution} since it can take on negative values. Even though the Wigner function is not a ``real'' probability distribution, it can still be used to compute actual probabilities. For example, integrating the Wigner function over $p$ yields the probability density (also referred to as the marginal distribution) over $q$,
\begin{align}
    \text{Pr}(q) = \int^\infty_{-\infty} W_\psi(q,p) \dd p
    &= \frac{1}{2\pi} \int^\infty_{-\infty} \dd p \int^\infty_{-\infty} \dd x
    \psi^*(q - \tfrac{1}{2} x) \psi(q + \tfrac{1}{2} x) e^{ixp} \nn \\
    &= \int^\infty_{-\infty} \dd x \psi^*(q - \tfrac{1}{2} x) \psi(q + \tfrac{1}{2} x) \delta(x) \nn \\
    &= \psi^*(q) \psi(q) = \abs{\psi(q)}^2,
\end{align}
where for simplicity of calculation we have considered the case of a pure state $\hat{\rho} = | \psi \rangle \langle \psi |$. Similarly, one can show that
\begin{equation}
	\text{Pr}(p) = \int^\infty_{-\infty} W_{\hat \rho}(q,p) \dd q = \abs{{\psi}(p)}^2
\end{equation}
is the probability density of $p$, where $\psi(p)$ is the wave function in $p$ representation (Exercise \ref{ex:integrating-W}). 
The Wigner function is normalised, i.e. 
\begin{equation}
\label{eq:Wigner-function-norm}
	\int^\infty_{-\infty} dx dp W_{\hat \rho}(q,p) =1.
\end{equation}
Through the Weyl transform, it is possible to associate a Wigner function to operators:
\begin{equation}
\label{eq:Wigner-function-operstors}
W_{\hat A}(q,p) \equiv \frac{1}{2\pi} \int^\infty_{-\infty}
    \bra{q + \tfrac{1}{2} x} \hat{A} \ket{q - \tfrac{1}{2} x} e^{ixp} \dd x.
\end{equation}
Expectation values of observables can be computed by means of the Wigner functions (as you can demonstrate in one of the exercises at the end of the chapter):
\begin{equation}
\label{eq:Wigner-function-exp-values}
	\langle \hat A \rangle = \text{Tr} [\hat A \hat \rho]= \int^\infty_{-\infty} W_{\hat \rho}(q,p) W_{\hat A}(q,p) dq dp.
\end{equation}

For instance, both the vacuum and the coherent state have Gaussian shapes which do not display any negativity, while the Wigner function for the single-photon Fock state shows clear indication of negativity. Indeed, states that are described by a Gaussian Wigner function are called {\it Gaussian states}. A cut of the Wigner function (as seen from the top) for a coherent state, a squeezed state, the vacuum, and a thermal state is represented in \figref{phase-space-representation}.

In \eqref{eq:Wigner-function}, the Wigner function is associated to a quantum state (possibly mixed) $\hat \rho$. However, it is also possible to associate a Wigner function to a process, by replacing the density matrix $\hat \rho$ in \eqref{eq:Wigner-function} with the operator describing the process\footnote{Technically, this requires a mapping associating a density matrix to quantum channel called the Choi Isomorphism.}, or also to a measurement, where in the latter case $\hat \rho$ is replaced by the projector associated with a given outcome, e.g., $| s \rangle_p  {_p\langle}s |$ for the outcome $p$ of the homodyne measurement $\hat p = \int ds | s \rangle_p  {_p\langle}s | s$. If the Wigner function is positive for all outcomes, then we say that the measurement is Wigner-positive. For a more formal definition, see the Appendix of Ref.~\citep{albarelli2018resource} and Ref.~\citep{Rahimi-Keshari2016}.  

\sse{Mari--Veitch theorem}
\label{sse: Mari-Eisert}

Any given CV quantum circuit is defined by (i) a specific input state $\rho_\text{in}$, (ii) a unitary evolution $\hat U$, and (iii) measurements $\hat \Pi$.  
The Mari--Veitch theorem, demonstrated independently in~\citep{Mari2012, veitch2012negative}, states that if:
\begin{itemize}
\item The input state $\hat \rho_\text{in}$ has positive Wigner function, and
\item The unitary operation transform states that are positively Wigner represented into states that are positively Wigner represented, and
\item The measurement operators $\hat \Pi$ has positive Wigner function,
\end{itemize} 
then there exists a classical algorithm able to efficiently simulate this circuit. This theorem is the analog of the Gottesman--Knill theorem seen for qubits in \chpref{chp:CircuitModel}. Intuitively, this happens because the Born rule can be expressed in terms of the Wigner functions of the evolved input state and the measurement operator, and if they are positive everywhere, they can be used to sample as from a regular probability distribution. Indeed, the Born rule can be expressed as:
\begin{equation}
\label{eq:Born-rule}
	P(\hat \Pi |\hat \rho_\text{out}) = \text{Tr} [\hat \Pi \hat \rho_\text{out}] = \int^\infty_{-\infty} W_{\hat \rho_\text{out}}(q,p) W_{\hat \Pi}(q,p) dq dp.
\end{equation}
where we have defined  $\hat{\rho}_\text{out} = \hat U \hat \rho_\text{in} \hat U^{\dagger}$, and where we have made use of Eq.(\ref{eq:Wigner-function-exp-values})\footnote{Using the Choi matrix formalism it is possible to formulate this theorem in a more general way, that involves the positivity of the Wigner function of Choi matrix of the process.}. 

Hence, including an element with negative Wigner function is mandatory in order to design a CV sub-universal quantum circuit that cannot be efficiently simulated by a classical device. By virtue of the Hudson theorem, 
this necessarily corresponds to the use of non-Gaussian resources. Indeed, the Hudson theorem states that the only pure states to possess a non-negative Wigner function are Gaussian states. Also, all Gaussian states (including their convex mixtures) have positive Wigner function. 
Previous criteria for the simulatability of CV circuits were given in terms of the Gaussianity of a circuit: if all elements in a CV circuits are Gaussian, then the circuit is classically efficiently simulatable~\citep{Bartlett2002}. This criterion is strictly included in the Mari--Veitch theorem, i.e., it recognises as classically efficiently simulatable a smaller class of CV circuits: there are indeed states and processes that are non-Gaussian, and yet for which the Wigner function is positive. Consider, for instance, the mixed state of the vacuum $\ket{0}$ and the single-photon state $\ket{1}$, $\rho = p \ketbra{0}{0} + (1 -p) \ketbra{1}{1}$ for any $p > 1/2$. 

An even more general criterion with respect to the Mari--Veitch theorem for the simulatability of CV circuits was given in Ref.~\citep{Rahimi-Keshari2016}, based on other quasi-probability distributions than the Wigner function.\\

\textit{Note: In class,~\nameref{tutorial4} is given after this section, but here it is placed after~\chpref{ch-CV} for convenience.}

\sse{Elementary operations and universal gate sets}
\label{univ-gate-sets-CV}

In the following, we are going to introduce the basic CV quantum operations, universality for such operations, and learn about several quantum computation models and protocols in CV.

\ssse{Definition of CV universality (1)}
\label{universality-def-CV}

The first definition of computational universality in CV that we will encounter in this course (and the first one historically) is the following: a CV quantum-computing system is universal if it can simulate the action of any Hamiltonian $H(\hat p_i, \hat q_j)$ consisting of a polynomial of the quadratures $\hat{q}_i$ and $\hat{p}_j$ in each mode $i,j$, to an arbitrary fixed accuracy~\citep{Lloyd1999,gu2009quantum}.

The whole construction relies on the following equality, for $H_A$ and $H_B$ arbitrary Hamiltonians and $\delta t$ a real number:
\beq
e^{i H_A \delta t} e^{i H_B \delta t} e^{-i H_A \delta t} e^{-i H_B \delta t} = e^{(H_A H_B - H_B H_A)\delta t^2} + O(\delta t^3).
\eeq
In other words, applying Hamiltonians $H_A$ and $H_B$ as specified above is equivalent to applying the Hamiltonian $i[H_A,H_B]$ in the short-time limit. More generally, a given set of Hamiltonians can generate any evolution within the algebra spanned by commutation. This property allows us to identify several classes of Hamiltonians:
\begin{itemize}
\item linear Hamiltonians like $\hat q$ and $\hat p$; they correspond to quadrature displacements in the case of the electromagnetic field;
\item quadratic Hamiltonians like $\hat q^2$ and $\hat p^2$; specific cases of this class of Hamiltonians are squeezing operators $\hat q\hat p + \hat p\hat q$ and the Fourier transform $\frac{\pi}{2} (\hat q^2 +\hat p^2)$ that rotates from one quadrature to the other;
\item entangling Hamiltonians, e.g., $\hat q_i \hat q_j$ for two modes $i$ and $j$;
\item a higher-order Hamiltonian like the cubic gate $\hat q^3$ or the Kerr Hamiltonian $(\hat q^2 + \hat p^2)^2$.
\end{itemize}
Evolutions with the first three classes are called Gaussian evolutions because they preserve Gaussian states. They are sufficient for certain quantum information tasks like CV quantum key distribution. 
The commutation of a polynomial in $\hat q$ and $\hat p$ with $\hat q$ and $\hat p$ themselves reduces the order of the polynomial by at least 1, commutation with quadratic Hamiltonians never increases the order, and commutation with a polynomial of order 3 or higher increases the order by at least 1. 
Generating Hamiltonians of order higher than 2 thus requires being able to implement at least one polynomial of degree greater or equal than 3. Let us now dwell on these operations and on the universality classes that they define more in details.

\ssse{Single-mode Gaussian transformations}

\begin{enumerate}

\item[\hspace{-0.15cm}a)] Quadrature displacements (linear in the quadratures):
\begin{align}
X(s) &= e^{- i s \hat p}, \hspace{0.25cm}  \text{such that} \hspace{0.5cm}  X(s) \ket{r}_q = \ket{r + s}_q 
\label{eq:CV-Xgate} \\
Z(s) &= e^{ i s \hat q}, \hspace{0.5cm}  \text{such that} \hspace{0.5cm}  Z(s) \ket{r}_p =  \ket{r + s}_p.
\label{eq:CV-Zgate} 
\end{align}

\item[\hspace{-0.15cm}b)] Rotations (quadratic in the quadratures)\footnote{
Note that the action of a rotation of the quadratures of a single mode looks like
\be
R(\theta)
\begin{pmatrix}
\hat{q} \\
\hat{p}
\end{pmatrix} 
=  
\begin{pmatrix}
\cos \theta & -\sin \theta \\
\sin \theta & \cos \theta
\end{pmatrix} 
\begin{pmatrix}
\hat{q} \\
\hat{p}
\end{pmatrix}
,
\ee
not to be confused with an equivalent matrix acting on the two-mode amplitude quadratures vector $\begin{pmatrix}
\hat{q}_1 \\
\hat{q}_2
\end{pmatrix}$,
realising the rotation of one quadrature $\hat{q}_1$ with respect to another one $\hat{q}_2$.}:
\be
R(\theta) = \exp \mleft[ \frac{i \theta \mleft( \hat q^2 + \hat p^2 \mright)}{2} \mright].
\ee
Particular example: Fourier transform $R(\pi/2) = F$, such that\footnote{
The corresponding rotation of the second mode quadrature is
\be
\begin{pmatrix}
q' \\
p'
\end{pmatrix}
= 
\begin{pmatrix}
0 & -1 \\
1 & 0
\end{pmatrix}
\begin{pmatrix}
q \\
p
\end{pmatrix}
=
\begin{pmatrix}
- p \\
q
\end{pmatrix}
,
\ee 
which corresponds to the transformations $F^\dag \hat q F = -\hat p$ and $F^\dag \hat p F = \hat q$, i.e., $F^\dag \hat a F = i \hat a$ and $F^\dag \hat a^\dag F = -i \hat a^\dag$.}
\begin{align}
F \ket{s}_q &= \frac{1}{\sqrt{2 \pi}} \int_{-\infty}^{\infty} \dd r e^{irs} \ket{r}_q  =  \ket{s}_p 
\label{eq:fourier} \\
F^{\dagger} \ket{s}_p &= \frac{1}{\sqrt{2 \pi}} \int_{-\infty}^{\infty} \dd r e^{-irs} \ket{r}_p = \ket{s}_q 
\end{align}
with $_{q}\braket{r}{s}_p = \frac{e^{irs}}{\sqrt{2 \pi}}$ and $\ket{s}_p, \ket{s}_q$ the eigenstates of the quadrature operators introduced in \eqref{eq:eigenstates-quadratures}.
The action of the Fourier transform can be verified by inserting the identity $\int_{-\infty}^{\infty} \dd s' \ket{s'}\hspace{-0.08cm}{_p} {_p}\hspace{-0.08cm}\bra{s'} = \mathcal I$ in \eqref{eq:fourier},
\be
\frac{1}{\sqrt{2 \pi}} \int_{-\infty}^{\infty} \dd r e^{irs} \ket{r}_q =
\frac{1}{\sqrt{2 \pi}} \int_{-\infty}^{\infty} \dd r \dd s' e^{irs} {_p}\braket{s'}{r}_q \ket{s'}_p = 
\frac{1}{2 \pi} \int_{-\infty}^{\infty} \dd r \dd s' e^{ir(s-s')} \ket{s'}_p
\ee
and by using that $\frac{1}{2 \pi} \int_{-\infty}^{\infty} \dd r e^{ir(s-s')} = \delta (s - s')$.

\item[\hspace{-0.15cm}c)] Squeezing (quadratic in the quadratures):
\be
S(s) = \exp \mleft[- \frac{i s \mleft(\hat q \hat p + \hat p \hat q \mright)}{2} \mright].
\ee

\item[\hspace{-0.15cm}d)] Shear (quadratic in the quadratures):
\be
D_{2,q} (s) = \exp \mleft[  i s \hat q^2 \mright].
\ee

\end{enumerate}

Any single-mode Gaussian operation can be decomposed into a) rotations; b) quadrature displacement; c) or d), i.e., squeezing or shear.
That is, 
\be
\lgr D_{1,q}(s) = Z(s), D_{2,q}(s), F = R(\pi/2)  \rgr  \mbox{   \bf{  universal set for single-mode Gaussian operations}}\nn
\ee
%

\ssse{Multimode Gaussian transformations}

The addition of any non trivial two-mode Gaussian gate, such as 

\begin{enumerate}
\item [\hspace{-0.15cm}e)] Controlled-Z gate (quadratic in the quadratures):
\be
C_Z = e^{i \hat{q}_1  \hat{q}_2},
\ee
\end{enumerate}

\noindent which acts as $e^{i \hat{q}_1  \hat{q}_2} |s\rangle_q |r\rangle_p = e^{i s  \hat{q}_2} |s\rangle_q |r\rangle_p  = |s\rangle_q |r + s\rangle_p $, yielding a conditional shift of the quadrature $\hat p$ in one mode, depending on the value of the quadrature $\hat q$ in the control mode.
Combined with the set of single-mode operations above allows for any general multi-mode Gaussian operation (see Ref.~\citep{menicucci2011graphical} for comments on the generalization to imperfect Gaussian operations and Gaussian measurements).
\begin{equation*}
\lgr D_{1,q}(s) = Z(s), D_{2,q}(s), F = R(\pi/2), C_Z  \rgr  \mbox{ \bf{ universal set for multi-mode Gaussian operations}}
\end{equation*}
They are enough to perform some algorithms, such as error-correcting codes regarding errors on single channels 
(however, non-Gaussian measurements are required to correct errors such as loss on all channels simultaneously). 
Yet, all the algorithms involving only Gaussian unitaries acting on Gaussian states with Gaussian measurements could be efficiently simulated on a classical computer, as we have seen in the previous section. 
\ssse{Single-mode universal transformations}

For a single mode, all Gaussian operations together with any non-Gaussian operation provide universality. For example, the set $D_{k,q} = \exp \mleft[  i s \hat q^k \mright]$ for $k = 1, 2, 3$ for all $s \in \R$ together with $F$ provides universal single-mode quantum computation (i.e., can be used to implement any single-mode unitary operation to arbitrary fixed accuracy):
\begin{equation*}
\lgr D_{1,q}(s) = Z(s), D_{2,q}(s), D_{3,q}(s), F = R(\pi/2)  \rgr  \mbox{   \bf{  universal set for single-mode quantum computing}}
\end{equation*}
In particular, we have the gate 

\begin{enumerate}
\item[\hspace{-0.15cm}f)] Cubic phase gate (cubic in the quadratures):
\be
D_{3,q}(s) = e^{i \hat{q}^3 s}.
\ee%
\end{enumerate}

\ssse{Multi-mode universal transformations}

Adding to this any nontrivial two-mode interaction provides universal quantum computation~\citep{Lloyd1999}. That is,
\begin{equation*}
\lgr D_{1,q}(s) = Z(s), D_{2,q}(s), D_{3,q}(s), F = R(\pi/2), C_Z   \rgr \mbox{   \bf{  universal set for multi-mode quantum computation}}
\end{equation*}

Summarizing:
\begin{equation*}
\mleft\{ e^{i \hat q s}, e^{i \hat{q}^2 s}, e^{i \frac{\pi}{4} (\hat{q}^2 + \hat{p}^2)}, \textcolor{red}{e^{i \hat{q}_1\otimes \hat{q}_2}}, \textcolor{blue}{e^{i \hat{q}^3 s}}  \mright\} \mbox{ \bf{ universal set for multi-mode quantum computation}}
\end{equation*}

\ssse{A sequence of Gaussian operations is also Gaussian}

We here show that from sequences of quadratic Hamiltonians, we will only be able to obtain quadratic Hamiltonians.
Let us define two arbitrary quadratic Hamiltonians:
\begin{align}
\label{eq:quadratic1}
    A &= a_0\hat{q} + a_1\hat{p} + a_2\hat{q}\hat{p} + a_3 \hat{p}\hat{q} + a_4\hat{q}^2 + a_5\hat{p}^2 , \\
    \label{eq:quadratic2}
    B &= b_0\hat{q} + b_1\hat{p} + b_2\hat{q}\hat{p} + b_3 \hat{p}\hat{q} + b_4\hat{q}^2 + b_5\hat{p}^2 .
\end{align}
We can calculate the product of their unitaries $e^{-iAt}e^{-iBt}$, by using the Baker--Campbell--Hausdorff (BCH) formula:
\be
e^{X}e^{Y} = e^{Z}, \qquad Z = X + Y + \frac{1}{2}[X,Y] + \frac{1}{12}[X,[X,Y]] - \frac{1}{12}[Y,[X,Y]] + \ldots
\ee
Up to second order, the BCH reads, in our case,
\be
e^{-iAt}e^{-iBt} = e^{-i(A+B)t -[A,B]t^2/2} + O(t^3, A, B),
\ee
where $A+B$ is quadratic in $\hat{q}$ and $\hat{p}$ as per Eqs.(\ref{eq:quadratic1}) and (\ref{eq:quadratic2}). By using the relation
\be
\comm{X+Y}{Z+T} = \comm{X}{Z} + \comm{X}{T} + \comm{Y}{Z} + \comm{Y}{T},
\ee
we can see that the commutator $\comm{A}{B}$ reads
\begin{align}
    \comm{A}{B} &= a_0 b_0 \comm{\hat{q}}{\hat{q}} + a_0 b_1 \comm{\hat{q}}{\hat{p}} + a_0 b_2 \comm{\hat{q}}{\hat{q}\hat{p}} + a_0 b_3 \comm{\hat{q}}{\hat{p}\hat{q}} + a_0 b_4 \comm{\hat{q}}{\hat{q}^2} + a_0 b_5 \comm{\hat{q}}{\hat{p}^2} \nn \\
    &\quad + a_1 b_0 \comm{\hat{p}}{\hat{q}} + \ldots + a_1 b_5 \comm{\hat{p}}{\hat{p}^2} \nn \\
     &\quad + \ldots \nn \\
    &\quad +  a_5 b_0 \comm{\hat{p}^2}{\hat{q}} + \ldots + b_5 a_5 \comm{\hat{p}^2}{\hat{p}^2}.
\end{align}
Now, let us calculate the commutators between $\hat{q}$ and $\hat{p}$ by using the canonical relations:
\begin{align}
\comm{\hat{q}}{\hat{q}} &= \comm{\hat{q}}{\hat{q}^2} = \comm{\hat{q}^2}{\hat{q}^2} = \comm{\hat{p}}{\hat{p}} = \comm{\hat{p}}{\hat{p}^2} = \comm{\hat{p}^2}{\hat{p}^2} = 0 , \\
\comm{\hat{q}}{\hat{p}} &= i, \quad \comm{\hat{p}}{\hat{q}} = -i , \\
\comm{\hat{q}}{\hat{p}\hat{q}} &= \comm{\hat{q}}{\hat{q}\hat{p}} = \comm{\hat{q}}{\hat{p}}\hat{q} + \hat{p}\comm{\hat{q}}{\hat{q}} = i\hat{q} , \\
\comm{\hat{p}}{\hat{p}\hat{q}} &= \comm{\hat{p}}{\hat{q}\hat{p}} = \comm{\hat{p}}{\hat{p}}\hat{q} + \hat{p}\comm{\hat{p}}{\hat{q}} = -i\hat{p} , \\
\comm{\hat{q}}{\hat{p}^2} &= \comm{\hat{q}}{\hat{p}}\hat{p} + \hat{p}\comm{\hat{q}}{\hat{p}} = 2i\hat{p} , \\
\comm{\hat{p}}{\hat{q}^2} &= \comm{\hat{p}}{\hat{q}}\hat{q} + \hat{q}\comm{\hat{p}}{\hat{q}} = -2i\hat{q} , \\
\comm{\hat{q}^2}{\hat{p}^2} &= \comm{\hat{q}^2}{\hat{p}}\hat{p} + \hat{p}\comm{\hat{q}^2}{\hat{p}} = 2i(\hat{q}\hat{p}+\hat{p}\hat{q}) .
\end{align}
Knowing that $\comm{X}{Y}=-\comm{Y}{X}$, we have all the commutation relations we need, and they are all at most quadratic in $\hat{q}$ and $\hat{p}$. Therefore, the product of two Gaussian operations is also Gaussian. It is straightforward to see that, by applying them repeatedly in sequence, we will never obtain a unitary with a higher-order polynomial.

Now we can understand why the choice of the second-order BCH is justified. The terms of the BCH formula that we omitted are nothing more than nested commutators. Here, we have seen that the commutator of two quadratic operators is also quadratic (at most). Therefore, nesting commutators of quadratic operators will never yield higher-order terms.

\ssse{Adding the cubic phase gate to the set of Gaussian operations yields all higher order polynomials}

We now show that, by adding the cubic phase gate to the set of Gaussian operations, we can reach all higher-order polynomials.
Let us consider the quadratic Hamiltonian $A$ from the previous paragraph, and the cubic phase gate $e^{i\hat{q}^3s}$. As we have seen before, it is enough if we can show that $\comm{A}{i\hat{q}^3s}$ contains higher-order terms in $\hat{q}$ and $\hat{p}$. Using again the distributive property of the commutators, it is enough to see that
\be
\comm{\hat{p}^2}{\hat{q}^3} = \comm{\hat{p}^2}{\hat{q}^2}\hat{q} + \hat{q}\comm{\hat{p}^2}{\hat{q}^2} = -2i \mleft( \hat{q}\hat{p} + \hat{p}\hat{q} \mright) \hat{q} - 2i \hat{q} \mleft( \hat{q}\hat{p} + \hat{p}\hat{q} \mright) = -2i \mleft( \hat{p}\hat{q}^2 + \hat{q}^2\hat{p} + 2\hat{q}\hat{p}\hat{q} \mright),
\ee
which is a third-order polynomial. What is more, one can see that $\comm{A}{i\hat{q}^3s}$ includes \textit{all} third-order monomials in $\hat{q}$. To get all third-order monomials in $\hat{p}$, one only needs to Fourier transform the cubic phase gate.

A simple inductive proof now shows that one can construct Hamiltonians that are arbitrary Hermitian polynomials in any order of $\hat{q}$ and $\hat{p}$. Suppose that one can construct any polynomial of order $M$ or less, where $M$ is of degree at least 3. Then, since 
\begin{align}
\comm{\hat{p}^3}{\hat{p}^m\hat{q}^n} &= i\hat{p}^{m+2}\hat{q}^{n-1} + \text{ lower-order terms}, \\
\comm{\hat{q}^3}{\hat{p}^m\hat{q}^n} &= i\hat{p}^{m-1}\hat{q}^{n+2} + \text{ lower-order terms},
\end{align}
one can, by commutation of $\hat{q}^3$ and $\hat{p}^3$ with monomials of order $M$, construct any monomial of order $M+1$. Since any polynomial of order $M+1$ can be constructed from monomials of order $M+1$ and lower, by applying linear operations and a single nonlinear operation a finite number of times, one can construct polynomials of arbitrary order in $\hat{q}$ and $\hat{p}$ to any desired degree of accuracy.

\section{Measurement-based quantum computation: the general paradigm in CV}
\label{se:q_comp_cv}

It is in principle possible to perform sequences of gates from the elementary gate set that we have identified above~\citep{hillmann2020universal}, and to thereby engineer quantum computations in CVs within the circuit model. However, CV quantum computation finds its most natural formulation within the measurement-based model. The reason for this is the availability of massively large cluster states, that can be deterministically generated (see experimental results by Furusawa, Pfister, Andersen, and Treps, as well as latest releases from the Canadian start-up Xanadu).

In this framework, we can reformulate the notion of universal quantum computation introduced in Sec.~\ref{universality-def-CV} as follows: for any CV unitary $U = e^{i H(\hat q_i, \hat p_j)}$ (corresponding to an arbitrary polynomial of $\hat q_i$ and $\hat p_j$) and any given input $\ket{\phi}$, there exists an appropriate graph state such that by entangling the graph state locally with $\ket{\phi}$ and applying an appropriate sequence of single-mode measurements, $U \ket{\phi}$ is computed. We are now going to retrace all the steps seen in \secref{sec-MBQC-dv}, where we introduced the measurement-based quantum computation model for qubits, but here in the framework of CVs. We start with the definition of CV cluster states.

\subsection{Cluster states in continuous variables}

Ideal cluster states are defined as follows: given a graph with $N$ vertices and a certain number of edges connecting these vertices according to a specific structure modeled by an adjacency matrix $V$, a CV cluster state is obtained starting from a collection of $N$ infinitely $p$-squeezed states, and applying $C_Z$ interactions according to the graph structure, i.e., the controlled-$Z$ gate $e^{i g \hat{q} \times  \hat{q}}$, with $g$ a real parameter, yielding
\be
\label{eq:def_clusters}
\ket{\psi_V} = \hat{C}_Z [V] \ket{0}_p^{\otimes N} = \prod_{j,k}^N e^{\fr{i}{2} V_{jk} \hat{q}_j \hat{q}_k} \ket{0}_p^{\otimes N} = e^{ \fr{i}{2} \hat{q}^T V \hat{q} } \ket{0}_p^{\otimes N} .
\ee
Here $V$ is a real and symmetric matrix, with finite elements.

Equation~(\ref{eq:def_clusters}) implies that a cluster state satisfies a nullifier relation, as detailed here below. 
Let us first recall the following definitions and property from \secref{sec:Stabilizers}: if an operator $K$ satisfies
\be
\label{eq: stabilizer}
K \ket{\phi} = \ket{\phi}   
\ee
for a state $\ket{\phi}$, we call it a ``stabilizer'' for $\ket{\phi}$. Equation~(\ref{eq: stabilizer}) implies that
\be
\label{eq: stabilizer2}
U K U^\dag \mleft( U \ket{\phi} \mright) =  U \ket{\phi},
\ee
i.e., that $U K U^\dag$ stabilizes $U \ket{\phi}$.
Note furthermore that
\be
\label{eq:stab0}
e^{-i s \hat p} \ket{0}_p \equiv X(s) \ket{0}_p = \ket{0}_p \forall s.
\ee

From Eqs.~(\ref{eq:stab0}) and (\ref{eq: stabilizer2}), it follows that the cluster state in \eqref{eq:def_clusters} is stabilized by the set
\begin{align}
K_i &= \hat{C}_Z [V] X_i(s) \hat{C}_Z [V]^\dag = e^{ \fr{i}{2} \hat{q}^T V \hat{q} } X_i(s) e^{- \fr{i}{2} \hat{q}^T V \hat{q} } \nn \\
&= \prod_{j,k} \prod_{l,m} e^{ \fr{i}{2} V_{j,k} \hat{q}_j \hat{q}_k } e^{ - i s \hat{p}_i } e^{ -\fr{i}{2} V_{l,m} \hat{q}_l \hat{q}_m } 
\end{align}
for each $i$. Using that $e^{i \hat{q}_1 \hat{q}_2} \hat{p}_1 e^{- i \hat{q}_1  \hat{q}_2} = \hat{p}_1 - \hat{q}_2$, we finally obtain that
\be
\label{eq:stabiizers_cluster}
K_i = e^{- i s \hat{p}_i} \prod_{k } V_{i,k} e^{ i s \hat{q}_k } = X_i (s) \prod_{k} V_{i,k} Z_k (s).
\ee
Equation~(\ref{eq:stabiizers_cluster}) is formally equivalent to its analog in the discrete-variable case [see Eq.~(20.11) in Ref.~\citep{Bruss2006}].

The $K_i$ form a group. The generators of the corresponding algebra are found by derivation since $K_i = e^{- i s H_i}$. 
Note that because of \eqref{eq: stabilizer} it follows that $H_i \ket{\psi_V} = 0 \forall i$. Hence the $H$ operators are called ``nullifiers'' for the state $\ket{\psi_V}$.
They can be easily calculated as
\begin{align}
H_i &= i \mleft. \fr{d K_i}{ds} \mright|_{s = 0} = i \fr{d}{ds} \mleft[ e^{- i s \hat{p}_i} \prod_{k} V_{i,k} e^{ i s \hat{q}_k} \mright]_{s = 0} \nn \\
&= i \mleft[ - i \hat{p}_i e^{- i s \hat{p}_i} \prod_{k} V_{i,k} e^{ i s \hat{q}_k} +  e^{- i s \hat{p}_i} \fr{d}{ds} \prod_{k} V_{i,k} e^{ i s \hat{q}_k} \mright]_{s = 0} \nn \\
&= \hat{p}_i + i \mleft[ e^{- i s \hat{p}_i} \sum_{k} V_{i,k} i \hat{q}_k \prod_{l} e^{ i s \hat{q}_l} \mright]_{s = 0} = \hat{p}_i - \sum_{k} V_{i,k} \hat{q}_k,
\end{align}
from which follows that
\be
\label{eq:nullifiers}
\mleft( \hat{p}_i - \sum_{k}  V_{i,k} \hat{q}_k \mright) \ket{\psi_V} = 0.
\ee
From Eq.~(\ref{eq:nullifiers}) then follows immediately that
\be
\bra{\psi_V} \Delta^2 \mleft( \hat{p}_i - \sum_{k} V_{i,k} \hat{q}_k \mright) \ket{\psi_V} = 0,
\ee
which for states with zero average also reads $\bra{\psi_V} \mleft( \hat{p}_i - \sum_{k}  V_{i,k} \hat{q}_k \mright)^2 \ket{\psi_V} = 0$.

\subsection{The CV MBQC paradigm}

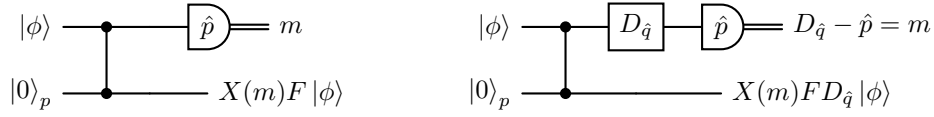
\begin{figure}[H]
\centering
\begin{quantikz}
\lstick{$\ket{\phi}$}  & \ctrl{1} & & \meterD{\hat{p}}\wire[r][1]{c} & \rstick{$m$}\wireoverride{n}  \\
\lstick{$\ket{0}_p$}  & \control{} & & \rstick{$X(m) F \ket{\phi}$}
\end{quantikz}\qquad\qquad
\begin{quantikz}
\lstick{$\ket{\phi}$}  & \ctrl{1} & \gate{D_{\hat{q}}} & \meterD{\hat{p}}\wire[r][1]{c} & \rstick{$D_{\hat{q}}-\hat{p}=m$}\wireoverride{n}  \\
\lstick{$\ket{0}_p$}  & \control{} & & \rstick{$X(m) F D_{\hat{q}}\ket{\phi}$}
\end{quantikz}
\caption{The basic setup for CV MBQC.}
\label{fig:scheme1}
\end{figure}

Here we follow Ref.~\citep{gu2009quantum}. Consider the scheme in Fig.~\ref{fig:scheme1}. 

\begin{itemize}

\item {\bf 1) Preparation:} One mode contains the initial state that we want to process, $\ket{\phi} = \int \dd s f(s) \ket{s}_q$; the other mode is initialized to $\ket{0}_p$.  The input state is hence $\ket{\phi} \otimes \ket{0}_p = \int \dd s f(s) \ket{s}_q \ket{0}_p$. Apply a $C_Z$ gate between the two, obtaining
\begin{align}
C_Z (\ket{\phi} \otimes \ket{0}_p) &= \int \dd s f(s) e^{i \hat q \otimes \hat q} \ket{s}_q \ket{0}_p = \int \dd s f(s) e^{i s \hat q_2} \ket{s}_q \ket{0}_p = \int \dd s f(s) \ket{s}_q \ket{s}_p \nn \\
&= \frac{1}{\sqrt{2 \pi}} \int \dd s f(s) \int \dd r e^{- i r s} \ket{r}_p \ket{s}_p,
\end{align}
where we have used that $e^{i s \hat q} \ket{0}_p = \ket{s}_p$.

\item {\bf 2--pre) Measure:} Measure the upper rail in the $\hat p$ basis with outcome $m$. This projects the second qumode into the state
\be
\ket{\psi}_\text{out} \propto \int \dd s f(s) e^{- i m s} \ket{s}_p = e^{- i m \hat p} \int \dd s f(s) \ket{s}_p = X(m) F \ket{\phi}.
\ee
The last equality is obtained since $F \ket{\phi} = \int \dd s f(s) F \ket{s}_q = \int \dd s f(s) \ket{s}_p$.
The effect of this circuit is to apply the identity (modulo a displacement and a rotation).

\item {\bf 2)} If we send as an input state the rotated state $D_{\hat{q}} \ket{\phi} = \int \dd s f(s) D_{\hat{q}} \ket{s}_q$, where $D_{\hat{q}} = e^{i f(\hat{q})}$ is an operator diagonal in the computational basis, then measuring $\hat{p}$ in the first qumode projects the second mode into
\be
\label{eq:comp_comp}
\ket{\psi}_\text{out} \propto X(m) F D_{\hat{q}} \ket{\phi}.
\ee
Since the $C_Z$ gate commutes with $D_{\hat{q}}$, the same result is obtained with $\ket{\phi}$ as an input state, and a rotation on the first mode after the $C_Z$ as in the left panel of \figref{fig:scheme1}. This is in turn equivalent to the situation in which no rotation is applied, but the first mode is measured in a rotated basis $D_{\hat{q}}^\dag \hat{p} D_{\hat{q}} \equiv \hat{p}_{f(\hat q)}$.
The extra displacement $X(m)$ depends on the outcome of the measurement on mode $1$, and can be compensated by choosing the measurement basis of the following steps (thus introducing, in general, time ordering).

\item {\bf 3) Universality of single-mode operations:} Repeat twice the previous protocol, for two different operators $D^1_{\hat{q}}$ and $D^2_{\hat{q}}$. The output state is
\begin{align}
\ket{\psi}_\text{out} &= X(m_2) F (D^2_{\hat{q}} X(m_1)) F D^1_{\hat{q}} \ket{\phi} \nn \\
&= X(m_2) F X(m_1) D^2_{\hat{q} + m_1} F D^1_{\hat{q}} \ket{\phi} \nn \\
&= X(m_2) F X(m_1) F D^2_{-\hat{p} + m_1} D^1_{\hat{q}} \ket{\phi} ,
\end{align}
where we have used the inequalities $X(-m) \hat{q} X(m) = \hat{q} + m$, $Z(-m) \hat{p} Z(m) = \hat{p} + m$, $F^\dag (-\hat{q}) F = \hat{p}$, $F^\dag \hat{p} F = \hat{q}$. If, instead of measuring the second mode on $\hat{p}_{f(\hat{q})}$, we would have measured it in the outcome-dependent basis $\hat{p}_{f(-\hat{q} - m_1)}$, we would have obtained as a result our deterministic desired output
\be
\label{eq:comp_cv}
\ket{\psi}_\text{out} = X(m_2) F X(m_1) F D^2_{\hat{p}} D^1_{\hat{q}} \ket{\phi}
\ee
(universal for single-mode operations if we repeat other times: we can obtain the desired transformation concatenating various $D_{\hat{q}}$ and $D_{\hat{p}}$), apart from by-product operations which do not need to be corrected.

\item {\bf 3)-4) Triviality of measurement adaptivity for Gaussian unitaries:} Let us focus on the building blocks of the universal set given above. For the Gaussian operations: 
\begin{itemize}
\item $F$ is obtained at each step of the computation.  
\item $D_{\hat{q}} = e^{i s \hat{q}}$ is obtained by measuring $\hat{p}_{s \hat{q}} = e^{-i s \hat{q}} \hat{p} e^{i s \hat{q}} = \hat{p} + s$ (measure $\hat{p}$ and add $s$ to the result). Note that $\hat{p}_{s (\hat{q} + m)} = \hat{p}_{s \hat{q}} = \hat{p} + s$ (no adaptation is required).
\item $D_{\hat{q}} = e^{i s \frac{\hat{q}^2}{2}}$ is obtained by measuring in the basis $\hat{p}_{s \hat{q}^2/2} = e^{-i s \frac{\hat{q}^2}{2}} \hat{p} e^{i s \frac{\hat{q}^2}{2}} = \hat{p} + s \hat{q}  = g (\hat{q} \sin \theta + \hat{p} \cos \theta)$ with $g = \sqrt{1 + s^2}$ and $\theta = \arctan s$. This is readily verified because the latter definition implies $\cos \theta = 1/\sqrt{1 + s^2}$ and $\sin \theta = s/\sqrt{1 + s^2}$. This corresponds to a rotated homodyne quadrature. Note that if we would have to adapt the basis we should measure according to $\hat{p}_{s (\hat{q} + m)^2/2} = \hat{p} + s \hat{q} + m s$.  This can be achieved by measuring in the same basis as without adaptation (i.e., $\hat{p} + s \hat{q}$) and adding $m s$ to the result
\end{itemize}
The adaptation required for these measurements is trivial and can be done afterwards (as a post-processing). Hence Gaussian operations can be implemented in any order or simultaneously (``parallelism'').

A cubic phase gate would instead require:
\begin{itemize}
\item $D_{\hat{q}} = e^{i s \frac{\hat{q}^3}{3}}$ is obtained by measuring in the basis $\hat{p}_{s \hat{q}^3/3} = e^{-i s \frac{\hat{q}^3}{3}} \hat{p} e^{i s \frac{\hat{q}^3}{3}} = \hat{p} + s \hat{q}^2$. If we have to measure according to $\hat{p}_{s (\hat{q} + m)^3/3} = \hat{p} + s \hat{q}^2 + 2 m s \hat{q} + m^2 s$, the term $2 m s \hat{q}$ requires a nontrivial adaptation of the measurement basis.
\end{itemize}

\item {\bf 5) Cluster states as a resource:} Given the fact that the $C_Z$ gates commute with the measurements, in practice the state used as the initial resource in the quantum computation protocol presented is a generalized cluster state, in which some of the modes (the input modes), also linked to the other nodes of the cluster, are initialized to code the modes of the input state. However, one can think of taking an initial cluster state (e.g., a square cluster state) and ``writing'' in some of its nodes the modes of the physical input state [e.g., by $C_Z$ gates and measurements of the input modes, or by teleportation~\citep{ukai2010universal}]. A state which allows this for each $U$ and each input state is said to be a universal resource. It has been demonstrated by Briegel that a square lattice graph (a cluster state) with unit weights is a universal resource for quantum computation. Depending on the specific kind of computation, other graphs than a square lattice could be more suitable for implementing the computation~\citep{horodecki2006lectures}.

\item {\bf 6) Two-mode interactions:} A sequence of single-mode operations can be implemented via the following measurements on a linear cluster. To achieve full universality, we have to add to the previous toolbox a two-mode interaction, e.g., the $C_Z$ gate. 
Such two-qubit gates can be constructed in a two-dimensional cluster state where two input qubits are entangled with a few other qubits, in analogy to the case of DV MBQC discussed in \secref{sec-MBQC-dv}. By a series of single-qubit measurements and rotations, we can end up with two of the other qubits representing the output state corresponding to the two-qubit gate having acted on the input state. 

\end{itemize}

In conclusion, note that the procedure above is an idealization: in real life, squeezed states will always have finite energy, i.e., squeezing degree. As a result, the output state of the computation --- even in the presence of ideal entangling gates and measurements --- will always be affected by Gaussian noise, caused by the finite squeezing. How to avoid accumulation of this (and other types of) noise is the subject of the following section. 

\section{Bosonic quantum error correction and GKP encoding}
\label{FT-CV-QC}

In classical informatics, when it comes to making sure that the errors that can occur during a computation can be corrected, it is convenient to resort to digitalized information, i.e., bits, as we saw in \secref{sec:ChallengesForQEC}. For this reason, combined with versatility, analog computers have been outperformed by digital computers in the 50s--60s, when the latter became sufficiently performant. Also note that from a computer-science perspective, the definition of computational models based on real numbers is problematic and less studied\footnote{If real computations were physically realizable, one could use them to solve NP-complete problems, and even $\#$P-complete problems, in polynomial time. Unlimited-precision real numbers in the physical universe are prohibited by the holographic principle and the Bekenstein bound.}.

Analogously, with quantum information, if the goal is to achieve fault-tolerant quantum computation, we must resort to qubit-like quantum information even when using CV hardware. In general, this means identifying two codewords $\ket{0}_L = \sum_n c^0_n \ket{n}$ and $\ket{1}_L = \sum_n c^1_n \ket{n}$, that are approximatively orthogonal. An example of qubit-like quantum information encoding in CV is based on the use of multicomponent cat states, where the qubit-like information is encoded in codewords $
\ket{0}_L \propto \mleft( \ket{\alpha} + \ket{i\alpha} + \ket{-\alpha} + \ket{-i \alpha} \mright)$ and $\ket{1}_L \propto \mleft( \ket{\alpha} - \ket{i} + \ket{-} - \ket{-i \alpha} \mright)$. This encoding is designed to correct for single-photon losses and phase errors up to $\pi/2$. Cat codes have first allowed for reaching the break-even point, in the sense that quantum information encoded in such cat states has been living longer than the one encoded in the corresponding qubit (within a transmon architecture)~\citep{ofek2016extending}. They are part of a family of bosonic codes that have discrete rotational symmetry~\citep{PhysRevX.10.011058}.  

In Ref.~\citep{Gottesman2001}, another way of encoding qubits in quantized harmonic oscillators, that exhibits instead a translational symmetry, was introduced by Gottesman, Kitaev, and Preskill, yielding the GKP encoding. This encoding has been shown to allow for the correction of arbitrary types of noise, below a certain threshold. Essentially, and without attempting to be rigorous, this is because GKP codes allows one to correct for single-mode displacements, and any noise-map can be decomposed into single-mode displacements~\citep{Gottesman2001}. GKP state have been generated experimentally, first with trapped ions~\citep{Fluhmann2018} and then in superconducting microwave cavities~\citep{campagne2019stabilized, kudra2021robust}, where they also allowed for the stabilization of quantum information with a lifetime gain of 2.2~\citep{sivak2023real}.

Note that both types of codes, at finite energy (i.e., finite amplitude for multimode cat codes and finite squeezing in GKP codes), will not allow on their own to reduce the error probability arbitrarily close to zero, as is the case in qubit codes. For this reason, one resorts to concatenation of bosonic codes with qubit error-correcting codes, such as the surface code we saw in \secref{sec:SurfaceCode}~\citep{Darmawan_2021, Noh_2022}. 

In what follows, we first give the basic ideas around rotationally symmetric bosonic codes. We then introduce the GKP encoding and the corresponding error-correcting scheme in detail. A comprehensive list of possible bosonic codes can be found here \url{https://errorcorrectionzoo.org}.

\subsection{Rotationally symmetric bosonic codes}

Here we briefly review rotationally symmetric bosonic codes~\citep{PhysRevX.10.011058}.
The simplest way of associating a qubit to a choice of levels of the quantized harmonic oscillator is to identify the qubit state zero with the vacuum, and the qubit state one with a single-photon state, i.e., $\ket{0}_L = \ket{0}$ and $\ket{1}_L = \ket{1}$. However, this encoding is not robust against losses: a single loss event translates into a bit flip at the logical level, i.e., $\hat a \ket{1} = \ket{0}$.

A more clever way of doing this mapping is through the so-called binomial encoding:
\be
\label{eq:binomial}
\ket{0}_L = \frac{\ket{0} + \ket{4}}{\sqrt{2}}; \qquad
\ket{1}_L = \ket{2}.
\ee
A loss event maps the encoded qubit $\ket{\psi}_L = \alpha \ket{0}_L + \beta \ket{1}_L$ into
\be
\hat a \ket{\psi}_L \propto \alpha \ket{3} + \beta \ket{1},
\ee
exhibiting a different parity than the encoded state. As such, a photon loss or gain can be detected by measuring the parity $\hat \Pi = (-1)^{\hat n}$, and the state can be restored by being brought back to the original code space with an appropriate recovery procedure~\citep{Joshi_2021}. This encoding has allowed for keeping the quantum information alive beyond the break-even point~\citep{ni2023beating}. The four-component cat state mentioned in the introduction is another example of a code that lives in the even-dimensional sub-space of the infinite-dimensional bosonic Hilbert space. 

\begin{figure}
  \centering 
  \includegraphics[width=\textwidth]{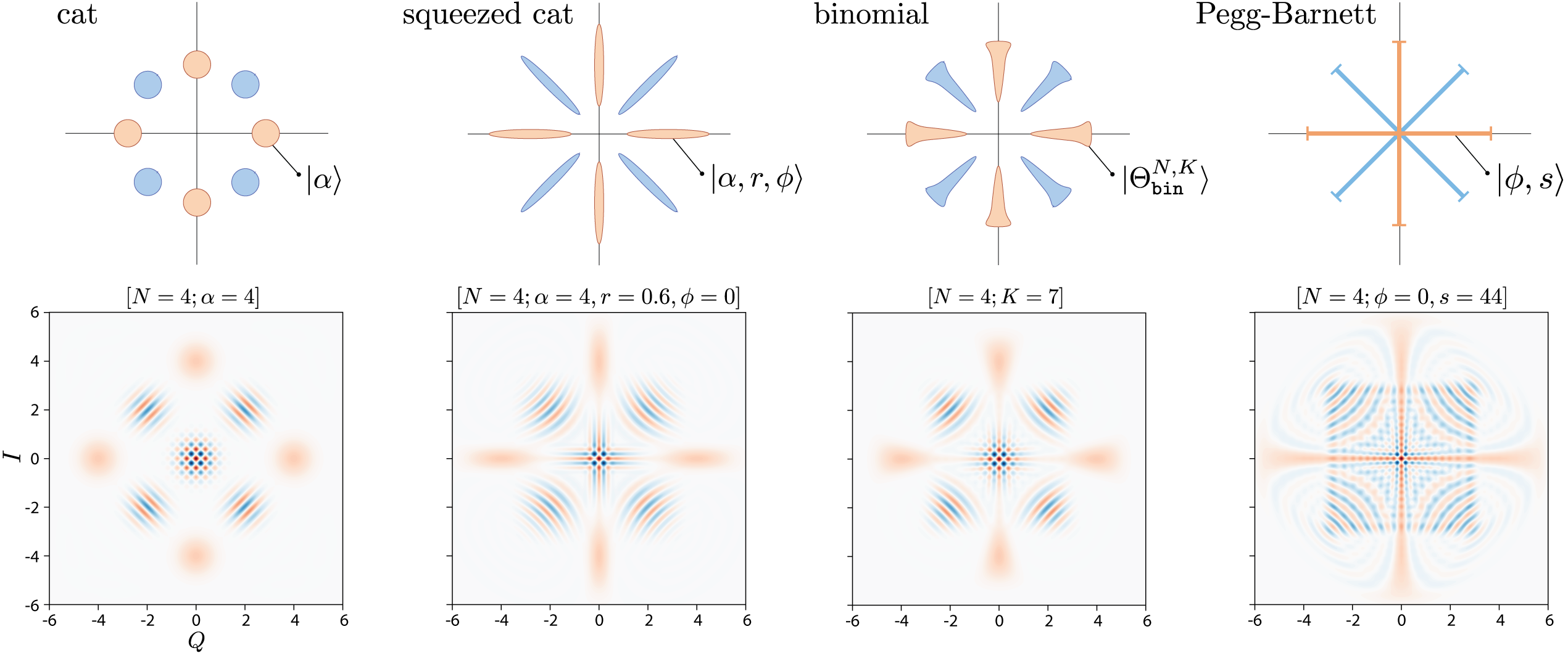}
  \caption{\label{fig:summary} From Ref.~\citep{PhysRevX.10.011058}.
  Graphical summary of several $N=4$ rotation codes: cat and squeezed cat, binomial, and Pegg--Barnett. Logical codewords are $+1$ eigenstates of the discrete rotation operator, and exhibit $N$-fold rotation symmetry. Top row: ball-and-stick diagrams illustrating $\ket{+_N}$ (orange) and $\ket{-_N}$ (blue). Indicated on each is the primitive $\ket{\Theta}$ for the code.
  Bottom row: Wigner functions for the $\ket{+_N}$ state, $W_{\ket{+_N} }(\alpha)$. Red/blue are positive/negative, the color scale on each plot is different, and $Q = \frac{1}{2} (\alpha + \alpha^*)$ and $I = \frac{1}{2i} (\alpha - \alpha^*)$ are the real and imaginary parts of $\alpha$.}
\end{figure}

In fact, both these codes are special examples of a family of bosonic codes that are invariant for discrete-rotational symmetry $\hat R_{N} = e^{i 2 \pi \hat n /N}$, and that can be written as
\begin{subequations} \label{eq:01codewords}
    \begin{align} \label{eq:comp0_rot}
        \ket{0_{N, \Theta}}
            &= \frac{1}{\sqrt{ \mathcal{N}_0} } \sum_{m=0}^{2N-1} e^{i \frac{m \pi}{N} \hat{n}} \ket{\Theta}, \\
        \ket{1_{N, \Theta}}
            &= \frac{1}{\sqrt{ \mathcal{N}_1} } \sum_{m=0}^{2N-1} (-1)^m e^{i \frac{m \pi}{N} \hat{n}} \ket{\Theta}, \label{eq:comp1_rot}
    \end{align}
\end{subequations}
where $\mathcal N_i$ are normalization constants, and $\ket{\Theta}$ is a primitive function. For instance, for the four-component cat code $\ket{\Theta} = \ket{\alpha}$ (see \figref{fig:summary}).

\begin{figure}
  \centering
  \includegraphics{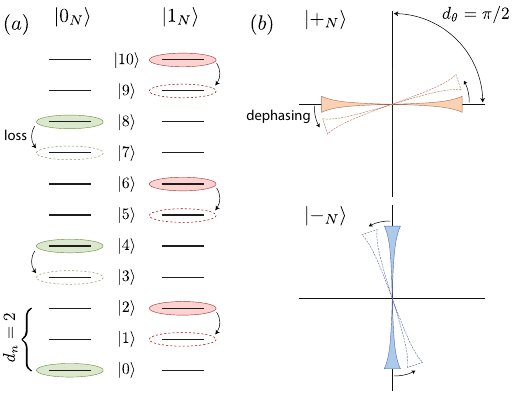}
  \caption{\label{fig:errorstructure} From Ref.~\citep{PhysRevX.10.011058}. Graphical summary of the Fock-space and phase-space structure of codewords for an $N=2$ rotation code.  
  (a) The computational-basis codewords $\ket{0_N}$ and $\ket{1_N}$ have support on every $2kN$ and $(2k+1)N$ Fock states for $k = 0, 1, 2, \dots$, respectively. Up to $N-1$ loss or gain errors can in principle be detected. 
  (b) The dual codewords $\ket{\pm_N}$ are related by a rotation in phase space by $\pi/2$. Rotation errors small compared to $d_\theta = \pi/2$ are detectable with a code-dependent uncertainty.}
\end{figure}

The rotational symmetry has important consequences for the Fock-space structure of the rotationally symmetric bosonic codes. Imposing further that the operator $\hat{R}_{2N} = e^{i \pi \hat{n} /N}$ acts as the logical operator $\hat{Z}$ allows for obtaining the Fock-space structure exemplified in \figref{fig:errorstructure}. Namely, given a state $\sum_n c_n \ket{n}$, the action of the operator $\hat{Z}$ is as follows:  
\be
e^{i \frac{ \pi \hat n}{N}} \sum_n c_n \ket{n} = \sum_n c_n e^{i \frac{ \pi n}{N}} \ket{n}.
\ee
It is clear that we need to impose:
\begin{align}
& \hat{Z} \ket{0_N} =  \ket{0_N} \Rightarrow c_n = 0 \text{ a part from those for which } e^{i \frac{ \pi \hat n}{N}} = 1 \Rightarrow n = 2 k N \\
& \hat{Z} \ket{1_N} = - \ket{1_N} \Rightarrow  c_n = 0 \text{ a part from those for which } e^{i \frac{ \pi \hat n}{N}} = -1 \Rightarrow n = (2 k + 1) N. 
\end{align} 
As a consequence, the Fock-space codewords for a rotationally symmetric code with rotational symmetry of $\hat{R}_N = e^{i 2 \pi \hat{n} /N}$ have Fock-space structure
\begin{subequations} \label{eq:01codewords-Fock}
    \begin{align} 
\ket{0_N} &= \sum_k c^0_{2kN} \ket{2kN} \\
\ket{1_N} &= \sum_k c^1_{(2k + 1)N} \ket{(2k+1)N}.
    \end{align}
\end{subequations}
 It is then clear that there is a trade-off between the phase and loss errors that rotationally symmetric bosonic codes are capable of correcting: the larger the rotational-symmetry parameter $N$, the smaller the phase shifts that can be corrected, and the larger the loss or gain events that can be corrected. More precisely, phase errors of up to $\pi/N$ and loss or gain of up to $N-1$ photons can be corrected.

Also note that it is possible to decompose arbitary noise processes into photon gain/loss and dephasing. As such, rotationally symmetric bosonic codes can be used as a universal error-correcting code. 

\subsection{GKP encoding}

We now turn to a code with a translational symmetry: the Gottesman--Kitaev--Preskill (GKP) encoding. 

\subsubsection{Ideal codespace}

\begin{figure}[H]
    \centering
    \includegraphics[width=0.9\textwidth]{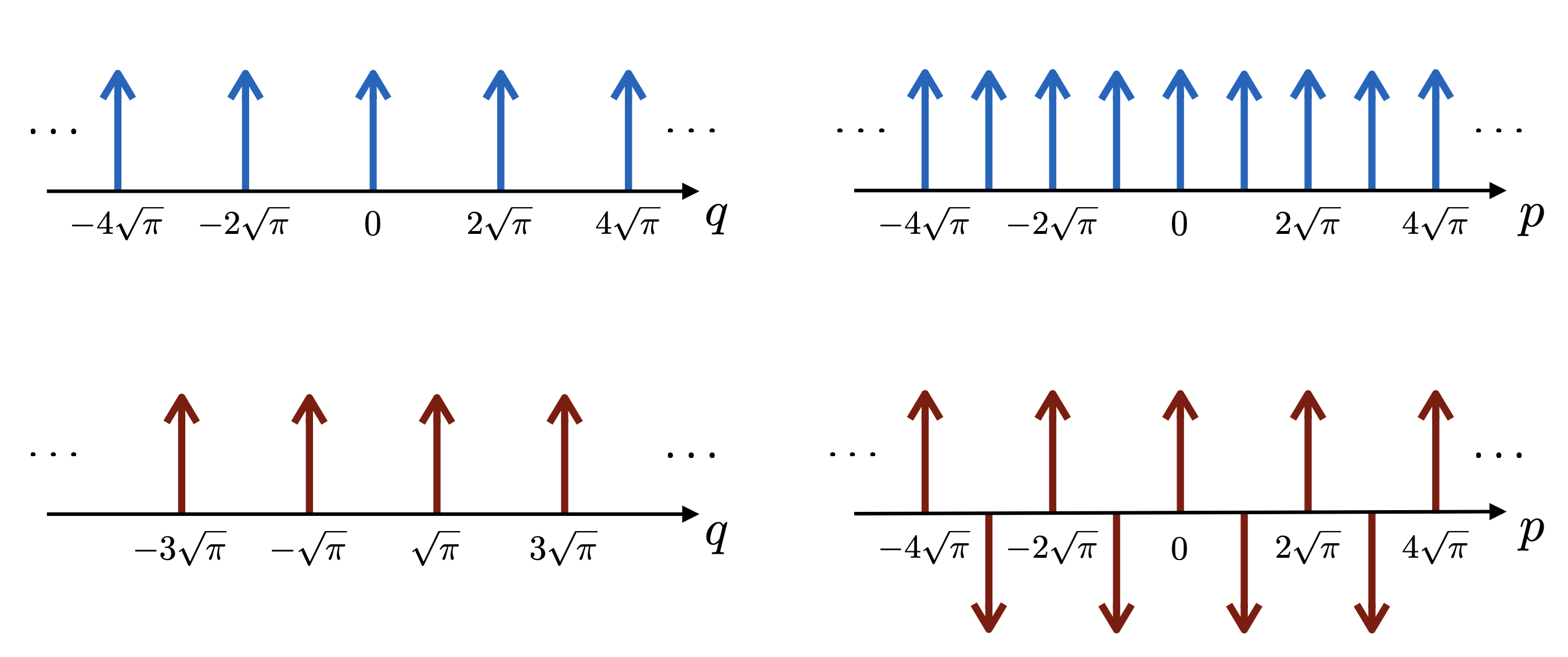}
    \caption{Figure from Ref.~\citep{DoucePhDThesis}. \underline{Left:} Wavefunction of perfect GKP states in the position representation. The upper figure, in blue, corresponds to $\ket{0_L}$, while the lower one, in red, corresponds to $\ket{1_L}$. Arrows stand for Dirac peaks. \underline{Right:} Same objects in momentum space (blue for $\ket{0_L}$ and red for $\ket{1_L}$).}
    \label{fig:GKP_ideal}
\end{figure}

Let us start by defining the basis of the GKP encoding.
Formally, the logical qubit states (codewords) are coherent superpositions of infinitely squeezed states, i.e., infinite combs of Dirac peaks~\citep{Gottesman2001}.
In the position basis, that is
\begin{equation}
\ket{0_L} = \sum_n \ket{2n\sqrt\pi}_q, \qquad
\ket{1_L} = \sum_n \ket{(2n+1)\sqrt\pi}_q.
\label{eq:GKP_ideal_q}
\end{equation}
The GKP codewords can equivalently be expressed in the position basis, as 
\begin{equation}
\ket{0_L} = \sum_n \ket{n\sqrt\pi}_p, \qquad
\ket{1_L} = \sum_n (-1)^n \ket{n\sqrt\pi}_p.
\label{eq:GKP_ideal_p}
\end{equation}
%
%
%
Note that we have omitted normalization because these ideal states (see \figref{fig:GKP_ideal}) are not normalizable.

Let us now find which continuous-variable gate implement desired logical operations on the codespace. Let us start with the Hadamard gate, $H_L$, which must act as follows: $H_L\ket{0_L}=\ket{+_L}$ and $H_L\ket{1_L}=\ket{-_L}$. Then\footnote{Note that technically, in the derivation above, one would have $|s\rangle_q = F^{\dagger}|s\rangle_p$ and hence obtain $H = F^{\dagger}$, but since $H = H^{\dagger}$, we can see that the Fourier transform and its Hermitian conjugate act equivalently on GKP code-words.},
\begin{equation}
\begin{aligned}
H_L \ket{0_L} &= \ket{+_L} = \ket{0_L} + \ket{1_L} \stackrel{\scriptsize\eqref{eq:GKP_ideal_q}}{=} \sum_n \ket{2n\sqrt{\pi}}_q + \sum_n \ket{(2n+1)\sqrt{\pi}}_q =  \sum_n \ket{n\sqrt{\pi}}_q = \\
&= F  \sum_n \ket{n\sqrt{\pi}}_p  \stackrel{\scriptsize\eqref{eq:GKP_ideal_p}}{=} F \ket{0_L}. 
\end{aligned}
\end{equation}
Note that we omit normalization because the states are not normalizable. A similar derivation can be done to see that $H_L\ket{1_L} = F \ket{1_L}$; we conclude that $H_L = F$.

Now, let us look at the logical Pauli matrices. We require the action $X_L \ket{0_L} = \ket{1_L}$ and vice versa. We obtain therefore: 
\begin{equation}
X_L \ket{0_L} = \ket{1_L} = \sum_n \ket{2n\sqrt{\pi} + \sqrt{\pi}}_q = e^{- i \hat p \sqrt{\pi}}\sum_n \ket{2n\sqrt{\pi}}_q = e^{- i \hat p \sqrt{\pi}}  \ket{0_L}. 
\end{equation}
A similar derivation shows that $e^{-i\sqrt{\pi} \hat{p}} \ket{1_L} = \ket{0_L}$; we conclude that $X_L = e^{-i\sqrt{\pi} \hat{p}}$.

Similarly, we can find $Z_L$, which must satisfy $Z_L \ket{0_L} = \ket{0_L}$ and $Z_L \ket{1_L} = -\ket{1_L}$. It is easy to verify that the operator $e^{i\sqrt{\pi} \hat{q}}$ does the trick: 
\begin{equation}
\begin{rcases}
    Z_L \ket{0_L} \stackrel{\scriptsize\eqref{eq:GKP_ideal_q}}{=} e^{i \sqrt{\pi}\hat{q}}\sum_n \ket{2n\sqrt{\pi}}_q \stackrel{*}{=}
    \sum_n e^{i2n \pi}\ket{2n\sqrt{\pi}}_q = \ket{0_L}\\
    Z_L \ket{1_L} = e^{i \sqrt{\pi}\hat{q}}\sum_n \ket{(2n + 1)\sqrt{\pi}}_q = \sum_n e^{i (2n + 1) \pi} \ket{(2n + 1)\sqrt{\pi}}_q = - \ket{1_L},
\end{rcases}
\end{equation}
where we used in step $*$ that $\ket{2n\sqrt{\pi}}_q$ are eigenstates of $\hat{q}$. 
The other Clifford logical operations corresponding to $C_Z$ gate, $S$ gate etc can be found in similar ways. 

At this point, one may wonder whether the recurring factor $\sqrt{\pi}$ is arbitrary or not.
In fact, one could define the GKP states in terms of any other factor than $\sqrt{\pi}$~\citep{Gottesman2001}, but we typically do not because of symmetry reasons. Since usually errors in $\hat{p}$ and $\hat{q}$ are comparable in magnitude, it makes sense to have logical operators that translate by the same amount. In our case, $X_L$ and $Z_L$ both translate $\hat{q}$ and $\hat{p}$, respectively, by $\sqrt{\pi}$. Thus, this GKP code can protect against shifts with $\abs{\Delta q}, \abs{\Delta p} < \sqrt{\pi}/2$.

\subsubsection{Definition of CV universality (2)}

A second definition of CV universality is based upon encodings, such as the GKP encoding. Consider qubits encoded in CV hardware. In this case, universality is achieved when one can implement at least one of the universal gate sets for DV quantum computation, that we introduced in \chpref{chp:CircuitModel}, encoded in the GKP encoding. 

On the non-normalizable states introduced in \eqref{eq:GKP_ideal_q}, Clifford operations correspond to the gates
\begin{equation}
\langle
\bar{H} =  F = e^{i\pi/4(\hat{q}^2+\hat{p}^2)}, \quad
\bar{C}_Z = \ e^{i\hat{q}_1 \hat{q}_2}, \quad
\bar{S} = e^{i\hat{q}^2/2}
\rangle.
\end{equation}
%
These gates are implemented by Gaussian CV operations, which is a feature (most likely) unique to the GKP encoding (other encodings have non-Gaussian operations as Clifford gates). 
To promote this set of operations to universality, we need a non-Clifford gate. This requires a non-Gaussian operation:
\begin{equation}
\bar{T} = \exp \mleft\{ \frac{i \pi}{4} \mleft[ \mleft( \frac{2 \hat q}{\sqrt{\pi}} \mright)^3 + \mleft( \frac{\hat q}{\sqrt{\pi}} \mright)^2 - \mleft( \frac{2 \hat q}{\sqrt{\pi}} \mright) \mright] \mright\}.
\end{equation}

\subsubsection{Realistic codespace}

\begin{figure}[h!]
\centering
\includegraphics[width=0.5\columnwidth]{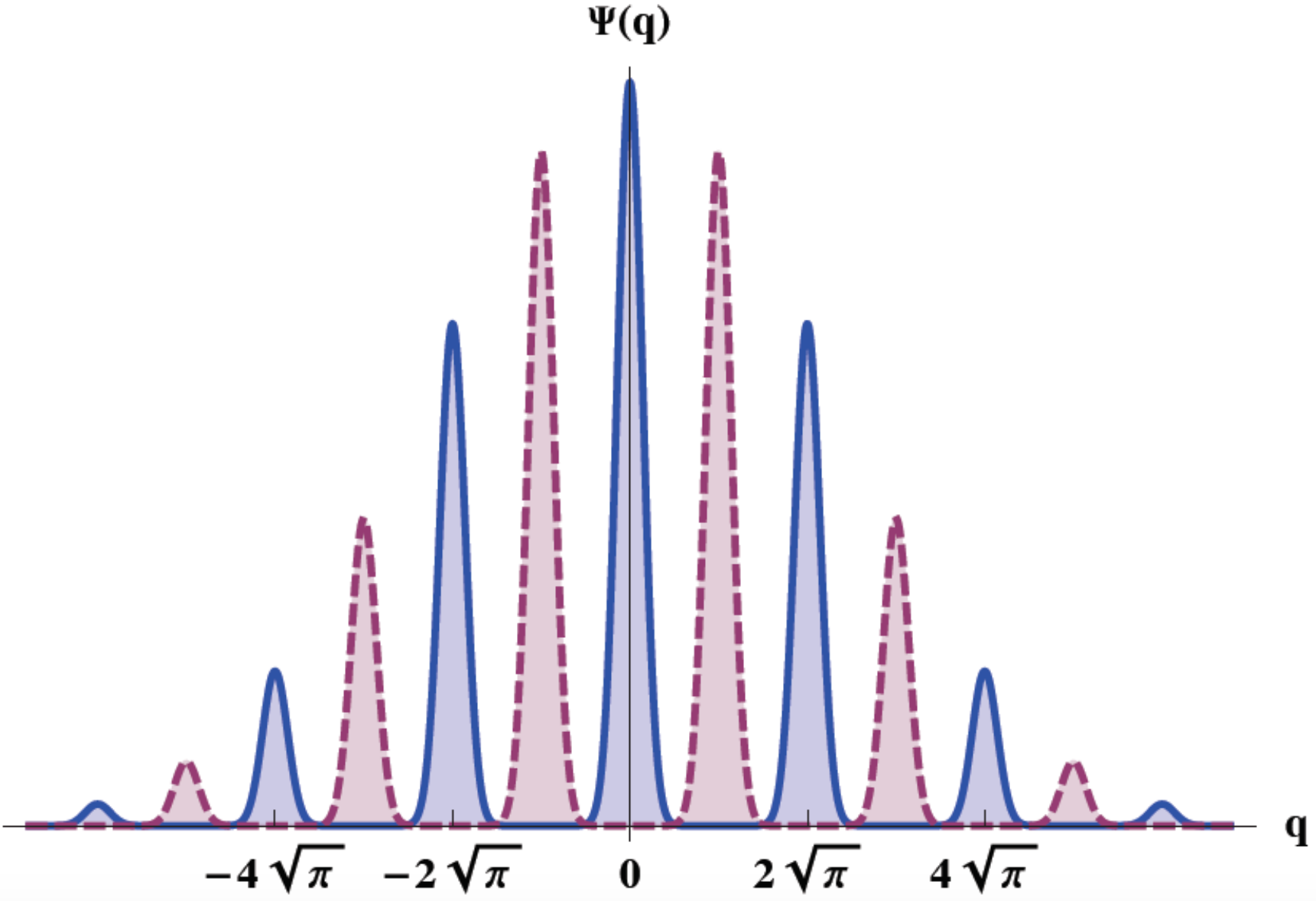}
\caption{\label{fig-GKP} Figure from Refs.~\citep{Douce2017, Douce2019}. Wavefunction in position representation of the GKP states $\ket{\tilde{0}_L}$ in solid blue and $\ket{\tilde{1}_L}$ in dashed red, with $\delta=\Delta=0.25$.}
\end{figure}

Realistic logical qubit states are normalizable finitely squeezed states, rather than non-normalizable infinitely squeezed states, and thus physically implementable. The Dirac peaks are hence replaced by a normalized Gaussian of width $\Delta$, while the infinite sum itself becomes a Gaussian envelope function of width $\delta^{-1}$ (see \figref{fig-GKP}).  When $\Delta = \delta$, the noise is symmetric in $\hat{q}$ and $\hat{p}$, and we refer to $\Delta$ as the squeezing parameter.

Formally, we can derive the wavefunctions of the realistic states $\ket{\tilde{0}_L},\ket{\tilde{1}_L}$ by introducing the following noise distributions:
\begin{equation}
G(u) = \frac{1}{\Delta \sqrt{2 \pi}} e^{-{u^2}/(2 \Delta^2)}, \qquad  F(v) = \frac{1}{\delta \sqrt{2 \pi}} e^{-{v^2}/(2 \delta^2)},
\label{eq:G(u)F(u)}
\end{equation}
which can be understood as the probability distributions of random displacements. Thus, the wavefunctions can be written as
\begin{align}
\braket{q}{\tilde{0}_L}
&= \bra{q} \int \dd u \dd v\, G(u) F(v) e^{-i u \hat{p}} e^{-i v \hat{q}} \ket{0_L} \nn \\
&= \bra{q} \int \dd u \dd v\, G(u) F(v) e^{-i u \hat{p}} \sum_n e^{-iv2n\sqrt{\pi}} \ket{2n\sqrt{\pi}}_q \nn \\
&= \bra{q} \int \dd u \dd v\, G(u) F(v) \sum_n e^{-iv2n\sqrt{\pi}}\ket{2n\sqrt{\pi} + u}_q \nn \\
&= \int \dd u \dd v\, G(u) F(v) \sum_n e^{-iv2n\sqrt{\pi}} \underbrace{\braket{q}{2n\sqrt{\pi} + u}_q}_{\delta(q - 2n\sqrt{\pi} - u)} \nn \\
&= \sum_n \int \dd v\, G(q-2n\sqrt{\pi}) F(v) e^{-iv2n\sqrt{\pi}} \; \stackrel{\scriptsize\text{Eq.}(\ref{eq:G(u)F(u)})}{=} \nn \\
&= \sum_n \exp{-\frac{(q-2n\sqrt{\pi})^2}{2\Delta^2}} \int \dd v\, \exp{-\frac{v^2}{2\delta^2}-iv2n\sqrt{\pi}} \stackrel{\text{\scriptsize Gauss. int.}}{=} \nn \\
&= N_0 \sum_n \exp \mleft\{ -\frac{(2n)^2\pi\delta^2}{2} \mright\} \exp \mleft\{ -\frac{(q-2n\sqrt\pi)^2}{2\Delta^2} \mright\},
\label{eq:GKP_real_0}
\end{align}
where we have used the Gaussian integral formula $\int \dd x\, e^{-(ax^2+bx+c)} = \sqrt{\pi/a}\: \exp{\frac{b^2}{4a}-c}$ and the normalization constant $N_0$.
Similarly, we can derive $\tilde{1}_L(q)$:
\begin{equation}
\braket{q}{\tilde{1}_L} = \bra{q} \int \dd u \dd v\, G(u) F(v) e^{-i u \hat{p}} e^{-i v \hat{q}} \ket{1_L} = N_1 \sum_n \exp \mleft\{ -\frac{(2n+1)^2\pi\delta^2}{2} \mright\} \exp \mleft\{ -\frac{(q-(2n+1)\sqrt\pi)^2}{2\Delta^2} \mright\},
\label{eq:GKP_real_1}
\end{equation}
with $N_1$ being a normalization constant.
Note that while these states are physically implementable, the logical basis states are no longer orthogonal. This introduces errors in the encoding that can be interpreted as qubit errors.

\subsection{Quantum error correction with the GKP encoding}

In this section, we prove that CV MBQC with finite squeezing and an additional supply of GKP states yields fault-tolerant quantum computation~\citep{Gottesman2001, menicucci2014fault}. In Ref.~\citep{gu2009quantum}, they showed how to implement standard quantum gates in CV MBQC, which would be sufficient for universal QC with GKP states~\citep{Gottesman2001}, i.e., relying on a DV encoding embedded in a CV hardware. What remains to prove is that these gates can be performed fault-tolerantly, admitting use of GKP auxiliary resource states.
This was achieved in Ref.~\citep{menicucci2014fault}, where it is shown that the noise in the $\hat p$ quadrature of a GKP-encoded quantum state can be replaced by the noise of the auxiliary $\ket{\tilde{0}_L}$ state following the procedure shown in~\figref{figerrcorr}. Repeating this gadget after a Fourier transform allows for correction of the other quadrature, thereby enabling fault tolerance. 

In order to explain this error-correction procedure, we follow a toy-model approach that has been developed by Glancy and Knill~\citep{Glancy2006}. This approach is based on a decomposition of the noise in several realizations of displacements, resulting in blurred ideal GKP states, and is referred to as the \textit{twirling approximation}~\citep{PhysRevA.101.012316}. Within this approach, we are going to show explicitly how GKP error correction allows one to correct for displacements, by analyzing the output of the circuit in \figref{figdisperr} with merely displaced perfect GKP states at the input. Since displacements form an operator basis, it follows that GKP states can correct any type of noise. Note that this works in principle for arbitrary noise, even when this is non-Gaussian.  

For physical states with finite squeezing there are subtleties, since error correction with coherent-envelope GKP states versus blurred ideal GKP states can, in principle, yield different results. These subtle differences are not completely captured by the Glancy--Knill picture, but for pedagogical reasons we will omit further details of this discussion. Furthermore, the twirling approximation has been shown to actually give good results, i.e., quantitatively similar to a fully-fledged treatment of coherent-envelope GKP states, when computing fidelities of recovered states in GKP quantum error-correction protocols [see Fig.~10 in Ref.~\citep{hillmann2021performance}]. 

\subsubsection{Noise model}

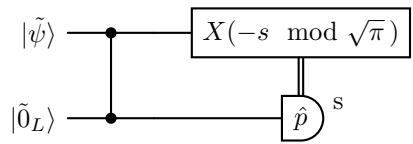
\begin{figure}[H]
\centering
\begin{quantikz}
\lstick{$\ket{\tilde{\psi}}$}  & \ctrl{1} & & \gate{X (-s \mod \sqrt{\pi}\,)} \\
\lstick{$\ket{\tilde{0}_L}$}  & \control{} & & \meterD{\hat{p}}\wire[u][1]{c}\rstick[label style={above right}]{s}
\end{quantikz}
\caption{\label{figerrcorr}Procedure to correct for errors in the $\hat{q}$ quadrature. $\ket{\tilde{0}_L}$ is a noisy GKP state and $\ket{\tilde{\psi}}$ is a noisy GKP-encoded CV state. $X(m)$ is a displacement operator $e^{-i m \hat p}$.}
\end{figure}

Let us say we have a noisy data qubit state $\ket{\tilde{\psi}}$ which is GKP-encoded. The scheme shown in \figref{figerrcorr} is a non-destructive $\hat{q}$-measurement on the data qubit: it gives information on what is the value of $\hat{q}$, so that we can identify noise in the form of a translation and correct it by displacing it back with an $X$ gate. 
Note that if the states were ideal, the protocol would act trivially: when measuring $\hat{p}$, we would obtain an outcome $s$ that would be a multiple of $\sqrt{\pi}$, and the correcting gate would have no effect.

\begin{figure}[H]
\centering
\begin{quantikz}
\lstick{$\ket{\psi}$} & \gate{e^{-iu_1\hat{p}_1}e^{-iv_1\hat{q}_1}} & \ctrl{1} & & \gate{X (v_2-u_1)} \\
\lstick{$\ket{0_L}$} & \gate{e^{-iu_2\hat{p}_2}e^{-iv_2\hat{q}_2}} & \control{} & & \meterD{\hat{p}}\wire[u][1]{c}\rstick[label style={above right}]{$n\sqrt{\pi}+u_1-v_2$}
\end{quantikz}
\caption{\label{figdisperr}Modeling the noise in the protocol. $\ket{0_L}$ is an ideal GKP state and $\ket{\psi}=\alpha\ket{0_L}+\beta\ket{1_L}$ is a perfect GKP-encoded CV state.}
\end{figure}
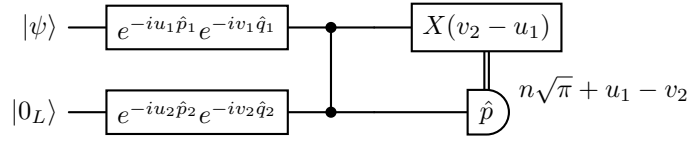

The noise in the scheme from \figref{figerrcorr} can be modeled as shown in \figref{figdisperr}, i.e., as displacements in $\hat{p}$ and $\hat{q}$ on both the data qubit and the auxiliary state.

Let us now compute the state obtained after measuring the $\hat{p}$ quadrature producing outcome $s$ in Fig.~\ref{figdisperr}, which we call $\ket{\Phi}$:
\begin{align} \label{eq:calculation-Tom-m}
\ket{\Phi}
&= \prescript{}{p}{\bra{s}}\hat{C}_Z e^{-iu_1\hat p_1}e^{-iv_1\hat q_1}e^{-iu_2\hat p_2}e^{-iv_2\hat q_2}\ket{\psi,0_L} \nn \\
&= \underbrace{\int\dd q_1\ket{q_1}_{q_1} \prescript{}{q_1}{\bra{q_1}}}_{\Id_1} \; \prescript{}{p_2}{\bra{s}}e^{i\hat{q}_1\hat{q}_2} e^{-iu_1\hat p_1}e^{-iv_1\hat q_1}e^{-iu_2\hat p_2}e^{-iv_2\hat q_2}\ket{\psi,0_L} \nn \\
&{\color{gray} \qquad\qquad\qquad\qquad \hookrightarrow e^{-i\hat{q}_1\hat{q}_2} \ket{q_1}_{q_1}\ket{s}_{p_2} = \underbrace{e^{-iq_1\hat{q}_2}}_{Z(-q_1)}\ket{q_1}_{q_1}\ket{s}_{p_2} = \ket{q_1}_{q_1}\ket{s-q_1}_{p_2}} \nn \\
&= \int\dd q_1\ket{q_1}_{q_1} \prescript{}{q_1}{\bra{q_1}} \; \prescript{}{p_2}{\bra{s-q_1}} e^{-iu_1\hat p_1}e^{-iv_1\hat q_1}e^{-iu_2\hat p_2}e^{-iv_2\hat q_2}\ket{\psi,0_L} \nn \\
&{\color{gray}\qquad\qquad\qquad\qquad  \hookrightarrow e^{iv_1\hat{q}_1} \underbrace{e^{iu_1\hat{p}_1}}_{X(-u_1)} \ket{q_1}_{q_1} = e^{iv_1\hat{q}_1} \ket{q_1-u_1}_{q_1} = e^{iv_1(q_1-u_1)} \ket{q_1-u_1}_{q_1}} \nn \\
&= \int\dd q_1\ket{q_1}_{q_1} \prescript{}{q_1}{\bra{q_1-u_1}} \; \prescript{}{p_2}{\bra{s-q_1}} e^{-iu_2\hat p_2}e^{-iv_2\hat q_2}\ket{\psi,0_L} e^{-iv_1(q_1-u_1)} \nn \\
&{\color{gray}\qquad\qquad\qquad\qquad  \hookrightarrow e^{iv_2\hat{q}_2} e^{iu_2\hat{p}_2} \ket{s-q_1}_{p_2} = \underbrace{e^{iv_2\hat{q}_2}}_{Z(v_2)} e^{iu_2(s-q_1)} \ket{s-q_1}_{p_2} =
e^{iu_2(s-q_1)} \ket{s-q_1+v_2}_{p_2} } \nn \\
&= \int\dd q_1\ket{q_1}_{q_1} \prescript{}{q_1}{\bra{q_1-u_1}} \; \prescript{}{p_2}{\braket{s-q_1+v_2}{\psi,0_L}} e^{-iu_2(s-q_1)}e^{-iv_1(q_1-u_1)} \nn \\
&=e^{i(v_1u_1-u_2 s)}\int\dd q_1 \ket{q_1}_{q_1} \prescript{}{q_1}{\bra{q_1-u_1}} \,\, \prescript{}{p_2}{\braket{s -q_1+v_2}{\psi,0_L}} e^{-i(v_1-u_2)q_1} = \{q_1\leadsto q_1+u_1\} \nn \\
&= e^{-iu_2(s-u_1)}\int\dd q_1 \ket{q_1 + u_1}_{q_1} \prescript{}{q_1}{\bra{q_1}} \,\, \prescript{}{p_2}{\braket{s-q_1-u_1+v_2}{\psi,0_L}} e^{-i(v_1-u_2)q_1}.
\end{align}
It is important to remark that, in this derivation, the bra/ket subscript $q_1$ ($p_2$) refers to the state being written in the $\hat{q}$ ($\hat{p}$) representation and acting on mode 1 (2). Do not confuse with the operators $\hat{q}_1,\hat{p}_2$ and their corresponding eigenvalues $q_1,p_2$.

Now, as we have seen in \figref{fig:GKP_ideal}, in the momentum representation, $\ket{0_L}$ is made up of Dirac peaks at every multiple of $\sqrt{\pi}$. Hence, 
\begin{equation}
\prescript{}{p_2}{\braket{s-q_1-u_1+v_2}{0_L}} = \delta(s-q_1-u_1+v_2 - l\sqrt{\pi}) \neq 0 \;\;\Leftrightarrow\;\; s-q_1-u_1+v_2 = l\sqrt{\pi}, \;\; l\in\Z.
\end{equation}
Moreover, in the position representation, the product $\prescript{}{q_1}{\braket{q_1}{\psi}}$ makes it so that $q_1$ must also be a multiple of $\sqrt{\pi}$, since both $\prescript{}{q_1}{\braket{q_1}{0_L}} = \delta(2n\sqrt{\pi}-q_1)$ and $\prescript{}{q_1}{\braket{q_1}{1_L}} = \delta((2n+1)\sqrt{\pi}-q_1)$ are multiples of $\sqrt{\pi}$.
Thus, the outcome of the measurement will be
\begin{equation}
    s = n\sqrt{\pi} + u_1 - v_2, \quad n\in\Z.
\end{equation}

Inserting this result into the post-measurement state $\ket{\Phi}$ yields
\begin{align}
    \ket{\Phi} &= e^{-iu_2(n\sqrt{\pi} - v_2)}\int\dd q_1 \ket{q_1 + u_1}_{q_1} \prescript{}{q_1}{\bra{q_1}} \,\, \prescript{}{p_2}{\braket{n\sqrt{\pi} - q_1}{\psi,0_L}} e^{-i(v_1-u_2)q_1} \nn \\
    &\qquad\qquad {\color{gray} \hookrightarrow \prescript{}{p_2}{\braket{n\sqrt{\pi}-q_1}{0_L}} = \sum_m \braket{n\sqrt{\pi}-q_1}{m\sqrt{\pi}}_p = \sum_m \delta ((n-m)\sqrt{\pi}-q_1) = \sum_m \delta (m\sqrt{\pi}-q_1), \quad m\in\Z} \nn \\
    &\qquad\qquad {\color{gray} \hookrightarrow \prescript{}{q_1}{\braket{q_1}{\psi}} =\delta (m'\sqrt{\pi}-q_1), \quad m'\in\Z} \nn \\
    &= e^{-iu_2(n\sqrt{\pi} - v_2)}\int\dd q_1 \ket{q_1 + u_1}_{q_1} e^{-i(v_1-u_2)q_1} \sum_m \delta (m\sqrt{\pi}-q_1)\delta (m'\sqrt{\pi}-q_1) \nn \\
    &\qquad\qquad {\color{gray} \hookrightarrow \delta (m\sqrt{\pi}-q_1)\delta (m'\sqrt{\pi}-q_1) = \delta_{m, m'}\delta (m\sqrt{\pi}-q_1)} \nn \\
    & = e^{-iu_2(n\sqrt{\pi} - v_2)}\int\dd q_1 \ket{q_1 + u_1}_{q_1} e^{-i(v_1-u_2)q_1} \sum_m \delta (m\sqrt{\pi}-q_1) = \{q_1 = m\sqrt{\pi}\} \nn \\
    & = e^{-iu_2(n\sqrt{\pi} - v_2)} \sum_m e^{-i(v_1-u_2)m\sqrt{\pi}} \int\dd q_1 \ket{m\sqrt{\pi} + u_1}_{q_1} \nn \\
    &\qquad\qquad {\color{gray} \hookrightarrow \ket{m\sqrt{\pi} + u_1}_{q_1} = e^{-iu_1\hat{p}_1} \ket{m\sqrt{\pi}}_{q_1} } \nn \\
    & = e^{-iu_2(n\sqrt{\pi} - v_2)} e^{-iu_1\hat{p}_1} \sum_m e^{-i(v_1-u_2)m\sqrt{\pi}} \ket{m\sqrt{\pi}}_{q_1} \nn \\
    & = e^{-iu_2(n\sqrt{\pi} - v_2)} e^{-iu_1\hat{p}_1} e^{-i(v_1-u_2)\hat{q}_1} \sum_m  \ket{m\sqrt{\pi}}_{q_1} \nn \\
    & = e^{-iu_2(n\sqrt{\pi} - v_2)} e^{-iu_1\hat{p}_1} e^{-i(v_1-u_2)\hat{q}_1} \ket{\psi}.
\end{align}
Finally, the error-corrected state is the post-measurement state $\ket{\Phi}$ displaced by \mbox{$-s \mod \sqrt{\pi} = v_2-u_1$} in the $\hat{q}$ quadrature:
\begin{equation}
    \ket{\Phi}_\text{EC} = X(v_2-u_1) \ket{\Phi} = e^{-i(v_2-u_1)\hat{p}_1} \ket{\Phi} = e^{-iu_2(n\sqrt{\pi} - v_2)} e^{-iv_2\hat{p}_1} e^{-i(v_1-u_2)\hat{q}_1} \ket{\psi}.
\end{equation}

Now, comparing the output state above with the input state $\ket{\tilde{\psi}} = e^{-iu_1\hat{p}_1} e^{-iv_1\hat{q}_1} \ket{\psi}$, we see that the noise in the $\hat{q}$ quadrature\footnote{Note that noise in the $\hat{q}$ quadrature refers to the term $e^{-i\hat{p}}$ and not to $e^{-i\hat{q}}$, since $e^{-i\hat{p}}=X$ induces a displacement in $\hat{q}$ and vice versa.} of the data qubit ($u_1$) has been replaced by the noise in the $\hat{p}$ quadrature of the auxiliary qubit ($v_2$).
On the downside, the noise from the auxiliary state has been added to the $\hat{p}$ quadrature of the data qubit ($v_1 \leadsto v_1-u_2$).
Overall, the error-correction protocol described here enables the replacement of noise in one quadrature of a data qubit with the noise of an auxiliary qubit, while increasing the noise in the orthogonal quadrature.

Due to the periodic nature of the GKP states, the protocol fails if the measurement outcome modulo $\sqrt{\pi}$ is so large that it gets displaced back to the wrong state, i.e., if $\abs{v_2-u_1} > \sqrt{\pi}/2$.
Above this error threshold, a logical bit-flip occurs, since the correcting translation leaves an additional $e^{\pm i \sqrt{\pi} \hat{p}}$ in the output state. 
In order to ensure that the noise parameters $u_1, v_2$ are small, the variance $\Delta$ in Eqs.~(\ref{eq:GKP_real_0})--(\ref{eq:GKP_real_1}) should also be small, i.e., the input states should resemble the ideal GKP states.
  
Note that even when the protocol succeeds, it is not enough, by itself, to correct random noise in GKP states. Since it replaces the noise in only one quadrature, it requires a Fourier transform and a second round of the protocol in the orthogonal quadrature.

\newpage
\section{Sampling models and sub-universal models in continuous variables}
\label{sse:sampling-in-CV}

\textit{Note: This section is no longer taught in the lectures, but it is left here for completeness.}\\

In one of the previous sections, we have dealt with MBQC, which is a model for (universal) quantum computation. In this section, we turn to sub-universal computational models in CV. In \secref{sse: Mari-Eisert}, we have seen that Wigner negativity is necessary in order to obtain quantum advantage --- at least of the exponential type.

However, if one aims at minimal extensions of Gaussian models, the boson sampling model that we have seen in Section~\ref{sec-Boson-sampling} appears as potentially ``over-kill'': both the input state and the measurement are described by negative Wigner functions, as they correspond, respectively, to single-photon states and photon counting measurement. Can we define other sub-universal models of quantum computation where only one of these elements is non-Gaussian, and show that they yield hard-to-sample output probability distributions?

\begin{figure}{H}
\centering
\includegraphics[width=\textwidth]{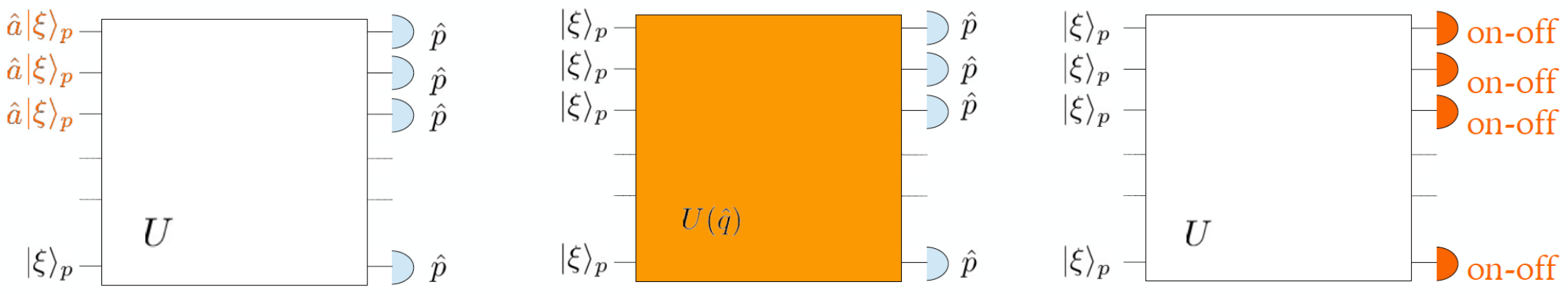}
\caption{Families of quantum circuits in continuous variables displaying minimal non-Gaussian character.}
\label{CV-sampling-models}
\end{figure}

Three different families of nontrivial sub-universal quantum circuits can be defined, depending on whether the element yielding the Wigner-function negativity is provided by the input state, the unitary evolution, or the measurement. This concept is exemplified in \figref{CV-sampling-models}.
Although Wigner negativity allows for stepping outside the range of applicability of the theorem in Ref.~\citep{Mari2012}, it is by itself not sufficient to imply classical hardness~\citep{garcia2020efficient}. 
The classical hardness of these circuits, therefore, has yet to be proven, for each circuit type. For the latter kind, corresponding to Gaussian boson sampling (GBS), this was done in Refs.~\citep{Lund2014, hamilton2017gaussian}. 
These circuits are composed of input squeezed states, passive linear optics evolution, and photon counters (rightmost panel in Fig.~\ref{CV-sampling-models}). Circuits of the second kind (central panel in Fig.~\ref{CV-sampling-models}) are, for instance, related to the CV implementation of instantaneous quantum computing --- another sub-universal model, where input states and measurements are Gaussian, while the evolution contains non-Gaussian gates~\citep{Douce2017, Douce2019}. First definitions of the former class of CV circuits, i.e., that display non-Gaussian input state and Gaussian operations and measurements (leftmost panel in Fig.~\ref{CV-sampling-models}), have been considered in Refs.~\citep{chakhmakhchyan2017boson, Chabaud2017}.

In this section, we will introduce the model CV-IQP (the CV version of the IQP model introduced in \secref{IQP-DV}) and sketch the proof of its computational hardness.

\subsection{Continuous-variable instantaneous quantum polytime}

\subsubsection{Definition of the model} 
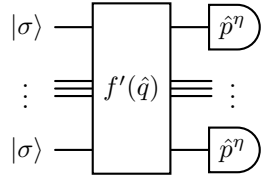
\begin{figure}[H]
\centering
\begin{quantikz}[wire types = {q, b, q}, classical gap = 0.1cm]
\lstick{$\ket{\sigma}$} & \gate[3]{f'(\hat{q})} & \meterD{\hat{p}^{\eta}} \\
\lstick{\vdots\;\;} & & \midstick{\vdots}\\
\lstick{$\ket{\sigma}$} & & \meterD{\hat{p}^{\eta}}
\end{quantikz}
\caption{\label{figCVIQP} IQP circuit in CVs. The states $\ket{\sigma}$ are finitely squeezed states with variance $\sigma$ in the $\hat p$ representation. The gate $f'(\hat q)$ is a uniform combination of elementary gates from the set in \eqref{eq:gates-sub2}. The finitely resolved homodyne measurement $\hat{p}^{\eta}$ has resolution $2 \eta$. }
\end{figure}

We define the model instantaneous quantum computing in CV (\figref{figCVIQP}).
This model is composed of the elementary gates
\begin{equation}
\label{eq:gates-sub2}
\mleft\{ e^{i d\hat{q}}, \, e^{i s \hat{q}^2}, \, e^{i c \hat{q}^3}, \, e^{i b \hat{q}_1 \hat{q}_2} \mright\}.
\end{equation}
All the gates in this model are diagonal in the position representation. As such, the Fourier transform is absent. Therefore, this model is not universal. For pedagogical purposes in these notes, we can assume in the definition that we are able to implement all the gates corresponding to all the possible choices of the real parameters in \eqref{eq:gates-sub2}. However, this has been shown to be unnecessary~\citep{Douce2019}.  We require momentum-squeezed states ${\ket{\sigma}}$ with $\sigma<1$ to be present at the input.

This model is a simpler version than the one introduced in Ref.~\citep{Douce2017}. 
Note the similarity with the IQP model defined in \secref{IQP-DV}. Here, too, we have a ``crossed'' structure of the type $p-q-p$ in input state, evolution and measurement, as it was $X-Z-X$ for the DV model.

For the proof of hardness, we proceed in the same way as seen in \chpref{ch-sub-univ}, when dealing with the IQP model: we show that adding postselection, the model becomes universal. In analogy to the Hadamard gadget of the DV model, we need therefore to develop a Fourier gadget.

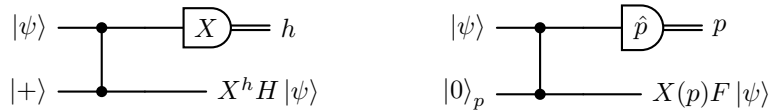
\begin{figure}[b]
\centering
\begin{quantikz}
\lstick{$\ket{\psi}$}  & \ctrl{1} & & \meterD{X}\wire[r][1]{c} & \rstick{$h$}\wireoverride{n} \\
\lstick{$\ket{+}$}  & \control{} & & \rstick{$X^h H \ket{\psi}$}
\end{quantikz}\qquad\qquad
\begin{quantikz}
\lstick{$\ket{\psi}$}  & \ctrl{1} & & \meterD{\hat{p}}\wire[r][1]{c} & \rstick{$p$}\wireoverride{n}  \\
\lstick{$\ket{0}_p$}  & \control{} & & \rstick{$X(p) F \ket{\psi}$}
\end{quantikz}
\caption{\label{Hadamard-gadget}Left: Hadamard gadget in a post-selected IQP circuit, where $h$ takes value $0$ if $+1$ is measured, while $h=1$ if the result is $-1$.
Right: Ideal Fourier gadget in CVs; an exact translation of the Hadamard gadget. ${\ket{0}_p}$ represents an infinitely $\hat p$-squeezed state with $\sigma = 0$, thus satisfying $\hat p {\ket{0}_p} = 0$.}
\end{figure}

In the idealised case of infinite-squeezing input states, this can be obtained fairly easily with the gadget of \figref{Hadamard-gadget}. Adding postselection allows one to recover the Fourier transform, which completes the set of gates in the model.

However, a realistic model departs from this idealized situation in at least two unavoidable aspects. First, in order to obtain a probability of success different from zero for the homodyne measurement, thereby giving meaning to postselection, we must introduce a binning of the real axis. This can still be equivalently modeled by using as a projective measurement. Instead of the ideal homodyne detector $\hat p = \int p \ketbra{p}{p}$, we are going to consider the finitely resolved homodyne detector $\hat p^{\eta}$ operator that we define as~\citep{Paris2003}
\be
\label{operator-proj}
\hat{p}^{\eta} = \sum_{k = - \infty}^{\infty} p_k \int_{- \infty}^{\infty} \dd p \, \chi^\eta_k(p) \ketbra{p}{p} \equiv \sum_{k = - \infty}^{\infty} p_k \hat{P}_k
\ee
with $\chi^\eta_k(p) = 1$ for $p \in [p_k - \eta, p_k + \eta]$ and $0$ outside, $p_k = 2 \eta k$ and $2 \eta$ the resolution, associated with the width of the detector pixels\footnote{Note that this model turns out to be equivalent to an ideal scheme with perfectly resolving homodyne detectors and a discretization (binning) of the measurement outcomes.}. It is easy to check that this is still a projective measurement, since $\sum_{k = - \infty}^{\infty} \hat{P}_k = \mathcal{I}$, and $\hat{P}_k \hat{P}_{k'} =  \hat{P}_k \delta_{k, k'}$\footnote{This result uses that $\int_{- \infty}^{\infty} \dd p' \chi^\eta_{k'}(p') \bra{p'} \delta{(p - p')} = \chi^\eta_{k'}(p) \bra{p}$ despite that $\chi^\eta_{k'}(p')$ is not a smooth function, which can be verified with Riemann-sum formalism.}. Note that this modelization is distinct from modeling imperfect detection efficiency~\citep{Leonhardt, Leonhardt1993, Paris2003}.
Using this operator for the measurements introduces errors in the Fourier transform applied with the Fourier gadget. In particular, the output state is a mixed state.
Second, realistic input states do have finite squeezing. Below, we are going to address the effect of these sources of noise, as well as how to deal with them.

\subsubsection{Fourier gadget for continuous variables}

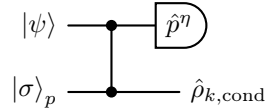
\begin{figure}[h]
\centering
\begin{quantikz}
\lstick{$\ket{\psi}$}  & \ctrl{1} &  \meterD{\hat{p}^\eta} \\
\lstick{$\ket{\sigma}_p$}  & \control{} & \rstick{$\hat{\rho}_{k,\text{cond}}$}
\end{quantikz}
\caption{A realistic Fourier gadget for continuous variables.
\label{figRealFTapp}}
\end{figure}

We consider in this subsection the actual Fourier gadget, provided by the circuit in \figref{figRealFTapp}, where we have removed the idealizations introduced for simplifying the discussion above. Namely, the auxiliary squeezed state is finitely squeezed, and the homodyne detection performed on the first mode possesses a finite resolution. This subsection is taken from the Supplementary Material of Ref.~\citep{Douce2017}.

\paragraph{Output state}

We compute the output state of the realistic Fourier-transform gate implementation. The circuit is reproduced in \figref{figRealFTapp}. By convention, the first (second) ket in the tensor product will refer to the upper (lower) arm.

We recall that we start from:
\be
\ket{\psi} \otimes \ket{\sigma}_p = \int \dd q\, \psi(q) \ket{q}_q \otimes \frac{1}{\pi^{1/4} \sqrt{\sigma}} \int \dd t \, e^{-\frac{t^2}{2\sigma^2}} \ket{t}_p.
\ee
Step by step we have first the $\hat{C}_Z$ gate:
\begin{align}
\label{eqintermediate}
\hat{C}_Z \ket{\psi} \otimes \ket{\sigma}_p &= \frac{1}{\pi^{1/4}\sqrt{\sigma}} \int \dd q \dd t \, e^{-\frac{t^2}{2\sigma^2}} \psi(q) \ket{q}_q \ket{q+t}_p \nn \\
&= \frac{1}{\pi^{1/4}\sqrt{\sigma}} \int \dd q \dd t \, e^{-\frac{(t-q)^2}{2\sigma^2}} \psi(q) \ket{q}_q \ket{t}_p \equiv \ket{\psi_{12}}.
\end{align}
We measure on the upper arm the finitely resolved $\hat{p}^{\eta}$ operator defined in \eqref{operator-proj}. 
When obtaining an outcome $p_k$, the measurement yields the conditional state on the lower arm
\begin{align}
\label{cond-state-full}
\hat{\rho}_{k,\text{cond}} &= \text{Tr}_1 \mleft[ \hat{P}_k \otimes \mathcal{I}_2 \ketbra{\psi_{12}}{\psi_{12}} \hat{P}_k \otimes \mathcal{I}_2 \mright] \nn \\
&= \int_{p_k-\eta}^{p_k+\eta} \dd s \hspace{0.1cm} \prescript{}{p,1}{\braket{s}{\psi_{12}}} \braket{\psi_{12}}{s}_{p,1} \nn \\
&= \frac{\eta}{\pi^{3/2} \sigma} \int \dd q \dd t \dd q' \dd t' \exp \mleft[-\frac{(t-q)^2}{2\sigma^2} \mright] \exp \mleft[-\frac{(t'-q')^2}{2\sigma^2} \mright] \psi(q)  \psi^*(q') \sinc[\eta (q-q')] e^{i p_k (q - q')} \ket{t}_p \prescript{}{p}{\bra{t'}} ,
\end{align}
where we have used
\be
\label{eq:sinc-integral}
\int_{-\eta}^{\eta} \dd s\, e^{i s (q - q')} = 2 \eta \, \sinc[\eta (q - q')].
\ee
We remark that the same expression as in \eqref{cond-state-full} is obtained if the homodyne detectors are perfectly resolving, and a discretization is performed after measurement by binning the measurement outcomes.

This state then has to be normalized by the probability of obtaining the outcome corresponding to the projection operator above. 
What really matters to us is $\hat{\rho}_{k=0,\text{cond}}$ corresponds to the outcome $p_k = 0$, because it is indeed the particular postselected state that corresponds to the implementation of the Fourier transform. For this specific outcome we have
\be
\label{eqpsi0}
\hat{\rho}_{k=0,\text{cond}} = \frac{\eta}{\pi^{3/2} \sigma} \int \dd q \dd t \dd q' \dd t' \exp \mleft[ -\frac{(t-q)^2}{2\sigma^2} \mright] \exp \mleft[ -\frac{(t'-q')^2}{2\sigma^2} \mright] \psi(q) \psi^*(q') \sinc[\eta (q-q')] \ket{t}_p \bra{t'}_p.
\ee
Note that in the limit of perfect resolution $\eta \rightarrow 0$ (upon normalization), we re-obtain the state that would be obtained in an MBQC implementation of the Fourier transform with a finitely squeezed auxiliary state. As can be seen in \eqref{eqpsi0}, finite squeezing means convolving the state with a Gaussian in the momentum representation, or equivalently multiplication with a Gaussian in the position representation. 

\paragraph{Probability of measuring $p_k = 0$, $\text{Prob}[k=0]$} 
\label{appB}

We evaluate here the probability of measuring an outcome $p_k = 0$ within a window function of width $2 \eta$, yielding the conditional state in \eqref{eqpsi0}. More precisely, we consider the expectation value of the following operator:
\be
\label{projector-0}
\hat{P}_0 = \int_{-\eta}^\eta \dd s \ket{s}_p \prescript{}{p}{\bra{s}} , 
\ee
taken in the state after the $\hat{C}_Z$ gate, that is [see \eqref{eqintermediate}],
\be
\ket{\psi_{12}} = \frac{1}{\pi^{1/4}\sqrt{\sigma}} \int \dd q \dd t \, \exp \mleft[ -\frac{(t-q)^2}{2\sigma^2} \mright] \psi(q) \ket{q}_q \ket{t}_p.
\ee
The calculation reads:
\begin{align}
\label{eqproba}
\text{Prob}[k=0] &= \bra{\psi_{12}} \hat{P}_0 \otimes \mathcal{I}_2 \ket{\psi_{12}} \nn \\ 
&= \frac{1}{\sigma\sqrt{\pi}} \int \dd q \dd q' \dd t \dd t' \dd s \, \exp \mleft[ -\frac{(t-q)^2}{2\sigma^2} \mright] \exp \mleft[ -\frac{(t'-q')^2}{2\sigma^2} \mright] \psi^*(q') \psi(q) \delta(t-t') _q\langle q' \vert s\rangle_p {_p\langle} s \vert q\rangle_q \nn \\ 
&= \frac{1}{2\sigma\pi^{3/2}} \int \dd q \dd q' \dd t \dd s \, \exp \mleft[ -\frac{(t-q)^2}{2\sigma^2} \mright] \exp \mleft[ -\frac{(t-q')^2}{2\sigma^2} \mright] \psi^*(q') \psi(q) e^{i s (q - q')} \nn \\
&= \frac{1}{2\pi} \int \dd q \dd q' \dd s \, \exp \mleft[ -\frac{(q-q')^2}{4\sigma^2} \mright] \psi^*(q') \psi(q) e^{i s (q - q')} \nn \\
&= \frac{2 \eta \sigma}{\sqrt\pi} \int \dd q \dd q' \frac{1}{2\sigma\sqrt{\pi}} \exp \mleft[ -\frac{(q-q')^2}{4\sigma^2} \mright] \psi^*(q') \psi(q) \sinc[\eta (q - q')] ,
\end{align}
where from the second to the third line we used that
\be
\label{identity}
\int_{-\infty}^{+\infty} \dd t \, \exp \mleft[ -\frac{(t-q)^2}{2\sigma^2} \mright] \exp \mleft[ -\frac{(t-q')^2}{2\sigma^2} \mright] = \sqrt{\pi} \sigma \exp \mleft[ -\frac{(q-q')^2}{4\sigma^2} \mright] ,
\ee
while in the last step we have used \eqref{eq:sinc-integral}. The probability can be Taylor-expanded in powers of $\eta$:
\be
\text{Prob}[k=0] = \frac{2 \eta \sigma}{\sqrt\pi} \mleft( \int \dd q \dd q' \frac{1}{2\sigma\sqrt{\pi}} \exp \mleft[ -\frac{(q-q')^2}{4\sigma^2} \mright] \psi^*(q') \psi(q) + O(\eta^2) \mright).
\ee
The first term in the parentheses is precisely the norm $\braket{\psi_{12}}{\psi_{12}}$ and hence is equal to 1. Consequently, the probability reads
\be
\text{Prob}[k=0] = \frac{2 \eta \sigma}{\sqrt\pi} + O(\eta^3).
\ee
The dominating order is thus proportional to the resolution $2 \eta$.

\paragraph{Large-squeezing limit} 

We note that Gaussian distributions obey the relation
\be
\frac{1}{2\sqrt{\pi} \sigma} \exp \mleft[ -\frac{(q-q')^2}{4\sigma^2} \mright] \underset{\sigma \rightarrow 0}{\longrightarrow} \delta(q-q') .
\ee
Based on this property, the integrals in \eqref{eqproba} actually yield
\begin{align}
& \int \dd q \dd q' \frac{1}{2\sigma\sqrt{\pi}} \exp \mleft[ -\frac{(q-q')^2}{4\sigma^2} \mright] \psi^*(q') \psi(q) \sinc[\eta (q - q')] \underset{\sigma \rightarrow 0}{\sim} 1 \\
& \int \dd q \dd q' \frac{1}{2\sigma\sqrt{\pi}} \exp \mleft[ -\frac{(q-q')^2}{4\sigma^2} \mright] \psi(q) \psi^*(q') \underset{\sigma \rightarrow 0}{\sim} 1.
\end{align}
Thus the probability of obtaining the outcome $p_k = 0$ becomes dominated by the pure-state contribution, and is determined by the expression
\be
\label{eqP0}
\text{Prob}[k=0] \underset{\sigma \rightarrow 0}{\sim} \text{Prob}^{(1)}[k=0] \underset{\sigma \rightarrow 0}{\sim} \frac{2 \eta \sigma}{\sqrt{\pi}}.
\ee
We notice that this probability is given as a function of the squeezed-state variance $\sigma$. Equation~(\ref{eqP0}) ensures that the postselection probability is nonzero, a necessary requirement to define it properly. This probability also needs to satisfy
\be
\text{Prob}[k=0] \gtrsim \frac{1}{2^n}.
\ee
This exponentially low probability is still compatible with the definition of the postselected class as explained in \secref{sse:complexity}.


\subsubsection{Dealing with errors} 

These sources of noise can, in principle, spoil the result of the computation and reduce the power of the postselected version of the circuit family, thereby preventing achieving arbitrary Post-BQP computations and spoiling the proof-of-hardness structure. How to solve this problem? 

The proof of hardness of CV-IQP with input finite-squeezing states makes use of GKP states. In Ref.~\citep{Douce2017}, it was assumed that input GKP states were present at the input. The universal set of CV operations given in \eqref{eq:gates-sub2} clearly includes a universal set of DV operations in GKP encoding:
\begin{equation}
\label{eq:gates-sub2-3}
\mleft\{ 
\bar{Z} = e^{i \hat{q} \sqrt{\pi}}, 
\bar{C_Z}= e^{i \hat{q}_1 \hat{q}_2}, 
\bar{T} = \exp \mleft( i \frac{\pi}{4} \mleft[ 2 \mleft( \frac{\hat q}{\sqrt{\pi}} \mright)^3 + \mleft( \frac{\hat q}{\sqrt{\pi}} \mright)^2 - 2 \frac{\hat q}{\sqrt{\pi}} \mright] \mright) \mright\},
\end{equation}
plus the Hadamard gate, which in the GKP encoding corresponds to the Fourier transform, that hence is obtained with postselection:
\begin{equation}
\mleft\{ \bar{H} = F \mright\}.
\end{equation}
Therefore, any postselected BQP computation can be simulated with a postselected instance of IQP circuits, i.e., Post-CV-IQP = PostBQP. In other words, for every computation in Post-BQP, we can find a CV-IQP circuit that,  with postselection corresponding to a nonzero probability (and consistent with the definition of the Post-BQP class), simulates that computation, despite the finite input squeezing and finite resolution.

However, it has later been shown in Ref.~\citep{Douce2019} that the circuit family is hard to sample even without input GKP states. The trick is to show that generation of GKP states can be subsumed in the gates of the circuit itself. The technicalities of that work are out of the scope of these notes.  

\newpage
\section*{Exercises}

\begin{enumerate}

\item Derive the commutation relation \eqref{eq:commutation-q-p} from \eqref{eq:commutation-a-ad} using the definition of the quadrature operators in terms of the annihilation and creation operators.

\item Derive the Fock coefficients of a coherent state from its definition as the eigenstate of the annihilation operator. In other words, demonstrate that \eqref{eq:CoherentState} implies \eqref{eq:CoherentStateFock}.

\item Different conventions are used in quantum optics. With the definition of the quadratures as in Eqs.~(\ref{eq:QuadratureOperatorQ}) and (\ref{eq:QuadratureOperatorP}), i.e., 
$$
\hat q = \frac{1}{\sqrt{2}}(\hat a + \hat a^\dag),\quad \hat p = \frac{1}{\sqrt{2}i}(\hat a - \hat a^\dag),
$$
the quadratures' variance (fluctuation) for the vacuum results in \eqref{eq:VarianceVacuum}, i.e., $\Delta_0^2 = 1/2$.
Show that if
$$
\hat q = (\hat a + \hat a^\dag), \quad \hat p = \frac{1}{i}(\hat a - \hat a^\dag),
$$
then $\Delta_0^2 = 1$, while if
$$
\hat q = \frac{1}{2}(\hat a + \hat a^\dag),\quad \hat p = \frac{1}{2i}(\hat a - \hat a^\dag),
$$
then $\Delta_0^2 = 1/4$.

\item \label{ex:integrating-W} Show for a pure state $\ket{\psi}$ that the marginal distribution obtained by integrating the Wigner function $W(q,p)$ in $q$ yields the probability density in $p$, $\abssq{\braket{p}{\psi}}$.

\item Show that the Wigner function is normalized to $1$.

\item Demonstrate Eq.(\ref{eq:Wigner-function-exp-values}).

\item In the section on CV MBQC, when explaining that the MBQC model needs to be adaptive in order to yield deterministic operations, we said that a) $\hat{p}_{s \hat{q}} = e^{-i s \hat q} \hat p e^{i s \hat q} = \hat p + s$; b) $\hat{p}_{s \hat{q}^2/2} = e^{-i s \frac{\hat{q}^2}{2}} \hat p e^{i s \frac{\hat{q}^2}{2}} = \hat p + s \hat q $; and c) $\hat{p}_{s (\hat{q}+m)^3/3} = \hat p + s \hat q^2 + 2 m s \hat q + m^2 s$, and we concluded that the case c) requires a nontrivial adaptation of the measurement basis.
Demonstrate expressions a), b), and c) using the Baker--Campbell--Hausdorff formula. 

\item Show that a cat state $\ket{\alpha} + \ket{- \alpha}$ has support only on even Fock states.

\item Show that if a state is stabilized by the stabilizer operators $e^{i 2 \sqrt{\pi} \hat p}$ and $e^{i 2 \sqrt{\pi} \hat q}$, then it is a GKP state. 

\item Show that $H_L\ket{1_L}=F\ket{1_L}$, where $\ket{1_L}$ is the ideal GKP state from \eqref{eq:GKP_ideal_q}--\eqref{eq:GKP_ideal_p}.

\item Verify \eqref{eq:GKP_real_1} by explicitly performing the double integral.

\end{enumerate}

\chapter*{Tutorial 4: CV\label{tutorial4}}
\addcontentsline{toc}{chapter}{Tutorial 4} 
\renewcommand{\thesection}{\arabic{section}} 
\setcounter{section}{0} 
\setcounter{figure}{0} 
\renewcommand{\thefigure}{T4.\arabic{figure}} 
\setcounter{tutorial}{4}
\setcounter{ex}{0}
\setlength{\parindent}{0pt} 

\setcounter{equation}{0} 
\renewcommand{\theequation}{T4.\arabic{equation}} 

\textit{Note: This tutorial is placed here for convenience, but during the course it is taught just after having introduced continuous-variable quantum computing (\secref{sse: Mari-Eisert}).}\\

This tutorial aims at strengthening the understanding of the basics of continuous-variable (CV) quantum computing. A large part of the material covered here is adapted from Refs.~\citep{Strandberg-licentiate} and~\citep{Strandberg-phd}.

\section{What is continuous in CV?}

In contrast to discrete-variable (DV) quantum computing, in the CV approach, we use observables characterized by a \emph{continuous} spectrum, such as the \emph{quadratures of an electromagnetic field} (a.k.a.~light).

\subsubsection*{Electromagnetic field}

An expression for a classical electromagnetic field, derived from Maxwell's equations under periodic boundary conditions, can be written as an expansion of plane waves:
\begin{equation}
    \vec{E} (\vec{r}, t) = \sum_n \sqrt{\frac{\hbar \omega_n}{2V\varepsilon_0}} \mleft( \alpha_n e^{i(\vec{k}_n\cdot\vec{r}-\omega_n t)} + \alpha_n^* e^{-i(\vec{k}_n\cdot\vec{r}-\omega_n t)} \mright),
    \label{eq:T4:E_classical}
\end{equation}
where $\sum_n$ denotes a sum over different frequency modes and $\alpha_n$ is dimensionless.
The field is presumed to occupy an empty volume $V$, $\varepsilon_0$ is the vacuum permittivity, and $\hbar$ is the reduced Planck constant ($\hbar=1$ from here on).
The frequency $\omega_n$ is related to the wave vector $\vec{k}_n$ via the linear dispersion relation $\abs{\vec{k}_n} = \omega_n/c$ with $c$ being the speed of light.

Now, \eqref{eq:T4:E_classical} suggests that the electromagnetic field is a wave.
However, while light can indeed behave as a wave, it can also be observed to have a particle-like nature.
For instance, light can be absorbed and emitted into discrete packets (or quanta) known as photons.
When the field is quantized, the coefficient $\alpha_n$ is promoted to an operator $\hat{a}_n$ called the photon-annihilation operator, while its complex conjugate, $\alpha^*_n$ becomes the photon-creation operator $\hat{a}_n^\dag$.
These operators obey the bosonic commutation relation
\begin{equation}
    \comm{\hat{a}_n}{\hat{a}_m^\dag} = \delta_{nm}.
\end{equation}
For a single mode of frequency $\omega$, this is the commutation relation $\comm{\hat{a}}{\hat{a}^\dagger} = 1$ of the ladder operators describing a quantum harmonic oscillator with Hamiltonian $H = \omega \hat{a}^\dag \hat{a}$.

\subsubsection*{Quadrature operators}

The number of photons ($\hat{n} = \hat{a}^\dag \hat{a}$) is a discrete observable, but the quantum state of an electromagnetic field also has observables with a continuous spectrum: the \emph{quadratures}.
The $\hat{q}$ and $\hat{p}$ quadratures (also denoted by $\hat{x}$ and $\hat{p}$, by $\hat{X}_1$ and $\hat{X}_2$, or by $I$ and $Q$) are conjugate operators defined as
\begin{equation}
    \hat{q} = \frac{1}{\sqrt{2}} \mleft( \hat{a}^\dag + \hat{a} \mright), \qquad
    \hat{p} = \frac{i}{\sqrt{2}} \mleft( \hat{a}^\dag - \hat{a} \mright).
    \label{eq:T4:quadratures}
\end{equation}
Therefore, the quantized version of \eqref{eq:T4:E_classical} for a single mode is
\begin{equation}
    \vec{E} (\vec{r}, t) = \xi_0 \mleft( \hat{a} e^{i(\vec{k}\cdot\vec{r} - \omega t)} + \hat{a}^\dag e^{-i(\vec{k}\cdot\vec{r} - \omega t)} \mright) \stackrel{\eqref{eq:T4:quadratures}}{=} \sqrt{2} \xi_0 \mleft[ \hat{q} \cos(\vec{k}\cdot\vec{r} - \omega t) + \hat{p} \sin(\vec{k}\cdot\vec{r} - \omega t) \mright],
\end{equation}
where $\xi_0$ contains the dimensional prefactors.
This provides an intuitive picture of the quadratures as the sine and cosine parts of the field.
In fact, the radiation field can be decomposed into one component \emph{in phase} with a reference $\cos(\vec{k}\vec{r} - \omega t)$, and one \emph{out of phase} by $\pi/2$:
\begin{figure}[H]
    \centering
    \includegraphics[width=0.7\textwidth]{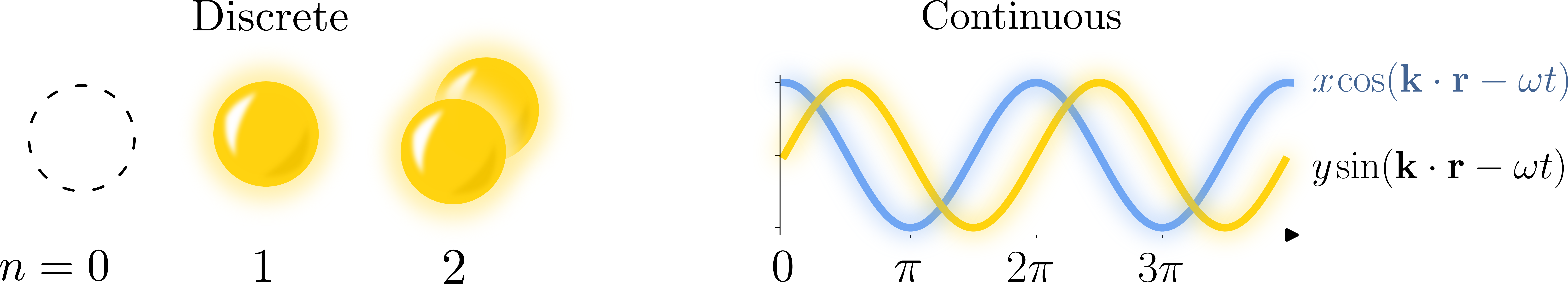}
\end{figure}
Since the components are shifted by one-quarter cycle, we say they are in \emph{quadrature phase}, thus giving name to the \emph{quadrature operators}.
As the notation implies, the quadratures are in a certain sense analogous to position and momentum, since they obey the same canonical commutation relation 
\begin{equation}
    \comm{\hat{q}}{\hat{p}}=i.
\end{equation}
However, they have nothing to do with the position and momentum of the electromagnetic oscillator.

\section{Phase-space representation: Wigner functions}

Classically, \emph{phase space} is the space in which all possible states of a system are represented by unique points.
For mechanical systems, the phase space usually has coordinates of position and momentum.
For instance:

\begin{figure}[H]
    \centering
    \includegraphics[width=0.7\textwidth]{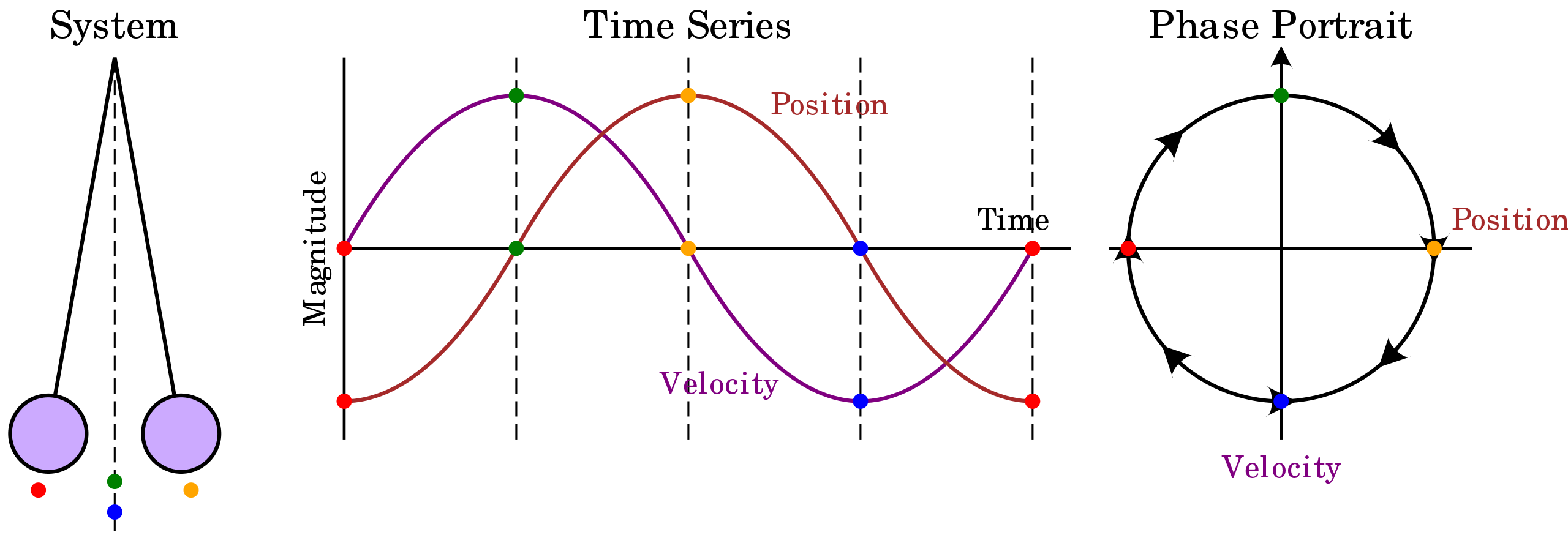}
    \caption{Figure by Krishnavedala, from \href{https://commons.wikimedia.org/w/index.php?curid=37054341}{Wikipedia - Phase space}.}
    \label{fig:enter-label}
\end{figure}

Since we have identified the quadratures $\hat{q}$ and $\hat{p}$ as the analogues of position and momentum variables, one could think that the classical phase-space representation works the same in the quantum realm, but that is not quite the case. 

A classical particle has a definite position and momentum, and hence it is represented by a point in phase space.
However, this cannot be the case for a quantum particle, due to the uncertainty principle.
Similarly, given an ensemble of classical particles, the probability of finding a particle at a certain position in phase space is specified by a probability distribution. 
However, for quantum particles, the uncertainty principle forbids the definition of a joint probability distribution at a point $(q, p)$ in phase space.
Fortunately, it is possible to define a \emph{quasiprobability}, such as the \emph{Wigner function}.

Quasiprobability distributions are similar to probability distributions in that they yield expectation values with respect to the weights of the distribution, but they violate some probability axioms.
For instance, the Wigner function is real-valued and integrating it along a direction in phase space gives the genuine probability of the orthogonal quadrature, but it can take on negative values.

\setcounter{footnote}{0}

The Wigner function may be formally defined in terms of the density operator\footnote{Remember that the density operator represents the quantum state of a physical system and is defined as $\hat{\rho} = \sum_j p_j \ketbra{\psi_j}$ for some states $\ket{\psi_j}$ that occur with probability $p_j$.} $\hat{\rho}$ like
\begin{equation}
	W(q, p) = \frac{1}{2\pi} \int^\infty_{-\infty}
    \bra{q-\frac{y}{2}} \hat{\rho} \ket{q+\frac{y}{2}} e^{ipy} \dd y \stackrel{\text{real}}{=} \frac{1}{2\pi}\int^\infty_{-\infty} \dd y \, e^{-ipy}
    \bra{q+\frac{y}{2}} \hat{\rho} \ket{q-\frac{y}{2}}.
\end{equation}
%

\subsection{Types of states, negativity, and nonclassicality}

As you have seen in the lectures, the Mari--Veitch theorem dictates that Wigner negativity is necessary (not sufficient) for quantum advantage in computation. Therefore, the Wigner function is a widely used tool in CV quantum computing. The volume of the negative part of the Wigner function has been suggested as a measure of non-Gaussianity, which for pure states is a inherently nonclassical feature, connected to resourcefulness for quantum computation. As we will now see, there are states, such as squeezed states, that albeit being represented by positive Wigner functions, are considered nonclassical by quantum optics, and they are resourceful states for tasks such as parameter estimation (see also Ref.~\citep{Strandberg-licentiate} for further explanation).\\

\underline{Note:} in all the Python codes below, we need the following preamble:
\begin{lstlisting}[language=Python]
import numpy as np
import qutip as qt

#N is the Hilbert space cutoff
\end{lstlisting}

\subsubsection{Fock states}

A Fock state is defined by the number of particles or quanta, $\ket{n}$. In DV quantum computing, $n$ usually denotes the number of photons or excitations, and we limit ourselves to the subspace spanned by $\{\ket{0}, \ket{1}\}$ (qubits).\\

Properties:
\begin{align}
\bullet \,&
\begin{cases}
\hat{a} \ket{n} = \sqrt{n} \ket{n-1} \\
\hat{a}^\dag \ket{n} = \sqrt{n+1} \ket{n+1}
\end{cases}
\implies\;
\hat{n} \ket{n} = \hat{a}^\dag \hat{a} \ket{n} = n \ket{n}
\;\implies\;
\expval{\hat{n}}_n = n \\
\bullet \,& \braket{n}{m} = \delta_{nm} \text{ (orthogonal)} \\
\bullet \,& \sum_n \ketbra{n}{n} = \Id \text{ (complete basis)}
\end{align}

Fock states are purely quantum mechanical and have no classical counterpart. 

\begin{figure}[H]
    \centering
    \includegraphics[width=0.4\textwidth]{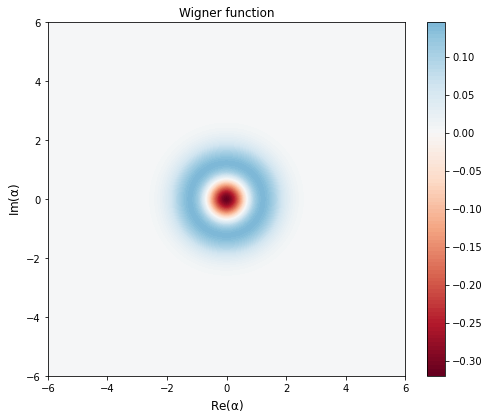}
    \includegraphics[width=0.4\textwidth]{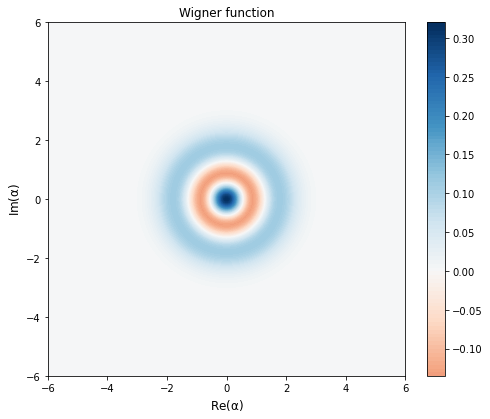}
    \caption{Wigner function of a Fock state $n=1$ (left) and $n=2$ (right). See the code for the figures below.}
\end{figure}

\begin{lstlisting}[language=Python]
N = 2 #we can cut off at N=2 because we only need {|0>, |1>}
fock1 = qt.fock(N, 1)
qt.plot_wigner(fock1,  figsize=(8, 6.5), alpha_max = 6, colorbar=True)

N = 3 #we can cut off at N=3 because we only need {|0>, |1>, |2>}
fock2 = qt.fock(N, 2)
qt.plot_wigner(fock2,  figsize=(8, 6.5), alpha_max = 6, colorbar=True)
\end{lstlisting}

\subsubsection{Coherent states}

A coherent state is the eigenstate of the annihilation operator:
\be
\hat{a}\ket{\alpha} = \alpha \ket{\alpha}, \quad \alpha=\abs{\alpha}e^{i\theta}\in\mathbb{C}.
\ee
In the Fock basis, a coherent state is 
\be
\ket{\alpha} = \sum_{n=0}^\infty \frac{\alpha^n}{\sqrt{n!}}e^{-\abs{\alpha}^2/2}\ket{n} \;\implies\;
\expval{\hat{n}}_\alpha = \abs{\alpha}^2.
\ee
Properties:
\begin{align}
\bullet \,&
P(n) = \abs{\braket{n}{\alpha}}^2 = e^{-\abs{\alpha}^2}\frac{\abs{\alpha}^{2n}}{n!} \text{ (Poisson distribution\footnotemark[2])}\\
\bullet \,& \braket{\alpha}{\beta} = e^{-\abs{\alpha-\beta}^2/2} \text{ (not orthogonal)}\\
\bullet \,& \frac{1}{\pi}\int \ketbra{\alpha} \dd^2 \alpha = 1,\; \alpha = \abs{\alpha} e^{i\theta} \text{ (overcomplete set)}
\end{align}

\footnotetext[2]{The Poisson distribution is a result of a process where the time (or an equivalent measure) between events has an exponential distribution, representing a memory-less process. A super-Poissonian distribution is a probability distribution that has a larger variance than a Poisson distribution with the same mean. Conversely, a sub-Poissonian distribution has a smaller variance. Thermal light is super-Poissonian (bunched), coherent light is Poissonian (random), and amplitude-squeezed light is sub-Poissonian (antibunched).}

We say that coherent states are quasi-classical, as some of their properties are classical and some are quantum. 

\begin{figure}[H]
    \centering
    \includegraphics[width=0.4\textwidth]{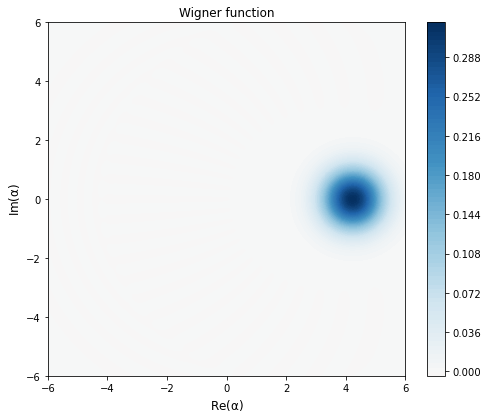}
    \includegraphics[width=0.4\textwidth]{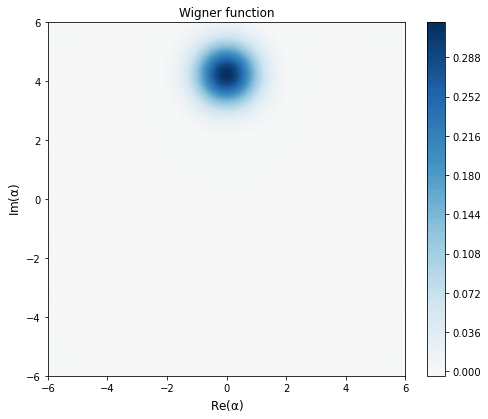}
    \caption{Wigner function of a coherent state $\alpha=3$ (left) and $\alpha=3j$ (right). See the code for the figures below.}
\end{figure}

\begin{lstlisting}[language=Python]
N = 32 #with N=32 we get a very decent-looking state
coherent1 = qt.coherent(N, 3)
qt.plot_wigner(coherent1,  figsize=(8, 6.5), alpha_max = 6, colorbar=True)
#We don't see the blob centered at Re(alpha)=3 because quTip renormalizes the axes by sqrt(2), for some reason. So instead, we see it at Re(alpha)=3*sqrt(2)

coherent2 = qt.coherent(N, 3j)
qt.plot_wigner(coherent2,  figsize=(8, 6.5), alpha_max = 6, colorbar=True)
\end{lstlisting}

\subsubsection{Vacuum state}

The vacuum state $\ket{0}$ is both a coherent state and a Fock state. A coherent state can be interpreted as a displaced vacuum:
\be
\ket{\alpha} = \hat{\mathcal{D}}(\alpha)\ket{0} = \exp{\alpha \hat{a}^\dag - \alpha^* \hat{a}} \ket{0}. 
\ee

\begin{figure}[H]
    \centering
    \includegraphics[width=0.4\textwidth]{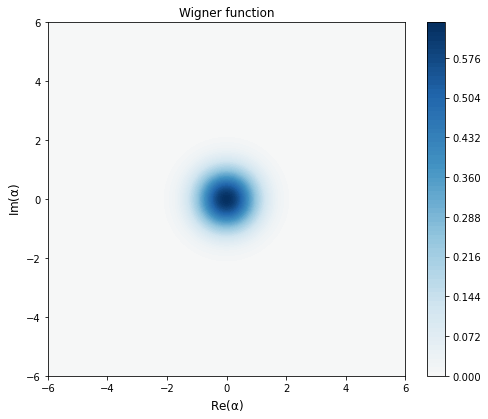}
    \includegraphics[width=0.4\textwidth]{Coherent.png}
    \caption{Wigner function of a vacuum state (left) and a displaced vacuum state (right). See the code for the figures below.}
\end{figure}

\begin{lstlisting}[language=Python]
N = 1 
vacuum = qt.fock(N, 0)
qt.plot_wigner(vacuum,  figsize=(8, 6.5), alpha_max = 6, colorbar=True)

N = 32 #since it's a coherent state, we need to increase the cutoff again
displaced_vacuum = qt.displace(N, 3) * qt.fock(N, 0)
qt.plot_wigner(displaced_vacuum,  figsize=(8, 6.5), alpha_max = 6, colorbar=True)
\end{lstlisting}

\subsubsection{Squeezed states}

A squeezed state is a coherent state with different fluctuations associated with the quadratures, i.e., different variances in $\hat{p}$ and $\hat{q}$:
\be
\ket{\alpha, \zeta} = \hat{\mathcal{D}}(\alpha) \hat{\mathcal{S}}(\zeta) \ket{0} = \exp{\alpha\hat{a}^\dag - \alpha^*\hat{a}} \exp{\frac{1}{2} \mleft( \zeta\hat{a}^2 - \zeta^*\hat{a}^{\dag 2} \mright)} \ket{0}, \qquad \zeta = \abs{\zeta} e^{i2\phi}.
\ee

Squeezed light exhibits sub-Poissonian statistics, which means that, according to quantum optics, squeezed light is nonclassical. However, the Wigner function is positive, which means it is classically efficiently simulatable together with Gaussian operations and Gaussian measurements.

\begin{figure}[H]
    \centering
    \includegraphics[width=0.4\textwidth]{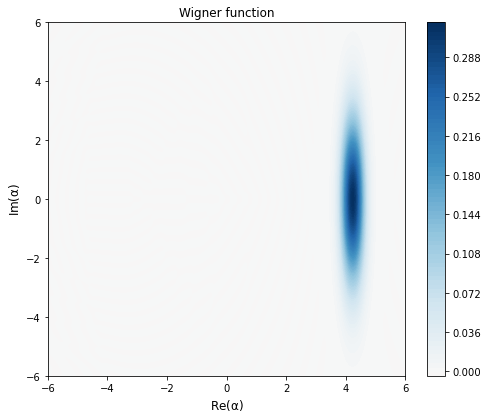}
    \includegraphics[width=0.4\textwidth]{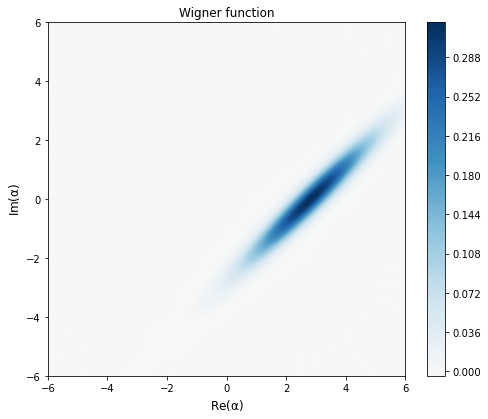}
    \caption{Wigner function of a squeezed state $\ket{\alpha, \zeta}$ with $\alpha=3, \zeta=1$ (left) and $\alpha=2, \zeta=-1i$ (right). See the code for the figures below.}
\end{figure}

\begin{lstlisting}[language=Python]
N = 64
squeezed1 =  qt.displace(N, 3) * qt.squeeze(N, 1) * qt.fock(N, 0)
qt.plot_wigner(squeezed1,  figsize=(8, 6.5), alpha_max = 6, colorbar=True)

squeezed2 =  qt.displace(N, 2) * qt.squeeze(N, -1j) * qt.fock(N, 0)
qt.plot_wigner(squeezed2,  figsize=(8, 6.5), alpha_max = 6, colorbar=True)
\end{lstlisting}

\subsubsection{Cubic phase states}

A cubic phase state is a achieved by implementing the cubic phase gate:
$$
\ket{\gamma, \zeta} =  e^{i\gamma \hat{q}^3}\hat{\mathcal{S}}(\zeta)\ket{0}, \qquad \hat{q} = (\hat{a} + \hat{a}^\dagger)/\sqrt{2}.
$$

At the end of 2021, at Chalmers, the generation of a cubic phase state was experimentally achieved for the first time --- paper published in 2022~\citep{kudra2021robust}. 

\begin{figure}[H]
    \centering
    \includegraphics[width=0.4\textwidth]{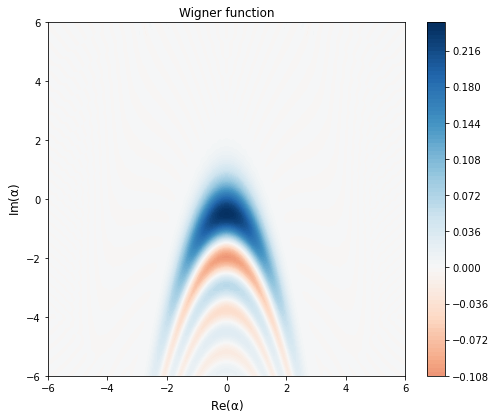}
    \includegraphics[width=0.4\textwidth]{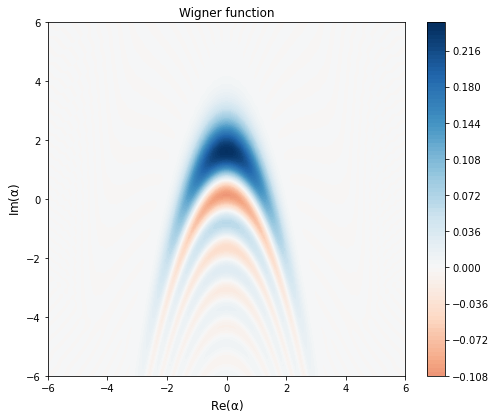}
    \caption{Wigner function of a cubic phase state $\ket{\gamma, \zeta}$ with $\gamma=-0.3, \zeta=-0.5$ (left) and a displaced cubic phase state with $\alpha=1.5j, \gamma=-0.3, \zeta=-0.5$ (right). See the code for the figures below.}
\end{figure}

\begin{lstlisting}[language=Python]
N = 128 #rule of thumb: the more complex the state, the higher the cutoff needs to be
q = (qt.create(N) + qt.destroy(N)) / np.sqrt(2)
cubic_gate = (-1j * 0.3 * q**3).expm()

cubic =  cubic_gate * (qt.squeeze(N, -0.5) * qt.fock(N, 0))
qt.plot_wigner(cubic, figsize=(8, 6.5), alpha_max = 6, colorbar=True)

cubic_displaced = qt.displace(N, 1.5j) * cubic_gate * (qt.squeeze(N, -0.5) * qt.fock(N, 0))
qt.plot_wigner(cubic_displaced, figsize=(8, 6.5), alpha_max = 6, colorbar=True)
\end{lstlisting}

\section{Cat states}

Cat states, named after Schr\"{o}dinger's cat, are the superposition of two coherent states:
\begin{align}
\ket{C_\alpha^+} &= N(\ket{\alpha} + \ket{-\alpha}) , \\
\ket{C_\alpha^-} &= N(\ket{\alpha} - \ket{-\alpha}) , \\
\ket{C_{i\alpha}^+} &= N(\ket{i\alpha} + \ket{-i\alpha}) , \\
\ket{C_{i\alpha}^-} &= N(\ket{i\alpha} - \ket{-i\alpha}).
\end{align}
We note that $\ket{C_\alpha^+}$ and $\ket{C_{i\alpha}^+}$ contain only \textit{even} Fock states, while $\ket{C_\alpha^-}$ and $\ket{C_{i\alpha}^-}$ contain only \textit{odd} Fock states. 

\subsection{Cat code}

In the cat code, we define logical states as follows:
\be
\ket{0_L} = \ket{C_\alpha^+} = N \mleft( \ket{\alpha} + \ket{-\alpha} \mright), \qquad
\ket{1_L} = \ket{C_{i\alpha}^+} = N \mleft( \ket{i\alpha} + \ket{-i\alpha} \mright)
\ee
Note that both $\ket{0_L}$ and $\ket{1_L}$ have only Fock states $0, 2, 4... (0\Mod 2)$, but are rotated one from the other in phase space (see \figref{fig:CatStates} below).\\

In most cavities, the most likely error is a single-photon loss ($\hat{a}$). When this occurs on the logical states defined above, we obtain
\begin{align}
\hat{a} \ket{0_L} &= \hat{a} \ket{C_\alpha^+} = \hat{a} N \mleft( \ket{\alpha} + \ket{-\alpha} \mright) = \alpha N \mleft( \ket{\alpha} - \ket{-\alpha} \mright) = \ket{C_\alpha^-} \\
\hat{a} \ket{1_L} &= \hat{a} \ket{C_{i\alpha}^+} = \hat{a} N \mleft( \ket{i\alpha} + \ket{-i\alpha} \mright) = i \alpha N \mleft( \ket{i\alpha} - \ket{-i\alpha} \mright) = \ket{C_{i\alpha}^-},
\end{align}
where $\hat{a}\ket{0_L}$ and $\hat{a}\ket{1_L}$ contain only Fock states $1, 3, 5, ...$. Now, it is clear that we can use a parity measurement (with stabilizer $e^{i\pi \hat{a}^\dag \hat{a}}$) to see if an error has occurred: no error for even photon number, and error for odd photon number. Very importantly, we can check for single-photon losses much faster than they occur. The cat code cannot correct two-photon losses or phase errors, but they are less likely to occur in a cavity. 

\begin{figure}[H]
    \centering
    \includegraphics[width=0.4\textwidth]{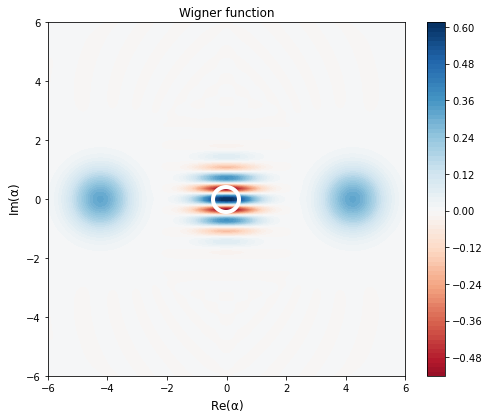}
    \includegraphics[width=0.4\textwidth]{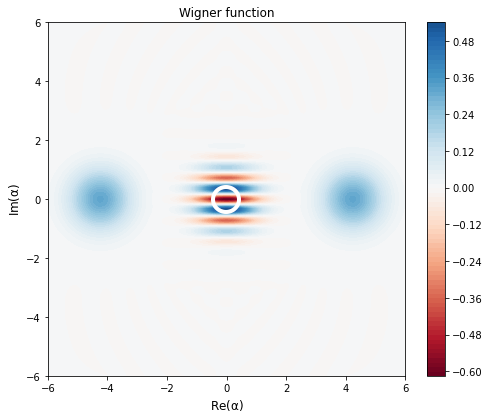}
    \includegraphics[width=0.4\textwidth]{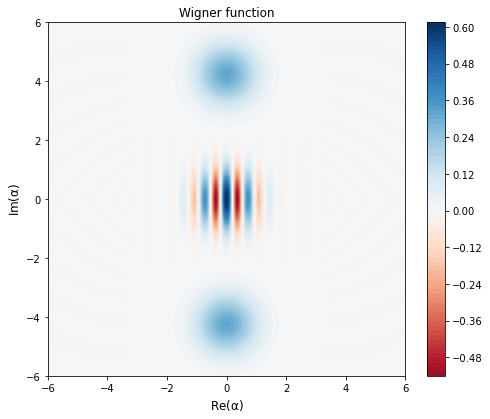}
    \includegraphics[width=0.4\textwidth]{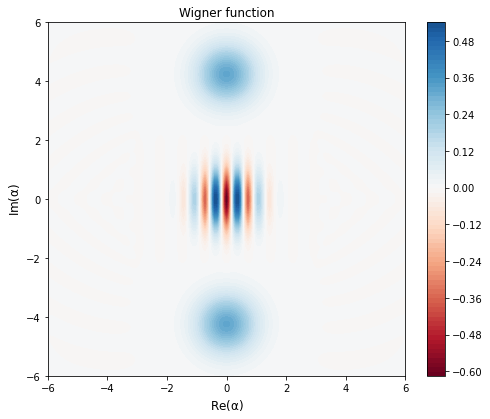}
    \caption{Wigner function of even (left) and odd (right) cat states. In particular, $\ket{C_{\alpha}^\pm}$ (top) and $\ket{C_{i\alpha}^\pm}$ (bottom). See the code for the figures below.}
    \label{fig:CatStates}
\end{figure}

\begin{lstlisting}[language=Python]
N = 32
cat_even = qt.coherent(N,3) + qt.coherent(N,-3)
qt.plot_wigner(cat_even, figsize=(8, 6.5), alpha_max = 6, colorbar=True)

cat_odd = qt.coherent(N,3) - qt.coherent(N,-3)
qt.plot_wigner(cat_odd, figsize=(8, 6.5), alpha_max = 6, colorbar=True)

cat_even_i = qt.coherent(N,3j) + qt.coherent(N,-3j)
qt.plot_wigner(cat_even_i, figsize=(8, 6.5), alpha_max = 6, colorbar=True)

cat_odd_i = qt.coherent(N,3j) - qt.coherent(N,-3j)
qt.plot_wigner(cat_odd_i, figsize=(8, 6.5), alpha_max = 6, colorbar=True)

#When an even cat loses a photon, it becomes an odd cat.
qt.plot_wigner(qt.destroy(N) * cat_even, figsize=(8, 6.5), alpha_max = 6, colorbar=True)

qt.plot_wigner(qt.destroy(N) * cat_even_i, figsize=(8, 6.5), alpha_max = 6, colorbar=True)
\end{lstlisting}

\subsubsection{The original cat code}

\textit{Note: We have not covered this in class -- this is only added for completeness.}\\

In the original cat code~\citep{Leghtas2013}, 
they create superpositions between cat states to define logical states:
\be
\ket{0_L} = \ket{C_\alpha^+} + \ket{C_{i\alpha}^+}, \qquad
\ket{1_L} = \ket{C_\alpha^+} - \ket{C_{i\alpha}^+}.
\ee
Note that $\ket{0_L}$ has only Fock states $0, 4, 8, ... (0\Mod 4)$ and $\ket{1_L}$ has states $2, 6, 10, ... (2\Mod 4)$.

In most cavities, the most likely error is a single-photon loss ($\hat{a}$). When this occurs on the logical states defined above, we obtain
\be
\hat{a}\ket{0_L} = \ket{0_L}_\perp = \ket{C_\alpha^-} + \ket{C_{i\alpha}^-},  \qquad
\hat{a}\ket{1_L} = \ket{1_L}_\perp = \ket{C_\alpha^-} - \ket{C_{i\alpha}^-},
\ee
where $\ket{0_L}_\perp$ contains only Fock states $3, 7, 11, ... (3\Mod 4)$ and  $\ket{1_L}$ has states $1, 5, 9, ... (1\Mod 4)$. Now, it is clear that we can use a parity measurement (with stabilizer $e^{i\pi \hat{a}^\dag \hat{a}}$) to see if an error has occurred: no error if ``even $\Mod 4$'', and error if ``odd $\Mod 4$''. Similarly as with the other cat code, one can check for single-photon losses much faster than they occur. This cat code still cannot correct two-photon losses or phase errors, but like we mentioned, they are less likely to occur in a cavity. \\

Note that both cat codes are equivalent up to a Hadamard transformation.\\

\begin{lstlisting}[language=Python]
N = 32
state_0L = cat_even + cat_even_i
qt.plot_wigner(state_0L, figsize=(8, 6.5), alpha_max = 6, colorbar=True)

state_1L = cat_even - cat_even_i
qt.plot_wigner(state_1L, figsize=(8, 6.5), alpha_max = 6, colorbar=True)

state_0L_perp = qt.destroy(N) * state_0L #same as cat_odd + cat_odd_i
qt.plot_wigner(state_0L_perp, figsize=(8, 6.5), alpha_max = 6, colorbar=True)

state_1L_perp = qt.destroy(N) * state_1L #same as cat_odd - cat_odd_i
qt.plot_wigner(state_1L_perp, figsize=(8, 6.5), alpha_max = 6, colorbar=True)
\end{lstlisting}

\begin{figure}[H]
    \centering
    \includegraphics[width=0.4\textwidth]{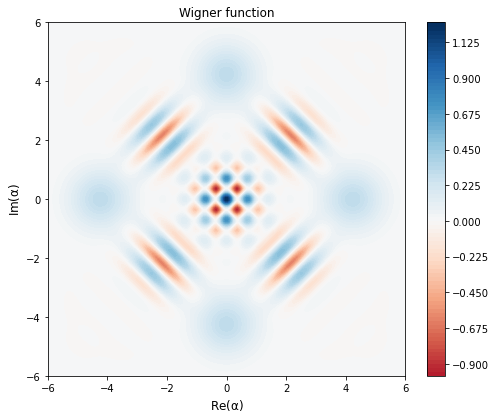}
    \includegraphics[width=0.4\textwidth]{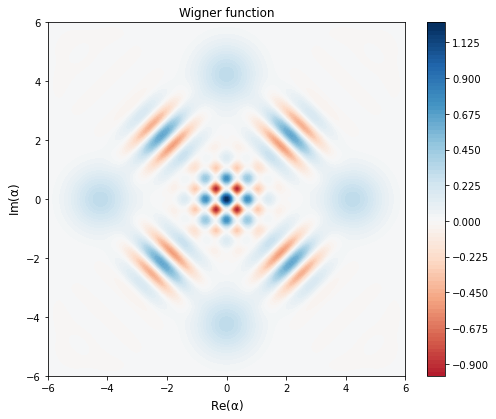}
    \includegraphics[width=0.4\textwidth]{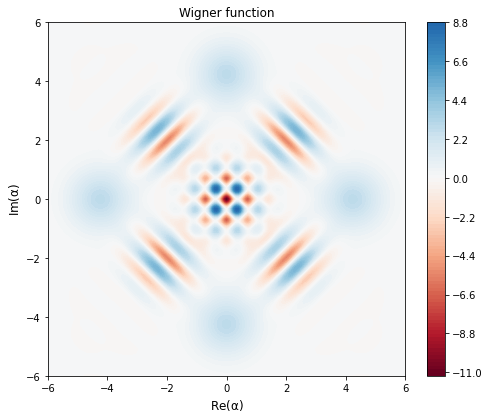}
    \includegraphics[width=0.4\textwidth]{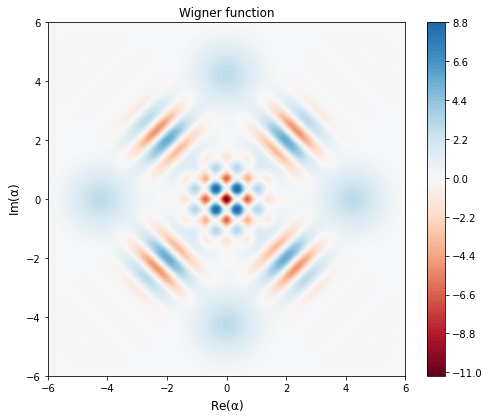}
    \caption{Wigner function of the superposition of cat states $\ket{0_L}$ (top left), $\ket{1_L}$ (top right), $\ket{0_L}_\perp$ (bottom left), and $\ket{1_L}_\perp$ (bottom right). See the code for the figures above.}
    \label{fig:my_label}
\end{figure}

\subsection{Wigner-function vs density-operator formalism}

Another feature of Wigner functions is that they are equivalent to the density-matrix formalism. This means that, in some cases, we can classically simulate a quantum CV circuit faster\footnote[3]{Note that faster does not mean that it is efficiently simulatable.} with Wigner functions than with density matrices~\citep{Bourassa2021}.
Let us see why with an example.\\

Cat states are linear superpositions of pure Gaussian states. We can write the density matrix of these states as
\be
\ketbra{C_\alpha^+} = N(\ketbra{\alpha} + \ketbra{-\alpha} + \ketbra{\alpha}{-\alpha} + \ketbra{-\alpha}{\alpha}).
\ee
The Wigner functions of the first two terms are Gaussian functions with the covariance matrix of the vacuum and a displacement (mean) proportional to $(\Re(\alpha), \Im{\alpha})$. The Wigner functions of the other two terms are complex-valued Gaussians with prefactor $e^{-2\abs{\alpha}^2}$ and vacuum covariance matrix. They are centered at $(i\Im(\alpha), -i\Re{\alpha})$ and at its complex conjugate.

\begin{figure}[H]
    \centering
    \includegraphics[width=\textwidth]{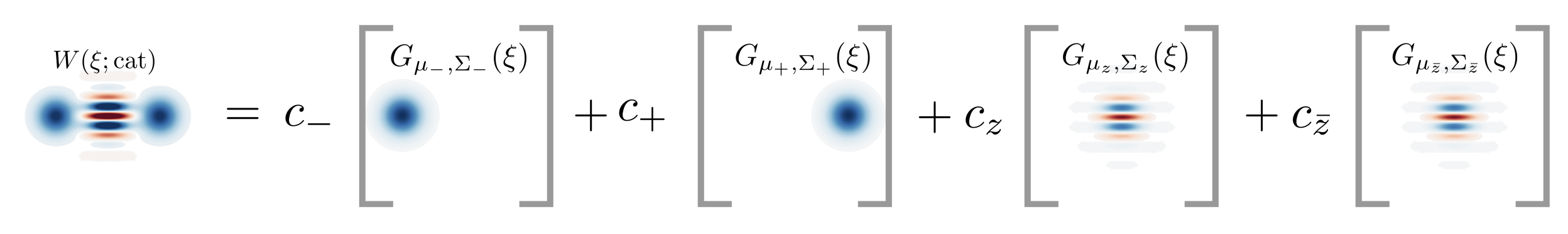}
    \caption{From Ref.~\citep{Bourassa2021}. Gaussian decomposition of the Wigner function of the (non-Gaussian) cat state. Note that the last two terms have complex coefficients and means, and are also complex conjugates of each other; thus the imaginary parts of the Gaussians cancel out and we only plot the real part. The cat-state negativity is produced by the sinusoidal oscillations of complex-valued Gaussian peaks.}
\end{figure}

Now, we said before that coherent states are not really orthogonal $\mleft( \braket{\alpha}{\beta} = \exp \mleft[ -\abs{\alpha-\beta}^2/2 \mright] \mright)$. However, we can make them almost orthogonal by using large $\alpha$ values, i.e., large photon numbers. This poses a problem when we realize that $\ket{\alpha}$ decomposes as an infinite sum of Fock states: high photon number means a huge Hilbert space. Therefore, when tracking the evolution of the density matrix, we need many dimensions and it gets hard to compute.
On the other hand, by tracking Wigner functions, for each cat state we only need to follow four Gaussians, which can be done much faster. 

\renewcommand{\thesection}{\thechapter.\arabic{section}} 
\setlength{\parindent}{15pt} 
\renewcommand{\theequation}{\thechapter.\arabic{equation}} 

\newpage
\addcontentsline{toc}{chapter}{Bibliography}

\bibliographystyle{apsrev4-1modified}

\bibliography{bibliography}
\end{document}